# ResFrac Technical Writeup

Mark McClure (mark@resfrac.com)
Charles Kang (charles@resfrac.com)
Chris Hewson (chris@resfrac.com)
Soma Medam (soma@resfrac.com)
Egor Dontsov (egor@resfrac.com)
Ankush Singh (ankush@resfrac.com)
Carlo Peruzzo (carlo@resfrac.com)
Elizaveta Gordeliy (lisa@resfrac.com)
Michael Nole (michaelnole@resfrac.com)
Jiacheng Wang (jiacheng@resfrac.com)
Jiaqi Du (jiaqi@resfrac.com)
Danial Zeinabady (danial@resfrac.com)

info@resfrac.com

# Table of Contents

# 1. Introduction

This document provides a detailed technical writeup for ResFrac. It covers the governing and constitutive equations, specification of initial and boundary conditions, and practical topics such as history matching. Section 15.

This document is complementary to the other documentation materials provided with ResFrac. The ResFrac Technical Summary provides a more succinct overview of ResFrac capabilities. The Getting Started Guide describes how to use the user interface to set up, run, and visualize simulations. We also provide step-by-step tutorials on performing ResFrac simulations for two applications: (1) hydraulic fracturing and production and (2) diagnostic fracture injection tests. We provide tutorial movies where the computer screen is recorded as we narrate and show how to use ResFrac.

For examples of ResFrac applications, refer to SPE 182593 (McClure and Kang, 2017), SPE 190049 (McClure and Kang, 2018), URTeC 123 (McClure et al., 2019), URTeC 608 (Kaufman et al., 2019), SPE 195980 (Fowler et al., 2019), and SPE 199716 (Cipolla et al., 2020). In particular, we recommend SPE 199726 (McClure et al., 2020), which gives answers to 'frequently asked questions' on practical aspects of model building. The most up-to-date list of technical papers written with ResFrac is available at:

<https://www.resfrac.com/library#resfractechnicalpapers>.

Finally, for information on particular input parameters, you can refer to the "Help" buttons built directly into the simulation builder tool. In document, references to the actual input parameters in the user-interface are provided in italics. In the "Index" section of the "Welcome" panel, there is a search bar that you can use to search through all input options.

The visibility of controls in the user-interface changes adaptively based on inputs. For example, depending on whether you select the black oil model or the compositional model, the input arguments shown in the "Fluid Model Options" panel are different. All user inputs are checked with a variety of 'validation' functions that provide error or warnings messages.

# 2. Problem geometry and meshing

## 2.1 Gridding overview

Figure *1* shows an example of a ResFrac simulation mesh. Matrix elements are meshed as cuboids; fracture elements are meshed as rectangles, and the well is meshed with line segments. You specify the fracture element length and aspect ratio (length divided by height). All fracture elements have the same length and height. The matrix elements are rectangular, but they do necessarily have uniform size, as shown in Figure *1*. The matrix mesh geometry is specified as part of the simulation setup; ResFrac does not use adaptive mesh refinement. You specify the wellbore element length.

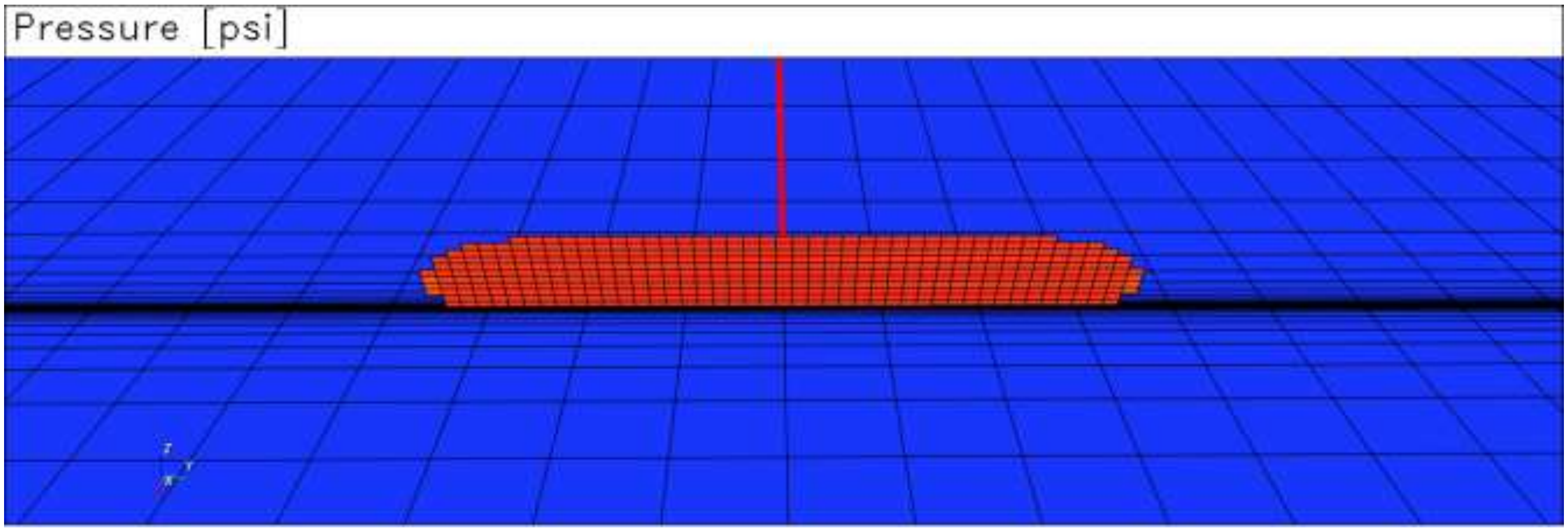

Figure 1: Example of a ResFrac simulation mesh with refinement towards the fracture.

Figure *1* shows an example of mesh that has been refined towards the fracture. In a conventional finite volume/difference simulator, mesh refinement towards the fracture is necessary in order to avoid numerical error. However, ResFrac uses the 1D subgrid (also called 1D submesh) method (McClure, 2017) to calculate fluid exchange between the fracture and matrix. The 1D subgrid method enables accurate calculations of fluid exchange between the fracture and the matrix, even with a coarse non-conforming mesh. A non-conforming mesh is a mesh where the edges of the fracture elements do not coincide with the edges of the matrix elements.

If you perform simulations of DFITs or other 'pressure transient' simulations, you probably want to refine the mesh towards the fracture, like shown in Figure *1*. This is because in pressure transients, you plot the derivative of pressure with respect to logarithmic time. This type of plot magnifies even small numerical imperfections. The 1D subgrid method is not perfect, and if you use it on a DFIT simulation, there can be a few odd wiggles in the derivative plot. But in most field scale simulations of hydraulic fracturing, production, etc., you can use a coarse nonconforming mesh with the 1D subgrid method, and the approximation error is very minor. The 1D submesh method defaults to be turned on. It can be toggled off using the parameter "1D submesh calculation for fracture-matrix connections" in the "Fracture Options" panel.

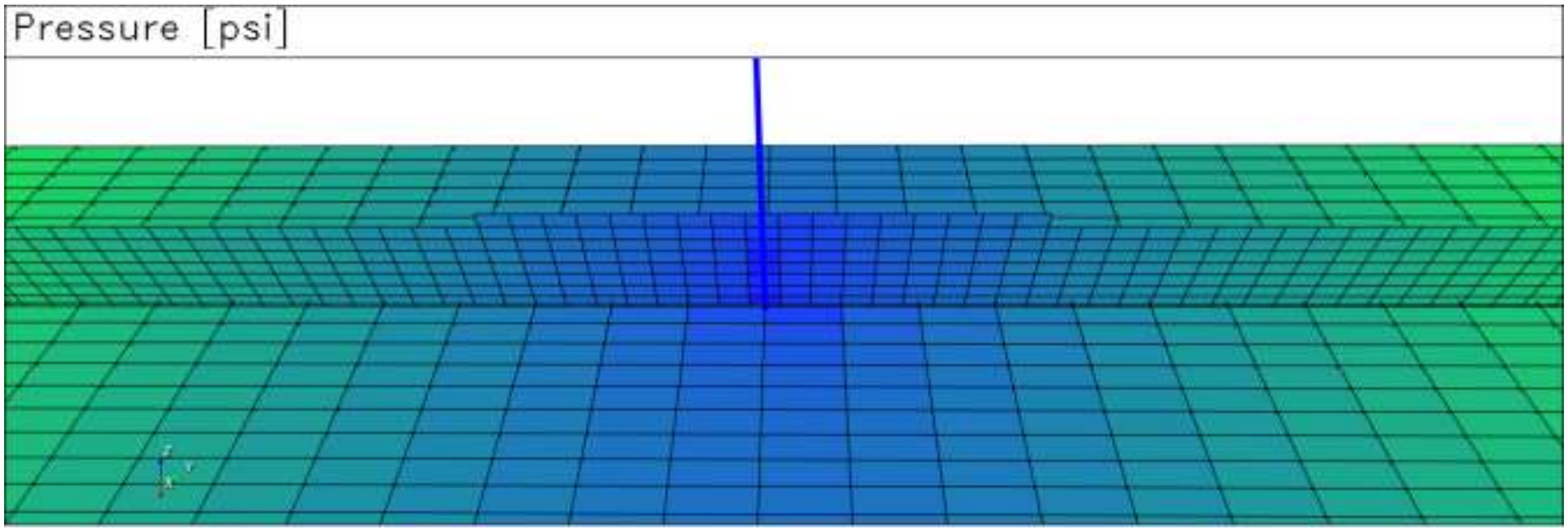

Figure 2: Example of a ResFrac simulation mesh that is relatively coarse and does not use refinement towards the fracture. Accurate results are still possible by using the 1D subgrid method (McClure, 2017).

To handle multiphase effects, the relative permeability for flow into the fracture is calculated using the composition of the matrix element and the pressure of the fracture element. Without this adjustment, the increase in GOR with depletion can easily be missed. The adjustment requires a new flash or black oil calculation, to determine fluid saturation at the pressure of the fracture element, but the composition of the matrix element. This adjustment defaults to be turned on. It can be turned off using the parameter “Adjust submesh for multiphase flow” in the “Fracture Options” panel.

We also introduced an option to make the rel perm for leakoff equal to no greater than the harmonic average of the leakoff calculated by this approach and the maximum possible rel perm of the phase in the matrix. For example, if the maximum possible rel perm of the water phase in the matrix element is set to 0 by the user, then water cannot leakoff. If the maximum possible rel perm to water phase is 0.01, then the rel perm for leakoff could be no more than the harmonic average of 0.01 and 1. This parameter is called “Limit leakoff rel perm to maximum possible matrix rel perm” and is located in the “Fracture Options” panel.

There are different options for specifying formation properties such as permeability and porosity. For each parameter, you can specify properties versus depth or properties on an element-by-element basis. Figure *3* shows an example of matrix properties viewed from the side. Formation properties are specified in the “Static Model and Initial Conditions” panel.

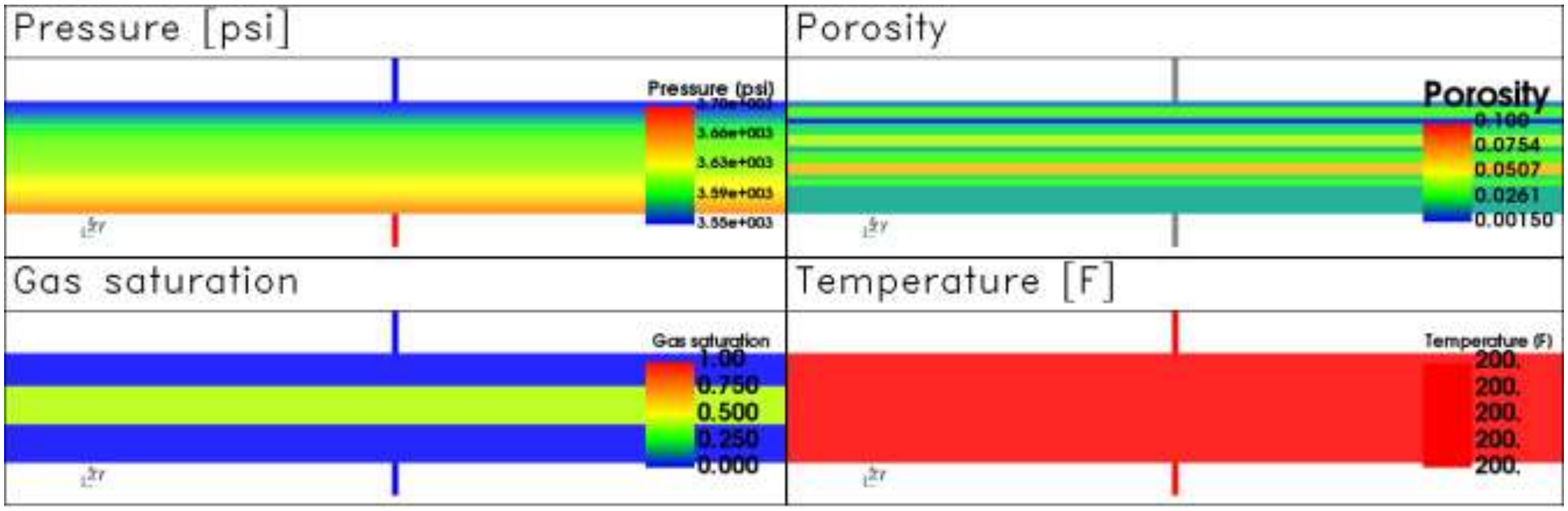

Figure 3: Side view of a matrix region showing formation properties. Porosity and gas saturation are defined by layer. By default, pressure is initialized at hydrostatic equilibrium (though it is possible to specify your own initial pressure profile). This is an isothermal simulation, so temperature is uniform.

You have the option to set up a rectilinear grid or import a corner point grid from a geomodelling package. If you set up a rectilinear grid, you need to manually specify the size and location of the matrix region, and the number and size of the matrix elements in each direction. Fractures cannot propagate out of the matrix region, and so you want to specify it to be large enough that fractures will never reach the edge. If you run a simulation and discover that fractures did, in fact, reach the edge, then you should discard that simulation and rerun with a larger mesh (or increase toughness to reduce the fracture size). The matrix mesh is set up in the “Meshing Options” panel.

Each side of the matrix region can be a ‘no flow’ boundary condition or a ‘constant pressure’ boundary condition. The default behavior is ‘no flow.’ With constant pressure boundary conditions, the pressure, composition, temperature, etc., adjacent to the matrix region is assumed to be equal to the initial conditions of each matrix element adjacent to the edge. This parameter is called “Matrix edge boundaries” and is specified in the “Static Model and Initial Conditions” panel.

In a rectilinear grid, to avoid upscaling issues, the user is heavily encouraged by a warning in the UI to use the ‘align z-mesh with layers’ wizard to make sure that matrix elements do not cut across geologic boundaries. In corner point grids, properties assigned by 'layer' (rather than element-by-element) are assigned based on the z-direction index of the elements, rather than by absolute depth intervals.

Because the fracture and matrix elements are nonconforming, fracture elements may cut across multiple depth intervals. This is handled by the code in different ways, depending on context. For flow, this is handled by handled directly by calculating fluid flow between each pair of fracture and matrix elements. For stress, the code takes the arithmetically thickness weighted average of the stress of each layer. For toughness, the code takes the highest toughness value encountered in any layer. For elastic moduli, the 'self-interaction coefficient' is calculated with a thickness weighted geometric average. Interaction coefficients between separate elements are calculated using the method in Section 10.

You specify the location and geometry of the well(s) with a series of vertices. You should specify the wellbore all the way to the surface. The matrix region will rarely, if ever, be specified all the way to the surface. In some problems, you will want to specify the matrix region to encompass a large region, such as the entire length of the lateral. But sometimes, you'll want to focus on one or a few stages along the well. In that case, you can specify the matrix region to be smaller, focusing only on the region of interest.

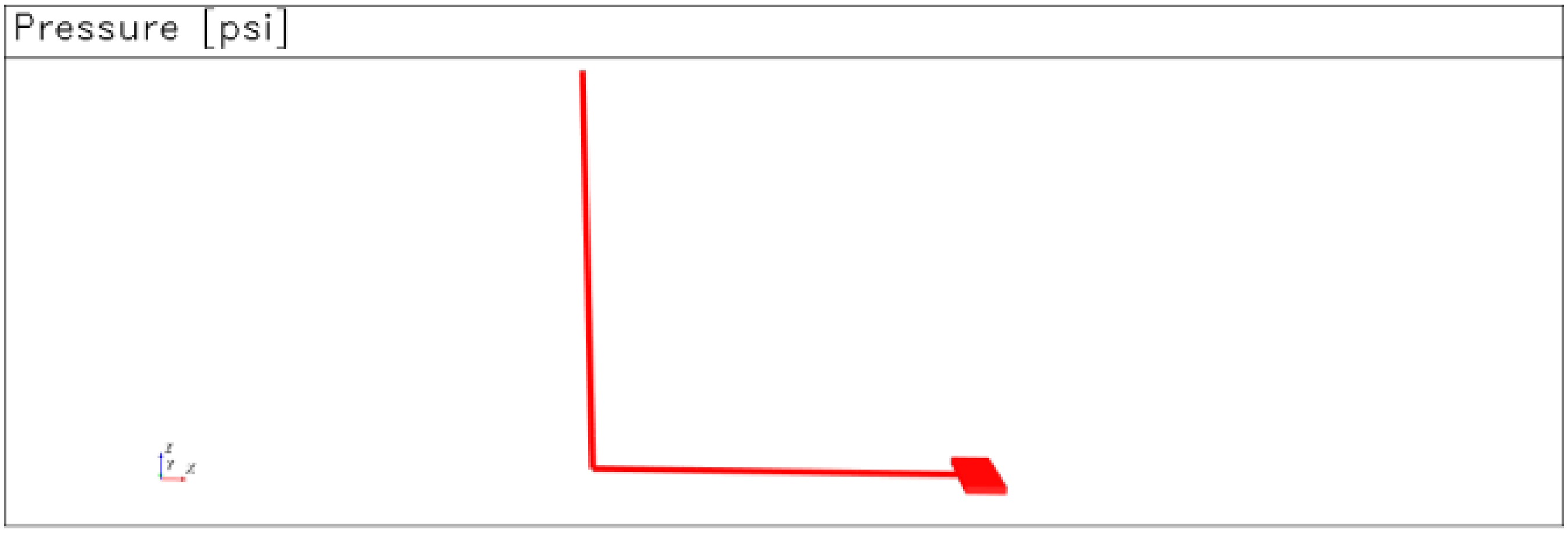

Figure 4: Example of a wellbore and matrix region. The wellbore is meshed to the surface. It has a long vertical section and then a long horizontal lateral. The matrix region is located at the end of the lateral section in the lower right.

ResFrac allows you to inject sequentially into stages. When you specify the well vertices, you can assign each vertex to a particular 'stage'. Then, when you specify the wellbore boundary conditions, you can opt to close off certain stages to flow over different periods of time. Wells vertices and stages are specified in the "Wells and Perforations" panel.

## 2.2 Corner point and general rectilinear grids

In addition to rectilinear, ResFrac supports two other gridding styles: corner point and 'general rectilinear.'

A corner point grid allows the user to maintain geological realism in their simulation. The subsurface does not generally follow straight lines and in many instances requires non-rectangular shapes in order to create a mesh that accurately represents the geology. Corner point (or pillar) grids are a widely used solution. To represent the geology, pillars are drawn that make up the skeletal structure of the grid and points are defined on these pillars in order to define the boundaries of the grid blocks. This type of geometry allows for the representation of pinchout, non-neighboring connected cells, and fault throw. Pinchout defines a grid block thinning at one end to the point of it collapsing on itself not allowing flow past the very small face. Through the use of pillars and separate z-coordinate defined for each grid block corner, non-neighboring connections and fault throws are model by connecting grid blocks that are not directly next to each other in their x, y, and z coordinate order. This allows for faces to have multiple other grid blocks that are connected to the face, or interior cells in the mesh that are not connected to another matrix element. The corner point grid is set up in the "Static Model and Initial Conditions" panel.

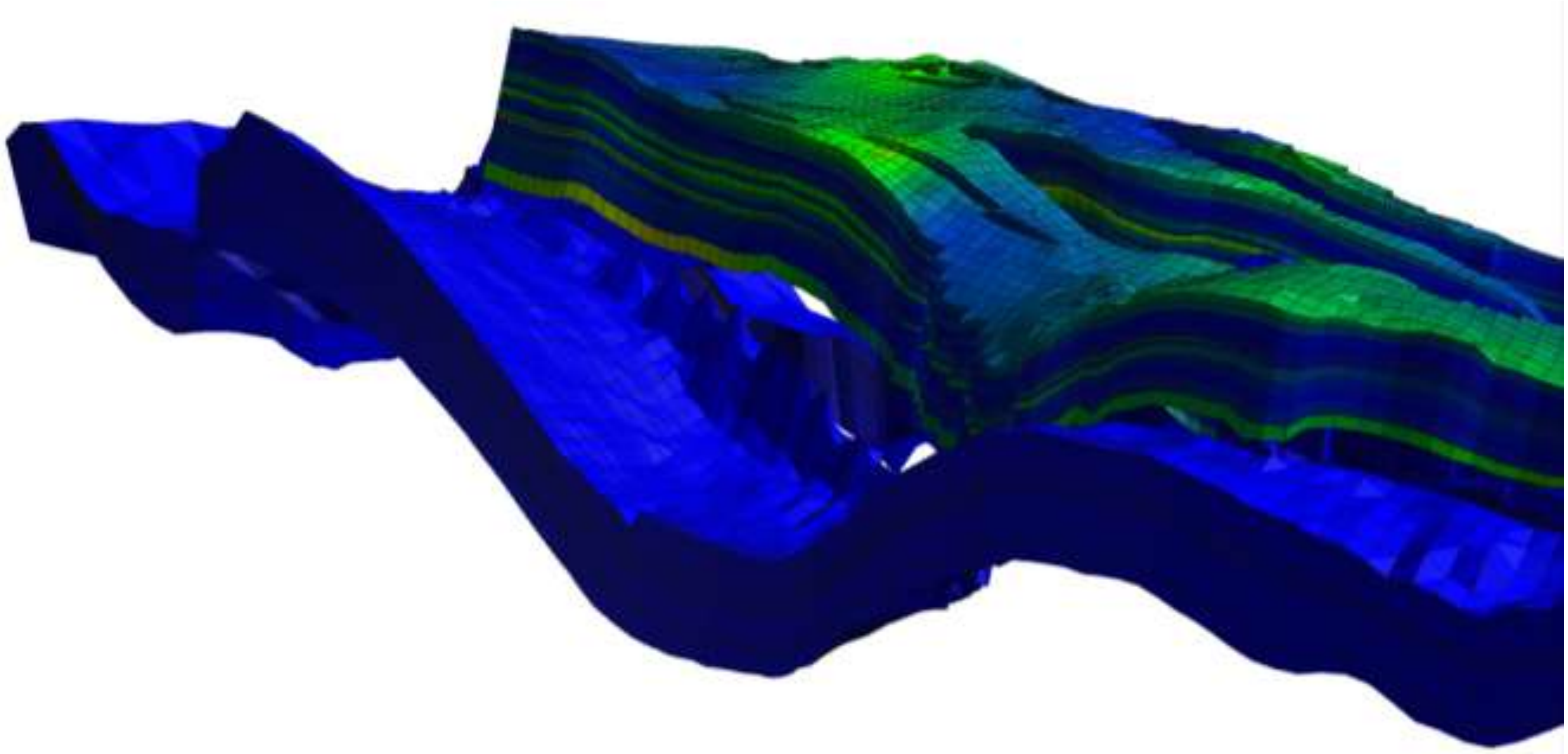

When importing a corner point grid from an externally created file, ResFrac uses the Eclipse/Petrel convention for file importation. To specify the mesh, three keywords are necessary to be defined: COORD, ZCORN, and DIMENS or SPECGRID. This importation wizard is called "Grid Import Wizard" and is specified in the "Static Model and Initial Conditions" panel.

COORD defines the top and bottom x, y, and z coordinates for each pillar of the grid and requires 6 * (NX + 1) * (NY + 1) entries where NX is the number grid blocks in the x-direction and NY is the number of grid blocks.

ZCORN defines the depth of the corners of each grid block. These corner depths are used in conjunction with the pillars defined in using the COORD keyword in order to determine the x and y coordinates of every corner in each separate grid block. ZCORN requires the input of 2 * NX * 2 * NY * 2 * NZ number of entries in order to completely define the grid.

In a general rectilinear grid, the overall mesh is similar to the corner point mesh with one difference, the pillars of the grid remain vertical. In the Eclipse format, these are defined by DX/DY/DZ and TOPS. DX/DY/DZ are the change in the X/Y/Z direction of the grid block and are assigned to the bottom left edge of each grid block, this assignment will create the sloping and slanted behavior of the grid. DX/DY/DZ are defined for each of the grid blocks. TOPS is the true vertical depth (TVD) to the top of the grid block. TOPS should be defined for each grid block in the model and works with DZ to define the depth of the grid blocks. Internally in ResFrac, a corner point grid and general rectilinear grid are represented the same way using the ".rfgrid" file format.

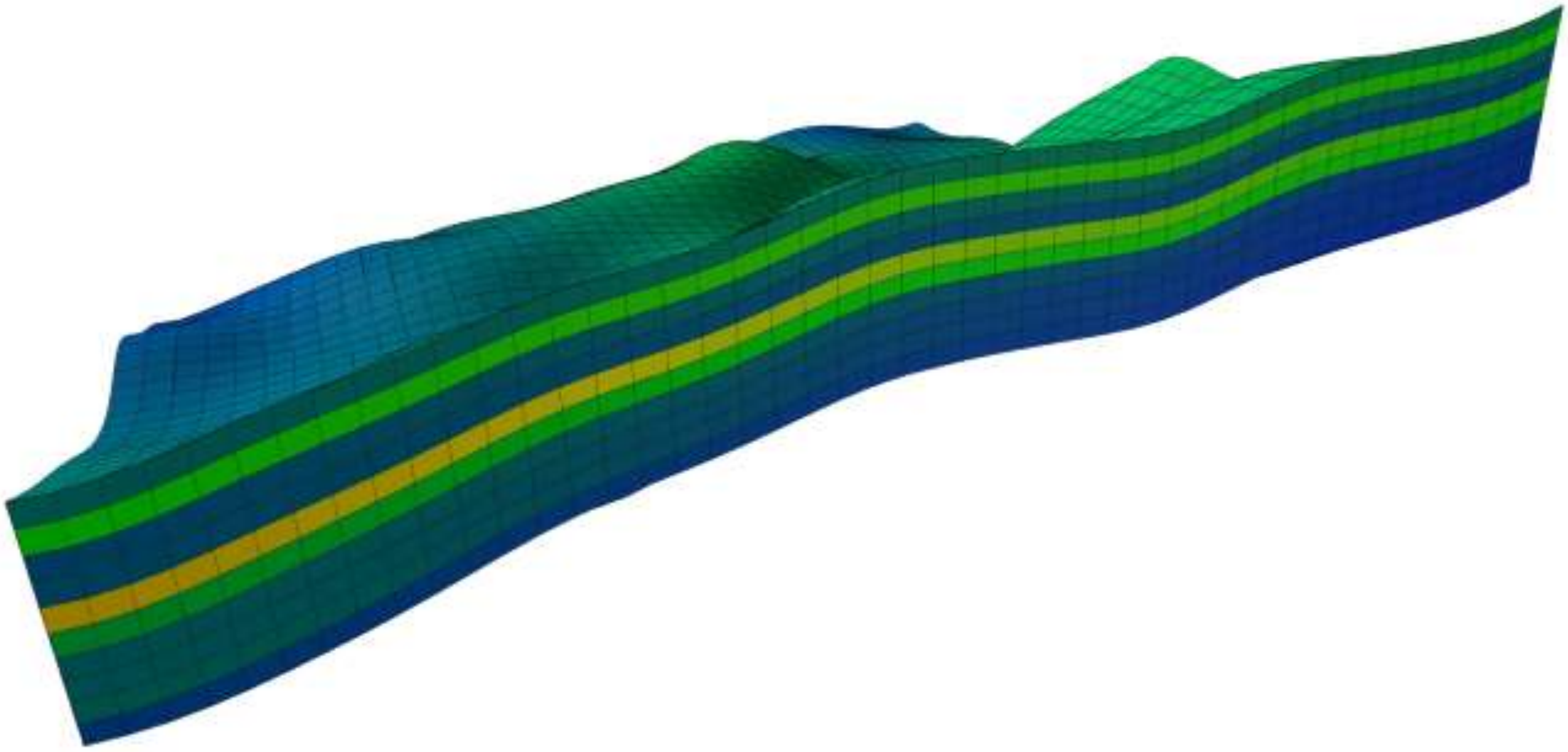

SPECGRID and DIMENS defines the size of the simulation grid. ResFrac does not support fully unstructured grids, and only one reservoir can be defined per simulation. Therefore the two keywords are defined in the same manner, any values defined after the first three inputs in SPECGRID are ignored. SPECGRID and DIMENS are defined with three entries, the first being NX, second NY, and third being NZ.

Once the grid is imported into ResFrac, a grid file is created. This grid file has the extension ".rfgrid" and is used for running the simulation. A grid file contains the overall size of the mesh “numberofmatrixelements”, "xycoordinates", and "zcoordinates". Similar to the external grid format, "xycoordinates" defines every pillar in the mesh, and "zcoordinates" defines the z coordinates of each grid block corner. The entry of “xycoordinates” and “zcoordinates”is done in sets of 6 and 8 respectively, and for each corner of the top layer and each grid block respectively in an increasing X, then Y, then Z manner meaning the entirety of the X grid blocks are cycled through before Y is increased. While you are able to import an .rfgrid file directly into ResFrac, it is highly recommended that you design your grid in Petrel, export to the eclipse format “.GRDECL”, and then import that file folder into ResFrac.

The non-uniform nature of the size and connection of the corner point faces results in a modified method to calculate the transmissibility between the faces. The general case of calculating the corner point or general rectilinear grid for the transmissibility between two blocks with a shared area that is not equal between the two faces is defined by (Cordazzo et al., 2002):

$$T_{12} = \frac{A_c}{\frac{A_1}{T_1}+\frac{A_2}{T_2}} = \frac{1}{\frac{A_1}{A_cT_1}+\frac{A_2}{A_cT_2}} = \frac{1}{\frac{1}{\frac{A_c}{A_1}T_1}+\frac{1}{\frac{A_c}{A_2}T_2}}, \quad 2\text{-}1$$

Where $T_{12}$ is the shared transmissibility between Blocks 1 and 2, $A_1$ and $A_2$ are the total area of the faces and $A_c$ is the shared contact area between Face 1 and 2.

## 2.3 1D submesh method

The 1D submesh method enables accurate calculation of fluid leakoff and production even if the mesh is relatively coarse. In a conventional reservoir simulator, accurate calculation of leakoff and production in very low permeability formation requires a highly refined mesh. This is inconvenient because it may require a very large number of elements to adequately mesh a system with a very large number of fractures. Also, with a corner point grid, it is not practical to perform mesh refinement around all the fractures.

To resolve this problem, the 1D submesh method creates a local submesh for each pair of matrix and fracture elements. The submeshes are constructed in such a way that they can be solved very efficiently, even with a large amount of mesh refinement. This allows good numerical accuracy for the leakoff and production calculations, even with very low permeability. The numerical details are described by McClure (2017).

Each submesh is defined by a function A(x), which describes the cross-sectional area for flow as a function of distance from the fracture. For 'linear' flow, A(x) is a constant. For 'radial' flow, A(x) increases linearly with distance. For 'spherical' flow, A(x) increases with the square of distance. The function A(x) is constructed numerically, in a fully general manner, so that the method can capture transitions between these end-member cases, and also represent the actual geometry of the neighboring fracture and matrix elements. A(x = 0) is always equal to the two-sided fracture surface area, and so at every early time, there is always 'linear' flow into the element. Within a matrix element, each small volume of rock is assigned to one and only one submesh, by estimating which fracture element it will drain into. Thus, if a fracture element is surrounded on its sides by other fracture elements, it will experience purely linear flow in the submesh, since streamlines from the matrix element will only flow linearly into that particular fracture element. However, if the element is not surrounded by any other elements and is in a large matrix element, the flow geometry will eventually transition to radial flow or (depending on vertical permeability) spherical flow.

It is helpful to be aware of the numerical approximations involved in the 1D submesh method. The 1D submesh method solves the single-phase flow equation on each mesh, and for simple single-phase flow problems, show very good accuracy. However, approximations are needed to handle effects from multiphase flow and pressure-dependent permeability.

Conventionally, for flow into the fracture, a simulator would use the relative permeability of the 'upwind' matrix element. However, with a relatively coarse mesh, this can lead to severe inaccuracy. At short and intermediate timescales, the pressure may be drawn down only within a few inches or feet of the fracture. If the matrix element is significantly larger than this 'radius of investigation,' the pressure and saturation in the overall element may have changed much less than the pressure drawdown in the near-fracture region. The result is that upwinding based on the overall saturation/rel perm of the matrix element may dramatically underestimate the impact of relative permeability changes during drawdown.

To resolve this problem, the simulator uses a 'multiphase flow adjustment.' The simulator needs to be able to calculate the effective flowing relative permeability, viscosity, and composition from the matrix element into the fracture element. The 1D submesh uses a single-phase calculation, and so the code only has access to pressure values from the submesh. It needs to be able to relate the pressure distribution in the 1D submesh and the composition of the overall matrix element to the effective flowing fluid properties into the fracture element.

The implementation of the 'multiphase flow adjustment' has evolved over time. Using the current most-up-to date 'multiphase 1D submesh version' (version #4), the simulator uses a simulated constant volume depletion to generate a spline function that allows flow parameters to be defined as a function of pressure.

At the start of each timestep, the code identifies all matrix elements containing a fracture element (and thus, containing a 1D submesh). In these elements, the code uses the composition of the overall matrix element and performs a constant volume depletion calculation. Starting at the current pressure, it calculates the relative permeability, viscosity, and composition of each phase. Then, it drains fluid from the 'cell' in proportions related to these phase mobilities and compositions. This is performed in a series of steps until reaching a low value of pressure. During the constant volume depletion, as the mixture goes below the bubble point, gas components tend to preferentially flow out of the 'cell,' resulting in a relative enrichment of the heavier components. This calculation mimics the process of depletion that occurs in the formation as fluid drains into the fracture(s). The values from each increment are fit with a spline to specify rel perm, viscosity, and composition as a function of pressure for each phase.

The code identifies a point roughly midway through the region of pressure depletion in the submesh and uses the pressure at that point in the interpolation functions to evaluate the properties of the flowing phases.

Section “19.4 Water banking” describes a special adjustment to the 1D submesh that is used to handle flowback of water that has leaked off from the fracture during fracturing.

The code uses a similar, though simpler, treatment for ‘pressure dependent permeability.’ Users have the option to specify a table of pressure dependent permeability multipliers to allow the formation permeability to decrease as pressure draws down. To avoid mesh effects, by default, the simulator uses the pressure in the fracture (not the matrix) to evaluate the pressure dependent permeability loss. Alternatively, there is an option to specify a ‘1D submesh PDP adjustment distance’ for evaluating the value in the PDP table. The code determines the pressure at that distance from the fracture in the submesh and evaluates the PDP multiplier at this value.

The figures below show a comparison between a simulation with a highly refined mesh (so that the 1D submesh method is not used) and a coarse mesh simulation (that uses the 1D submesh method). No effort has been made to calibrate numerical parameters so that the ‘coarse’ simulation will match the results from the ‘refined’ simulation. We can see that there are moderate differences in the results. The refined simulation has lower GOR and more water production than the coarse mesh simulation. Oil production is fairly similar in both simulations.

The moderate mismatch with the refined mesh solution is not ideal. However, in most practical applications, it has minor significance. In practical shale applications, we do not have knowledge of the relative permeability curves that should be used. Without an ability to know relative permeability prior to performing a history match, precise numerical accuracy is not needed. Regardless of mesh refinement, we would be varying the relative permeability curves to match field data. This would be necessary using either a fine-meshed solution or a (much faster-running) coarse mesh solution using the 1D submesh approach.

There are some applications where high-resolution calculations of multiphase flow definitely *are* needed. For example, this is necessary if simulating gas-injection EOR, where compositional mixing in the near-fracture region is critical to consider. In these cases, users should use the ‘mesh refinement wizard’ in the meshing panel, which automatically refines the mesh along the areas where fractures may form.

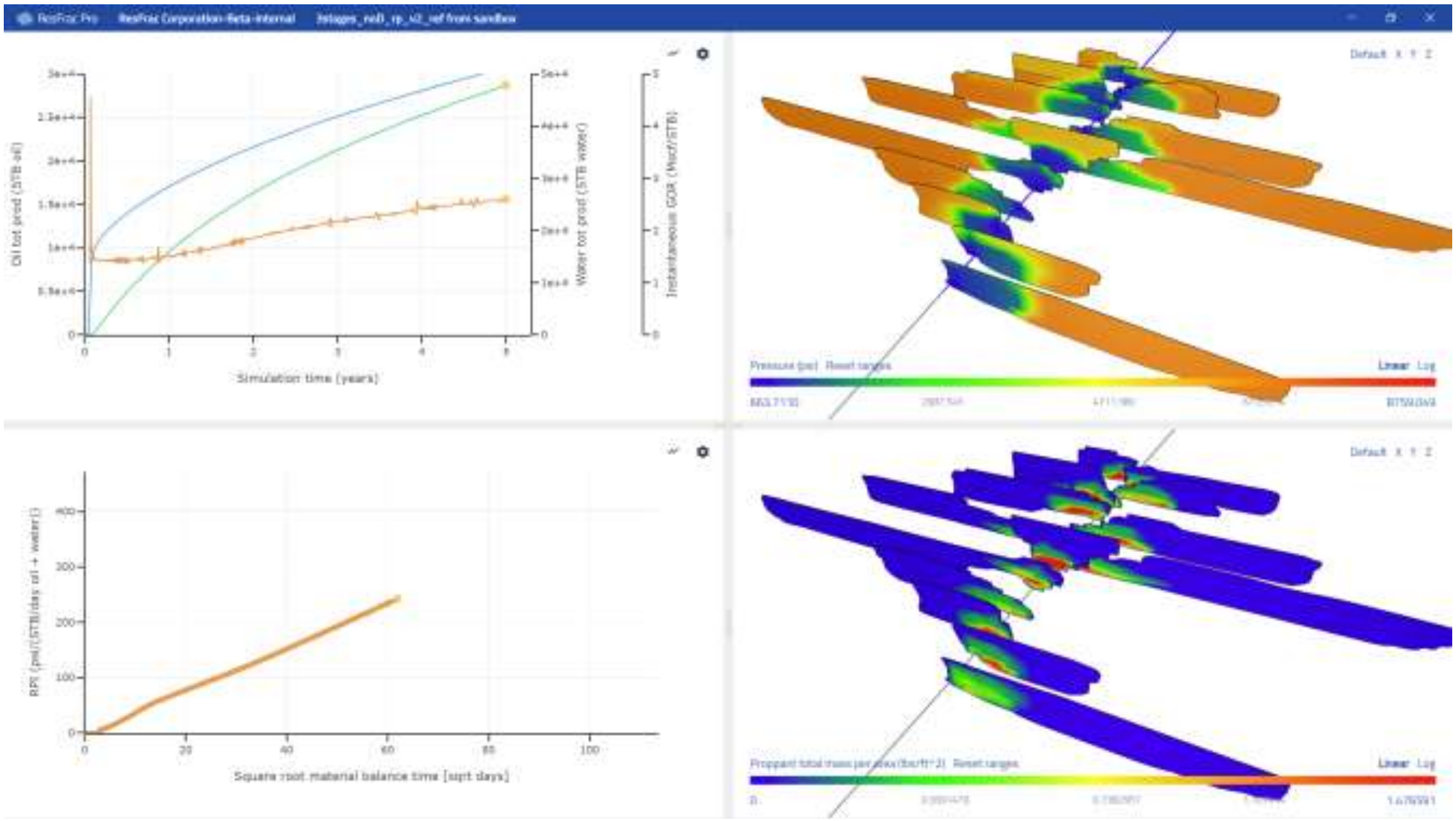

*Figure 5: Production simulation using a highly refined mesh.*

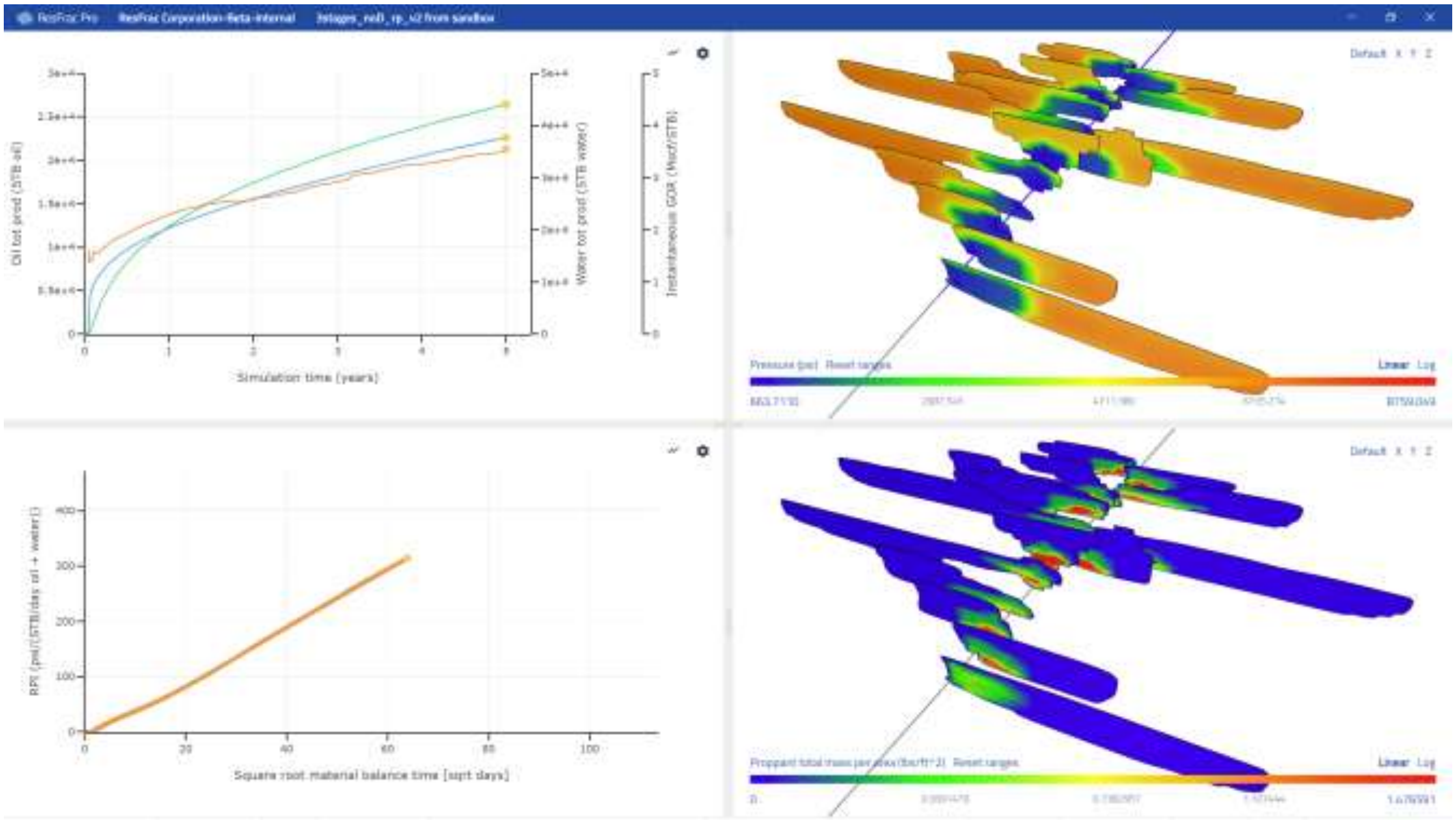

*Figure 6: Production simulation using a coarse mesh, and the 1D submesh method.*

## 3. The overall system of equations

ResFrac uses the finite volume method for transport. In each timestep, the simulator enforces $N_c$ component molar balance equations, one energy balance equation, $N_{pr}$ proppant mass balance equations, and $N_s$ water solute mass balance equations. The total number of equations in each element is $N_c + N_{pr} + N_s + 1$. In addition to these transport equations, wellbore elements have one additional equation, momentum balance. Matrix and fracture elements each have additional equations related to mechanical deformation, as described in Sections 10 and 11. Boundary condition elements have equations enforcing boundary conditions, described in Section 12. The simulator provides the option to turn off thermal transport (making an isothermal simulation) and neglect poroelastic stress response to pressure change.

Calculations of fluid properties are discussed in Section 14. One of the $N_c$ components is water. In the black oil model, there are two other components: 'oil' and 'gas.' In the compositional fluid model, the other components are 'flash' components that form one or two phases that are assumed to be immiscible with the water phase. The flash components are defined by molar mass, pseudocritical temperature and pressure, acentric factor, and binary interaction coefficients. In the "Startup" panel, you specify whether you want to use a black oil treatment or a compositional treatment. Once selected, you specify the details of the fluid model in the "Fluid Model Options" panel.

The $N_s$ water solutes are defined by their molar mass and parameterized in terms of mass fraction within the water component. Hence, the molar mass of the water 'component' is not constant; it is a function of water solute mass fraction. The water solutes may be inert tracers such as salts or they may be gel molecules that impart non-Newtonian flow characteristics, as described in Section 9. Water solutes are defined in the "Water Solutes" panel.

Proppant types are defined by grain diameter, density, and parameters related to their permeability in a packed bed, described in Section 8. The proppant grains are assumed to have constant density. Proppants are defined in the "Proppants" panel.

Momentum balance is used to calculate flow rate in each wellbore element. The calculation includes friction, hydrostatic head, pressure gradient, momentum advection, and momentum accumulation.

The geomechanics calculations consider stress shadowing from fracture opening and poroelastic stress change due to pressure change in the matrix. You have the option to turn on or off poro (and thermo) elastic stress changes using controls in the startup panel.

The boundary condition equations can be specified at the wellhead or as bottomhole constraints. If you specify a bottomhole constraint, the wellbore is removed from the simulation during the period of time when the bottomhole constraints is specified. Well controls are specified in the "Well Controls" panel.

# 4. Fluid flow in the matrix

## 4.1 Overall

The molar flow rate of component c from matrix element i to matrix element j is calculated using Darcy's law (Aziz and Settari, 1979):

$$q_{c,ij} = T_{ij} \sum_{p=1}^{p=3} \frac{z_{c,p,ij} k_{rp,ij} \rho_{M,p,ij}}{\mu_{p,ij}} \left(\Phi_{p,i} - \Phi_{p,j}\right), \qquad 4\text{-}1$$

where p is the phase, $z_{c,p,ij}$ is flowing molar fraction, $k_{rp,ij}$ is relative permeability, $\rho_{M,p,ij}$ is flowing molar density, and $\mu_{p,ij}$ is viscosity. The parameter $T_{ij}$ includes the geometric effect of element shape and harmonically averages permeability (Karimi-Fard et al., 2004). The fluid potential of each phase, $\Phi_p$, drives flow based on fluid pressure and hydrostatic head. Capillary pressure and non-Darcy flow in the matrix are neglected. There is an option to include the effect of non-Newtonian fluid rheology in the matrix.

There are four ways of specifying relative permeability (in decreasing level of complexity):

1. Tables of drainage and imbibition relative permeability curves for water-hydrocarbon and oil-gas. Three phase relative permeabilities are calculated by interpolating between the two-phase curves using the method of Baker (1988). Hysteresis effects are calculated using the methods of Land (1968) and Killough (1976). This option is called 'BakerThreePhase.'
2. Tables of kw and krow versus Sw and kg and krog versus Sg. This option is called 'ThreePhaseRelPerm'.
3. Tables of relative permeability versus phase saturation.
4. Power law Brooks-Corey parameters for each phase.

The power law Brooks-Corey relative permeability is calculated as:

$$k_{rp} = k_{rp,muliplier}\left(\frac{S_p - S_{pr}}{1 - S_{pr}}\right)^{n_p} \quad 4\text{-}2$$

Regardless of the value of $k_{rp,muliplier}$, the relative permability is not permitted to exceed 1.0.

When using the 'ThreePhaseRelPerm' options (ie, tables of kw and krow versus Sw and kg and krog versus Sg), there are four rel perm options available. In all cases, krw and krg are calculated directly from the function of rel perm versus their own saturation. However, the four options vary in their methods of calculating kro.

The option 'BakerOnlykroI' is:

$$k_{ro} = \frac{S_g k_{rog} + (S_w - S_{wco}) k_{row}}{S_g + S_w - S_{wco}} \quad 4\text{-}3$$

The option 'BakerOnlykroII' is:

$$k_{ro} = \frac{(S_g - S_{gco}) k_{rog} + (S_w - S_{wco}) k_{row}}{S_g + S_w - S_{wco}} \quad 4\text{-}4$$

These two Baker options used the Baker rel perm interpolation method, but only for the oil relative permeability.

The option StoneI is:

$$k_{ro} = k_{rocw} SS_o F_w F_g \quad 4\text{-}5$$

$$SS_o = \frac{S_o - S_{om}}{1 - S_{wco} - S_{om}} \quad 4\text{-}6$$

$$SS_w = \frac{S_w - S_{wco}}{1 - S_{wco} - S_{om}} \quad 4\text{-}7$$

$$SS_g = \frac{S_g}{1 - S_{wco} - S_{om}} \quad 4\text{-}8$$

$$F_w = \frac{k_{row}}{k_{rocw}(1-SS_w)}$$ 4-9

$$F_g = \frac{k_{rog}}{k_{rocw}(1-SS_g)}$$ 4-10

The option StoneII is:

$$k_{ro} = k_{rocw}\left(\left(\frac{k_{row}}{k_{rocw}} + k_{rw}\right)\left(\frac{k_{rog}}{k_{rocw}} + k_{rg}\right) - k_{rw} - k_{rg}\right)$$ 4-11

$S_{wco}$ is the connate water saturation. $k_{rocw}$ is the oil rel perm at the connate water saturation. The alternative rel perm models are summarized by Baker (1988). When using the 'ThreePhaseRelPerm' model, the default in ResFrac is 'BakerOnlykroII.'

Hysteresis is not supported with the 'ThreePhaseRelPerm' option. Hysteresis is only included with the full 'BakerThreePhase' rel perm option. The full details of the 'BakerThreePhase' option are given by Baker (1988).

Relative permeability curves in the matrix are specified in the "Curve Sets" panel. You define different sets of relative permeability relations, each called a 'curve set,' and then in the table called "Geologic Units (Facies List)" in the "Static Model and Initial Conditions" panel, you assign each layer to a curve set.

## 4.2 Enforcing initially static fluid

In shale, capillary pressure allows complex vertical distributions of fluid saturation and pore pressure to develop. ResFrac does not include these capillary effects. So, if you specify these as an initial condition, and also specify a nonzero vertical permeability, then at the start of the simulation, fluid may crossflow between layers (even if you don't inject or produce from a well). One strategy to avoid this problem is to specify zero vertical permeability.

Alternatively, you can use a special option to subtract a constant from the hydrostatic potential differences. At the start of the first timestep, the simulator calculates the hydrostatic potential difference for each phase and each element-element connection in the reservoir. This quantity is stored. Then, for flow, over the entire duration of the simulation, the hydrostatic potential difference is calculated as:

$$\Delta\Phi_{eff} = \Delta\Phi_{actual} - \Delta\Phi_{init}$$ 4-12

At the start of the simulation, $\Delta\Phi_{actual} = \Delta\Phi_{init}$, and so there is no effective potential difference and no flow. The term '$-\Delta\Phi_{init}$' captures the effect of the capillary effects. This method is useful for corner point grids, since elements may not be aligned vertically. This option is specified by the parameter "Enforce initial equilibrium in the matrix" in the "Numerical Options" panel.

## 4.3 Dual porosity

ResFrac allows you to use a dual porosity treatment in the reservoir. In the dual porosity model, each matrix element is divided into two collocated elements. One element, the 'fracture' element, has low pore volume. The other element, the 'matrix' element, has high pore volume but is only hydraulically connected to its collocated fracture element. The flow between the matrix element and the collocated fracture element is controlled by a 'shape factor' and the 'matrix' element permeability.

There are two dual porosity options (Zimmerman et al., 1993). Pseudo-steady state dual porosity models the volumetric fluid flow rate (per element volume) between the matrix and fracture elements as:

$$Q_d = \frac{-\alpha k_m}{\mu}\left(P_f - P_m\right), \quad 4\text{-}13$$

where $\alpha$ is a shape factor (units of inverse length squared), $k_m$ is the permeability of the matrix, and $Q_d$ is the volumetric flow rate per element volume. The permeability of the fracture element is used to calculate flow between adjacent elements.

The transient dual porosity model accounts for transient flow between the 'matrix' and 'fracture' element. Rather than calculating fluid exchange using the 'average' pressure in the matrix element, the transient model attempts to account for the pressure gradient that develops due to fluid flow in/out of the 'fractures.' One way to model this transient behavior is through a MINC (multiple interacting continua) model, in which there are multiple nested collocated matrix elements. However, this can be computationally intensive because of the need for a large number of elements. Instead, ResFrac uses the semi-analytical approach proposed by Zimmerman et al. (1993). This approach uses only a single matrix element and approximates transient behavior with a semi-analytical modification to Equation 4-4. The "Startup" panel has an option to turn on dual porosity and select "Transient" or "PSS" (Pseudo-Steady State). Then, the "Geologic units" table in the "Static Model and Initial Conditions" panel allows you to turn on or off dual porosity in each individual layer and specify the dual porosity parameters.

ResFrac supports pressure dependent matrix permeability: reversible, irreversible, or both. In reversible pressure dependent permeability, you input a table of permeability multipliers as a function of pressure change. These multipliers are applied directly as a function of element pressure at every timestep.

With irreversible pressure dependent permeability, each element remembers the highest and lowest pressure reached in the element. You input a table of irreversible pressure dependent permeability multipliers versus maximum historical pressure. Irreversible multipliers greater than 1.0 are intended to capture processes such as shear stimulation of natural fractures or formation of secondary unpropped hydraulic fractures. Multipliers less than 1.0 are intended to capture processes such as collapse of microcrack permeability (Heller et al., 2014; King et al., 2018). Both may be applied simultaneously. For example, if the maximum 'dP' is 2000 psi, and the minimum 'dP' is -4000 psi, the code will apply two separate irreversible multipliers – the multiplier > 1 that corresponds to 2000 psi, and the multiplier < 1 that corresponds to -4000 psi.

Pressure dependent permeability curves are specified in the "Curve Sets" panel. You define different sets of relative permeability relations, each called a 'curve set,' and then in the table called "Geologic Units (Facies List)" in the "Static Model and Initial Conditions" panel, you assign each layer to a curve set. As an alternative to the parameterized reversible permeability increase, the permeability increase can be calculated from a pressure-dependent leakoff (PDL) relation, described in Section 4.7.

The porosity of matrix elements is calculated assuming a constant porosity compressibility:

$$\phi = \phi_{init} exp\left(c_\phi (P - P_{init})\right) \quad 4\text{-}14$$

Following the common convention used in petroleum reservoir engineering, the porosity compressibility, $c_\phi$, is the derivative of porosity with respect to internal fluid pressure given zero lateral strain (Palmer and Mansoori, 1998; Zimmermann, 2017). Initial porosity can be specified on an element-by-element basis or by layer in the table called "Geologic Units (Facies List)" in the "Static Model and Initial Conditions" panel.

## 4.4 Adsorption/desorption

For the Black Oil model, adsorbed gas is calculated according to a Langmuir isotherm (Yu et al., 2014):

$$v_a(P) = \frac{v_{a,L}P}{P+P_L}, \qquad \text{4-15}$$

where $v_a$ is the gas volume of adsorption at pressure P, $v_{a,L}$ is the Langmuir volume, and $P_L$ is the Langmuir pressure. Adsorption is only available as an option in isothermal models. Adsorption is turned on using an option in the "Startup" panel. The adsorption model parameters are specified in the table called "Geologic Units (Facies List)" in the "Static Model and Initial Conditions" panel.

In the compositional model, adsorbed gas is calculated using a multicomponent extension of the Langmuir isotherm model (Wang et al., 2019):

$$\omega_i = \frac{\omega_{i,max}B_i x_i P}{1+P\sum_j B_j x_j}, \qquad \text{4-16}$$

where $\omega_i$ are the moles of adsorbed component *i* per unit mass of rock, $\omega_{i,max}$ is the max number of moles per unit mass of rock that can be adsorbed for a given component *i* (or component-wise "Langmuir volume"), $B_i$ is the inverse Langmuir pressure of component *i* ($MPa^{-1}$), $x_i$ is the mole fraction of component *i* in the fluid phases, and *P* is the system pressure (MPa). A Langmuir isotherm is defined by both a Langmuir volume and a Langmuir pressure for each component in the system, and different isotherms can be applied to different facies.

## 4.5 Diffusion

The compositional model simulates molecular diffusion of each component in each phase due to gradients in component concentrations. A multicomponent extension of Fick's law is used:

$$J_{i\alpha} = -c_\alpha D_{mi\alpha} \nabla x_{i\alpha} \qquad \text{4-17}$$

where $J_{i\alpha}$ is the flux of component *i* in phase $\alpha$ [mol/m$^2$-sec], $c_\alpha$ is the molar density of phase $\alpha$ [mol/m$^3$], $D_{mi\alpha}$ is the effective mixture diffusion coefficient of component *i* in phase $\alpha$ [m$^2$/sec], and $x_{i\alpha}$ is the mole fraction of component *i* in phase $\alpha$ [mol/mol].

The effective diffusion coefficient is a function of bulk binary diffusivities, the pore structure of the matrix, and phase saturation. A common representation of the effective diffusion coefficient as a function of porosity, phase saturation, and bulk diffusivity is as follows (Ho and Webb, 2006):

$$D_{mi\alpha} = \phi^m S_\alpha^n D_{i\alpha} \qquad \text{4-18}$$

where $D_{i\alpha}$ is the bulk diffusivity of component *i* in phase $\alpha$. The cementation exponent, *m*, and the saturation exponent, *n*, are functions of the volume of pore space, the converging/diverging nature of the pore throats, and flow path topology. Millington and Quirk (1961) derived *m* and *n* for unconsolidated porous media and proposed values of 4/3 and 10/3, respectively. Millington and Quirk note that $D/D_0$ in a saturated medium is closely related to 1/F, where F is the formation factor and is proportional to $\frac{1}{\phi^m}$. In nanoporous media, the cementation exponent in the formation factor can be 3 or even higher. Thus, a typical range for both *m* and *n* is usually between 1 and 3.

For the Fickian diffusion model, bulk diffusivity of each component in each phase is calculated using a pseudo-binary diffusivity approximation via the Wilke (1950) equation:

$$D_{i\alpha} = \frac{1-x_{i\alpha}}{\sum_{j\neq i}\frac{x_{j\alpha}}{D_{ij\alpha}}} \quad \text{4-19}$$

where $D_{ij\alpha}$ is the Maxwell-Stefan bulk diffusion coefficient between component *i* and *j* in phase $\alpha$.

The Maxwell-Stefan bulk diffusion coefficients can be specified in one of two ways. First, the user can specify matrices of constant bulk diffusion coefficients for each phase. Matrices must be $N_c$ x $N_c$, where $N_c$ is the number of components, and diffusion coefficient matrices must be specified for both the oil and gas phases. Alternatively, bulk diffusion coefficients can be computed internally as a function of pressure, temperature, and composition using the extended Sigmund correlation (Sigmund, 1976; Da Silva and Belery, 1989; Yang et al., 2022):

$$\frac{\rho_M D_{ij}}{\rho_M^0 D_{ij}^0} = 0.99589 + 0.096016\rho_{pr} - 0.22035\rho_{pr}^2 + 0.032874\rho_{pr}^3 \quad \text{4-20}$$

where $\rho_M$ is the phase molar density, $\rho_{pr}$ is the pseudo-reduced molar density, and $\rho_M^0 D_{ij}^0$ is the low-pressure density-diffusivity product between components *i* and *j*. For $\rho_{pr}$ greater than 3.0, this equation is modified as follows (Da Silva and Belery, 1989):

$$\frac{\rho_M D_{ij}}{\rho_M^0 D_{ij}^0} = 0.18839 exp\left(3 - \rho_{pr}\right) \quad \text{4-21}$$

Diffusive fluxes are computed individually for each phase. This is important because gas phase diffusivities can be several orders of magnitude larger than liquid phase diffusivities. When approaching the critical point, an abrupt change in fluid phase state classification from “oil” to “gas” in ResFrac can cause discontinuous changes in diffusive flux through each phase even if composition is not abruptly changing. To handle this transition, ResFrac interpolates diffusivity, fluid composition, and fluid density as a function of distance from reference oil and gas phase compositions. Then, diffusive fluxes through the oil-like and gas-like fractions of the fluid are separately computed.

## 4.6 Well matrix connection

For open hole well simulations, it is important to accurately calculate connection properties between a given well and the surrounding matrix. Specifically, given a well and matrix elements, we first calculate the intersection length of the well segment that is inside the matrix element. The second step is to use Peaceman’s equation, which allows us to capture the logarithmic singularity of pressure near the well.

Peaceman’s equation is derived for single phase flow, but it accounts for permeability anisotropy, as well as mesh anisotropy (Peaceman, 1982). The flow rate at the well $q_w$ is related to the pressure difference between the well and matrix elements as

$$q_w = T_w(p_0 - p_w),$$

where $p_w$ is the pressure of the well element, while $p_0$ is the corresponding pressure of the matrix element. The transmissibility $T_w$ is calculated as

$$T_w = \frac{2\pi k\, h}{\mu log(r_0/r_w)},$$

where $h$ is the length of the well segment inside the matrix element, $\mu$ is fluid viscosity, $r_w$ is the radius of the wellbore. Assuming that the well is oriented along the $z$ direction, the permeability is $k = \sqrt{k_x k_y}$, while the length scale is

$$r_0 = 0.28 \frac{\sqrt{(k_y/k_x)^{1/2}\Delta x^2 + (k_x/k_y)^{1/2}\Delta y^2}}{(k_y/k_x)^{1/4} + (k_x/k_y)^{1/4}}.$$

Here $k_x$ and $k_y$ are permeabilities in the $x$ and $y$ directions, respectively, while $\Delta x$ and $\Delta y$ are the matrix element dimensions. Wells that are oriented in different directions are treated in a similar fashion.

For the case of a perforation connected directly to a matrix, we need to use a three-dimensional solution for flow. Specifically, pressure solution around a perforation can be written as

$$p(r) = p_w + \frac{q_w \mu}{4\pi k}\left(\frac{1}{r_w} - \frac{1}{r}\right).$$

Here $p_w$ is well pressure, while $r_w$ is well radius. Alternatively, it can be written as

$$p(r) = p_0 + \frac{q_w \mu}{4\pi k}\left(\frac{1}{r_0} - \frac{1}{r}\right),$$

where $p_0$ is the pressure in the matrix element containing well and $r_0$ is the effective length scale. Flow rates to neighboring cells can be computed as

$$q_j = \frac{k A_j}{\mu \Delta x_j}(p_j - p_0).$$

While the sum of all flow rates should be equal to the total rate, i.e. $\sum q_j = q_w$. This leads to

$$\sum \frac{k A_j}{\mu \Delta x_j}(p_j - p_0) = \sum \frac{k A_j}{\mu \Delta x_j} \frac{q_w \mu}{4\pi k}\left(\frac{1}{r_0} - \frac{1}{\Delta x_j}\right) = q_w.$$

This result can be simplified to calculate the effective length as

$$r_0 = \frac{\sum \frac{k A_j}{4\pi \Delta x_j}}{1 + \sum \frac{k A_j}{4\pi \Delta x_j^2}}.$$

Since transmissivity is defined as $q_w = T_w(p_0 - p_w)$, it can be calculated as

$$T_w = \frac{4\pi k}{\mu}\left(\frac{1}{r_w} - \frac{1}{r_0}\right).$$

Note that the summation is done for all six neighbors, while $\Delta x_j$ are the cell sizes in different directions and $A_j$ are the corresponding flow areas.

## 4.7 Pressure-dependent leakoff (PDL)

In naturally fractured formations, fluid loss can increase sharply once the fluid pressure exceeds the pressure at which natural fissures open mechanically. This behavior is called pressure-dependent leakoff (PDL). It is commonly identified from the characteristic hump in the superposition derivative, $G\, dP/dG$, during the pressure falloff after a diagnostic fracture injection test (Barree and Mukherjee, 1996). Fracture models based on Carter leakoff represent PDL by increasing the leakoff coefficient exponentially with pressure above the fissure opening pressure. ResFrac does not use a leakoff coefficient. Leakoff is calculated as transient flow from the fracture into the matrix (Section 2.3). Therefore, PDL is implemented as a pressure-dependent multiplier on matrix permeability, with inputs expressed as a leakoff coefficient multiplier so that they correspond to the parameters used in conventional PDL analysis.

The leakoff coefficient multiplier, $C_P$, is a function of fluid pressure, $P$:

$$C_P = \begin{cases} 1, & P \le P_{fis} \\ \min\left(\exp\left(b\left(P - P_{fis}\right)\right), C_{P,max}\right), & P > P_{fis} \end{cases} \quad 4\text{-}22$$

where $b$ is the "PDL leakoff coefficient exponent," $P_{fis}$ is the fissure reopening pressure, and $C_{P,max}$ is the "PDL maximum leakoff coefficient multiplier," which must be greater than or equal to one.

The fissure reopening pressure is calculated for each matrix element from its initial pore pressure, $P_{init}$, its initial minimum horizontal stress, $\sigma_{hmin,init}$, and the "PDL fissure reopening pressure fraction," $f_{fis}$. If "PDL fissure fraction applies to delta" is true (the default):

$$P_{fis} = P_{init} + f_{fis}\left(\sigma_{hmin,init} - P_{init}\right), \quad 4\text{-}23$$

so that $f_{fis} = 0$ places fissure reopening at the initial pore pressure, and $f_{fis} = 1$ places it at the initial minimum horizontal stress. If "PDL fissure fraction applies to delta" is false:

$$P_{fis} = \max\left(f_{fis}\sigma_{hmin,init}, P_{init}\right). \quad 4\text{-}24$$

If a per-element stress array is provided, $\sigma_{hmin,init}$ is taken from its Sxx values (the x-direction of the simulation mesh is aligned with the minimum horizontal stress). Otherwise, it is calculated from the minimum horizontal stress inputs at the depth of the element. If "Shift PDP table / PDL fissure pressure with stress depletion" is true, $P_{fis}$ is also shifted by the poroelastic and thermoelastic stress change in the direction of the minimum horizontal stress (Section 11), so that the fissure reopening pressure follows the stress change caused by depletion or injection.

The reservoir-controlled leakoff coefficient is proportional to the square root of permeability. Therefore, the leakoff coefficient multiplier is applied to permeability as its square. The three principal permeabilities of the matrix element are multiplied by:

$$F_{PDL} = C_P^2, \quad 4\text{-}25$$

so the permeability multiplier never exceeds $C_{P,max}^2$. Because $C_P$ is calculated from the current pressure, PDL is reversible. The leakoff enhancement disappears when the pressure falls below $P_{fis}$, for example during shut-in or flowback, and it returns if the pressure rises above $P_{fis}$ again during a later injection.

Because $P_{fis}$ is above the initial pore pressure, the multiplier is mostly active where injection raises the pressure: at the fracture face and in the formation near the fracture. For a fracture–matrix connection that uses the 1D submesh (Section 2.3), the multiplier is evaluated at both the fracture element pressure, $P_f$, and the matrix element pressure, $P_m$, and the two values are averaged:

$$F_{PDL,con} = \frac{1}{2}\left[F_{PDL}\left(P_f\right) + F_{PDL}\left(P_m\right)\right]. \quad 4\text{-}26$$

If "Pressure dependent permeability arithmetic averaging" is false, the harmonic average is used instead. If "Pressure dependent permeability not greater than one if fracture is closed" is true, $F_{PDL,con}$ is damped toward one when the adjacent fracture element is mechanically closed. For connections between matrix elements, the multiplier of each element is evaluated at its own pressure, as for other pressure-dependent permeability multipliers.

Optionally, the same factor can be applied to the filtercake. If "Apply PDP/PDL multiplier to filter cake permeability" is true in the curve set of the matrix element, the filtercake permeability at the connection (Equations 9-3 and 9-4) is multiplied by the pressure-dependent permeability factor of the connection:

$$k_{fc,con,eff} = F_{PDL,con} k_{fc,con}. \qquad \text{4-27}$$

This represents fissures that open above the fissure reopening pressure and bypass the filtercake. The option applies to fracture–matrix and well–matrix connections, and it also works with the 'Reversible PDP' option and with table-based pressure-dependent permeability, in which case the corresponding pressure-dependent permeability factor of the connection is used.

PDL is specified for each curve set in the "Curve Sets" panel. Set "Use parameterized PDP/PDL curves" to true, set "Pressure dependent permeability/leakoff option" to 'Reversible PDL,' and specify the exponent, the fissure reopening pressure fraction, and the maximum multiplier. With 'Reversible PDL,' the parameterized reversible permeability increase parameters are not used, but the parameterized reversible and irreversible permeability loss parameters and the irreversible permeability increase parameters still apply. Each layer is assigned to a curve set in the table called "Geologic Units (Facies List)" in the "Static Model and Initial Conditions" panel. Large values of $b$ change permeability by orders of magnitude over a small pressure range, which can slow convergence. The simulator issues a warning if the permeability multiplier increases more than tenfold over less than 350 psi, which corresponds to $b$ greater than about 0.0033 psi$^{-1}$.

The property "Permeability (x-direction)" in the 3D visualization shows the matrix permeability multiplied by the factor evaluated at the matrix element pressure only. Because Equation 4-26 also includes the factor evaluated at the fracture pressure, PDL can increase leakoff substantially even where the visualized permeability of the matrix element is unchanged. Leakoff rates are a more direct measure of the effect of PDL.

# 5. Fluid flow in the fractures

This section describes general-purpose constitutive equations for calculating fracture aperture and transport properties. The equations consider Darcy and non-Darcy flow, Newtonian and non-Newtonian fluid, multiphase flow, are valid for mechanically open or closed fractures, and are valid for any value of proppant volume fraction.

The literature contains a variety of well-validated equations for describing fracture aperture and flow properties under different conditions. For example, the cubic law describes flow through an open fracture (Witherspoon et al., 1980); joint stiffness laws describe flow through a closed proppant-free fracture (Barton et al., 1985); and flow through a proppant-filled fracture can be modeled as flow through a packed bed of particles. Because simulation codes usually focus on either fracturing or production, they do not need to realistically describe the transition from mechanically open to closed. Because our objective is to simulate fracturing and production within a single framework, we require general-purpose equations that handle transitions between limiting cases.

Dontsov and Peirce (2014; 2015) developed a set of constitutive equations for capturing the transition from slurry flow through an open fracture to fluid flow through packed particles in a closed fracture. They neglected the ability of proppant-free fractures to retain aperture after closure due to roughness. McClure and Horne (2013) developed a framework for modeling the transition from mechanically open to closed, applying a nonlinear joint closure law after closure. However, they did not include the effect of proppant. Their algorithm was applied by McClure et al. (2016b) to achieve very close matches to field data from diagnostic fracture injection tests (DFITs), including before, during, and after closure. Shiozawa and McClure (2016a) combined the approaches of Dontsov and Peirce (2014; 2015) and McClure and Horne (2013). However, the constitutive equations developed by Shiozawa and McClure (2016a) do not guarantee that the proppant volume fraction cannot exceed unity at high closure stress and are only applicable for single-phase, Darcy, Newtonian flow. The relations developed in ResFrac follow an approach similar to this prior work, but with modifications to ensure physically plausible results under all conditions. Our relations include multiphase flow, non-Darcy flow, and non-Newtonian flow.

In limiting cases, our relations reduce to well-validated and widely used equations for fracture flow (such as the cubic law for a mechanically open fracture). In intermediate cases, the relations smoothly (without discontinuity) transition between the different constitutive laws. The equations always yield results that satisfy physical constraints (such as the requirement that proppant volume fraction cannot exceed unity). Within the overall framework, we apply specific constitutive equations to calculate flow properties. Specific constitutive equations could be replaced, if desired, within the overall framework.

Before providing the equations for flow through the fracture, it is necessary to describe the equations for calculating aperture. A fracture is defined as being mechanically open if the fracture walls have come out of contact because fluid pressure exceeds the normal stress. A fracture is defined as being mechanically closed if the fracture walls are in contact because the fluid pressure is less than the normal stress.

The aperture of a mechanically closed element is calculated as:

$$E(\sigma_n > P) = E_{res} + \frac{1-C_{pr,c}}{1-C_{pr}}[E_{cr} + E_b], \quad (5\text{-}1)$$

where:

$$E_{cr} = E_0(1-\gamma_b)\left(\frac{1}{1+\frac{9(\sigma_n - P)}{\sigma_{n,ref}}}\right), \quad (5\text{-}2)$$

$$E_b = \left(E_0\gamma_b + E_{pr}\right)exp\left(-c_{b,\phi}(\sigma_n - P)\right), \quad (5\text{-}3)$$

$$\gamma_b = min\left(1.0, \frac{C_{pr,c}}{C_{p,max}}\right). \quad (5\text{-}4)$$

When the fracture transitions from open to closed, the aperture is equal to: $E_{res} + E_0 + E_{pr}$, or equivalently, $E_{res} + E_{cr} + E_b$.

The $E_{cr}$ term represents the contribution of the unpropped fracture roughness to the aperture (Barton et al., 1985; Willis-Richards et al., 1995). The parameter $\sigma_{n,ref}$ is the effective normal stress at which the $E_{cr}$ term reaches 10% of its maximum value. $E_{res}$ is the irreducible minimum aperture, the fracture aperture at very high effective normal stress.

$C_{pr}$ is the volume fraction of proppant. $C_{pr,max}$ is the maximum possible proppant volume fraction in a packed bed. For random packing and moderately heterogeneous particles, $C_{pr,max}$ is around 0.5 - 0.66. $\gamma_b$ is the fraction of the fracture "roughness" part of the aperture ($E_0$) that is filled with proppant. $C_{pr,c}$ is the effective proppant volume fraction at closure, equal to the volume of proppant per area divided by $E_0 + E_{pr} + E_{res}$ (discussed further below). Ideally, $C_{pr,c}$ should be less than or equal to $C_{p,max}$. Because $E_{pr}$ is updated explicitly (discussed below), $C_{pr,c}$ may exceed $C_{p,max}$ slightly. In this case, $\gamma_b$ is not permitted to exceed unity. $E_{pr}$ is excess aperture at closure above $E_0 + E_{res}$; it is only nonzero if the fracture contains a large volume of proppant.

The $\frac{1-C_{pr,c}}{1-C_{pr}}$ factor stiffens the fracture as it is compressed (which increases $C_{pr}$) and guarantees the physical constraint that the proppant volume fraction never exceeds unity. As the fracture compresses after closure, $C_{pr}$ may exceed $C_{pr,max}$. This is a physically valid result that corresponds to the case of proppant crushing and embedment into the fracture walls. The values of $c_{b,\phi}$ and $\sigma_{n,ref}$ are chosen such that $E_{cr}$ is much more compliant than $E_b$, consistent with the idea that an unpropped fracture is much more sensitive to changes in closure stress than a propped fracture.

Equation 5-1 is an implicit equation for E because of the dependence of $C_{pr}$ or $C_{pr,c}$ on aperture. This dependence arises from the definition that the mass of proppant per area at closure is equal to the current mass of proppant per area:

$$m_{pr,a} = EC_{pr}\overline{\rho}_{pr} = C_{pr,c}(E_0 + E_{pr} + E_{res})\overline{\rho}_{pr}, \qquad 5\text{-}5$$

where $m_{pr,a}$ is the proppant mass per area. The average proppant density $\overline{\rho}_{pr}$ is constant because the grains are assumed incompressible. If $C_{pr}$ is known, Equation 5-5 is solved for $C_{pr,c}$ (in terms of E), which is plugged into Equation 5-1. If $m_{pr,a}$ is known, then Equation 5-5 is solved for $C_{pr}$ (in terms of E) and plugged into Equation 5-1. In either case, after the substitution, Equation 5-1 is a quadratic equation for E. For physically realistic input parameters, the quadratic equation always has exactly one positive real root.

$C_{pr,c}$ is equal to the concentration of proppant that would be present if the closed fracture was unloaded to effective normal stress equal to zero (holding all proppant in-place); it is not necessarily equal to the value of $C_{pr}$ when closure actually occurred. The distinction is necessary because it is possible for proppant to flow through a mechanically closed fracture. This could occur if the proppant grains have very low diameter (see Section 8 for a discussion of proppant bridging), the fracture is rough (relatively large value of $E_0$), the effective normal stress is low, and the proppant volume fraction is low (so that $C_{pr,c}$ is less than $C_{pr,max}$).

The aperture of a mechanically open fracture is calculated as:

$$E(\sigma_n = P) = E_0 + E_{res} + E_{pr} + E_{open}. \qquad 5\text{-}6$$

Within each timestep, $E_0$, $E_{res}$, and $E_{pr}$ are constant. If the element is closed, then E is calculated according to Equation 5-1. If the element is open, then E is calculated according to Equation 5-6, with $E_{open}$ as an additional unknown in the system of equations.

It would not be realistic to initiate fracture elements with significant $E_0$ and $E_{res}$. Instead, when a fracture element is initiated, all aperture terms are equal to zero except $E_{res}$, which is given a tiny initial value, 0.00000328 ft. Strictly, this does not conserve mass, but this initial aperture is so tiny that the error is negligible. As fluid flows into the element, pressure increases, causing the aperture to increase. As the fracture opens, an algorithm is used to progressively update $E_0$ and $E_{res}$ until they reach user-defined maximum values. This mimics the process of roughness formation as the fracture surfaces form and separate for the first time. Effectively, elements with $E_0$ and $E_{res}$ less than their maximum value are considered fracture surfaces that have yet fully formed. The vast majority of elements in a simulation have $E_0$ and $E_{res}$ greater than their maximum value.

If $E_0$ and $E_{res}$ are less than their maximum allowed value, they are updated at the end of each timestep. The code determines whether $E_{open}$ exceeds 10% of the total aperture. If so, $E_{open}$ is decreased, and $E_0$ and/or $E_{res}$ are increased by the same amount, in order to enforce that $E_{open}$ equals 10% of E. The variables are updated so that the adjustment does not change E overall. $E_0$ is equal to zero until $E_{res}$ reaches its maximum value, $E_{res,max}$. Then, $E_0$ is increased until it reaches its maximum possible value, $E_{0,max}$. The values $E_{0,max}$ and $E_{res,max}$ are user-defined within each facies; typical values are 0.00082 ft and 0.0000328 $ft$, respectively (Barton et al., 1985). Once $E_0$ and $E_{res}$ have reached their maximum value, they do not increase further. Because roughness generation is an irreversible process (the breaking apart of the rock), the adjustment only increases $E_0$ and $E_{res}$; it never decreases them.

The maximum allowed values of $E_0$ and $E_{res}$ and the value of $\sigma_{n,ref}$ (90% closure stress) are specified by layer in the table called "Geologic Units (Facies List)" in the "Static Model and Initial Conditions" panel.

Next, $E_{pr}$ is updated. $E_{pr}$ is defined such that if fluid was drained from the element until closure (holding the volume of proppant in the element constant), then at the point of closure (when $C_{pr}$ reached $C_{pr,max}$), aperture would be equal to $E_{res}$ + $E_0$ + $E_{pr}$ (satisfying Equation 5-1 because $E_{open}$ reaches zero at the point of mechanical closure). Therefore, if:

$$\frac{EC_{pr}}{C_{pr,max}} \leq E_0 + E_{res}, \quad (5\text{-}7)$$

then $E_{pr}$ is set equal to zero. This corresponds to the case where there is so little proppant in the element that the roughness dominated part of the aperture, $E_{res}$ + $E_0$, is capable of containing all the proppant in the element at closure at a volume fraction less than or equal to $C_{pr,max}$. If the inequality in Equation 5-7 is not satisfied, then $E_{pr}$ is greater than zero and calculated as:

$$E_{pr} = \frac{EC_{pr}}{C_{pr,max}} - E_0 - E_{res}. \quad (5\text{-}8)$$

This method of updating $E_{pr}$ satisfies the physical constraint that $C_{pr}$ must be less than or equal to $C_{pr,max}$ at closure. Because $E_{pr}$ is updated explicitly at the end of the timestep, $C_{pr}$ may slightly exceed $C_{pr,max}$ at closure (because closure occurs implicitly during the nonlinear solve within each timestep), but the difference is slight and so is considered acceptable. If the overshoot greater than a few percent, the timestep is discarded and repeated with a smaller timestep duration.

$q_p$, the volumetric flow rate of phase p, is written as a weighted average of the volumetric flow rate of fluid/proppant slurry flowing through an open fracture, $q_{p,o}$, and a value that is similar to the flow rate of pure fluid through a closed fracture, $q_{p,c}$ (defined below). A weighting factor $\gamma_f$ is defined that equals zero when $E_{open}$ is equal to zero (so that $q_p$ is equal to $q_{p,c}$) and is asymptotically equal to one when $E_{open}$ is large (so that $q_p$ is equal to $q_{p,o}$). If $\gamma_f$ is equal to 0.5, then $q_{p,c}$ and $q_{p,o}$ are weighted equally. Equal weighting occurs when $E_{open}$ is equal to 10% of $E_0 + E_{pr} + E_{res}$ (the aperture at closure). As discussed below, if there is no proppant in the fracture, $q_{p,c}$ is identically equal to $q_{p,o}$ and so the value of the $\gamma_f$ is irrelevant because it is a weighting between two identical values. $q_p$ and $\gamma_f$ are calculated as:

$$\gamma_f = \frac{E_{open}}{0.1(E_0+E_{pr}+E_{res})+E_{open}}. \quad (5\text{-}9)$$

$$q_p = \gamma_f q_{p,o} + (1-\gamma_f) q_{p,c}. \quad (5\text{-}10)$$

The flow rate of fluid/proppant slurry through an open fracture in one-dimensional flow over a distance $\Delta x$, across width W, and driven by difference in flow potential $\Delta\Phi_p$ is calculated as:

$$\frac{\Delta \Phi_p}{\Delta x} = \frac{-q_{p,o}}{W}\left(\frac{12\mu_{p,s}}{\left(E_{adj}E\right)^3 k_{rp,cro}}\right) + sgn\left(\Delta \Phi_p\right)\left(\frac{q_{p,o}}{WE}\right)^2 \beta_{cro}\beta_{rp,cro}\rho_{p,s}. \quad \text{5-11}$$

The fracture conductivity is equal to $E^3/12$, the classical cubic law (Witherspoon et al., 1980). In this equation, the parameter $W$ is not aperture; instead, it is the width of the flowing region (ie, cross-sectional area is equal to aperture times width of the flowing region). The parameter $E_{adj}$ is the 'Effective fracture aperture conductivity factor.' It is used to account for the effect of multistranded fracture propagation, roughness, pinchpoints, and other causes of nonideality. Practically, it may be given values from 0.2-1.

The effective phase viscosity, $\mu_{p,s}$, includes an adjustment for the effect of proppant concentration (Section 8) and so depends on both fluid properties and proppant concentration. As proppant volume fraction approaches the maximum possible, effective viscosity goes to infinity, representing the immobilization of the slurry into a packed bed of particles (Section 8.6; Dontsov and Peirce, 2014; 2015). The viscosity calculation accounts for non-Newtonian fluid properties caused by fluid additives. The phase relative permeability, $k_{rp,cro}$, is calculated as $S_p^{n_{p,cr}}$, where $n_{p,cr}$ is an exponent. This calculation assumes there is not any residual saturation in mechanically open fractures. $\beta_{rp,cro}$ is calculated as the reciprocal of the relative permeability, following the recommendation from Fourar and Lenormand (2000; 2001). The phase slurry density $\rho_{p,s}$ includes the density of the phase and the density of the proppant flowing in the fluid. It is calculated assuming that the proppant partitions in equal volume fractions into each phase. The operator sgn returns -1.0 if the value in the parentheses is negative and 1.0 otherwise.

$\beta_{cro}$ is calculated as a function of aperture from the correlation developed for non-Darcy fracture flow by Chen et al. (2015). Chen et al. (2015) performed a large number of fracture flow experiments and correlated $\beta$ with aperture:

$$\beta = 0.022\left(\frac{\xi}{2E}\right)^{.66} E^{-1}, \quad \text{5-12}$$

where $\xi$ was defined as the "maximum asperity height" and effectively acted like a tuning parameter.

At large E in a mechanically open fracture, flow becomes analogous to flow through an open duct. Flow through an open duct is described by the Darcy-Weisbach equation (the equation used for calculating frictional pressure gradient for flow through a pipe). Analogy between Equation 5-11 and correlations for the Fanning friction factor used in the Darcy-Weisbach equation indicates that $\beta$ should not be given a value lower than about 0.0194/E, which corresponds with Fanning friction factor for fully developed turbulent flow. Therefore, we use 5-12 to calculate $\beta_{cro}$, but use 0.0194/E as a minimum value.

The various constants related to the calculation of flow through the fractures, such as the value of $\xi$, are defined by different parameters in the "Fracture Options" panel.

Phase potential $\Phi$ is the sum of fluid pressure and gravitational head. The same value of $\Delta\Phi_p$ needs to be used in the calculations of both $q_{p,o}$ and $q_{p,c}$. However, for flow through an open fracture, proppant is suspended in slurry with the fluid, and the density of the phase should be calculated including the effect of proppant on mixture density. The effect of proppant on slurry density causes gravitational slurry convection, in which proppant-laden slurry injected into a proppant-free fracture tends to sink to the bottom of the fracture (Section 8). On the other hand, when the proppant is wedged between the walls of the fracture and is not supported by the fluid, the phase density in the calculation of $\Phi$ should be equal to the pure fluid density. To define a single, unique value of $\Phi_p$ that can be used in either calculation, we average the pure fluid density and proppant-laden slurry density using $\gamma_f$:

$$\Phi_p = P_p - gz\left(\gamma_f \rho_p + (1-\gamma_f)\left(C_{pr}\overline{\rho}_{pr} + (1-C_{pr})\rho_p\right)\right), \quad \text{5-13}$$

where z is depth and $\overline{\rho}_{pr}$ is the average density of the proppant grains. $P_p$ is the fluid pressure of phase p (equal to simply P because capillary pressure is neglected in the present work).

If the phase is non-Newtonian (aqueous phase containing gel), the apparent viscosity in Equation 5-11 is calculated using the analytical solution for flow between parallel plates. We have not found an analytical solution for flow of an MPL fluid between parallel plates. However, a solution for an Ellis fluid is available, and the Ellis fluid model yields a rheology curve that is very similar to the MPL. The apparent viscosity of an Ellis fluid flowing through parallel plates is (Matsuhisa and Bird, 1965):

$$\mu_a = \frac{\mu_0}{2\left[1+\frac{3}{\alpha+2}\left(\frac{dP}{dx}\frac{E}{\tau_{1/2}}\right)^{\alpha-1}\right]}, \quad \text{5-14}$$

where $\alpha$ and $\tau_{1/2}$ are Ellis model parameters chosen such that rheology curve is consistent with the equivalent MPL curve. Water solute properties that impart non-Newtonian fluid properties are defined in the “Water Solutes” panel.

The relation for $q_{p,c}$ is:

$$\frac{\Delta\Phi_p}{\Delta x} = -\left(\frac{q_{p,c}}{W}\right)\frac{1}{M_{p,cr}+M_{p,b}} + sgn(\Delta\Phi_p)\left(\frac{q_{p,c}}{WE}\right)^2 \rho_p \frac{M_{p,cr}+M_{p,b}}{\frac{M_{p,cr}}{\beta_{cr}\beta_{rp,cr}}+\frac{M_{p,b}}{\beta_b\beta_{rp,b}}}. \quad \text{5-15}$$

In contrast to Equation 5-11, the fluid density used in the non-Darcy term in Equation 5-15 is equal to the pure phase density and does not include the effect of proppant on slurry density.

The parameter $M_{p,cr}$ is defined as:

$$M_{p,cr} = \frac{(E_{res}+E_{cr}+E_{open})^3 k_{rp,cr}}{12\mu_{p,cr}}, \quad \text{5-16}$$

where $k_{rp,cr}$ is the relative permeability for crack flow, and $\mu_{p,cr}$ is the phase viscosity for crack flow (Equation 5-14). The fracture conductivity is calculated using the cubic law, but neglecting the $E_{pr}$ part of the aperture. If there is no proppant, then $E_{pr}$ is zero and $E_{cr}$ is equal to $E_0$ and so 5-15 uses an identical conductivity as 5-11. If the fracture is closed and completely full of proppant, then the conductivity used in Equation 5-16 is based only on $E_{res}$ and is very small. In this case, Equation 5-15 is dominated by $M_{p,b}$ (defined below), and the equation reduces to Forchheimer’s law for flow through porous media. $\beta_{cr}$ is calculated from Equation 5-12, and $\beta_{crp,cr}$ is calculated as the reciprocal of $k_{rp,cr}$ (Fourar and Lenormand, 2000; 2001).

The parameter $M_{p,b}$ is defined as:

$$M_{p,b} = \frac{E_b k_b k_{rp,b}}{\mu_{p,b}}, \quad \text{5-17}$$

where $E_b$ is the proppant bed part of the aperture, $k_b$ is the proppant bed permeability (Equation 5-21), $k_{rp,b}$ is the relative permeability for flow through the proppant bed, and $\mu_{p,b}$ is the phase viscosity for flow through the proppant bed (Equation 5-22). In the case of a proppant-free fracture, $M_{p,b}$ is equal to zero, Equation 5-15 is dominated by $M_{p,cr}$, and the relation simplifies to Forchheimer's law for flow through a fracture.

When $E_b$ and $E_{cr}$ are both nonzero (indicating a fracture that contains proppant, but not enough to entirely fill the space between the asperities at closure), Equation 5-15 calculates flow assuming fluid is transported in parallel through the "bed" and "crack" parts of the aperture. The effective conductivity for flow in parallel is the sum of the conductivity of each layer. Consequently, the M terms are summed in Equation 5-15. Next, the $\beta$ terms are averaged to find an equivalent $\beta$ for the combined flow. While high M terms correspond to high flow capacity, high $\beta$ terms correspond to low flow capacity. Thus, the appropriate average of $\beta$ for flow in parallel is harmonic, not arithmetic. This is analogous to how conductance and resistance are averaged to find effective values in an electric circuit in series or parallel. In the harmonic average, the terms are weighted according to their corresponding M term because the fraction of flow in the bed and crack part of the aperture is approximately equal to their respective M terms divided by the sum of the M terms.

$k_{rp,cr}$ is calculated from the power-law form of the Brooks-Corey model:

$$k_{rp,cr} = k_{rp,cr,multiplier} \left(\frac{S_p - S_{pr,cr}}{1 - S_{pr,cr}}\right)^{n_{p,cr}}, \qquad \text{5-18}$$

where $k_{rp,cr,multiplier}$ is a multiplying factor, $n_{p,cr}$ is a curvature parameter, and $S_{pr,cr}$ is the residual phase saturation. Regardless of the value of $k_{rp,cr,multiplier}$, the relative permeability is not allowed to exceed 1.0.

To ensure consistency with the relative permeability calculation used for mechanically open fractures (which assumes zero residual phase saturations and $k_{rp,max}$ equal to 1.0), the values of $S_{pr,cr,max}$ and $k_{rp,cr,max}$ are defined as functions of effective normal stress. When the fracture walls first contact at $\sigma_n{}'$ (effective normal stress) equal to zero, consistency requires that $S_{pr,cr,max}$ and $k_{rp,cr,max}$ must be equal to zero and 1.0, respectively. As effective normal stress increases, they asymptotically approach user-defined limiting values, $S_{pr,cr,max}(\sigma_n' \gg 0)$ and $k_{rp,cr,max}(\sigma_n' \gg 0)$. The transition is controlled by a hyperbolic relationship, mimicking the relationship for aperture developed by Barton et al. (1985):

$$k_{rp,cr,max}(\sigma_n{}') = k_{rp,cr,max}(\sigma_n' \gg 0) + \left(1 - k_{rp,cr,max}(\sigma_n' \gg 0)\right)\frac{1}{1+9\sigma_n'/\sigma_{n,ref}}, \qquad \text{5-19}$$

$$S_{pr,cr,max}(\sigma_n') = S_{pr,cr,max}(\sigma_n' \gg 0)\left(1 - \frac{1}{1+9\sigma_n'/\sigma_{n,ref}}\right). \qquad \text{5-20}$$

The bed relative permeability is calculated using Equation 5-18. For bed flow, values of $k_{rp,b,max}$ and $S_{pr,b,max}$ are constant, not functions of effective normal stress (unlike Equations 5-19 and 5-20).

The Brooks-Corey parameters for flow through propped and unpropped fractures are specified in the "Fracture Options" panel. The parameter "X-curve for open fractures" in that panel allows you to toggle on or off whether open fractures are X-curve (or have the same relative permeability behavior as closed unpropped fractures), and the parameter "Maximum relative reduction in residual saturation for open elements" provides additional ability to tune this behavior.

There are two ways to calculate proppant bed permeability. If you leave the parameter "proppant pack permeability compressibility" blank when you specify the proppants, then the proppant bed permeability is calculated using a modified Kozeny-Carmen equation (Krauss and Mays, 2014):

$$k_b = k_{0,b} \frac{d^2(\phi - f_b)^3}{(1-\phi+f_b)^2}, \quad (5\text{-}21)$$

where d is the proppant diameter, $f_b$ and $k_{0,b}$ are user-defined constants, and $\phi$ is the porosity (1.0 – $C_{pr}$). If there are multiple different types of proppant present, then d, $f_b$, and $k_{0,b}$ are equal to the volume fraction weighted average of the properties of each type.

If you specify the "proppant pack permeability compressibility," then the permeability of the proppant pack is calculated with the equation:

$$k_b = k_{b,sn0} exp\left(-c_{b,k}(\sigma_n - P)\right), \quad (5\text{-}22)$$

where $k_{b,sn0}$ is the permeability of the proppant bed at normal stress equal to zero (calculated from Equation 5-21), and $c_{b,k}$ is the "proppant pack permeability compressibility."

All parameters related to proppant pack permeability and porosity are defined in the "Proppants" table in the "Proppants" panel.

Figure *7* shows fracture conductivity versus effective normal stress for different amounts of proppant mass per area. The calculation is performed using the values specified in the caption. When ResFrac is run, it automatically calculates the table of conductivity versus effective normal stress and outputs them to the comments file. Note that 'fracture conductivity' is not a parameter used directly by ResFrac. Instead, it is uses 5-11 and 5-15 to calculate flow. Figure *7* shows the equivalent fracture conductivity is non-Darcy pressure gradient were neglected.

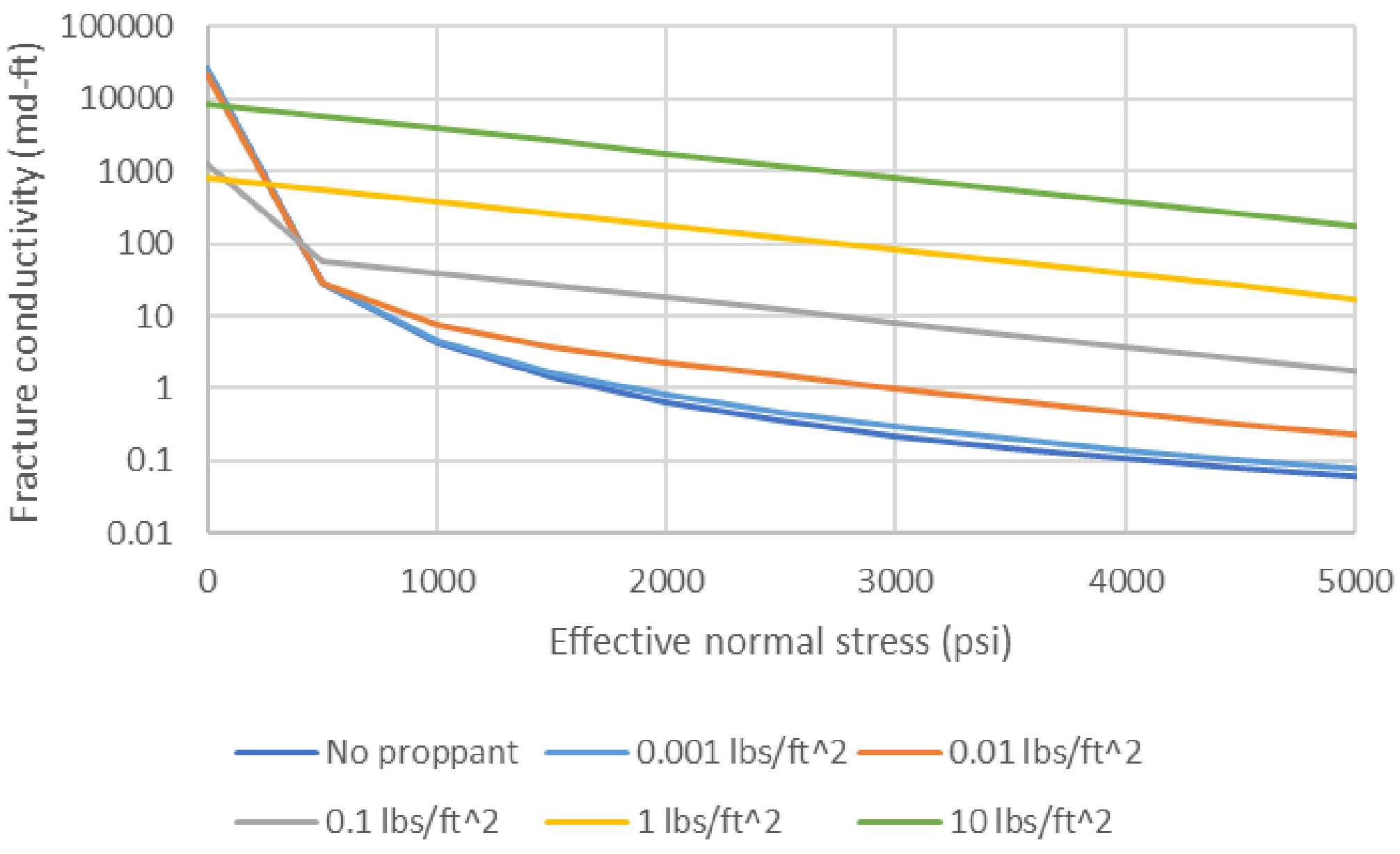

Figure 7: Fracture conductivity versus effective normal stress for different amounts of proppant mass per area. Calculated using $E_0$ equal to 0.0015 ft, $\sigma_{n,ref}$ equal to 500 psi, $k_{0,b}$ equal to 0.007, f equal to 0, and $c_{b,k}$ equal to 7e-4 $psi^{-1}$.

The user has the option to customize the relationship between effective normal stress and conductivity. They can: (a) specify a table of propped conductivity multipliers as a function of effective normal stress (overriding the standard exponential conductivity function; (b) specify different tables of proppant conductivity multipliers as a function of effective normal stress at different values of proppant concentration (and the code interpolates between them based proppant concentration); (c) a table of unpropped conductivity multipliers as a function of effective normal stress (overriding the standard cubic law relation).

If the phase is non-Newtonian, the shear rate for flow in the proppant bed is calculated as (Cannella et al., 1988):

$$\dot{\gamma}_{p,b} = \frac{\alpha_c u_p}{\sqrt{k_b k_{rp,b} S_p \phi}}, \qquad 5\text{-}23$$

where $\alpha_c$ is a constant (around 1.0 to 15.0). The shear rate is related to viscosity, as discussed in Section 9. The parameter "Alpha in the Cannella equation" in the "Fracture Options" panel allows you to specify the value of $\alpha_c$.

A wide variety of equations are available for predicting $\beta_b$ (Ergun and Orning, 1949; Geertsma, 1974; Martins et al., 1990; Lopez-Hernandez et al., 2004;). Lopez-Hernandez et al. (2004) compared equations and recommended the Martins et al. (1990) equation because it is based on experiments from a variety of types of proppant under different conditions. Based on this recommendation, we adopt the Martins et al. (1990) equation:

$$\beta_b = \frac{0.0000077638}{k_b^{1.036}}, \qquad 5\text{-}24$$

where the units of $\beta_b$ are $m^{-1}$ and the units of $k_b$ are $m^2$. The correlation is not dimensionally consistent.

Geertsma (1974) proposed a widely-used an expression for $\beta_{rp}$ for flow through porous media. However, it was developed based on experiments with gas/water flow at water saturation less than 0.5. It scales $\beta_{rp}$ according to phase saturation to the -5.5 power, which implies unrealistically large values of $\beta_{rp}$ at low phase saturation. Also, it does not consider the effect of residual phase saturation, which is unphysical because $\beta$ should approach infinity as relative permeability approaches zero. Fourar and Lenormand (2000) reviewed experiments on multiphase flow through packed bed reactors used in chemical engineering. They found that $\beta_{rp}$ can well-approximated by setting $\beta_{rp}$ equal to the inverse of $k_{rp}$. This is equivalent to the Lockhart and Martinelli (1949) approach for flow through a packed bed (Sáez and Carbonell, 1985). We follow the recommendation of Fourar and Lenormand (2000) and set $\beta_{rp,b}$ to be the inverse of $k_{rp,b}$. The parameters "Fracture relative beta model" and "Proppant bed relative beta model" in the "Fracture Options" panel allow you to control how the relative beta parameters are calculated.

# 6. Fluid flow in the wellbore

Flow velocity is calculated by solving the momentum balance equation (in addition to the component, thermal, water solute, and proppant conservation equations that are solved in all elements). Wellbore flow is calculated using a homogeneous model that assumes that the superficial velocity of each phase is the same (except for gravitational settling of proppant, discussed below). The multiphase viscosity is calculated as the mass fraction weighted average of the viscosity of the individual components (Cicchitti et al., 1959). The homogeneous model is a strong simplification of multiphase flow because it neglects buoyancy and viscosity differences between phases. During multiphase production, it is recommended that you specify the boundary conditions as "bottomhole" boundary conditions, instead of wellhead conditions, which removes the wellbore from the model.

The momentum balance equation in a well is written as (Hasan and Kabir, 2002):

$$\frac{d(\overline{\rho}v)}{dt} - \frac{dP}{dz} - v\frac{d(\overline{\rho}v)}{dz} + g sin(\theta_w)\overline{\rho} + \frac{fv^2\overline{\rho}}{D} = 0, \quad 6\text{-}1$$

where $\overline{\rho}$ is the average density in the element (including all fluid phases and proppants), D is wellbore diameter, v is superficial velocity, f is the Fanning friction factor, g is the gravitational constant, and $\theta_w$ is the angle from vertical.

The hydrocarbon phases are Newtonian, but if the aqueous phase contains gel, it is non-Newtonian and described with the modified power law (Section 9). Correlations for f are available for either Newtonian or power law fluids, but not modified power law fluids (which model the fluid as Newtonian at low shear rate and power law at high shear rate). We calculate f by taking a weighted average of $f_N$ and $f_{PL}$, the friction factors for Newtonian and power law fluids, respectively. The Newtonian friction factor is calculated from the Chen (1979) correlation. The power law friction factor is calculated as the weighted average of the laminar flow power law friction factor (Keck et al., 1992) and the turbulent flow power law friction factor (Dodge and Metzner, 1959). The weighting factor in the average is based on Reynolds number, uses a sigmoidal function, and is centered at Reynolds number of 2000, approximately the transition value to turbulent flow (Keck et al., 1992). The values of $f_N$ and $f_{PL}$ are averaged with weighting calculated from a sigmoidal function of the logarithm of shear rate, with a transition at $\dot{\gamma}_{1/2}$. The effect of proppant on the slurry viscosity in the friction factor calculations is calculated using the method developed by Keck et al. (1992). The parameters "Wellbore friction adjustment factor" and "Wellbore proppant friction adjustment factor" allow tuning of wellbore friction. Typically, because of friction reducers that aren't accounted for in the standard correlation, we significantly tune down the effect of friction. You also have the option to specify friction adjustment factors for each type of water solute in the "Water solutes" table in the "Water Solutes" panel.

Proppant flows through the wellbore due to both convection and gravitational settling. Proppant settling in the vertical section of the wellbore is neglected. Proppant settling in the lateral section of the wellbore – due to inadequate fluid velocity – can be included by turning on the parameter "Include proppant settling in laterals" in the "Proppants" panel.

Perforation pressure drop between wellbore elements and fracture elements is modeled using the equation (Cramer, 1987):

$$\Delta P_{pf} = \frac{0.808Q^2\overline{\rho}}{C_{pf}^2 N_{pf}^2 D_{pf}^4}, \quad 6\text{-}2$$

where Q is total volumetric flow rate, $N_{pf}$ is the number of perforations in the cluster, $D_{pf}$ is the diameter of the perforations, and $C_{pf}$ is the coefficient of discharge. Note that despite the 0.808 factor, Equation 6-2 is written in consistent units, not field units. Perforation pressure drop between the wellbore and matrix elements is neglected. The parameters in Equation 6-2 are specified for each perforation cluster. Perforation properties can be specified for each individual cluster or on an average 'per stage' basis using the controls in the "Wells and Perforations” panel.

ResFrac implements the perforation erosion correlation from Long and Xu (2017). This correlation models perforation erosion as being caused by proppant flow. Early on, the discharge coefficient is rapidly increased to a limiting value. Much more slowly, the perforation diameter increases. Both processes are parameterized by coefficients that the user can input or leave at the default values recommended by Long and Xu (2017). We typically initialize simulations with relatively large values of discharge coefficient, so that the discharge coefficient ‘erosion’ is minor. The erosion parameter can be calibrated empirically from downhole measurements performed after the completion of the job (Cramer et al., 2019). Perf erosion is controlled by the parameters “Perforation erosion alpha” and “Perforation erosion beta” in the “Wells and Perforations” panel. You can also specify these erosion parameters separately for each individual proppant type in the “Proppants” panel.

To account for nonuniform perforation diameter, you can input the standard deviation of perforation diameter. ResFrac assumes that the perforations in each cluster are normally distributed with a standard deviation given by this user-input parameter. Because perforation pressure drop scales with the fourth power of diameter, variability in perforation erosion results in an increase in the effective diameter above the mean value. This parameter is “Perforation diameter standard deviation” in the “Wells and Perforations” panel.

Proppant may bridge off at perforations, as described in Section 8.

In addition to perforation pressure drop, there may be additional pressure drop due to near wellbore tortuosity. This is modeled in the code with the following empirical equation (Wright, 2000):

$$\Delta P_{nw} = A_{nw} Q^{\alpha_{nw}}. \qquad \text{6-3}$$

In openhole wellbore sections, perforation pressure drop is zero. However, the near wellbore complexity pressure drop applies to all connections between wellbore and fracture elements. The parameters in Equation 6-3 are specified at each wellbore vertex, allowing the near wellbore pressure drop to be different at different points along the wellbore. Near wellbore tortuosity can be empirically measured with a step down test (Cramer et al., 2019). ResFrac optionally includes a simplified model for ‘erosion’ of the near wellbore tortuosity. Near-wellbore tortuosity coefficients and exponents can be specified along sections of the wellbore or at each individual perforation in the “Well trajectories and perforation locations” section of the “Wells and Perforations” panel. Near-wellbore tortuosity erosion is controlled by the parameter “Near-wellbore complexity erosion factor” in the “Other Physics Options” panel.

During production, when using the compositional model, flash calculations are performed to simulate surface separation facilities. The separator is held at 200 psia and 120˚F. The water and gas phases are removed at the separator. The liquid phase from the separator is then flashed to stock tank conditions, 14.7 psia and 68˚F. The stock tank gas is separated and collected with the separator gas. After these separations, the volumes of the produced water, oil, and gas are calculated at standard conditions. “Standard pressure” and “Standard temperature” can be modified in the “Other Physics Options” panel. “Separator pressure” and “Separator temperature” are specified in the “Well Controls” panel.

# 7. Thermal transport

Convective energy transport is calculated assuming constant heat capacity for the water phase and hydrocarbon phases. Heat conduction between elements is calculated using Fourier’s law. Thermal conductivity and the heat capacity and density of the solid grains in the matrix are specified within each facies. Thermal conductivity and heat capacity are specified in the “Geologic units (facies list)” table in the “Static Model and Initial Conditions” panel.

In addition to being used for fluid flow, the 1D subgrid method (McClure, 2017) is also implemented for heat conduction. Therefore, you can get accurate heat conduction calculations between the matrix and fracture elements, even if you use a relatively coarse mesh. A simplification is that the heat conduction calculations with the 1D subgrid method do not account for the effect of heat convection on the temperature distribution in the element at the ‘subgrid’ scale.

The formation surrounding the well is not meshed all the way to the surface. However, conductive heat exchange between the wellbore and the surrounding formation may be significant and so is included in the calculation. Wellbore heat exchange with the surrounding formation is calculated using the boundary element technique described by Zhang et al. (2011). The code calculates the convolution integral of the temperature derivative with respect to time multiplied by the Green’s function for radial heat conduction from a cylinder at constant temperature. This technique requires storing temperatures from all previous timesteps at each element and reevaluating the convolution integral at each timestep. Nevertheless, because the calculation is performed independently for each well element (conduction is assumed radial), the calculation is fast and does not have a significant effect on simulation runtime. Heat transfer between the fluid in the wellbore and the exterior of the wellbore (outside the casing and cement) is described by a user defined total heat transfer coefficient, $U_{w,tot}$. The code simultaneously calculates the temperature on the outside of the wellbore and the total heat conduction transfer rate.

# 8. Proppant transport

In this section, we describe the relations implemented in the simulator for proppant transport in the fracture and wellbore. The simulator calculates transport parameters (such as settling rate) for each of the $N_{pr}$ proppant types in the simulation. Some transport properties depend on the overall proppant volume fraction, $C_{pr}$. In this case, the total proppant volume fraction (considering all types of proppant present) is used in the calculation.

## 8.1 Proppant bridging/screenout

Proppant is unable to flow through a fracture if grain diameter is greater than aperture or if bridging occurs. It is generally accepted that proppant bridging occurs when E/d (aperture divided by grain diameter) falls below a certain factor. However, there is disagreement about the value of the factor. Gruesbeck and Collins (1982) performed experiments on proppant bridging across perforations. They found that bridging occurs if perforation diameter is less than about six particle diameters. Based on these results, it is often assumed that bridging occurs at values of E/d in the range of three to six (Dontsov and Peirce, 2014; 2015; Smith and Montgomery, 2015). On the other hand, Barree and Conway (2001) performed proppant bridging experiments for flow through slots, rather than perforations. They found that bridging did not occur until the particle size was close to the slot aperture and recommend using a bridging factor of E/d close to one.

To avoid numerical problems, it is useful to implement the bridging transition over a range of aperture values. This can be accomplished with a blocking function $\chi$, equal to 1.0 if proppant can flow, 0.0 if it is immobile due to bridging, and between 0 and 1 for values of E/d near the bridging factor.

The blocking function is defined as:

$$\chi = 1 if \frac{E}{d} > \left(\frac{E}{d}\right)_{upper} \qquad \text{8-1}$$

$$\chi = 0 if \frac{E}{d} < \left(\frac{E}{d}\right)_{lower}$$ 8-2

$$\chi = \frac{\frac{E}{d} - \left(\frac{E}{d}\right)_{lower}}{\left(\frac{E}{d}\right)_{upper} - \left(\frac{E}{d}\right)_{lower}} if \left(\frac{E}{d}\right)_{lower} < \frac{E}{d} < \left(\frac{E}{d}\right)_{upper}$$ 8-3

The values of $\left(\frac{E}{d}\right)_{lower}$ and $\left(\frac{E}{d}\right)_{upper}$ are inputs to the simulator specified by the properties "Proppant screenout minimum ratio" and "Proppant screenout maximum ratio" in the "Proppants" panel. We choose to use default values equal to 1.25 and 1.75, respectively. These values are close to the values recommended by Barree and Conway (2001). Their experiments were performed for slot flow and were more representative of the conditions in a fracture than the experiments of Gruesbeck and Collins (1982).

ResFrac also considers the possibility that proppant could bridge off at the perforations. It is commonly assumed in the industry that proppant may bridge out once perforation diameter becomes smaller than six times, or perhaps 8-10 times the proppant diameter. Similar to fracture bridging, ResFrac allows you to specify upper and lower bridging ratios for perforation bridging with the parameters "Proppant screenout minimum ratio for flow through a perforation" and "Proppant screenout maximum ratio for flow through a perforation" in the "Proppants" panel.

## 8.2 Viscous drag

Due to viscous drag, the proppant moves along with the flowing fluid. Proppant tends to accumulate in the center of the aperture (away from the fracture walls), where the velocity is greatest. This allows the proppant to flow at a greater superficial velocity than the carrying fluid (Barree and Conway, 1995; Liu and Sharma, 2005; Dontsov and Peirce, 2014; 2015).

Barree and Conway (1995) performed slot flow experiments with proppant flowing with Newtonian fluid and matched the results with the equation:

$$\frac{Q_{pr}}{Q} = C_{pr}\left(1.27 - \left|\left(C_{pr} - 0.1\right)^{1.5}\right|\right)$$ 8-4

where $Q_{pr}$ is the volumetric flow rate of proppant, and Q is the volumetric flow rate. Equation 8-4 indicates that at volumetric fraction equal to 0.1, the proppant flows at a superficial velocity 27% greater than the bulk fluid.

Equation 8-4 is numerically inconvenient because there is a discontinuity in the derivative at $C_{pr}$ equal to 0.1. A very close match to Equation 8-4 with continuous derivative is possible with a polynomial regression to the equation (Equation 8-4).

At concentrations close to $C_{pr,max}$, the grains physically interfere with each other, causing screenout, and the particle velocity decreases, reaching zero at $C_{pr,max}$. This can be modeled by multiplying $\frac{Q_{pr}}{Q}$ by a factor: $1 - \left(C_{pr}/C_{p,max}\right)^{s}$. The sharpness of the screenout transition can be adjusted by varying s. In the limit of very large s, the transition is abrupt – proppant flows freely until $C_{pr}$ reaches its maximum, and then immediately becomes immobile. A very sharp screenout transition would cause numerical problems and isn't intuitively reasonable. Therefore, we use s equal to 20, which yields a sharp transition starting at around $0.9C_{pr,max}$.

The combined equation is:

$$\frac{Q_{pr}}{Q} = C_{pr}\left(1-\left(\frac{C_{pr}}{C_{pr,max}}\right)^{20}\right)(-3.74578663C_{pr}^4 + 6.18208260C_{pr}^3 - 4.35820344C_{pr}^2 + 0.660118234C_{pr} + 1.23837722) \qquad 8\text{-}5$$

## 8.3 Gravitational settling

Because proppant is (nearly always) denser than the carrying fluid, gravity causes it to settle downward. Stokes' law is the simplest and most commonly-used expression for calculating settling rate. However, it is greatly overestimates settling rate in low viscosity fluids. Ferguson and Church (2006) developed an equation for isolated particle settling velocity, $V_{t,\infty}$, that is valid over the full range of practical values for grain size, density, and viscosity:

$$V_{t,\infty} = \frac{Rgd^2}{\frac{18\mu}{\rho_f}+\sqrt{0.75Rgd^3}}, \qquad 8\text{-}6$$

where R is equal to $(\rho_{pr} - \rho_f)/\rho_f$, $\rho_{pr}$ is the density of the grains, and $\rho_f$ is the density of the fluid. The ∞ subscript indicates that the equation is valid for isolated particles. Particle shape has an effect on coefficients in Equation 8-6. Equation 8-6 uses the coefficients recommended by Ferguson and Church (2006) for natural sand grains.

Equation 8-6 is not applicable for non-Newtonian fluids. Chien (1994) developed a general equation for settling velocity that can be used for non-Newtonian fluids and at all practical values of grain size, density, and viscosity:

$$V_{t,\infty} = 12\left(\frac{\mu_a}{d\rho_f}\right)\left[\sqrt{1+7.27dR\left(\frac{d\rho_f}{10\mu_a}\right)^2}-1\right]. \qquad 8\text{-}7$$

where apparent viscosity is calculated from a fluid model such as the modified power law.

For Newtonian fluids, Equation 8-7 yields predictions within 5% of Equation 8-6. For non-Newtonian fluids, Equation 8-7 is implicit because the apparent viscosity must be calculated from the settling shear rate. The settling shear rate can be calculated as (Chien, 1994):

$$\dot{\gamma}_S = \frac{V_t}{d}. \qquad 8\text{-}8$$

Equation 8-8 assumes fluid shear around the particle is only caused by gravitational settling. This is a simplification because fluid flow through the fracture induces shear in the flowing fluid (Novotny, 1977). Nevertheless, the effect of horizontal flow on particle shear rate is nearly always neglected in fracturing simulators. We believe this is justified for the following reasons. In Newtonian fluids, viscosity is not a function of shear rate and so the effect of flow rate on viscosity is zero. In shear thinning fluids, the particles tend to migrate to the center of the fracture, away from the walls (Tehrani, 1996; Lecampion and Garagash, 2014), where the shear rate is lowest, much lower than near the walls (Novotny, 1977). Therefore, in shear thinning fluid, the effect of convective flow on shear rate around flowing particles is relatively low. In either case, the horizontal velocity of the flowing particles should not have a strong effect on $\dot{\gamma}_S$.

In viscoelastic fluids such as HVFR (high viscosity friction reducer), these equations may overestimate settling velocity. Elastic properties of the polymer molecules impart additional drag (Murch et al., 2017). While this is not typically used, the user has the option to use an adjustment factor to modify the particle settling velocity.

## 8.4 Hindered settling

At high proppant volume fraction, settling is hindered due to hydrodynamic interaction between particles. Richardson and Zaki (1954) discovered that hindered settling is equivalent to the process of particle fluidization. Fluidization occurs when fluid flows upward through a bed of unconsolidated particles. At a threshold fluidization velocity, the particles are lifted by the flow, causing expansion of the bed and a drop in the solid volume fraction. Richardson and Zaki (1954) found that correlations designed to describe fluidization are also valid for hindered settling. Fluidization is an important process in reaction beds used in chemical engineering, and so it is possible to draw on a large and rich literature describing this process.

A large number of correlations are available in the literature to describe fluidization/hindered settling. Chhabra (2007) performed a critical review of these correlations, comparing correlations and experimental data from a variety of sources. Chhabra (2007) recommended the correlation from Garside and Al-Dibouni (1977):

$$V_{t,hind} = V_{t,\infty} \frac{(1-C_{pr})^Z}{1+2.35\left(\frac{d}{D_t}\right)}, \tag{8-9}$$

where $D_t$ is the diameter of the tube (the correlation was developed for settling in a cylinder) and Z is equal to:

$$Z = \frac{5.09+0.2839 Re_t^{0.877}}{1+0.104 Re_t^{0.87}}, \tag{8-10}$$

where $Re_t$ is equal to:

$$Re_t(V_{t,hind}) = \frac{\rho_f V_{t,hind} d}{\mu}. \tag{8-11}$$

The $1 + 2.35\left(\frac{d}{D_t}\right)$ term is effectively a 'wall effect' adjustment on settling velocity. $D_t$ is set to the 'hydraulic diameter' for slot flow, which is two times the aperture. Thus, if the aperture is 4x the particle diameter, then the settling velocity will be reduced by a factor of 1/(1 + 2.35*1/8) = 1.3 (ie, reduced by 23%).

For non-Newtonian fluids, Chhabra (2007) found that Equations 8-9 and 8-10 can still be used, as long as Equation 8-11 is replaced with the relevant Reynolds number. For power law fluid, the settling Reynolds number is:

$$Re_t'(V_{t,hind}) = \frac{\rho V_{t,hind}^{2-n} d^n}{K}. \tag{8-12}$$

The Reynolds number for a Modified Power Law fluid is:

$$Re_{MPL}'(V_{t,hind}) = \frac{\rho V_{t,hind} d}{\mu_0}\left(1 + \frac{\mu_0}{K}\left(\frac{V_{t,hind}}{d}\right)^{1-n}\right). \tag{8-13}$$

Similar to the equation for viscous drag (Equation 8-5), Equation 8-13 can be multiplied by $1 - \left(C_p/C_{p,max}\right)^s$,

with s equal to 20, to enforce a sharp drop in settling velocity to zero at $C_{p,max}$. This jamming adjustment is simultaneously applied to both the settling velocity and the horizontal proppant velocity, and so does not greatly extend the distance that proppant can flow before settling into the bed.

## 8.5 Clustered settling

In quiescent, shear thinning fluid, particle grains tend to cluster together, agglomerating into effectively larger particles, which accelerates settling (Clark et al., 1977; Kirkby and Rockefeller, 1985; McMechan and Shah, 2001; Daugan et al., 2004; Liu and Sharma, 2005; Mora et al., 2005). Clustered settling only occurs in quiescent fluid (Liu and Sharma, 2005) and shear thinning fluid (Kirkby and Rockefeller, 1985; Daugan et al., 2004). Clustered settling occurs because of shear thinning in the wake behind settling particles (Daugan et al., 2004; Mora et al., 2005).

We have been unable to find a well-validated general-purpose equation from the literature for predicting clustered settling. Therefore, in this section, we propose an equation that qualitatively and quantitatively matches observations in the literature.

Clustered settling requires time to initiate, as particles cluster into columns. However, these clusters are able to form over settling distances measured at the lab scale (Kirkby and Rockefeller, 1985; Daugan et al., 2004). Therefore, at the field scale, where settling distances are much greater, it is acceptable to assume that clustered settling begins effectively instantaneously, and it is reasonable to use an equation for clustered settling that is not time-dependent.

Figures 10 and 11 in the paper by Kirkby and Rockefeller (1985) show settling velocity as a function of proppant concentration for several different fluids and proppants. Figure 7 in the paper by Kirkby and Rockefeller (1985) provides the fluid rheological information. The Xanthan gum solutions followed power law behavior with n equal to around 0.25. The HPG solutions followed Newtonian behavior at low shear rates and power law fluid at high shear rates with n equal to around 0.5. The Xanthan solutions exhibited power law behavior over the full experimental range (the minimum experimental shear rate is around 0.1 $s^{-1}$). The HPG solutions transitioned towards a low shear rate plateau at around 1 $s^{-1}$.

The following generalizations can be made from the Kirkby and Rockefeller (1985) results: fluids with lower n show greater tendency for clustered settling; clustered settling has a relatively stronger effect for smaller proppant; the peak clustered settling velocity occurs at volume fraction around 0.1; settling velocity decreases as volume fraction increased above 0.1; and settling velocity is lower than the isolated particle settling velocity at volume fractions greater than around 0.3. Kirkby and Rockefeller (1985) also performed experiments with a Newtonian glycerol solution. The Newtonian glycerol solution had monotonically decreasing settling velocity as a function of proppant concentration, consistent with conventional hindered settling in a Newtonian fluid. This confirms that that clustered settling only occurs in shear thinning fluids. For 45 mesh particles, HPG peak settling velocity was around 10 times larger than isolated settling velocity. In Xanthan, it was 20 times larger than isolated settling velocity. For 20 mesh particles, HPG peak settling velocity was around 5 times larger than isolated settling velocity, and for Xanthan, peak settling velocity was l0 times larger than isolated settling velocity. In contrast, for Newtonian hindered settling, settling velocity should have been reduced by about 50% from isolated settling velocity at a volume fraction of 0.1. The low concentration settling velocities in the HPG solutions were consistent with calculated settling velocities from Equations 8-6 or 8-7 using the zero shear rate viscosity.

Based on these results, we propose to model clustered settling by multiplying settling velocity by the following empirical adjustment factor:

$$V_{t,adj,clust} = 1 + \frac{0.065}{d}\frac{n+1}{2n}C_{pr}exp\left(-40C_{pr}^{2.2}\right)\frac{exp(-u*50)}{1+0.1\dot{\gamma}_{1/2}\frac{d}{V_{t,\infty}}}, \quad 8\text{-}14$$

where u is the slurry Darcy velocity, $\dot{\gamma}_{1/2}$ is the shear rate at the transition to power law behavior (Section 9), and $\frac{V_{t,\infty}}{d}$ is the shear rate for a settling isolated particle. $V_{t,adj,clust}$ is designed to be multiplied by settling velocity after the hindered settling adjustment.

The $0.1\frac{V_{t,\infty}}{d}$ term is used because if the shear rate is so low that the particles are settling in the zero shear rate plateau, then the settling is effectively Newtonian and clustered settling will not occur. In a general -purpose simulator, elements may contain very tiny concentration of gel. Under these conditions, $\dot{\gamma}_{1/2}$ is very large because the fluid is effectively Newtonian, and clustered settling should not occur. Thus, the $0.1\frac{V_{t,\infty}}{d}$ term enables the equation to smoothly handle situations with all possible concentrations of gel – from very low (effectively Newtonian) to high (modified power law). For a fluid with a non-negligible amount of gel, $0.1\frac{V_{t,\infty}}{d}$ will be significantly smaller than $\dot{\gamma}_{1/2}$ and clustered settling will occur.

Equation 8-14 is equal to 1.0 in the limit of large horizontal velocity, n equal to 1.0 (Newtonian), $C_{pr}$ equal to zero, and $\frac{V_{t,\infty}}{d} \ll \dot{\gamma}_{1/2}$ (low gel concentration). Equation 8-14 assumes that clustered settling ceases to be significant at horizontal velocity around 0.3-0.5 ft/s; this is only a rough estimate. The literature indicates that clustered settling only occurs in quiescent fluid, but we are not aware of experiments that have quantitatively determined the threshold velocity at which clustered settling no longer occurs. The clustering adjustment may scale with particle and fluid density, but there is not sufficient data to include this effect in Equation 8-14.

Figure 7 shows the combined clustered and hindered velocity adjustment from Equations 8-5 and 8-14 for a 20 mesh particle, $\frac{V_{t,\infty}}{d} \gg 0.1\dot{\gamma}_{1/2}$, n equal to 0.25, and at horizontal velocity of 0, 0.0656, 0.164, and 0.492 ft/s. The trend is very similar to shown in Figure 11 from Kirkby and Rockefeller (1985).

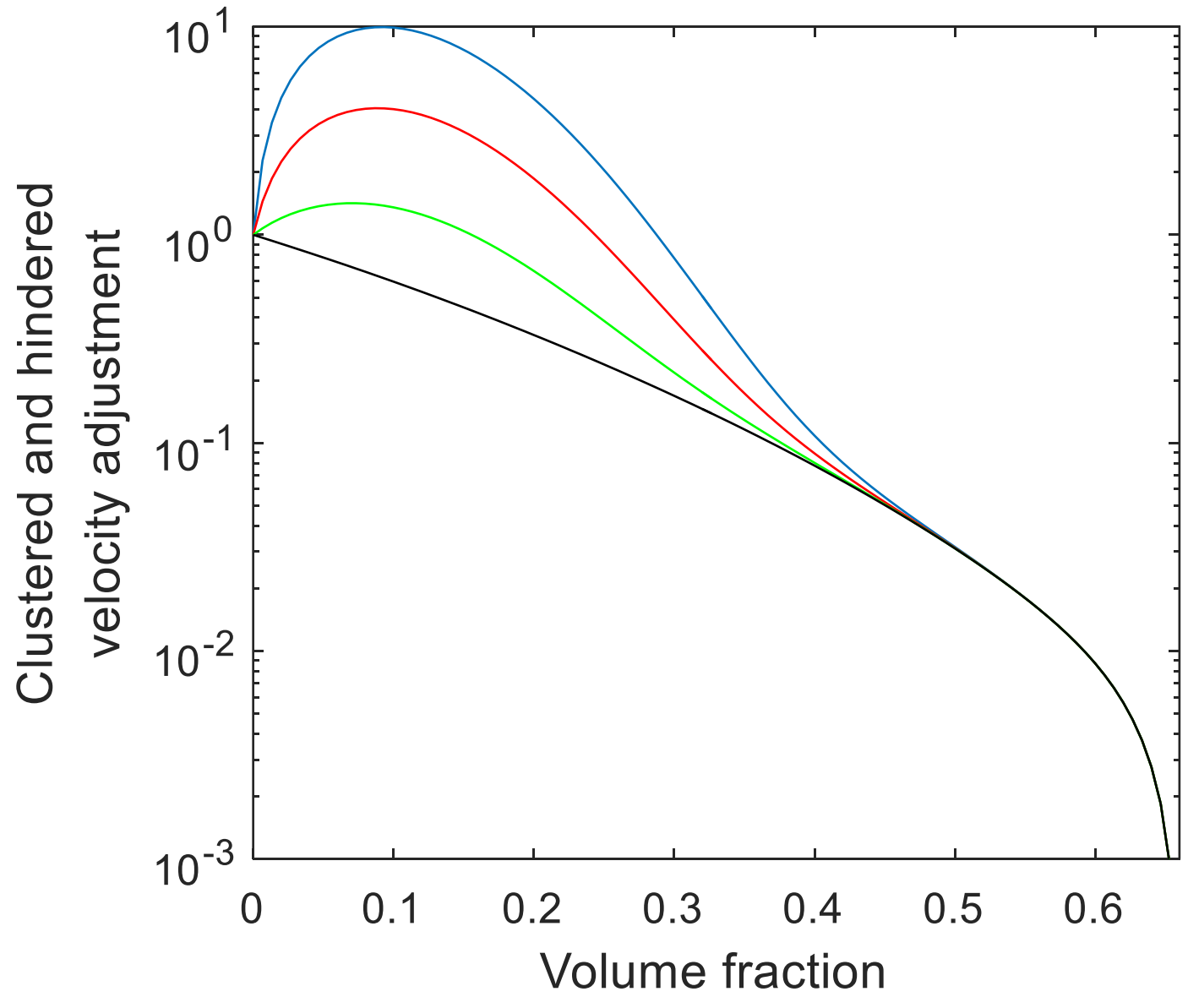

Figure 8: Combined clustered and hindered velocity adjustment from Equations 8.4-1 and 8.5-1 for a 20 mesh particle, $\frac{V_{t,\infty}}{d} \gg 0.1\dot{\gamma}_{1/2}$, and at horizontal velocities of 0, 0.0656, 0.164, and 0.492 ft/s.

## 8.6 Slurry viscosity

The effective viscosity of solid/liquid slurry increases as a function of concentration. A large number of correlations are available in the literature to describe this process. One commonly used expression is Eiler's equation (Keck et al., 1992):

$$\mu_{r,N} = \left\{1 + 1.25\left[\frac{C_{pr}}{1-\frac{C_{pr}}{C_{pr,max}}}\right]\right\}^2, \quad \text{8-15}$$

where $\mu_r$ is the viscosity divided by the viscosity at solute concentration of zero. The subscript N indicates that Equation 8-15 is only applicable for Newtonian fluid.

Keck et al. (1992) developed the following equation for power law slurry:

$$\mu_{r,PL} = \left\{1 + \left[0.75(e^{1.5n} - 1)e^{-(1-n)\dot{\gamma}/1000}\right]\frac{1.25C_{pr}}{1-C_{pr}/C_{pr,max}}\right\}^2. \quad \text{8-16}$$

Equation 8-16 is designed so that the shear rate $\dot{\gamma}$ should be calculated from the Newtonian shear rate, equal to $12\,Q/(E^2W)$ for slot flow. Equations 8-15 and 8-16 predict that the slurry viscosity goes to infinity as the proppant volume concentration approaches $C_{pr,max}$, reflecting the immobilization of the proppant slurry as it jams into a packed bed.

There is not a correlation available in the literature for the effective viscosity of proppant slurry in modified power law fluid. A pragmatic choice is to use a weighted average between the values in Equation 8-15 and 8-16 based on $\dot{\gamma}$ and $\dot{\gamma}_{\frac{1}{2}}$. A sigmoidal averaging function h can be defined as:

$$h(f(x), g(x), x, x_t) = \frac{f(x)}{1+\frac{x}{x_t}} + \left(1 - \frac{1}{1+\frac{x}{x_t}}\right)g(x), \quad \text{8-17}$$

where f(x) and g(x) are the functions being averaged, x is the value input to the functions, and $x_t$ is the transition value. For $x \ll x_t$, h is equal to f(x), and for $x \gg x_t$, h is equal to g(x). Equation 8-17 performs an interpolation scaled over the logarithm of the function values. The transition between the functions begins at about $0.1x_t$ and is nearly finished at around $10x_t$.

Therefore, the proppant concentration viscosity adjustment of a modified power law fluid can be approximated as:

$$\mu_{r,MPL} = h\left(\mu_{r,N}(\dot{\gamma}), \mu_{r,PL}(\dot{\gamma}), \dot{\gamma}, \dot{\gamma}_{\frac{1}{2}}\right). \quad \text{8-18}$$

## 8.7 Multiphase flow

Fracturing is most often performed with a single phase liquid (usually water-based). However, gas/liquid flow can occur during pumping of energized fracturing fluids and other unconventional fracturing fluids (Jacobs, 2014; Ribeiro et al., 2015). Fracture multiphase fluid flow with proppant transport reaches the limits of the literature, and so it is only practically possible to implement simple first-order approximations. In this section, we briefly review the literature and propose simple and pragmatic approaches.

Hindered particle settling in gas/liquid mixtures is similar to the process of bed fluidization by gas/liquid mixtures (Chapter 8 from Chhabra, 2007). In chemical process reactors using multiphase bed fluidization, fluidized columns are operated with either the liquid or gas as the continuous phase and in either concurrent or countercurrent flow (Muroyama and Fan, 1985). As with multiphase flow in pipes, different flow patterns may be apparent: coalesced bubble flow, dispersed bubble flow, slug flow, etc.

In fluidized bed reactors with a continuous liquid phase, the particles are contained in and supported by the liquid phase. If gas flows upward with the liquid, the bed can be fluidized at a lower liquid superficial velocity, and greater bed expansion occurs at the same liquid superficial velocity. The rapidly flowing gas imparts kinetic energy to the liquid phase and increases the in-situ liquid phase velocity, which creates additional lift (Zhang et al., 1998). The reduction in fluidization velocity is less significant for higher liquid viscosity and for shear thinning fluids (Miura and Kawase, 1997; 1998). For particles with diameter less than a few mm (typical of proppants used during fracturing), the flowing gas phase can actually decrease the bed expansion during fluidization. Larger particles tend to break up gas bubbles. Smaller particles do not break up bubbles, and so they tend to aggregate (Muroyama and Fan, 1985).

These processes may be less significant during flow through a fracture, compared to flow through a packed bed reactor. Viscous forces are relatively stronger during flow in a fracture because of the close proximity of the fracture walls (because Reynolds number scales with aperture). On the other hand, Pan (1999) and Chen and Horne (2006) showed that complex flow patterns can emerge during multiphase flow in fractures, depending on viscous, inertial, capillary, and gravitational forces. These are very complex processes, and significant research efforts would be necessary to develop and validate relations that can be used in a general-purpose hydraulic fracturing simulator.

The simplest treatment of particle settling in multiphase flow is to calculate the settling velocity with the mass fraction weighted average of the phase viscosities. This method is used in homogenous pipe flow models of multiphase flow (Cicchitti et al., 1959; Hasan and Kabir, 2002).

Estimating the shear rate experienced by a non-Newtonian aqueous phase during multiphase flow and particle settling in a fracture would be very challenging and depend on flow pattern. We have chosen to calculate shear rate by scaling the phase superficial velocity by the square root of the product of saturation and relative permeability, following the scaling of the Cannella et al. (1988) equation for multiphase flow in porous media (Equation 5-22; Sharma et al., 2011).

For the effect of proppant concentration on viscosity, we have chosen to apply the viscosity adjustment factors summarized in Section 8.6 to each phase separately and assume the particle volume fraction is the same in each phase. In reality, the solid particles may tend to concentrate into the wetting phase, but there is no literature available investigating this potential effect.

The equations for hindered settling are not functions of fluid properties, and so the simplest treatment is to leave them unchanged. The discussion of fluidization above indicates that hindered settling is not unaffected by multiphase flow. However, because the gas phase can cause either an increase or decrease in the bed expansion, there is not a simple way to even qualitatively include these phenomena in a general-purpose simulator.

## 8.8 Gravitational convection

Fluid convection driven by density differences accelerates downward movement of proppant (Clifton and Wang,

1988; Cleary and Fonseca, 1992; Barree and Conway, 1995; Unwin and Hammond, 1995; Hammond, 1995; Clark and Zhu (1996); Shah and Asadi, 1998; Mobbs and Hammond, 2001; Clark, 2006; Shokir and Al-Quraishi, 2009). Slurry mixture density is significantly increased by the presence of proppant. When proppant-laden slurry enters a fracture that is initially filled only with water, the denser slurry tends to gravitationally convect downward, which increases the overall downward particle velocity. Clark and Zhu (1996) and Clark (2006) demonstrated that the role of convection can be assessed with a dimensionless number. For Newtonian fluids, the dimensionless number is:

$$N_{gc} = \frac{12Q\mu}{gE^3(\rho_{pr} - \rho_f)h_f}, \qquad 8\text{-}19$$

where $h_f$ is fracture height. If $N_{gc}$ is less than one, gravitational slurry convection is significant.

This process is included ResFrac through the inclusion of proppant in the calculation of phase potential (Equation 5-13).

## 8.9 Bed load transport

Some investigators have claimed that bed load transport is a dominant process for proppant transport in slickwater fracturing. Patankar et al. (2002) and Wang et al. (2003) performed laboratory scale experiments of proppant transport in a slot and used them to derive bed load transport correlations. These correlations have been applied directly to field scale hydraulic fracturing (Woodworth and Miskimins, 2007; Weng et al., 2011; Shiozawa and McClure, 2016b). Mack et al. (2014) proposed to evaluate proppants on the basis of their bed load transport properties and perform hydraulic fracturing design on the basis of correlations for bed load transport.

In contrast, Biot and Medlin (1985) stated that "sand transport in the bed load does not scale up with fracture height … bed load transport is a significant factor in laboratory-scale experiments but not on the scale of field treatments." In other words, the rate of bed load proppant transport is the same at the lab scale and at the field scale. In the lab, when a small amount of proppant is circulated through a slot, bed load transport can have a dominant effect. But in the field, when the volumes of proppant are much larger, bed load transport has a minimal effect. The distance that proppant can be transported in viscous drag before settling to the bottom scales directly with the size of the fracture. Therefore, settling distances in small-scale laboratory experiments are small, but in the field, viscous drag can transport proppant significant distances, even in slickwater.

McClure (2018) reviewed the discussion in the literature and applied a variety of literature correlations to published laboratory data on slot flow from Patankar et al. (2002) and Medlin et al. (1985). McClure (2018) concluded that bed load transport is not a significant transport mechanism at the field scale. McClure (2018) reviewed transport correlations from the civil engineering and sedimentary geology literature on the rate of sand transport along flowing rivers.

In fracture elements that are located at the top of the proppant bed in the fracture, the element is submeshed vertically in order to accurately calculate the percentage of proppant in the element that has settled into an immobile bed at the bottom of the element. This submeshing calculation avoids discretization dependence that can lead to overestimation of proppant transport along the top row of fracture elements in the proppant bed.

For bed load transport, ResFrac implements the Wiberg and Smith (1989) correlation.

$$Q_s = W\left(d\sqrt{gdR}\right)\left(9.64N_{Sh}^{0.166}\right)\left(N_{sh} - N_{Sh,c}\right)^{3/2} \qquad 8\text{-}20$$

where $Q_s$ is the solid volumetric flow rate, W is the width available for flow (equal to the aperture in the case of a fracture), $N_{Sh,c}$ is the critical Shields number for the onset of bed load transport (we use the recommended value of 0.047).

## 8.10 Bed slumping

The angle of repose, $\theta_r$, is defined as the steepest angle that can be supported by a pile of particles without slumping. Some authors have suggested bed slumping may be an important process during hydraulic fracturing (Sahai et al., 2014; Mack et al., 2014). A typical angle of repose for sand is 33° (Bolton, 1986). Mack et al. (2014) describes proppants designed to have a lower angle of repose, as low as 23°.

ResFrac uses a simplified treatment of bed slumping. The code identifies whether or not the proppant in each element is 'supported' from below. Supported elements are either at the bottom of the fracture or have an underlying element that is entirely full of proppant or closed. In elements that are 'supported,' the calculation in the previous section is used to determine the amount of proppant in the element that has settled into an immobile bed. If there is an element with 'immobile bed' horizontally adjacent to an 'unsupported' element, then proppant is allowed to slump out of the immobile bed into the adjacent unsupported element. Effectively, this makes the angle of repose mesh dependent. If fracture element height equals length, the angle of repose is effectively 45°. If fracture element length is double height, the angle of repose is effectively 26.5°.

## 8.11 Wellbore proppant dynamics

Based on computational fluid dynamics simulations, Wu et al. (2017) argued that proppant may have difficulty 'turning the corner' into perforations as it flows at high velocity down the wellbore. To include this effect in ResFrac, a proppant holdup factor is used to calculate proppant flowrate when proppant is flowing out of the well into the formation. The mass rate of proppant transport is multiplied by an adjustment factor:

$$m_{p,adj} = exp\left(\frac{-v}{v_{pt}}\right), \quad \text{8-21}$$

where v is the flow velocity in the wellbore and $v_{pt}$ is a scaling velocity inputted by the user. The default value of $v_{pt}$ in ResFrac is a very high value so that this adjustment does not have an effect. This behavior is controlled by the parameter "Proppant outflow from well turning scaling velocity" in the "Proppants" panel.

As discussed by Weddle et al. (2018), proppant can settle into an immobile bed at the bottom of a horizontal wellbore, especially nearer to the toe in each stage, where fluid velocity is lowest. As reviewed by McClure (2018), there is a robust literature describing the settling of particles suspended in a slurry during pipe flow. Correlations from this literature are implemented in ResFrac. In each wellbore element, ResFrac tracks the fraction of each proppant type that has settled into an immobile bed at the bottom of the well. Proppant that has settled into an immobile bed is not permitted to flow. Proppant settling is turned on using the parameter "Include proppant settling in laterals" in the "Proppants" panel.

ResFrac only permits proppant settling in laterals that have dip less than or equal to 30° (horizontal wells have dip of 0°).

The correlation used for predicting proppant settling has been updated from an earlier version. Older ResFrac versions used correlations from Wicks (1968), Thomas (1962), and Wasp et al. (1977). Now, ResFrac uses the correlation from Oroskar and Turian (1980). Literature review indicates that the Oroskar and Turian (1980) correlation is believed to be the most reliable, and it is the most widely used today in a diverse set of engineering applications. The correlation is used to calculate $v_D$, the deposition velocity. If the flow rate (above the immobile proppant bed) is less than the deposition velocity, then proppant will begin to deposit into an immobile bed along the bottom of the pipe. The velocity for homogeneous flow – at which proppant should be entirely suspended through the pipe with no perceptible vertical concentration gradient – is assumed to be double the deposition velocity.

At velocities between $v_D$ and $v_D + 0.5(v_D + v_{c,homogeneous})$, both the settling rate and the erosion rate are assumed to be zero. At velocities between $v_D + 0.5(v_D + v_{c,homogeneous})$ and $v_{c,homogeneous}$, erosion rate is linearly interpolated between 0 at $v_D + 0.5(v_D + v_{c,homogeneous})$ and 100% per second at $v_{c,homogeneous}$.

The equations for settling and erosion are applied separately for each proppant type.

## 8.12 Updated wellbore proppant dynamics

An updated version of proppant dynamics in the wellbore is introduced in 2024. The model is based on the recent paper Dontsov (2023b), which synthesized the correlation for proppant transport and distribution between perforations based on the available computational and experimental data Ahmad and Miskimins (2019a), Ahmad and Miskimins (2019b), Liu et. al (2021), Snider et. Al (2022), Wu and Sharma (2016). The amount of proppant that flows into perforations depends on three primary physical phenomena: (a) the distribution of slurry into each shot (controlled by limited entry), (b) particle settling in the wellbore (noticeable at low rates), and (c) proppant turning efficiency (caused by particle inertia). The combination of these phenomena can be used to perform phasing optimization, see e.g. Dontsov et. al (2023), which would promote more uniform distribution of proppant.

The distribution of slurry between perforations is controlled by limited entry and perforation cluster design. While this is crucial for the final proppant distribution, we also need to consider particle turning that is characterized by ‘turning efficiency’ and is related to the percentage of proppant that is able to complete the turn into a perforation. This efficiency primarily depends on particle size, and typically varies from 0.7 for bigger particles to 0.9 for smaller particles, see Dontsov (2023b) for more details.

The flow rate in the wellbore decreases from heel to toe due to slurry outflow through perforations. Consequently, the ability of fluid to suspend particles reduces and toe clusters tend to be more affected by particle settling than the heel clusters.

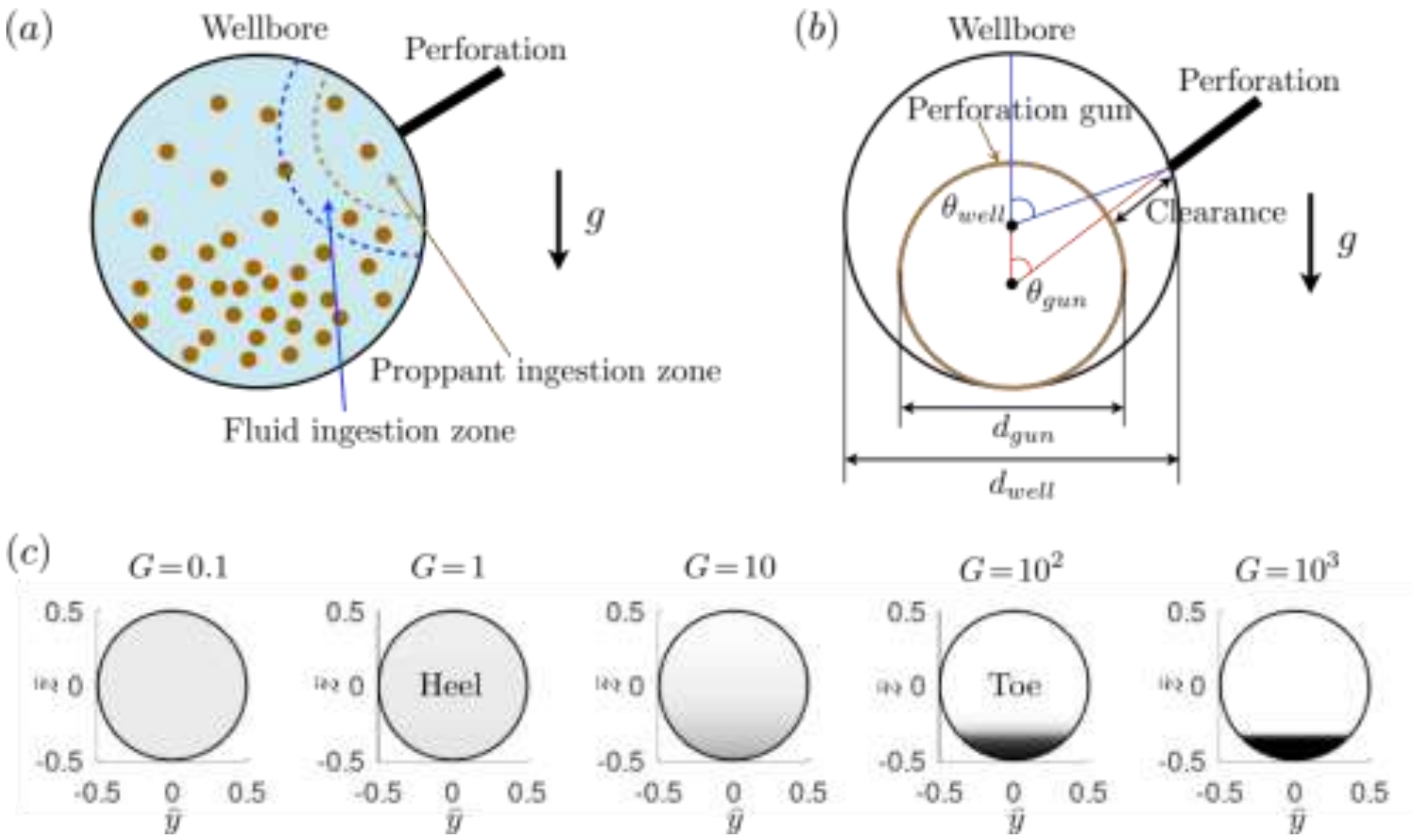

*Figure 9: (a) Schematics of proppant distribution in the wellbore cross-section. (b) Illustration of perforation gun clearance. (c) Calculation of proppant distribution for different values of the dimensionless gravity G.*

To illustrate the effect of particle settling, Figure 9 shows a wellbore cross-section and schematically shows non-uniform particle distribution. The fluid and proppant 'ingestion zones' denote the regions in the well that end up flowing into a particular perforation shot. In the limiting case of a uniform velocity distribution within the well's cross-section, the area of the fluid ingestion zone relative to the area of the total wellbore's cross-section is simply equal to the slurry flow rate through the perforation divided by the total flow rate through the wellbore. Note that for typical parameters, size of the ingestion zone is relatively small. As a hypothetical example, let's assume 10 clusters with 3 perforations and 90 bpm injection rate that is distributed uniformly between the shots. This corresponds to 3 bpm per shot. For the heel-most perforation, the ratio between the flow rate per shot to total flow rate is 3/90 = 0.033. This ratio then increases slightly along the stage, but is still small. In the middle of the stage, the ratio is 3/45 = 0.067. Right before the last cluster, this ratio is 3/9 = 0.33. Comparing proppant ingestion and fluid ingestion, the area associated with proppant ingestion is smaller by an amount determined by the turning efficiency that is discussed earlier. As can be seen from Figure 9, particle settling starts playing an important role because if the perforation is located at the bottom of the well and the ingestion zone is small, then this perforation takes more proppant due to higher local concentration of particles. A similar process, albeit with an opposite outcome, applies for perforations located at the top.

To quantify particle settling, we define a parameter *G*, for dimensionless gravity. This parameter defines particle volume fraction distribution in the well, see figure 9. In S.I. units, the parameter *G* is defined as:

$$G = \frac{8\phi_m(\rho_p-\rho_f)g cos(\theta_w) d_{well}}{f_D \rho_f v_w^2}, \qquad 8\text{-}22$$

where $\phi_m = 0.585$ is the maximum flowing volume fraction of particles, $\rho_p$ is particle mass density, $\rho_f$ is fluid mass density, $g = 9.81\, m/s^2$ is gravitational constant, $d_{well}$ is wellbore diameter, $f_D = 0.04$ is a fitting parameter that can also be interpreted as a friction factor in the pipe, and $v_w$ is the average wellbore velocity or the total wellbore flow rate divided by the cross-sectional area. To account for dipping wells, the dip angle $\theta_w$ is introduced that simply reduces the gravitational constant. The only parameter that varies along the stage is the average wellbore velocity. At the stage heel, the injection rate is high, the velocity is high, and therefore the parameter $G$ is low (on the order of 1 for typical parameters). This leads to nearly uniform proppant distribution. At the same time, velocity is much lower for toe clusters and the dimensionless gravity $G$ can exceed 100. This results in significantly asymmetric particle distribution, in which particle flow resembles a 'flowing bed' state, as indicated in figure 9.

Figure 9 also illustrates how the perforation gun is often noticeably smaller than the inner wellbore diameter. This creates a gap or clearance between the gun and the wellbore. Consequently, the initial perforation diameter can vary with respect to the clearance or phasing of perforations. The relations between the gun-centered and well-centered azimuths can be determined from geometry as:

$$\theta_{well} = \theta_{gun} + sin^{-1}\left(sin\left(\theta_{gun}\right)\right), \theta_{gun} = tan^{-1}\left(\left(r + cos(\theta_{well})\right)^{-1} sin(\theta_{well})\right), r = 1 - \frac{d_{gun}}{d_{well}}. \quad 8\text{-}23$$

Figure 10 shows the relation that is implemented to adjust perforation diameter as a function of gun clearance Bell et al. (1995).

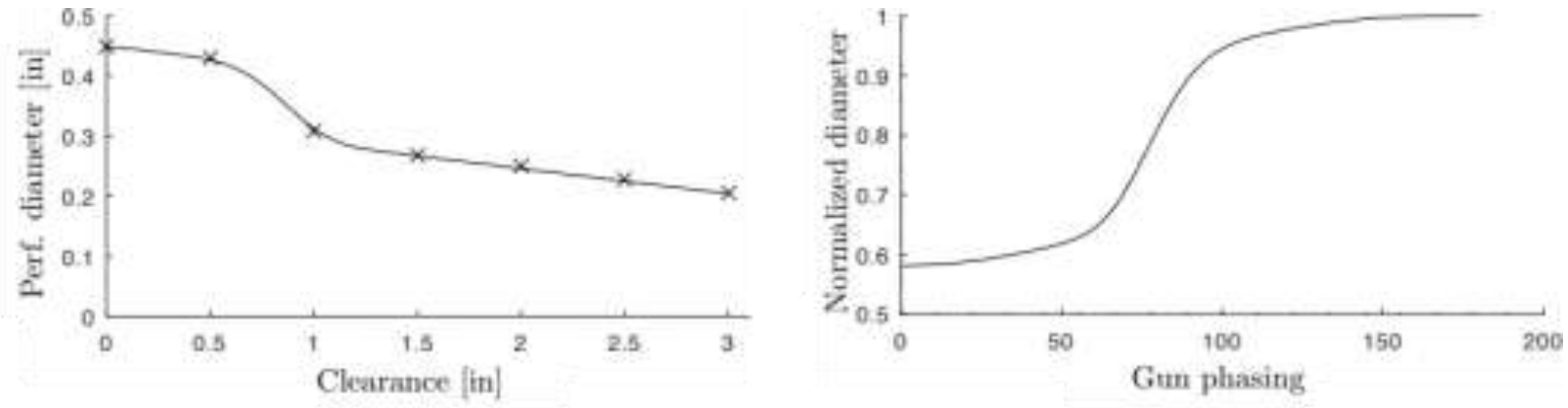

*Figure 10: Variation of perforation diameter versus gun clearance.*

Perforation erosion has a major effect on proppant distribution because it causes the slurry flow distribution to evolve over time. It is usually observed that heel-side perforations erode preferentially in the downstream direction. To capture this behavior, we introduce $D_u$ as the effective diameter in the upstream direction, $D_s$ as the diameter in the direction perpendicular to fluid flow, and $D_d$ as the downstream diameter, see also Dontsov et. al (2024). It is also useful to introduce equivalent diameter as

$$D_{eq} = \sqrt{\frac{4A}{\pi}}, A = \frac{\pi}{8} D_s(D_u + D_d), \quad 8\text{-}24$$

which essentially represents area of the perforation. The erosion model is the modification of Long et al. (2017) erosion model and reads:

$$\frac{dD_{cir}}{dt} = \alpha C v_p^2, \frac{dD_{ax}}{dt} = \alpha C\left(v_p^2 + \frac{1}{2}\gamma v_w^2\right), \frac{dC_d}{dt} = \beta C v_p^2\left(1 - \frac{C_d}{C_{d,max}}\right). \quad 8\text{-}25$$

Here $\gamma$ is a new parameter that is related to erosion associated with lateral fluid velocity, $C_d$ is the coefficient of discharge, $C_{d,max}$ is the maximum possible value of coefficient of discharge, $C$ is proppant concentration, $v_p$ is the velocity through the perforation, $v_w$ is the average fluid velocity is the wellbore, and $t$ is time.

A cartoon of an eroded perforation is shown in figure 11. The perf shape is a combination of a circular part in the upstream direction and an elliptical part in the downstream direction (caused by the term proportional to $\gamma$).

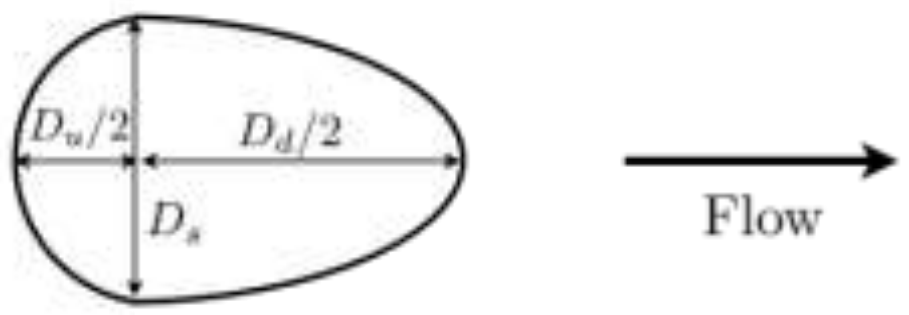

*Figure 11: Schematics of an eroded perforation.*

We use $\alpha = 3x10^{-13} m^2\, s/kg\,, \beta = 10^{-9}\, ms/kg\,, \gamma = 100$ for default values of the erosion parameters.

Perforation erosion, combined with slurry distribution controlled by limiting entry, turning efficiency, and particle settling primarily determine proppant distribution between perforation clusters. However, there are a few other corrections that are also important. First of all, complex fluid rheology affects the level of particle suspension. To provide a pragmatic way of modeling this, we introduced a suspension multiplier, which allows proppant to be better suspended in the fluid. The second important thing is so-called inline correction, which is relevant mostly for oriented perforations. Particles that were unable to complete a turn into first perforation, tend to accumulate in the vicinity of this perforation. And if the following perforation has the same azimuth, then it receives more proppant. The inline correction multiplier allows to change magnitude of this effect. Finally, a strong erosion uncertainty is often observed in field data. To account for such behavior, we have added a spatially-correlated erosion uncertainty for the parameter $\alpha$.

## 8.13 Proppant flowback

Proppant flowback can occur if a sufficiently strong pressure gradient is applied during production. It is mostly likely to occur if there is a thick proppant pack, and the grains can resettle as proppant squeezes out. Note this can occur even if the fracture is mechanically closed onto the proppant! Based on a literature review, the correlations from Canon et al. (2003) appear to be the most realistic. Their correlation has been implemented in ResFrac and can be optionally turned on in the 'advanced' section of the proppants panel. The correlation is used to calculate a pressure gradient (psi/ft) above which flowback can occur. If the actual pressure gradient exceeds that threshold, flowback is allowed to occur. If the correlation predicts that flowback should occur, the blocking parameter $\chi$ is modified. Typically, in a mechanically closed fracture, $\chi$ would be zero. But if flowback is predicted, it is increased above zero to allow proppant to entrain in the flowing fluid. The figure below shows the distribution of proppant and aperture before and after flowback. Proppant flowback is turned on using the parameter "Proppant flowback from closed fractures" in the "Proppants" panel. Garbino et al. (2024) provide practical case studies using the ResFrac proppant flowback capability.

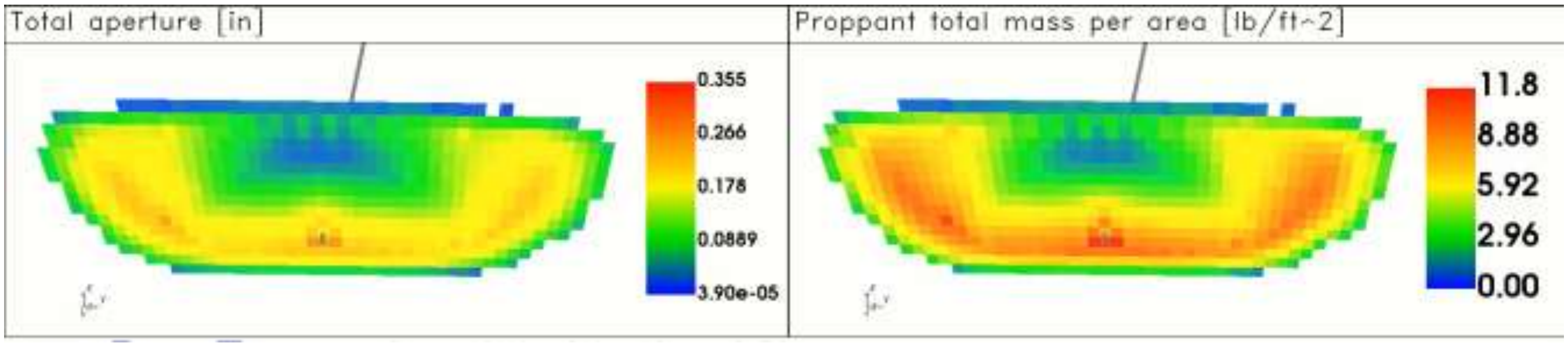

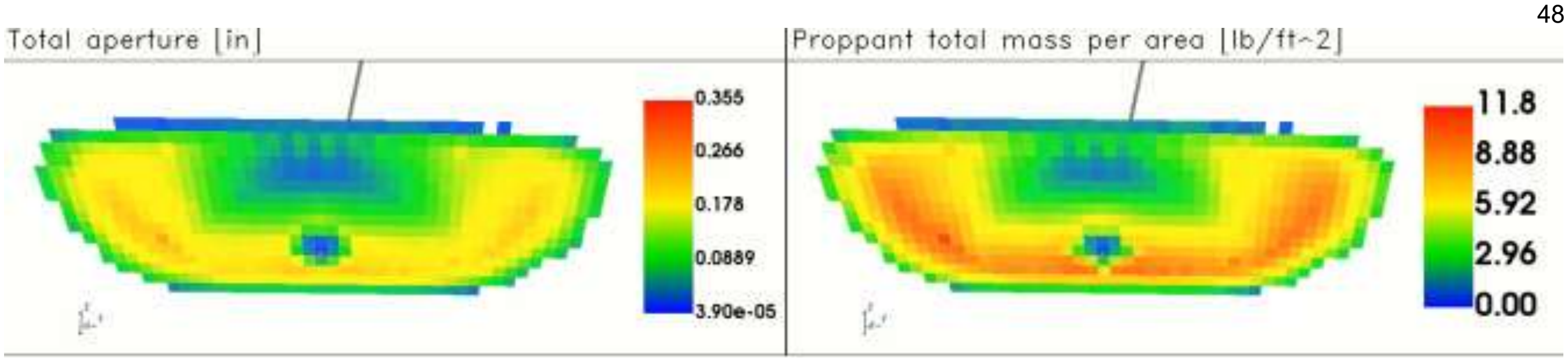

Figure 12: Distribution of proppant and aperture before and after flowback.

## 8.14 Proppant trapping

Fracture closure occurs slowly in unconventionals, giving the proppant a long time to settle. As a result, proppant often settles out all the way to the bottom of the fracture, leaving large areas entirely unpropped. Is this realistic? Proppant definitely settles downward due to gravity, so we should model this effect. But perhaps processes cause some residual trapping of proppant. This could be concentration at points of high leakoff, settling onto ledges or other branch points, etc. Gale et al. (2018) and Maity and Ciezobka (2020) show core-across results from the Wolfcamp shale that appear to demonstrate these effects. Singh et al. (2021) describe this process in laboratory experiments with rough walled fractures.

Radioactive proppant tracers have been reported to show the first proppant injected being observed near the wellbore, suggesting some is trapped, rather than being transported far away. For example, Figure 10 from Weddle et al. (2018) shows three separate proppant tracers injected at the start, middle, and end of the stage are all observed at the well in approximately equal quantities.

To address these issues, some users of conventional frac simulators stop the simulation at shut-in, and assume proppant is 'frozen in place' after shut-in.

We handle this effect by allowing some proppant to be trapped as if flows through the fracture. A certain amount of each proppant type i can be immobilized, or trapped, in the fracture. The change in immobilization is calculated as:

$$\frac{dm_{i,imm}}{dt} = K_{imm}(m_i - m_{i,imm}), \qquad \text{8-26}$$

where:

- $m_{i,imm}$ is mass/area of immobile proppant for proppant type i
- $K_{imm}$ is a rate constant (min^-1)
- $m_i$ is mass/area for proppant type i
- $m_{tot,max}$ is the maximum allowed mass/area for all proppant types

The user specifies two parameters: $m_{tot,max}$ and K. Typical values are K = 0.1 min^-1 and $m_{tot,max}$ and 0.15 lbs/ft^3. By default, $m_{tot,max}$ is set to zero, so that there is not any trapping. The value of $m_{tot,max}$ is specified with the parameter "Maximum immobilized proppant mass per area," and K is specified with the parameter "Proppant immobilization rate," both in the "Proppants" panel.

Running simulations with proppant trapping leads to more proppant being deposited near the wellbore. Proppant can still settle out to the bottom of the fracture, and large regions of the fracture(s) may remain unpropped. But trapping reduces the tendency for extreme settling to the bottom of the fractures and improves proppant placement near the well. This causes more realistic overall results when comparing with field-scale data. With extreme settling, production between clusters becomes extremely erratic unless you set the unpropped fracture conductivity to be a high value, but if you do that, then the simulations (incorrectly in the great majority of cases) will suggest that you can reduce proppant mass without seeing impact on production. Singh et al. (2025) provides practical examples of calibrating the proppant trapping algorithm to field data.

## 8.15 Proppant trapping multipliers by type

'Maximum proppant trapping per area' can be specified globally or customized by layer. In addition, trapping can be customized by proppant type by specifying a trapping 'multiplier' for each type. The multipliers must be less than or equal to 1.0.

The trapping algorithm works as following:

(1) The local value of 'maximum immobilized proppant' is determined from either the specified 'global' value or from the specified value for that particular layer.
(2) If the overall amount of proppant (of all types) trapped in the element exceeds the local value of 'maximum immobilized proppant,' then no additional proppant may be trapped, of any type.
(3) The maximum amount of trapping for each individual type is calculated as the product of its multiplier and the local amount of 'maximum immobilized proppant.' If the amount of trapped proppant of that type reaches the limit, then that type of proppant is not permitted to trap further, even if the overall amount of trapped proppant does not exceed the local maximum.

Example 1:
maximumimmobilizedproppant = 0.2 lbs/ft^2.
Prop1 trapping multiplier is 0.5 and Prop2 trapping multiplier is 0.25.

In this case, no more than 0.5*0.2 = 0.1 lbs/ft^2 of Prop1 could be trapped, and no more than 0.25*0.2 = 0.05 lbs/ft^2 of Prop2 could be trapped. Both types of proppant would be limited by their 'individual proppant' type multipliers, and no element would ever have more than 0.15 lbs/ft^2 of proppant trapped.

Example 2:
maximumimmobilizedproppant = 0.2 lbs/ft^2.
Prop1 trapping multiplier is 0.75 and Prop2 trapping multiplier is 0.75.

In this case, the maximum amount of proppant trapped of Prop1 and 2 is individually 0.15 lbs/ft^2. However, it would not be possible to trap 0.15 lbs/ft^2 of both types of proppant, because that would add up to 0.3 lbs/ft^2, which would exceed the local value of 'maximum immobilized proppant.' Thus, both types of proppant will be able to immobilized until they each individually reach 0.15 lbs/ft^2 or they both collectively reach 0.2 lbs/ft^2.

The process is irreversible and path dependent. For example, if Prop1 was pumped first and Prop2 second, then the Prop1 would trap to 0.15 lbs/ft^2 and Prop2 would then trap to 0.05 lbs/ft^2.

## 8.16 Proppant washout

The proppant washout model simulates the remobilization of immobilized proppant due to high fluid velocity in the fracture. ResFrac implements the washout model developed by Kryvenko (2025). If invoked, this model appends the proppant trapping equation (8-26) with a remobilization term:

$$\frac{dm_{i,imm}}{dt} = K_{imm}\left(m_i - m_{i,imm}\right) - K_{washout}u_D^2 m_{i,imm}, \quad 8\text{-}27$$

where in the remobilization term $K_{washout}$ is the immobilization rate (s$^{-1}$) and dimensionless superficial velocity, $u_D$ is computed as follows:

$$u_D = \frac{u}{u_c}, \quad 8\text{-}28$$

where $u$ is the superficial velocity in the fracture and the $u_c$, the critical washout velocity, is defined as:

$$u_c = \sqrt{\frac{8N_{sh,c}d_p g(\rho_p - \rho_f)}{f\rho_f}},$$

where $N_{sh,c}$ is the critical Shields number, $d_p$is the proppant particle diameter, $g$ is gravitational acceleration, $\rho_p$ is the proppant particle density, $\rho_f$ is the density of the fluid, and $f$ is a friction factor.

# 9. Water solutes

## 9.1 Fluid viscosity

The $N_s$ water solutes are convected within the water phase. The water solutes can be inert tracers, or they can impart non-Newtonian rheological characteristics based on the modified power law. Another type, 'Modified Power Law vT,' specifies the modified power law parameters from laboratory measurements of $K'$ and $n'$ versus aging time and temperature (Section 9.5).

The apparent viscosity of the aqueous phase is modeled with the modified power law model (Capobianchi and Irvine, 1992):

$$\mu_{a,adj} = \frac{\mu_{0,adj}}{1+\frac{\mu_{0,adj}}{K}(\dot{\gamma})^{1-n}} = \frac{\mu_{0,adj}}{1+\left(\frac{\dot{\gamma}}{\dot{\gamma}_{1/2}}\right)^{1-n}}, \quad 9\text{-}1$$

where n is the exponent from the equivalent power-law model, $K$ is the constant from equivalent power law model, $\dot{\gamma}$ is the shear rate, and $\dot{\gamma}_{1/2}$ is the shear rate at the transition from Newtonian to power law behavior. The viscosity adjustment parameter $\mu_{0,adj}$ is multiplied by the viscosity of the pure water phase to calculate viscosity. At low shear rate, $\mu_{0,adj}$ is constant. At high shear rate, the apparent viscosity asymptotically approaches power law behavior parameterized by K and n. The modified power law model is very similar to the Ellis fluid, though not identical. The two models are identical at low and high shear rate, but differ slightly at the transition. The modified power law model is more convenient for most applications because it is defined in terms of shear rate. The Ellis model is defined in terms of shear stress. The "Water solutes" table in the "Water Solutes" panel is used to specify the flow parameters for each type of water solutes.

A formulation is needed to calculate $\mu_0$, $\dot{\gamma}_{1/2}$, and n as a function of water viscosity and the concentration of fluid additives. To accomplish this, the user specifies three parameters for each modified power law water solute: viscosity multiplier per 0.001 mass fraction, $\dot{\gamma}_{1/2}$, and n. If multiple water solutes are present, the mixture values of $\dot{\gamma}_{1/2}$ and n are calculated as the mass fraction weighted average. As an example, if viscosity multiplier per 0.001 is equal to 5, and the mass fraction of the solute is 0.002, then $\mu_{0,adj}$ is equal to 10. $\mu_{a,adj}$ is not permitted to be less than one.

ResFrac can model reactions between water solutes. You define a first-order reaction rate constant $X_{ws1,ws2}$. Water solute 1 converts into water solute 2 at the rate:

$$\frac{1}{\overline{m}}\frac{d\overline{m}}{dt} = X_{ws1,ws2} \qquad 9\text{-}2$$

You can use these reactions to capture processes like gel crosslinking and breaking. For example, you could specify three types of water solutes: 'not-yet' cross-linked gel, cross-linked gel, and broken gel. Then define two water solute reactions: a reaction from 'not-yet' cross-linked to cross-linked, and a reaction from cross-linked to broken. Reactions are specified in the "Reactions between water solutes" table in the "Water Solutes" panel.

## 9.2 Filtercake

In each facies, the user inputs a maximum water solute molar mass permitted to flow into the rock. If the water solute molar mass is greater than the maximum, then water solute is not permitted to flow into the matrix. Instead, it forms a filtercake on the fracture walls. The "Maximum flowing molar mass" is specified in the "Geologic units (facies list)" table in the "Static Model and Initial Conditions" panel. The molar masses of the solutes are defined in the "Water solutes" table in the "Water Solutes" panel.

ResFrac uses a simple model for formation of filtercake. The thickness of the filtercake, $w_{fc}$, is calculated as the cumulative mass of filtercake screened out per area (on each side of the fracture), divided by filtercake density, $\rho_{fc}$. The permeability of the filtercake is assumed to be a constant, $k_{fc}$. The filtercake modifies the effective permeability for flow between fracture element i and adjacent matrix element j. The effective permeability is calculated using the solution for effective permeability for flow in series:

$$\left(k_{eff}\right)_{ij} = \frac{l_j + w_{fc}}{\frac{l_j}{k_j} + \frac{w_{fc}}{k_{fc}}}, \qquad 9\text{-}3$$

where $k_j$ is the permeability of element j and $l_j$ is the distance from the fracture wall to the center of element j. It is assumed that once the filtercake has formed, it does not erode. The "Filtercake permeability" is specified in the "Water Solutes" panel.

Equation 9-3 assumes that the filtercake permeability $k_{fc}$ is a single constant. In practice, different water solutes form filtercake of different permeability. ResFrac allows the user to assign an individual filtercake permeability $k_{fc,i}$ to each water solute $i$ in the "Water Solutes" panel.

The simulator tracks the cumulative mass of each water solute at every fracture-to-matrix and well-to-matrix contact. The filtercake at each connection is treated as a stack of sublayers, one per water solute, with sublayer thickness proportional to mass fraction. The effective filtercake permeability at each contact is calculated as the mass-fraction-weighted harmonic mean:

$$k_{fc,con} = \frac{1}{\sum \frac{f_i}{k_{fc,i}}} \xi_L \qquad 9\text{-}4$$

where $k_{fc,con}$ is the effective filtercake permeability at a single connection; $f_i$ is mass fraction of water solute $i$; $\xi_L$ is a layer-specific multiplier (default 1.0) specified in the "Water Solutes" panel.

In naturally fractured formations, fissures that open at high pressure can bypass the filtercake. If "Apply PDP/PDL multiplier to filter cake permeability" is true for a curve set, $k_{fc,con}$ is multiplied by the pressure-dependent permeability factor of the connection, as described in Section 4.7 (Equation 4-27).

Per-solute values of $k_{\text{fc,i}}$ are rarely known directly. Typically, what is available from core-scale leakoff tests or a history match is the bulk filtercake leakoff coefficient $C_{\text{w}}$ of a reference fluid mixture, along with a relative sense of how each water solute contributes to filtercake permeability. The "Filtercake Permeability Wizard" converts these inputs into the set of $k_{\text{fc,i}}$. ResFrac allows the user to provide $C_{\text{w}}$, the fracture propagation pressure minus pore pressure $\Delta p_{\text{cake}}$, the reference fluid mixture, and a dimensionless relative filtercake permeability multiplier $\eta_i$ for each water solute. The bulk filtercake permeability $k_{\text{fc,mix}}$ of the reference mixture is solved from:

$$C_w = \sqrt{\frac{k_{fc,mix}\alpha\Delta p_{cake}}{2\mu_w}} \qquad 9\text{-}5$$

where $\mu_w$ is the water viscosity; $\alpha = \frac{1}{\sum f_i}$ is a normalization over the reference mixture mass fractions. Each water solute is assigned a filtercake permeability of the form:

$$k_{fc,i} = k_{scaler}\eta_i \qquad 9\text{-}6$$

where $k_{\text{scaler}}$ is a scaling constant; evaluated at the reference mixture composition:

$$k_{scaler} = k_{fc,mix}\sum\frac{f_i}{\eta_i} \qquad 9\text{-}7$$

Combining Equations 9-6 and 9-7 gives the resulting $k_{\text{fc,i}}$ used during simulation. The $k_{\text{fc,i}}$ values populate to "Water Solutes" panel and may be edited manually after the "Filtercake permeability wizard" runs.

## 9.3 Spurt loss of water solutes

Spurt loss of hydraulic fracturing fluid can occur in certain formations, most commonly in conventional reservoirs. During spurt loss, fracturing fluid leaks into the formation without its water solutes being filtered out to form a cake. The formation possesses a finite spurt loss capacity; only after this capacity is fully exhausted does a filtercake begin to form. Consequently, experiencing spurt loss increases both total fluid leakoff and the risk of a screenout.

When specified, each water solute has an inherent spurt loss mass per unit area. Because all water solutes concurrently consume the overall spurt loss capacity of the formation, a dimensionless spurt loss metric ($D_{spurt}$) is utilized to track the cumulative consumption across all solutes. $D_{spurt}$ is calculated as

$$D_{spurt} = \sum_1^N (m_{leakoff} * \frac{m_{solute,i}}{A_{leakoff} * m_{spurt,i} * \lambda}), \qquad 9\text{-}8$$

where $m_{leakoff}$ is the leakoff mass of the fluid mixture; $m_{solute,i}$ and $m_{spurt,i}$ are the mass fraction and spurt loss mass per unit area of water solute $i$, respectively; $N$ is the total number of water solutes; $A_{leakoff}$ is the leakoff area; $\lambda$ is the layer specific spurt loss multiplier.

Spurt loss terminates as soon as the accumulated dimensionless spurt loss reaches 1.0, at which point filtercake begins forming with any subsequent leakoff.

Conventionally, spurt loss severity is defined by the spurt loss volume per unit area of the overall injecting fluid mixture. In some cases, individual water solutes are also assigned a "spurt loss potency." Using the water solute mass fractions, the potencies of each solute, and the fluid mixture's total spurt loss volume per area, the individual spurt loss mass per area for each water solute can be determined as:

$$m_{spurt,i} = V_{leakoff} * \rho_{\text{mixture}} * m_{solute,i} * p_i * \sum_1^N (\frac{1}{p_i}), \quad \text{9-9}$$

where $V_{leakoff}$ is the spurt loss volume per unit area of the fluid mixture; $\rho_{\text{mixture}}$ is the fluid mixture density; $p_i$ is the spurt loss potency of water solute $i$.

Conversely, if the spurt loss mass per area for each individual water solute is known, the overall spurt loss volume per area of a fluid mixture can be back-calculated using the mass fractions of its constituent solutes as

$$V_{leakoff} = \frac{1}{\rho_{\text{mixture}} \sum_1^N (\frac{m_{solute,i}}{m_{spurt,i}})} \quad \text{9-10}$$

## 9.4 Using pipe friction tables for water solutes

For some fluids, the standard Fanning friction factor calculations combined with scalar adjustment factors may not adequately capture the non-linear rheological behavior across varying pipe diameters and injection rates. To address this, the simulator allows users to define custom pipe friction tables and assign them to specific water solutes. When a table is assigned, the frictional pressure drop of the fluid mixture is governed by the tabulated empirical data rather than the built-in Fanning calculations.

A pipe friction table defines the frictional pressure gradient as a function of three independent variables: fluid load (water solute concentration), pipe inner diameter, and volumetric flow rate. To evaluate the pressure gradient for a specific simulated state, the model employs a multi-dimensional interpolation and extrapolation strategy. For a given fluid load and pipe diameter, the pressure gradient is interpolated along the flow rate axis using a shape-preserving spline. To interpolate between tabulated diameters, the model performs linear interpolation with respect to 1/D^5. If the simulated pipe diameter falls outside the tabulated bounds (or if only a single diameter is provided), the model extrapolates the pressure gradient using the inverse fifth-power relationship derived from standard pipe friction scaling as

$$\left(\frac{\Delta P_{fric}}{\Delta L}\right)_{actual} = \left(\frac{\Delta P_{fric}}{\Delta L}\right)_{table} \left(\frac{ID_{table}}{ID_{actual}}\right)^5, \quad \text{9-11}$$

where $\left(\frac{\Delta P_{fric}}{\Delta L}\right)_{actual}$ and $\left(\frac{\Delta P_{fric}}{\Delta L}\right)_{table}$ are the actual and closest table recorded pressure drop gradient due to friction, respectively; $ID_{actual}$ and $ID_{table}$ are the actual and closest table recorded pipe inner diameters, respectively.

For interpolation between two bounding tabulated diameters ($ID_{low}$ and $ID_{high}$), the pressure gradient is calculated as

$$\left(\frac{\Delta P_{fric}}{\Delta L}\right)_{actual} = \left(\frac{\Delta P_{fric}}{\Delta L}\right)_{low} + \left(\left(\frac{\Delta P_{fric}}{\Delta L}\right)_{high} - \left(\frac{\Delta P_{fric}}{\Delta L}\right)_{low}\right) \frac{\left(\frac{1}{ID_{actual}^5} - \frac{1}{ID_{low}^5}\right)}{\left(\frac{1}{ID_{high}^5} - \frac{1}{ID_{low}^5}\right)}, \quad \text{9-12}$$

where $\left(\frac{\Delta P_{fric}}{\Delta L}\right)_{high}$ and $\left(\frac{\Delta P_{fric}}{\Delta L}\right)_{low}$ are the immediate upper and lower bounding table recorded pressure drop gradient due to friction.

The model linearly interpolates the pressure gradient between tabulated fluid loads. If the simulated fluid load exceeds the highest tabulated value, the pressure gradient is clamped to the maximum available fluid load (no upward extrapolation). If the fluid load falls below the lowest tabulated value, a blending fraction is utilized to smoothly transition back to standard water friction calculated by Fanning method. Moreover, since not all water solutes in the fluid mixture may be assigned friction tables, the model also blends friction of all water solutes into a single effective friction factor for the momentum solver.

First, the interpolated pressure gradient from a table is reverse-engineered into an equivalent Fanning friction factor ($f_{table}$) to maintain compatibility with the fundamental momentum equations as

$$f_{table} = \frac{\left(\left(\frac{\Delta P}{\Delta L}\right)_{actual} ID_{actual}\right)}{2\,\rho_{mixture} * v_{true}^2}, \quad \text{9-13}$$

where $v_{true}$ is the true mixture velocity.

To account for fluid mixtures where the mass fraction of a tabulated solute ($c_{acc}$) is lower than the table's minimum defined concentration ($c_{min}$), a transition multiplier $\alpha$ is calculated as

$$\alpha = min\left(1.0, \frac{c_{acc}}{c_{min}}\right), \quad \text{9-14}$$

This ensures a continuous physical transition to pure water behavior as the solute concentration approaches zero.

Finally, the overall blended friction factor for the well element is calculated using a mass-fraction-weighted average of the active solutes. Solutes without a designated friction table, as well as the sub-minimum transition portion of tabulated solutes, default back to the baseline Fanning friction factor ($f$) as

$$f_{blend} = f x_{base} + \sum_k^M x_k \left[\alpha_k f_{table,k} + (1 - \alpha_k) f\right], \quad \text{9-15}$$

where $x_{base}$ is the mass fraction share of all water solutes without a designated friction table; $x_k$ is the mass fraction share of water solutes utilizing friction table $k$; $\alpha_k$ is the low-concentration transition multiplier for table $k$; $f_{table,k}$ is the equivalent Fanning friction factor of water solutes utilizing friction table $k$; $M$ is the total number of pipe friction tables.

## 9.5 Time- and temperature-dependent rheology (‘Modified Power Law vT’ water solutes)

The rheology of crosslinked gels changes as the fluid ages at temperature. Viscosity typically builds as the gel crosslinks and then declines as the polymer degrades or the breaker acts. Laboratories report this behavior as tables of the power law consistency index, $K'$, and the flow behavior index, $n'$, versus time, measured at a given gel concentration and temperature. The ‘Modified Power Law vT’ water solute type (‘ModifiedPowerLawvT’ in the settings file) uses these tables directly. Instead of the constant viscosity multiplier and exponent of a ‘Modified Power Law’ solute (Section 9.1), the modified power law parameters of a ‘Modified Power Law vT’ solute are interpolated from the tables as functions of the age of the fluid, the local temperature, and the local solute concentration.

A ‘Modified Power Law vT’ water solute is defined in the “Water solutes” table in the “Water Solutes” panel, like other water solutes. Its shear rate for transition to the zero-shear rate plateau, $\dot{\gamma}_{1/2}$, is specified there, but its viscosity multiplier per 0.001 mass fraction and power law exponent are not used. The laboratory data are entered in the “Parameter versus time tables” in the “Water Solutes” panel. Each table holds one measurement for a single water solute, concentration, and temperature, and contains rows of aging time (the time since the fluid was mixed), $K'$, and $n'$; a read-only column shows the corresponding viscosity at 100 s$^{-1}$. A water solute can have several tables at different concentrations and temperatures, and every 'Modified Power Law vT' solute requires at least one table. $K'$ and $n'$ do not need to vary monotonically with aging time, so a viscosity that first rises and then falls can be represented.

At shear rates well above $\dot{\gamma}_{1/2}$, the apparent viscosity reproduces the measured power law:

$$\mu_a = K'(t_a, T, w)\dot{\gamma}^{n'(t_a,T,w)-1}, \qquad 9\text{-}16$$

where $t_a$ is the age of the fluid, $T$ is the local temperature, and $w$ is the local mass fraction of the water solute.

To achieve this, the interpolated $K'$ and $n'$ are converted into the parameters of the modified power law (Equation 9-1). The exponent is set to $n = n'$, and $\dot{\gamma}_{1/2}$ is held at its input value at all ages. At high shear rate, Equation 9-1 approaches $\mu_w \mu_{0,adj}\left(\dot{\gamma}/\dot{\gamma}_{1/2}\right)^{n-1}$, where $\mu_w$ is the viscosity of pure water. Setting this equal to Equation 9-16 gives the target zero-shear viscosity multiplier:

$$\mu_{0,target} = K'/\left(\mu_w \dot{\gamma}_{1/2}^{1-n'}\right). \qquad 9\text{-}17$$

The contribution of the solute to $\mu_{0,adj}$ is set so that a fluid containing only this solute has $\mu_{0,adj} = \mu_{0,target}$, and a 'Modified Power Law vT' solute is never allowed to reduce the viscosity below that of pure water. With these substitutions, 'Modified Power Law vT' solutes enter the mixture rules of Section 9.1 and the apparent viscosity calculations for flow in fractures (Equation 5-14) and in the wellbore (and in the matrix, if non-Newtonian flow in the matrix is enabled) in the same way as 'Modified Power Law' solutes. In the wellbore, the zero-shear multiplier is also derived from the interpolated $K'$, so the parameter "Viscosity multiplier per 0.001 mass fraction (in the well)" is not used.

Between measured concentrations, $K'$ and $n'$ are interpolated linearly in mass fraction. Above the highest measured concentration, they are held at the values of that concentration (no extrapolation). Below the lowest measured concentration at the local temperature, $w_{min}$, holding the table values would make dilution have no effect. Instead, $\mu_{0,target}$ is interpolated linearly between its value at $w_{min}$ and pure water:

$$\mu_{0,target}(w) = 1 + \left[\mu_{0,target}(w_{min}) - 1\right]\frac{w}{w_{min}}, \text{ for } w < w_{min}, \qquad 9\text{-}18$$

with $n'$ held at its value at $w_{min}$. As the solute is diluted toward zero, its effect on viscosity vanishes smoothly.

The simulator must know the age of the fluid in every element. Age is not a conserved quantity, so it cannot be transported directly, and fluids of different ages mix at every connection in every timestep. ResFrac therefore tracks age with conserved companion water solutes. At the start of the simulation, each 'Modified Power Law vT' solute automatically creates two companions: an age tracer, named "<solute name> (age tracer)," and a degraded solute, named "<solute name> (degraded)." The age tracer is an inert solute with no effect on viscosity or wellbore friction. It keeps the other properties of its parent (such as molar mass, filtercake permeability, and spurt loss mass per area) so that it moves with the parent. The degraded solute is a 'Modified Power Law' solute whose rheology is fixed at the fully aged state of its parent. Both companions are solved with the water solute mass balance equations (Section 3), like any other water solute.

At the start of each timestep, in every fracture, matrix, and wellbore element, mass is transferred from pure water to the age tracer at a rate proportional to the local mass fraction of the parent solute, $w_p$:

$$\Delta w_{tr} = \frac{w_p \Delta t_{age}}{\tau_{tr}}, \quad \text{9-19}$$

where $w_{tr}$ is the mass fraction of the age tracer, $\Delta t_{age}$ is the aging increment, and $\tau_{tr}$ is an internal timescale equal to 100 times the longest aging time in the tables of the solute, which keeps the tracer mass small compared with the parent mass. The aging increment is equal to the timestep duration, $\Delta t$, unless a breaker is specified. A breaker is represented by the optional "Breaker kick-in time," $t_{kick}$, and "Breaker aggressiveness factor," $F_{brk}$, in the "Water solutes" table; both must be specified. Once the age of the fluid in an element reaches $t_{kick}$, aging accelerates:

$$\Delta t_{age} = \begin{cases} \Delta t, & t_a < t_{kick} \\ F_{brk}\Delta t, & t_a \geq t_{kick} \end{cases} \quad \text{9-20}$$

Because the parent and its tracer are transported together, dilution by water does not change their ratio, and when fluids of different ages mix, the ratio becomes the parent-mass-weighted average age. The age of the fluid in each element is recovered from the ratio:

$$t_a = \tau_{tr}\frac{w_{tr}}{w_p}. \quad \text{9-21}$$

This approach requires no additional governing equation, and the age is preserved through restarts because it is stored in the water solute mass fractions.

For interpolation, the age is normalized by the maximum aging time at the local temperature, $t_{max}(T)$:

$$t_D = \min\left(\frac{t_a}{t_{max}(T)}, 1\right). \quad \text{9-22}$$

At each measured temperature, $t_{max}$ is the longest aging time among all tables at that temperature. Between measured temperatures, it is interpolated linearly, and outside the measured range, it is held at the nearest value.

For each combination of temperature and concentration, $n'$ and $\ln K'$ are fit versus $t_D$ with a monotonicity-preserving spline (Steffen, 1990). Outside the range of aging times measured for that combination, the nearest measured values are held (no extrapolation). At the local conditions, values are interpolated linearly between the bracketing concentrations at each bracketing temperature, and then linearly between the bracketing temperatures; outside the measured temperature range, the nearest temperature is used. $K'$ is interpolated in logarithmic space because it often decreases by orders of magnitude as the fluid breaks. Interpolating $K'$ linearly while $n'$ is interpolated independently can produce intermediate states that are more viscous than the measurements on either side.

When the age of the fluid in an element reaches $t_{max}(T)$, the remaining parent solute mass is transferred to the degraded companion, and the tracer mass is returned to pure water. The degraded companion is always evaluated at $t_D = 1$. Retiring fully aged fluid prevents old resident fluid from dominating the average age, which would otherwise assign an artificially old age to newly injected fluid when it arrives. If the parent solute disappears from an element, for example through a water solute reaction, any remaining tracer mass is returned to pure water. Because the parameter versus time tables already describe how the rheology changes with time, the simulator issues a warning if a 'Modified Power Law vT' solute is also consumed or produced by a water solute reaction (Equation 9-2), since this may double count the time dependence.

The 3D visualization includes the properties "Fluid n'," "Fluid K'," and "Fluid age" for fracture and matrix elements. "Fluid n'" and "Fluid K'" are the blended values for the water phase. "Fluid age" is the parent-mass-weighted age of all ‘Modified Power Law vT’ solutes in the element, with degraded solute counted at its maximum aging time. In the “Water Solutes” panel, the viscosity curves of a fluid mixture that contains a ‘Modified Power Law vT’ solute are plotted at dimensionless ages of 0, 0.5, and 1.

Laboratory reports often give $K'$ and $n'$ at only a few aging times, along with a denser record of apparent viscosity at a fixed shear rate. The “Parameter versus time infill wizard,” located below the parameter versus time tables in the “Water Solutes” panel, uses the denser viscosity record to add rows to a selected table. For each row of aging time, $t_a$, apparent viscosity, $\mu_a$, and shear rate, $\dot{\gamma}$, entered in the wizard, $n'$ is interpolated through the measured rows of the table with the same monotonicity-preserving spline used by the simulator (and held at the end values outside the measured range), and $K'$ is calculated so that the power law reproduces the entered viscosity:

$$K' = \mu_a \dot{\gamma}^{1-n'(t_a)}. \quad \text{9-23}$$

Rows at aging times that already exist in the table are skipped, and measured rows are never changed. Generated rows are marked “Generated by wizard,” and they are replaced the next time the wizard is applied to the same table.

# 10. Fracture stress shadows, geometry, initiation, and propagation

## 10.1 Fracture initiation, opening, and stress shadowing

Hydraulic fractures can initiate at perforation clusters, the edges of preexisting fractures, or at ‘initiation points’ placed along openhole wellbore sections. The spacing of the openhole initiation points are a user input. By default, hydraulic fractures do not initiate in mode III from the top and bottom of preexisting fractures (which leads to an en echelon array of cracks, but there is an option to turn this on. To initiate, the fluid pressure at the initiation point must exceed the locally calculated Shmin, plus an effective tensile strength. Further, for discrete propagation algorithm, the crack does not form unless the fluid pressure is high enough to propagate the fracture once it has formed. Thus, the fracture toughness has an impact on initiation in this case. On the other hand, for continuous propagation algorithm, the fracture is always initiated once the fluid pressure exceeds the local Shmin plus tensile strength. See Section 10.3 for the description of continuous algorithm. The fracture toughness and effective tensile strength are input for each layer in the “Geologic units (facies list) table in the “Static Model and Initial Conditions” panel).

To avoid mesh dependence, the user specifies a “Minimum fracture initiation radius” in the “Fracture Options” panel (by default, 7.5 ft). If the fracture element size is greater than the minimum initiation size, then the crack is initiated as a single element. If the element size is less than the minimum initiation size, then multiple elements are created in order to make the initial crack radius approximately equal to the minimum initiation radius.

When fracture elements are mechanically open, force balance requires the fluid pressure to equal the normal stress (Crouch and Starfield, 1983). Fracture opening induces backstress that increases normal stress. When pressure begins to exceed normal stress, the fracture aperture increases, increasing normal stress and maintaining the system in equilibrium. If normal stress begins to exceed fluid pressure, the fracture aperture decreases and eventually the walls come into contact at mechanical closure. After closure, the fracture retains aperture and conductivity due to roughness of the fracture walls and/or proppant. The aperture of mechanically closed fractures calculated using nonlinear joint closure laws (Section 5)

The stresses induced by fracture opening are calculated according to the three-dimensional displacement discontinuity (DD) method described by Shou et al. (1997) using constant displacement rectangular elements.

The method assumes that the fracture is contained within in an infinite, homogeneous, linearly elastic medium. The stresses induced by normal displacements of mechanically closed elements are small, and so are neglected for simplicity and computational efficiency.

The Shou et al. (1997) method is based on the Green's function for an opening mode displacement discontinuity. The Green's function for opening mode displacement discontinuity yields solutions that are consistent with analytical solutions from fracture mechanics, such as the solution from Sneddon (1946).

Optionally, ResFrac can use an approximate method to account for heterogeneity of elastic moduli. Consider the interaction coefficient predicting the impact of opening at element i on element j. The Voigt-Reuss-Hill average of the moduli between these two points is calculated to determine the modulus used in the Shou et al. (1997) technique. This behavior can be modified using the parameter "Modulus averaging technique" in the "Other Physics Options" panel.

For example, Figure *13* shows a ResFrac solution to a problem with layered elastic moduli. The problem is set up to mimic the validation problem described in Section 5.1 from Peirce and Siebrits (2001). They perform a more rigorous and accurate solution. The ResFrac simulation is set up using special options such that a circular crack forms with uniform internal fluid pressure. Because there is extreme contrast in Young's modulus, and a thin layer, this is an uncommonly severe example of modulus heterogeneity. Nevertheless, the ResFrac simulation does a reasonable job of reproducing the more rigorous result from Peirce and Siebrits (2001). The max apertures in the upper and lower lobes from Peirce and Siebrits (2001) are 1.8 mm and 0.9 mm. In the ResFrac simulation, the max aperture in the lobes is 2.3 mm and 1.18 mm.

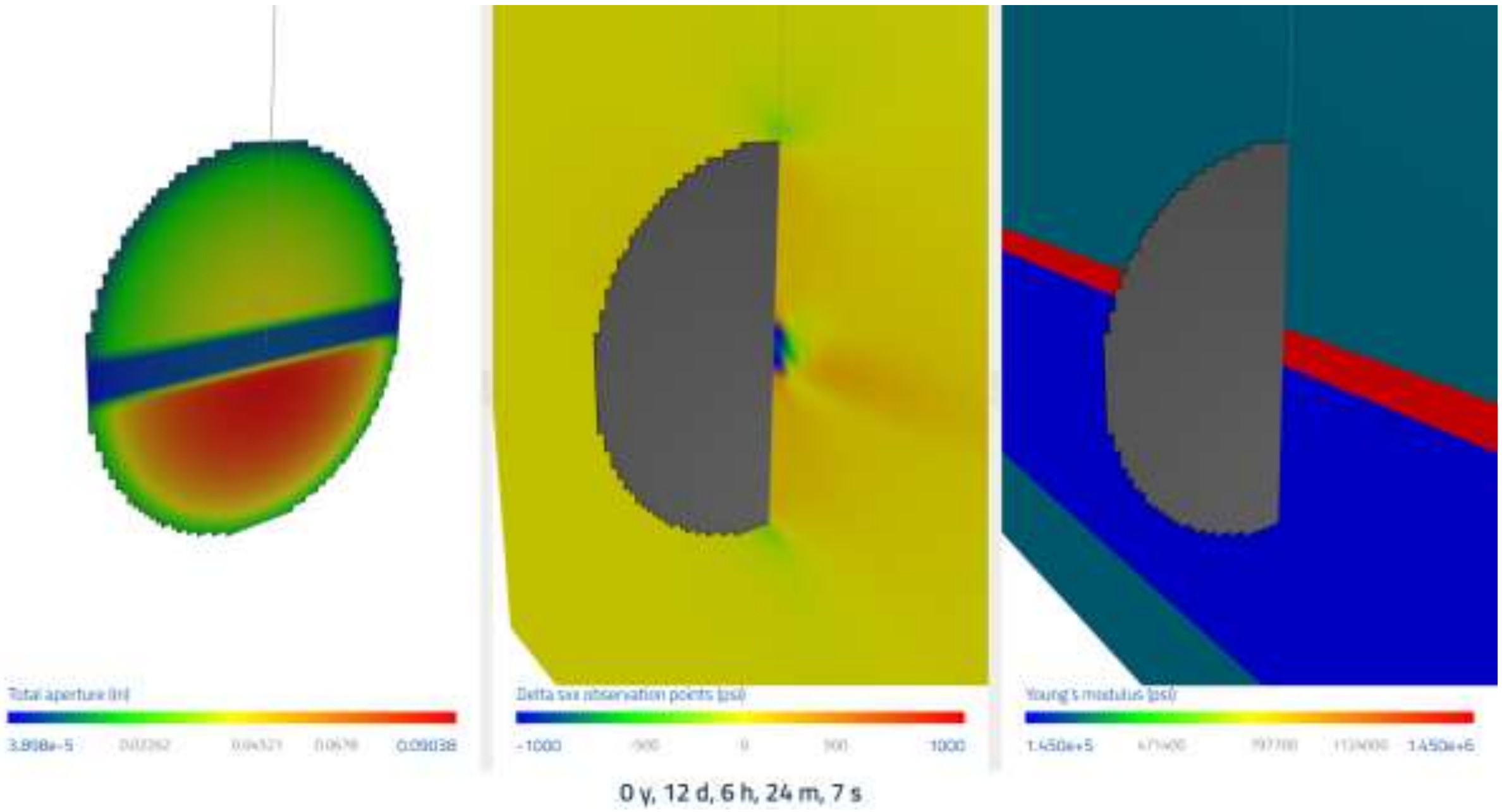

Figure 13: ResFrac approximate solution to the problem described in Section 5.1 from Peirce and Siebrits (2001).

With the boundary element method, numerical artifacts cause the technique to be inaccurate within a certain distance of the elements. This can be problematic when modeling fractures with very close spacing. To resolve this problem, ResFrac adjusts the calculation when calculating stress shadow within a distance of one element length from the plane cutting tangent through the fracture element. The adjustment is performed by moving the 'observation point' outward perpendicular to the crack until it is a distance equal to one element length. This behavior can be modified with the parameter "BEM close proximity adjustment distance" in the "Fracture

Options" panel.

Figure *14* shows an example. The lower panel shows a fine-mesh solution the Sneddon (1946) problem. The top left panel shows a coarse solution without any adjustment to the calculation. Numerical artifacts are visible near the fracture. The right panel shows the same result with the adjustment, which has smoothed the artifacts.

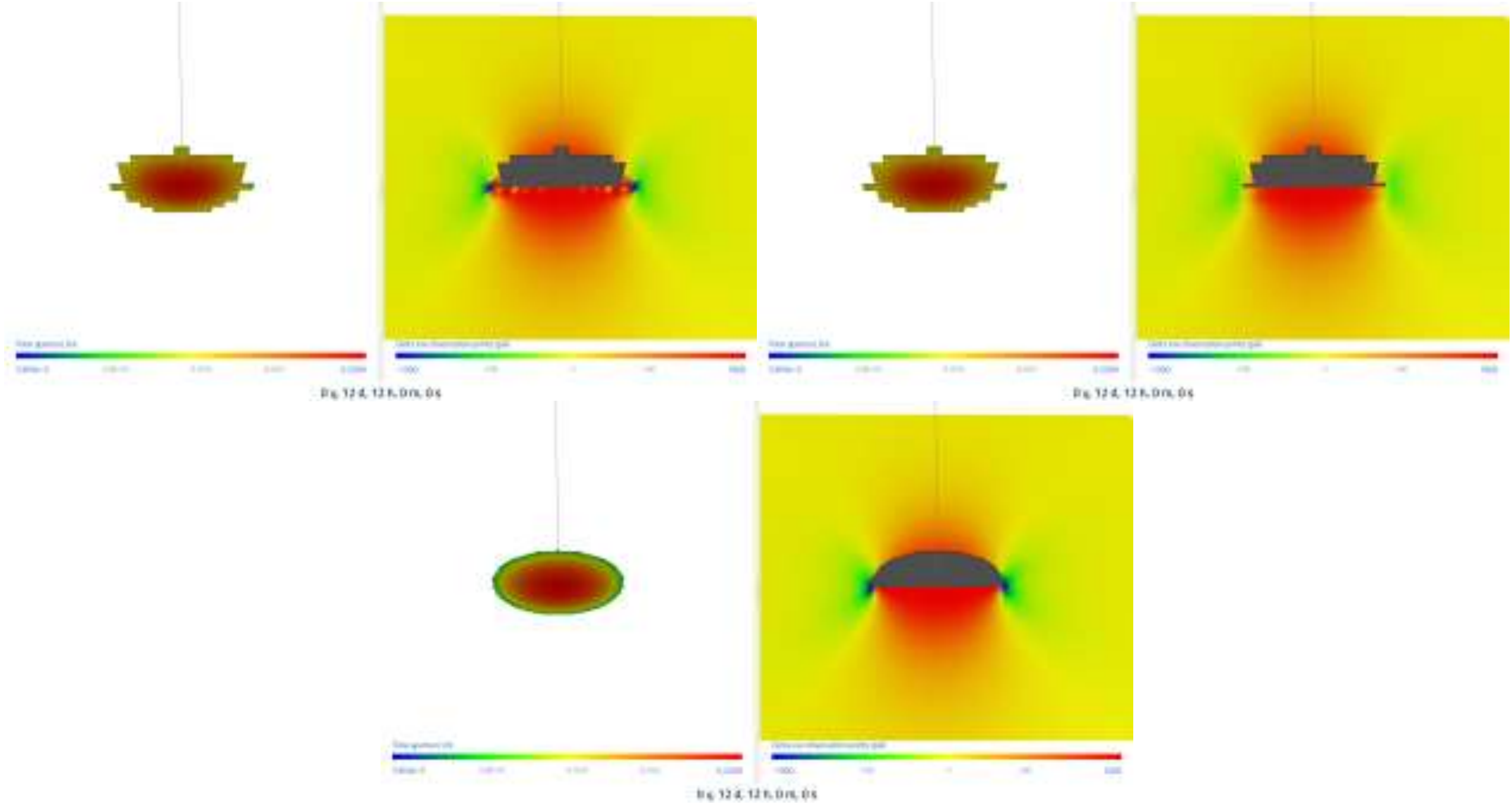

Figure 14: Three solutions to the Sneddon (1946) problem. The top left uses a coarse mesh and no adjustment for proximity to elements. The top right uses a coarse mesh and does include the adjustment for proximity to elements. The bottom figure uses a fine mesh.

ResFrac allows you to place stress observation planes and points. At these points, the DD method is used to calculate stress changes, and these are outputted for visualization. ResFrac also provides an option to output strain and displacement values at these points. All are calculated using the method from Shou et al. (1997).

ResFrac does not calculate fracture sliding or calculate the stresses induced by sliding. Hydraulic fractures form perpendicular to the minimum principal stress and so do not initially bear any shear stress (or experience shear displacement discontinuity). Because of stress interaction with neighboring fractures, it is possible for shear stress to be induced on hydraulic fractures and for them to experience shear displacement. However, this is a second order effect that is not usually included in hydraulic fracturing simulators (some simulators have included this effect, such as the work by Wu and Olson, 2016, and McClure and Horne, 2013, but it is uncommon). Fracture sliding is more important in discrete fracture network simulations that include preexisting natural fractures because natural fractures are usually oriented at an angle to the principal stresses and initially bear shear stress than is relieved by slip when fluid pressure increases. Even among fracturing simulators that use a DFN of the natural fractures, the stresses induced by sliding are often neglected (such as in the code described by Weng et al., 2011).

## 10.2 Fracture propagation

Fractures propagate according to the theory of linear elastic fracture mechanics. The fracture tip extends when the stress intensity factor reaches the fracture toughness. Fracture toughness is permitted to be anisotropic, different in the horizontal and vertical directions, and is permitted to vary by facies. Anisotropic apparent toughness may be caused by bedding plane slip or by stress layering (Fu et al., 2019).

The stress intensity factor is calculated numerically (Sheibani and Olson, 2013). By default, the fracture is assumed to propagate linearly. If you set the parameter "Straight fractures" in the "Fracture Options" panel to false, ResFrac will allow curving fracture paths. The fracture propagates in the direction of the locally calculated maximum horizontal stress. Fracture propagation is implemented by adding a new element ahead of the crack tip; remeshing is never performed. McClure et al. (2016a) validated this approach by implementing it in a 3D hydraulic fracturing simulator and matching analytical solutions for fracture propagation. Also, the approach is validated by matching ResFrac simulations to analytical solutions in our automated test suite, described below.

Fracture height confinement is primarily controlled by differences in stress between layers (Warpinski et al., 1982). The user can specify different stress in different layers. Height confinement also occurs due to shear along bedding planes (Chuprakov and Prioul, 2015). To mimic this process, our code allows the user to specify the location of delamination planes that are mechanical barriers to fracture propagation using the "Frictional interfaces" table in the "Static Model and Initial Conditions" panel. At these layers, the user specifies an elevated fracture toughness that must be exceeded to cross the barrier.

ResFrac allows fracture toughness to scale with fracture size. Field evidence strongly suggests that fracture toughness scales with size (Shylapobersky, 1986). Empirically, you'll find that if you use a laboratory-derived toughness to simulate fracturing in very low permeability rock, the fracture will be unrealistically long. Elevated, scale-dependent fracture toughness has also been described in the geology literature (Delaney et al., 1986; Scholz, 2010). Following Delaney et al. (1986) and Scholz (2010), scale-dependent fracture toughness is modeled with the equation:

$$K_{Ic} = K_{Ic,init}\left(1 + K_{Ic,fac}\sqrt{L_{eff}}\right), \qquad 10\text{-}1$$

where $L_{eff}$ is the length-scale of the fracture, and $K_{Ic,fac}$ is a scaling parameter. By default, $L_{eff}$ is defined as either height or length, whichever is smaller. However, there is an option to define $L_{eff}$ to be equal to the larger of the two quantities. These parameters are "Relative toughness per square root fracture size" and "Scale toughness by larger dimension" in the "Fracture Options" panel.

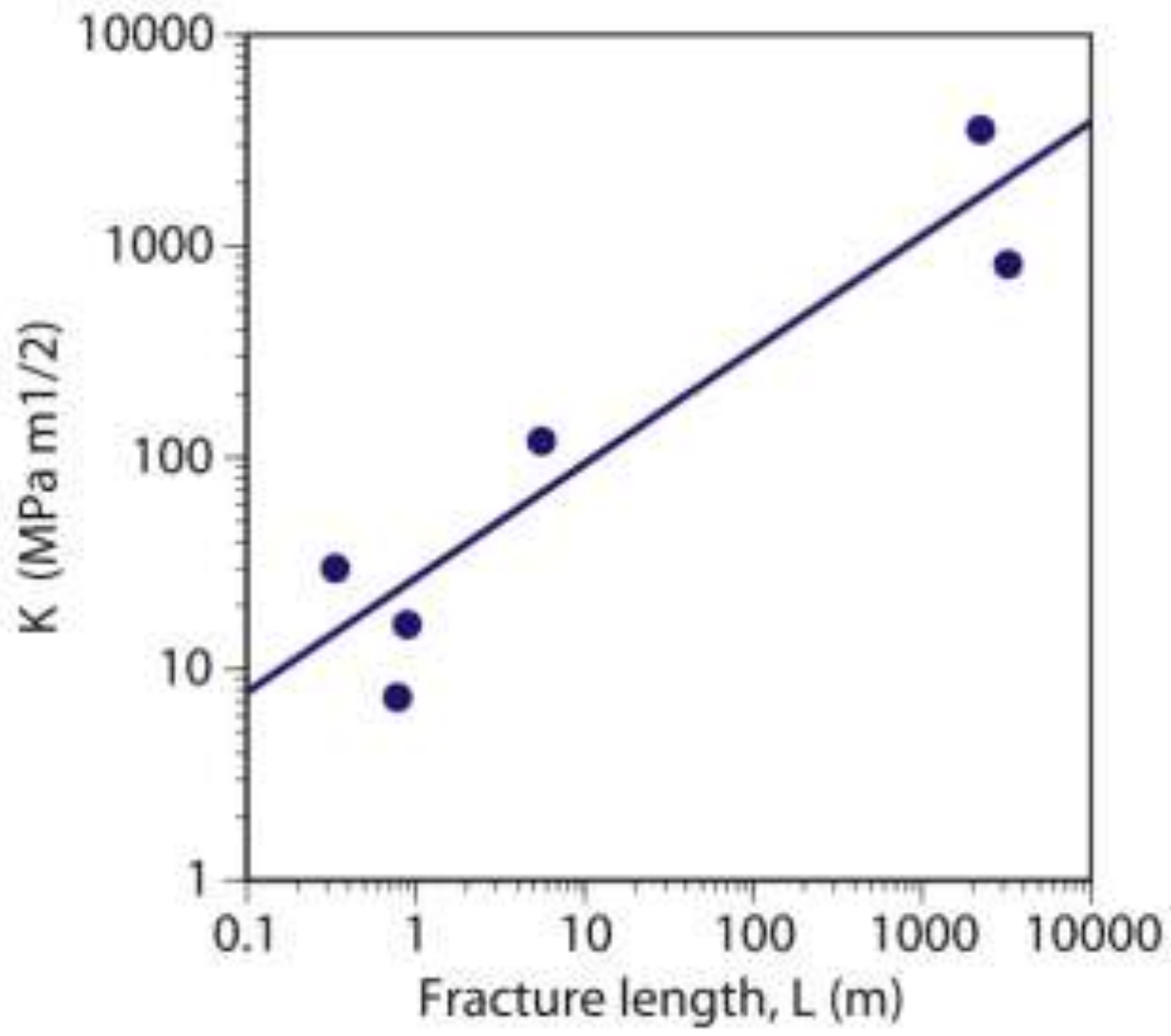

Figure 15: Figure reproduced from Scholz (2010). Illustrates observed relationship between apparent fracture toughness and fracture dimension.

## 10.3 Continuous front tracking algorithm with Multi Layer Tip Elements (MuLTipEl)

Numerical algorithm that allows to continuously track fracture front using Multi-Layer Tip Elements (MuLTipEl) was first introduced for the case of a plane strain hydraulic fracture in (Dontsov, 2021b). Its extension to fully three-dimensional fractures is summarized in (Dontsov et. al, 2022). This section briefly outlines the concept and presents a few examples, while readers are referred to the aforementioned papers for more detailed information.

To understand the primary concept, consider a uniformly pressurized plane strain fracture, such as the one shown in Figure *16(a)* (marked as “Before propagation” in the legend). If the fracture incrementally propagates by fully opening new elements, then the fracture length instantly increases. Consequently, the maximum width decreases, see the black dashed line in Figure *16(a)* (marked as “After propagation” in the legend). This implies that the fracture state changes abruptly and this corresponds to “discrete” propagation. In contrast, continuous propagation features an additional stress applied to the tip elements see in Figure *16(b)*. Initial magnitude of this stress is sufficient to keep the element fully closed. The value of this stress is then adjusted gradually, which, in turn, allows the element to open continuously. Such an approach allows to reduce mesh dependence and also to account for the effects of layers within each fracture element. Thus, this eliminates the restriction of having sufficiently fine meshes that resolve all the geological layers.

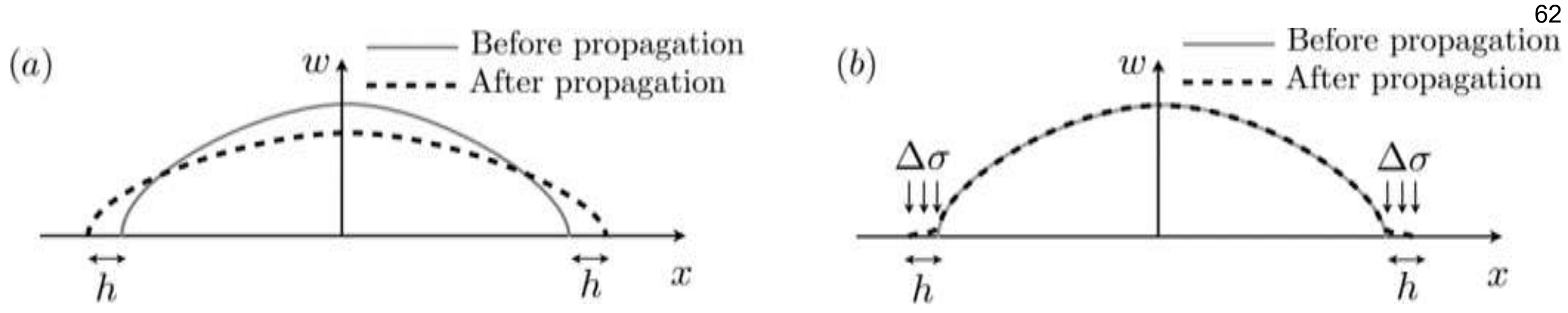

Figure 16: Schematics of incremental propagation for a plane strain hydraulic fracture for the discrete case (panel (a)) and for the continuous case (panel (b)).

Several numerical examples are considered next. Namely, the case of a radially symmetric fracture and a constant height or PKN fracture. Input parameters are summarized in Tab.1 and Tab.2. In these tables, $K_{Ic}$ is fracture toughness, $E$ is Young's modulus, $\nu$ is Poisson's ratio, $\mu$ is fluid viscosity, $Q_0$ is the injection rate, $t$ is the injection time, $\tau$ is the dimensionless time, and $H$ is the fracture height (applies for the constant height fracture only). No leak-off is assumed in these examples for simplicity. Three considered sets of parameters for each geometry approximately correspond to the domination of viscosity (*M* case), toughness (*K* case), and the transition (*MK* case).

Table 1: Input parameters for numerical simulations of propagation of a radial hydraulic fracture in the viscosity M, transition MK, and toughness K regimes.

| Property | M | MK | K |
|---|---|---|---|
| $K_{Ic}\left[psi \cdot in^{\frac{1}{2}}\right]$ | 2000 | 4000 | 6000 |
| $E[Mpsi]$ | 6 | 6 | 6 |
| $\nu$ | 0.25 | 0.25 | 0.25 |
| $\mu[cp]$ | 3 | 25 | 80 |
| $Q_0[bpm]$ | 20 | 20 | 20 |
| $t[hrs]$ | 1 | 1 | 1 |
| $\tau$ | $9 \times 10^{-3}$ | 8 | $6 \times 10^{6}$ |

Table 2: Input parameters for numerical simulations of propagation of a constant height hydraulic fracture in the viscosity M, transition MK, and toughness K regimes.

| Property | M | MK | K |
|---|---|---|---|
| $K_{Ic}\left[psi \cdot in^{\frac{1}{2}}\right]$ | 4000 | 8000 | 15000 |
| $H[ft]$ | 200 | 200 | 200 |
| $E[Mpsi]$ | 6 | 6 | 6 |
| $\nu$ | 0.25 | 0.25 | 0.25 |
| $\mu[cp]$ | 3 | 30 | 80 |
| $Q_0[bpm]$ | 20 | 20 | 20 |
| $t[hrs]$ | 1 | 1 | 1 |
| $\tau$ | $4 \times 10^3$ | 30 | 0.2 |

The competition between fluid viscosity and toughness for a radial fracture is controlled by the dimensionless time, defined as

$$\tau = \left(\frac{K'^{18}t^2}{\mu'^5 E'^{13} Q_0^3}\right)^{1/2}, E' = \frac{E}{1-\nu^2}, K' = \sqrt{\frac{32}{\pi}} K_{Ic}, , \mu' = 12\mu. \quad 10\text{-}2$$

According to (Dontsov, 2016), if $\tau \leq 5 \times 10^{-2}$, then viscosity dominates. At the same time, if $\tau \geq 3 \times 10^6$, then the limit of toughness domination is reached. Similar competition for a constant height hydraulic fracture is also controlled by the dimensionless time, but its expression is different.

$$\tau = \frac{2\pi^{\frac{1}{2}} E'^4 \mu Q_0^2 t}{H^{7/2} K_{Ic}^5}. \quad 10\text{-}3$$

According to the analysis of a constant height fracture (Dontsov, 2021a), $\tau \leq 0.1$ corresponds to toughness domination, while $\tau \geq 2 \times 10^3$ implies viscosity domination.

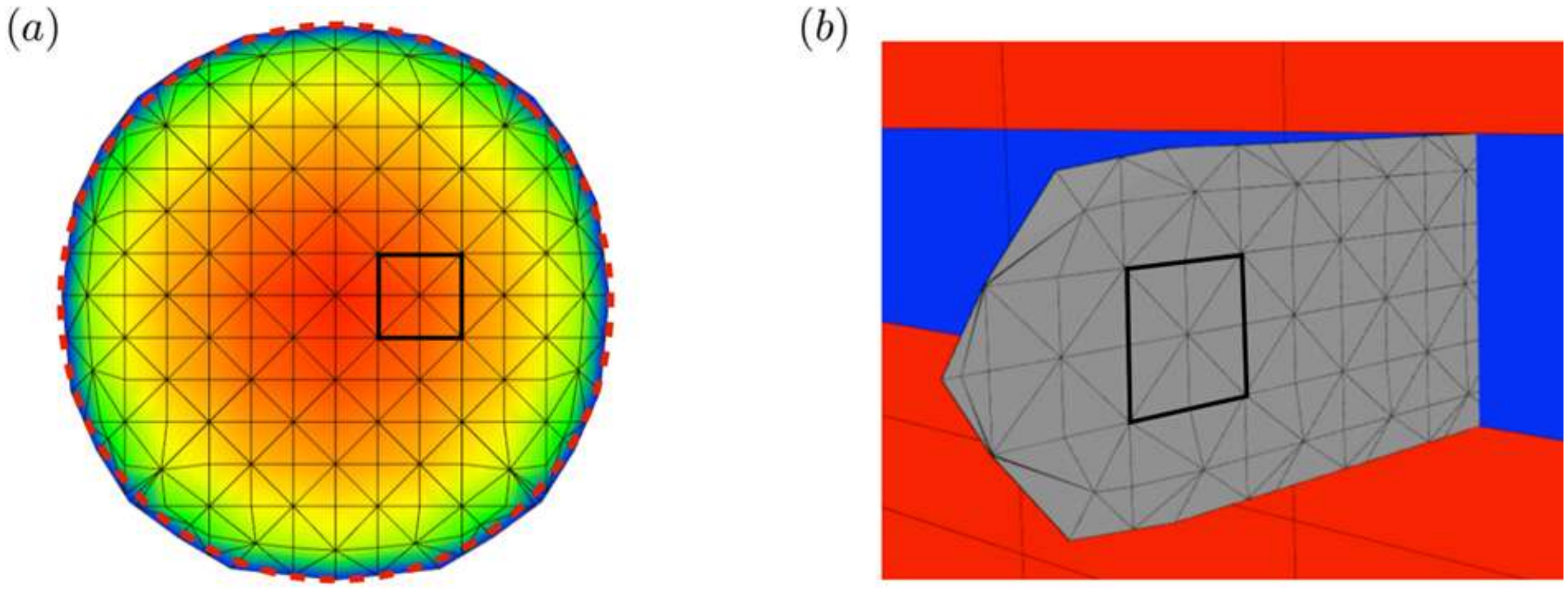

Figure 17: (a): Result of numerical simulation for a radial hydraulic fracture for the parameters corresponding to the transition region, i.e. MK. Numerical mesh is highlighted and the color fill corresponds to fracture aperture. Thin lines indicate triangle boundaries for visualization, while thick black square shows the element size used in calculations. The dashed red line plots a circle to visually estimate the degree of circularity of the fracture. (b): Result of numerical simulation for a constant height hydraulic fracture for the M K parameters. The fracture element is highlighted by the thick black square. The red and blue regions correspond to matrix elements and visualize the confining stress, where the red zones correspond to barriers.

Figure *17* illustrates results of numerical simulations for the radial (panel *(a)*) and constant height (panel *(b)*) geometries. Thin black lines outline triangles that are used for visualization, while the thick black squares show the element size. Red matrix elements indicate fracture barriers. It is clear that the fracture front tracking allows to precisely arrest the fracture at the location of the fracture barriers and also to capture circularity of the radial fracture.

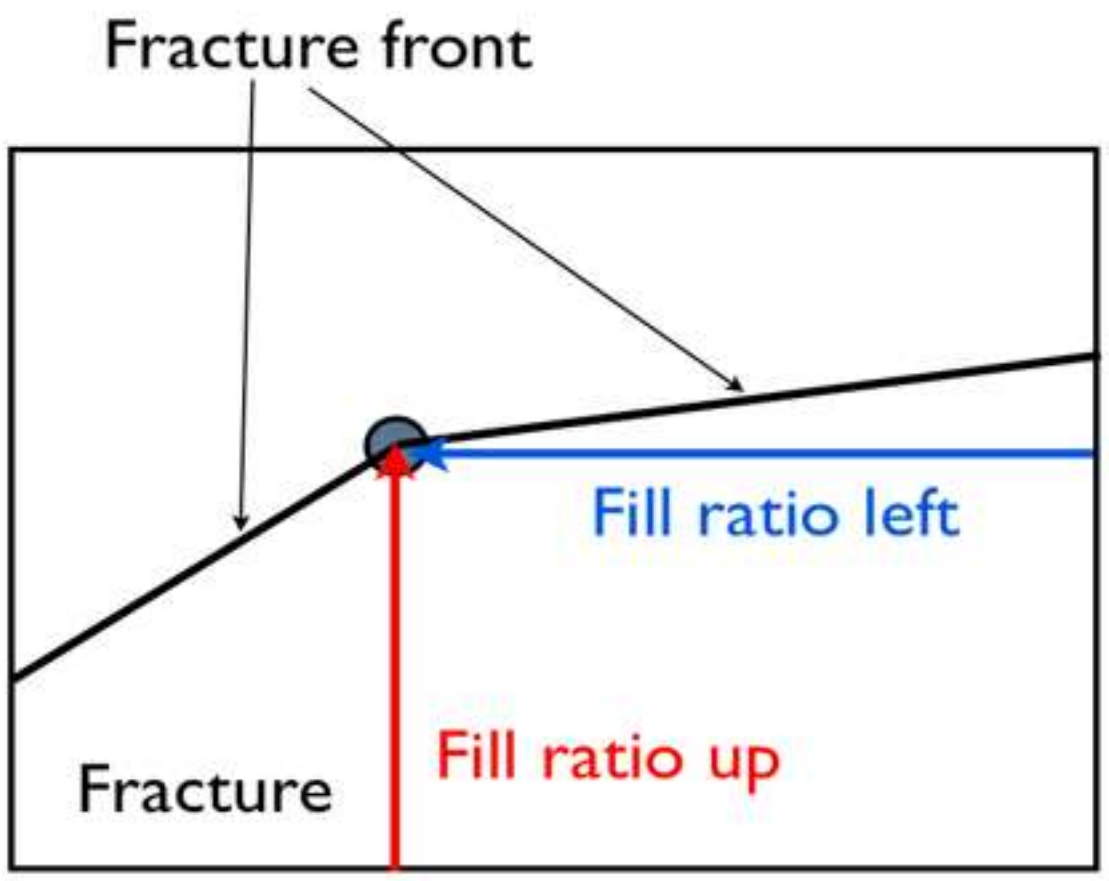

Figure 18: Schematics of calculating fracture front.

Figure 18 shows schematics for calculating fracture front. Each fracture element containing the front has quantities 'fill ratio up', 'fill ratio down', 'fill ratio left', and 'fill ratio right'. These parameters are used to determine location of the fracture front. Typically, only two out of four values are active, as shown in . The fill ratio left determines distance to the front point from the right edge of the element. The fill ratio up determines the distance from the lower edge of the element to the fracture front. These parameters vary from 0 to 1, i.e. represent the fraction of the element length or height. Each tip element has at least one front point. These points are then connected by straight lies to obtain the overall fracture front. To calculate fracture height or length, we use the coordinates of the fracture points for all fracture elements. Thus, the fracture dimensions account for partially filled elements. The parameter 'effective fill ratio', which is a function of the fill ratios in active directions, provides a viable approximation for the geometric area fraction. As a result, the ‘true area’ of a fracture element is the product between the effective fill ratio and the total area of a rectangular numerical element. Finally, proppant concentration within each tip element is assumed to be constant and is distributed uniformly over the filled element area. Consequently, proppant mass is calculated as the product between proppant concentration and true area of the element.

To investigate mesh dependence of the algorithm, simulations were performed for six different meshes (ranging from 5 ft to 160 ft) for three sets of parameters for each fracture geometry. Results are depicted in the figures below. The crosses correspond to MuLTipEl algorithm, the hollow circular markers show the results of the discrete algorithm, while the dashed black lines show the reference solutions from (Dontsov, 2016) and (Dontsov, 2021a). Results demonstrate that MuLTipEl algorithm reaches the correct solution much sooner (i.e. with coarser mesh). The discrete algorithm, on the other hand, requires finer meshes to reach similarly accurate results. Note that the reference solution for PKN fracture is approximate and is calculated assuming a flat fracture tip, see (Dontsov, 2021a). As a result, there is a slight discrepancy between the analytical solution and the converged numerical result.

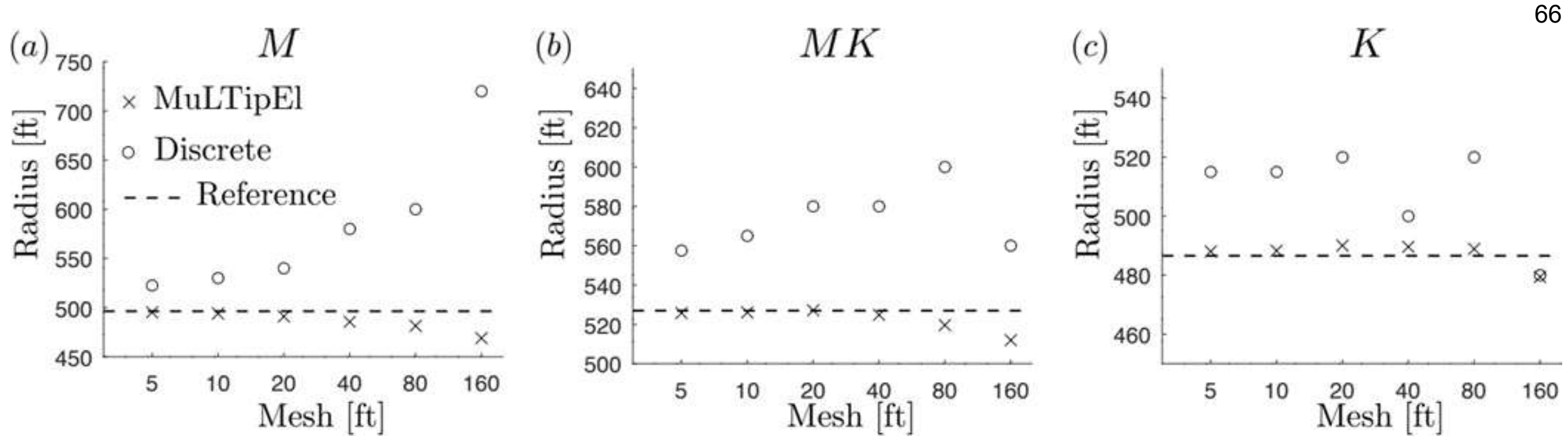

Figure 19: Mesh sensitivity results for the radial geometry for three sets of parameters summarized in Tab. 1. Crosses indicate the values of the final fracture radius calculated using MuLTipEl algorithm, circular markers correspond to the results from the discrete algorithm, while the dashed black lines show the reference solution.

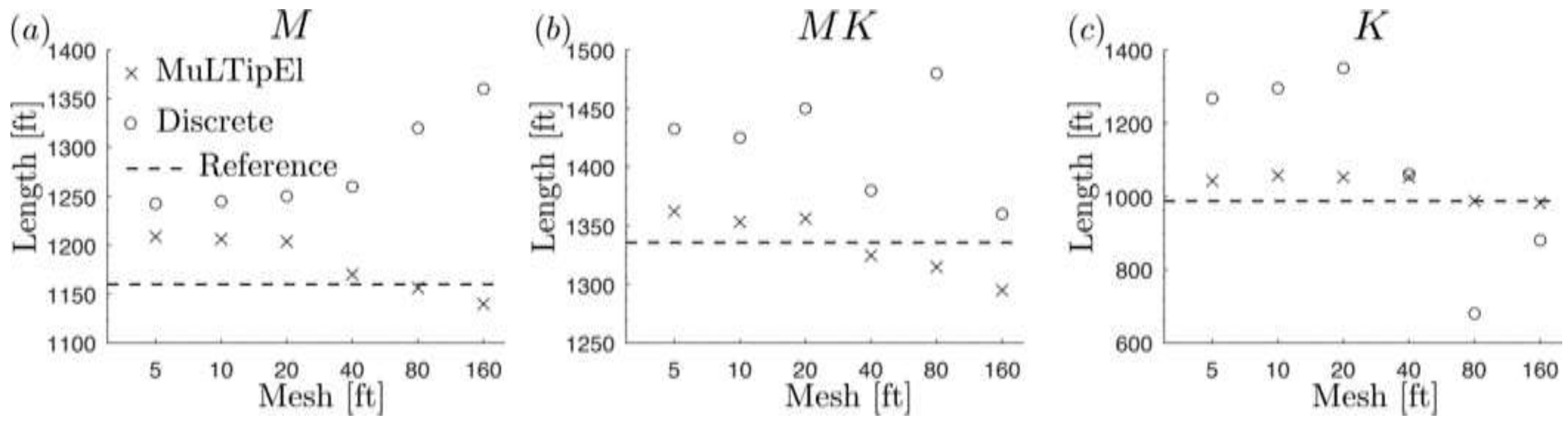

Figure 20: Mesh sensitivity results for the constant height geometry for three sets of parameters summarized in Tab. 2. Crosses indicate the values of the final fracture length calculated using MuLTipEl algorithm, circular markers correspond to the results from the discrete algorithm, while the dashed black lines show the reference solution.

Finally, Figure *21* shows comparison for a single fracture for a field case. Again, simulations are performed using both discrete and continuous approaches. Several thin stress barriers are present above the wellbore. As can be seen from the results, the MuLTipEl algorithm is able to account for these layers and to arrest the fracture growth even on coarse meshes. In contrast, its discrete counterpart requires a fine mesh to fully capture the barriers. This permits using coarser meshes with MuLTipEl algorithm and to calculate accurate result much quicker. Note, however, that in practical applications the fracture geometry is often calibrated to field data, which effectively reduces the issue of mesh dependence for the discrete approach.

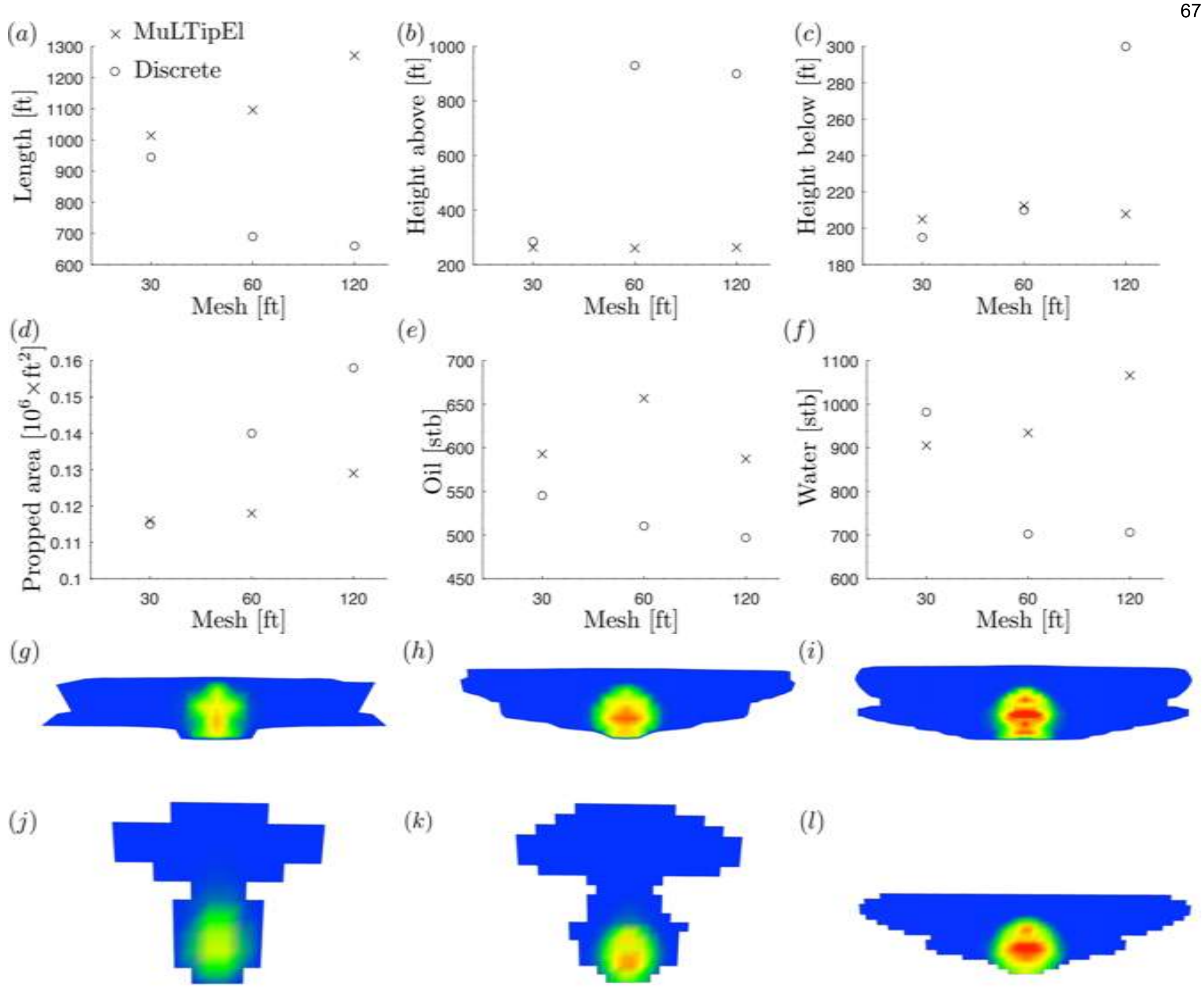

Figure 21: Mesh sensitivity results for a single hydraulic fracture for the example field data. Crosses indicate the values of the final fracture characteristics calculated using MuLTipEl algorithm, circular markers correspond to the results from the discrete algorithm. (a): Comparison of fracture length. (b): Comparison of height above. (c): Comparison of height below. (d): Comparison of propped surface area. (e): Comparison of oil production. (f): Comparison of water production. (g) − (i): Fracture geometries for 120 ft, 60 ft, and 30 ft meshes for MuLTipEl. (j) − (l): Fracture geometries for 120 ft, 60 ft, and 30 ft meshes for discrete.

## 10.4 Propagation from preexisting fractures

This section describes propagation from preexisting fractures focusing on continuous fracture front tracking algorithm.

First of all, it is important to outline a distinction between preexisting hydraulic fractures and preexisting natural fractures. For the purpose of this document, preexisting hydraulic fractures are preexisting fractures that are aligned with Shmax direction. Such fractures always propagate in mode I. At the same time, preexisting natural fractures are the fractures that are not aligned with the Shmax direction. Propagation from such fractures can occur only in mode II or in mode III. Please keep in mind that the fractures are still enforced to be straight for ‘continuous’ propagation, so that all newly initiated elements propagating from preexisting fractures are aligned with the Shmax direction. The criterion of initiation of mode II and mode III fractures is similar to that for the initiation from a well.

The natural fracture collision relative distance is used for collisions between hydraulic and natural fracture and between natural fractures. Collisions between preexisting hydraulic fracture and other hydraulic fractures are calculated using ‘Fracture collision relative distance’. Moreover, preexisting hydraulic fractures are also assigned an ‘initiating well’ if they are connected to a well. Depending on the situation, either the same well or different well collision distance is used. Depleted collision applies to all types of collisions. If the natural fracture collision relative distance is not specified, then it is set equal to fracture collision relative distance.

Note that the prescribed fracture dimensions are reproduced precisely and partially filled elements are used at the fracture edges to ensure this. This applies to both preexisting hydraulic fractures as well as to preexisting natural fractures. Note that in addition, the preexisting hydraulic fractures are always snapped to grid in the Shmax direction. Snapping in the vertical direction applies to all preexisting fractures.

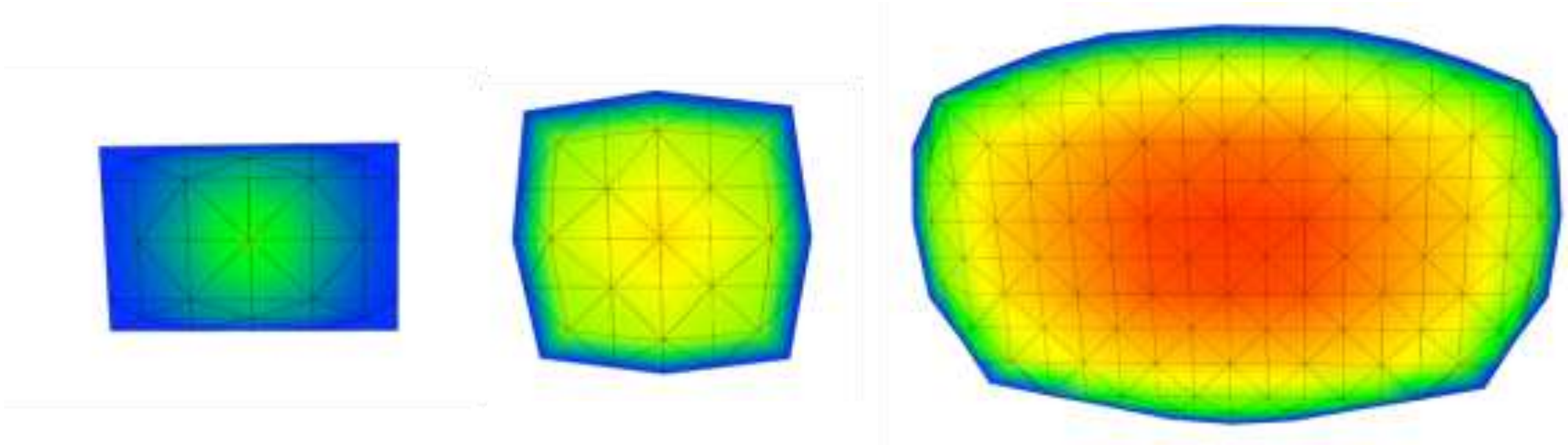

Figure 22: Example of propagation from preexisting hydraulic fracture.

shows an example of mode I propagation from a preexisting hydraulic fracture. The fracture inflates first, then starts to propagate, and eventually becomes affected by the stress barriers above and below.

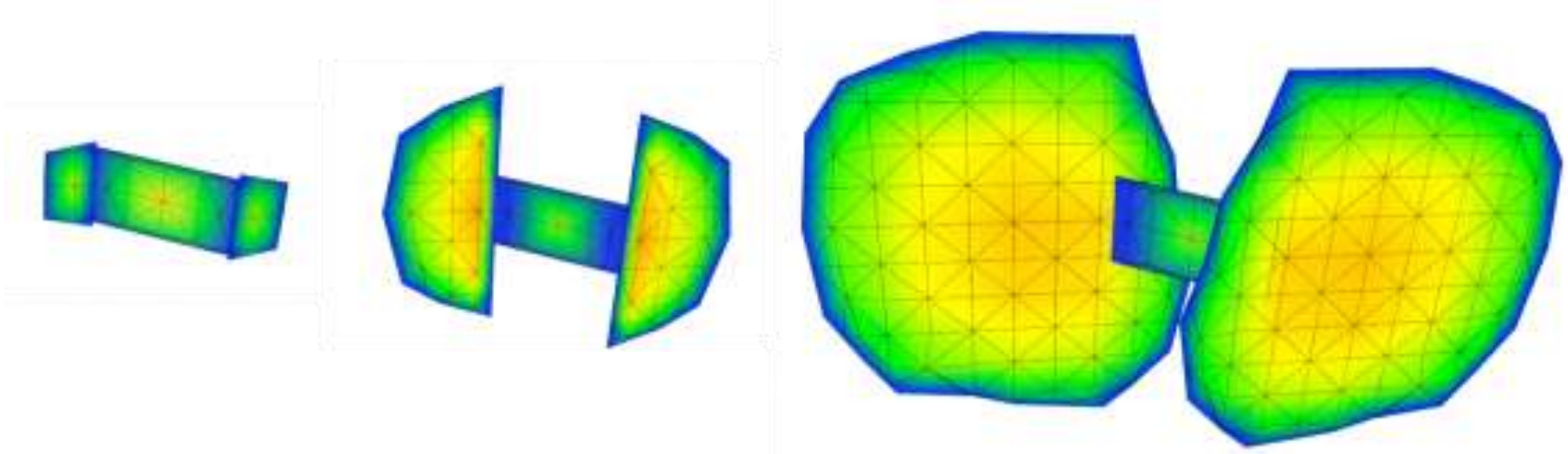

Figure 23: Example mode II propagation from a preexisting natural fracture.

The figure shows an example of mode II propagation from a preexisting natural fracture. The injection point is located in the middle of the natural fracture. Note that mode III propagation is disabled in this example. As can be seen, mode II fractures are initiated from the edges of the natural fracture and start propagating away from each other. Note that the newly generated fractures are restricted from propagating backwards. However, as the injection continues, the fractures grow in height and eventually start to propagate backwards by circumventing the natural fracture.

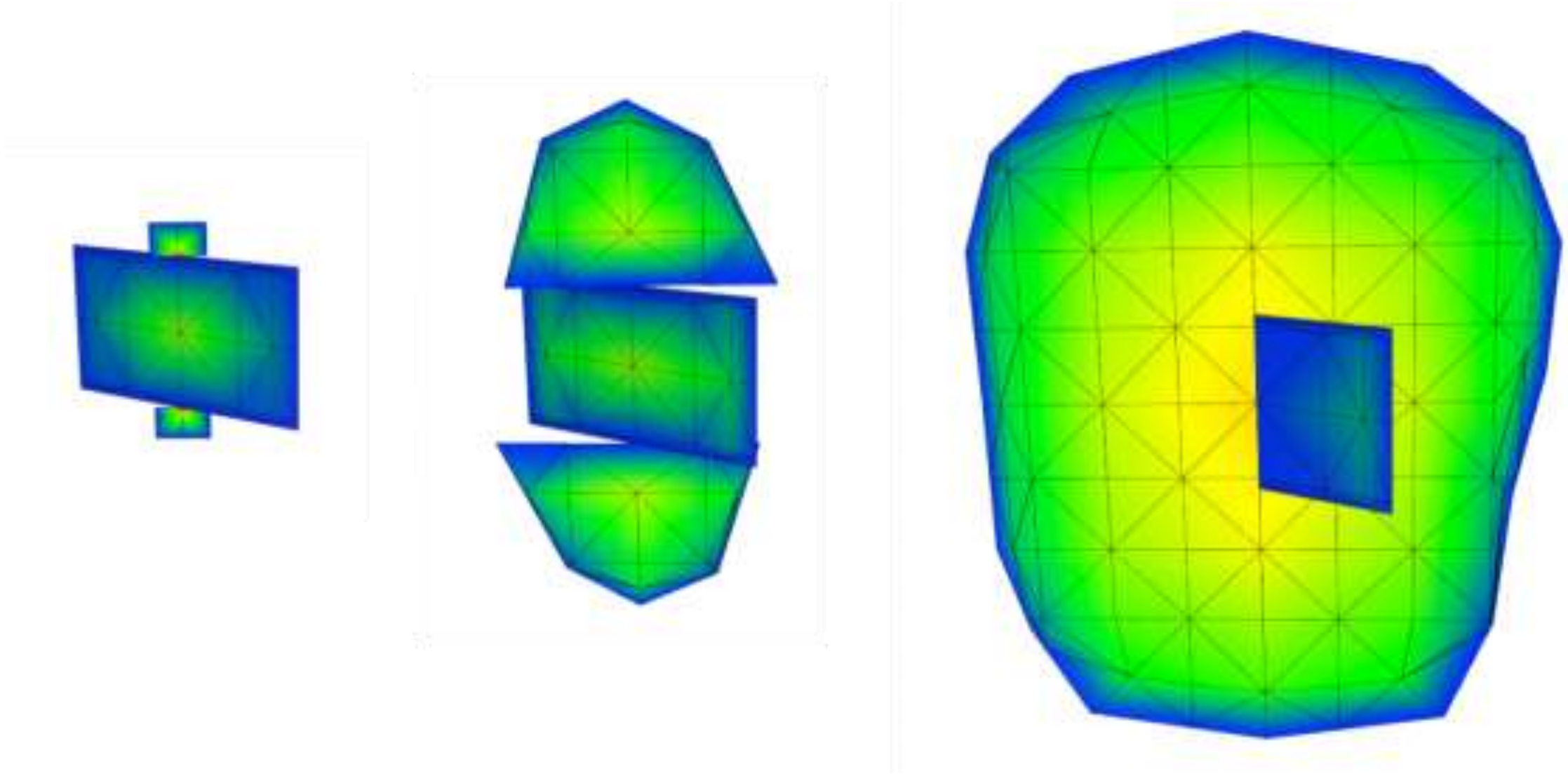

Figure 24: Example mode III propagation from a preexisting natural fracture.

The figure illustrates mode III propagation from a preexisting natural fracture. Mode III fractures represent 'vertical' propagation from natural fractures and are initiated in the middle of each fracture element. Thus, this inevitably introduces some mesh dependence since the number of generated fractures varies with mesh size. Also note that the corner elements do not initiate mode III fractures. Once the mode III fractures grow further, they eventually circumvent the natural fracture and merge together, forming a single mode III fracture.

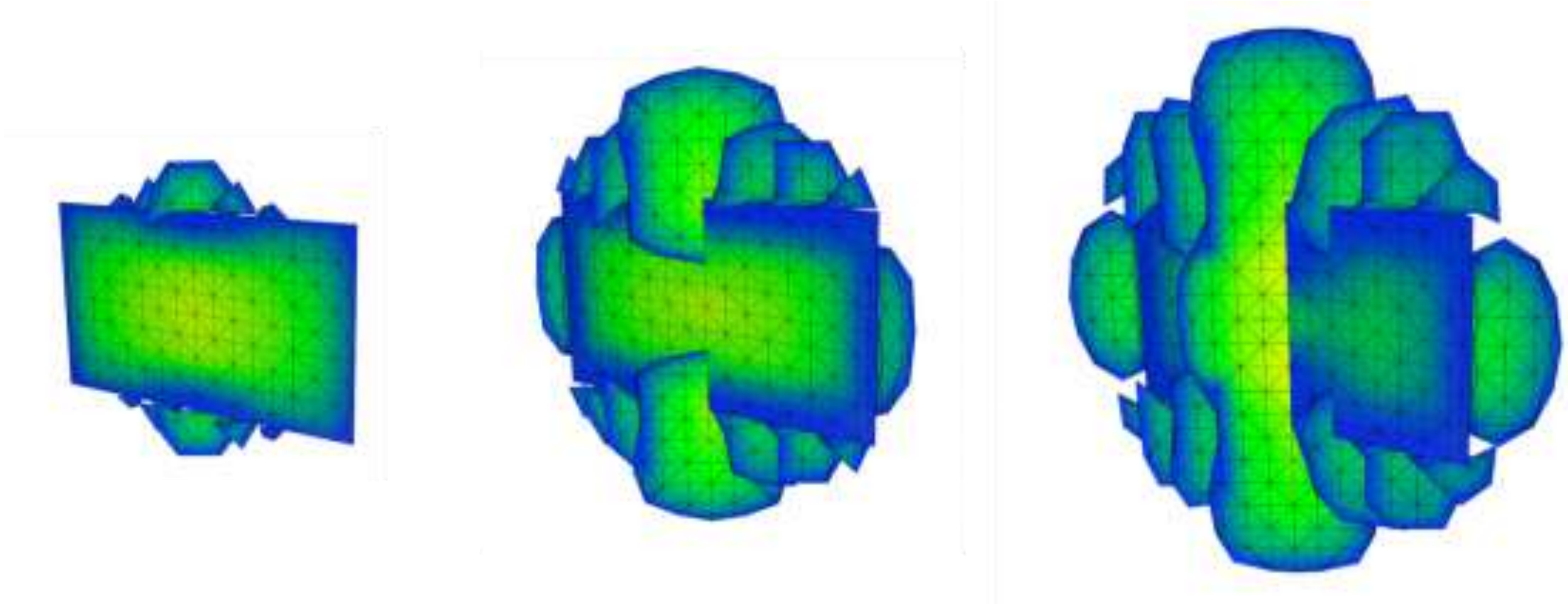

Figure 25: Example mode II and mode III propagation from a preexisting natural fracture.

For completeness, the figure shows an example simulation for a finer mesh and when both mode II and mode III propagation types are allowed. As can be seen from the figure, mode III fractures are initiated first, then more mode III fractures and mode II fractures are initiated. In this example, the injection point is in the middle of the natural fracture and hence the flux disproportionally benefits the centrally located mode III fractures, which eventually combine into one.

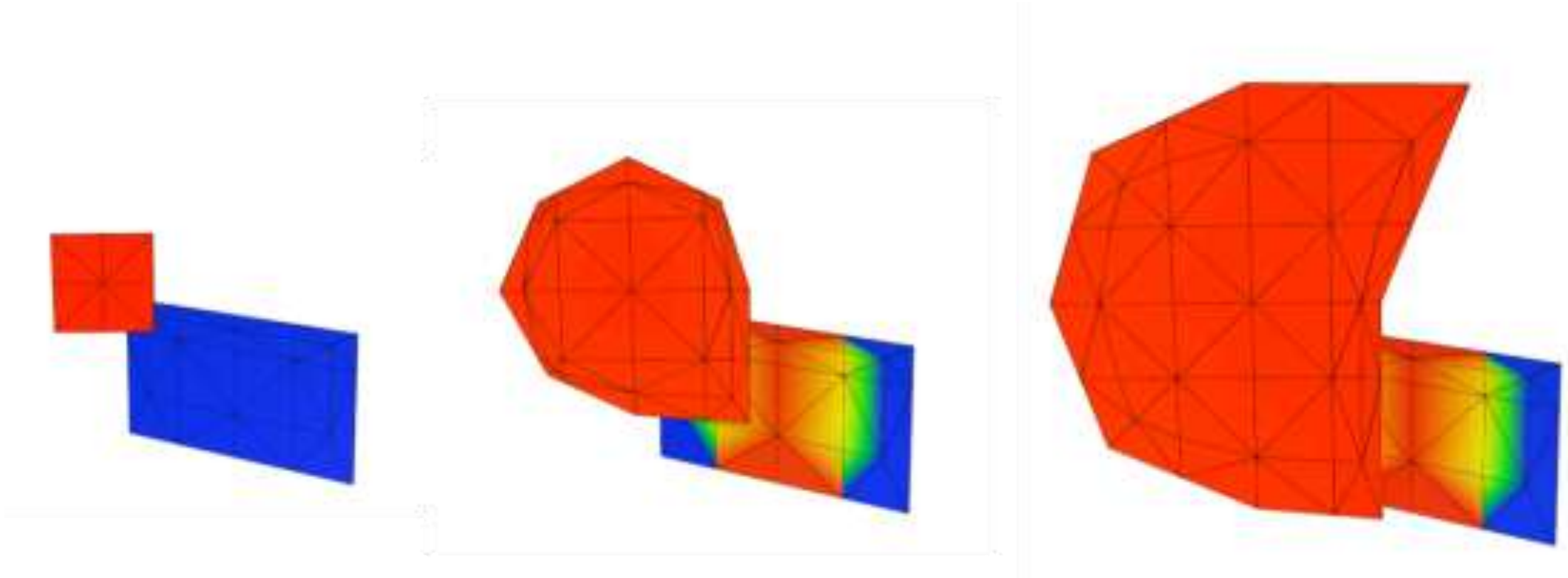

Figure 26: Collision between hydraulic and preexisting fractures from the side.

Another important topic is fracture collisions. This example shows a collision between a hydraulic fracture and a natural fracture. There are different crossing mechanisms, namely 'Always', 'Random', 'Crossing frequency', 'Never', and 'Extended Renshaw-Pollard'. For this example, the crossing strategy is set to 'never', i.e. the hydraulic fracture is arrested once it reaches the natural fracture. The fluid pressure is shown by color. As can be seen from the figure, the hydraulic fracture is indeed arrested by the natural fracture. But then it starts growing vertically and eventually circumvents the natural fracture. A larger preexisting fracture would have an ability to prevent such a behavior and to arrest propagation in one direction.

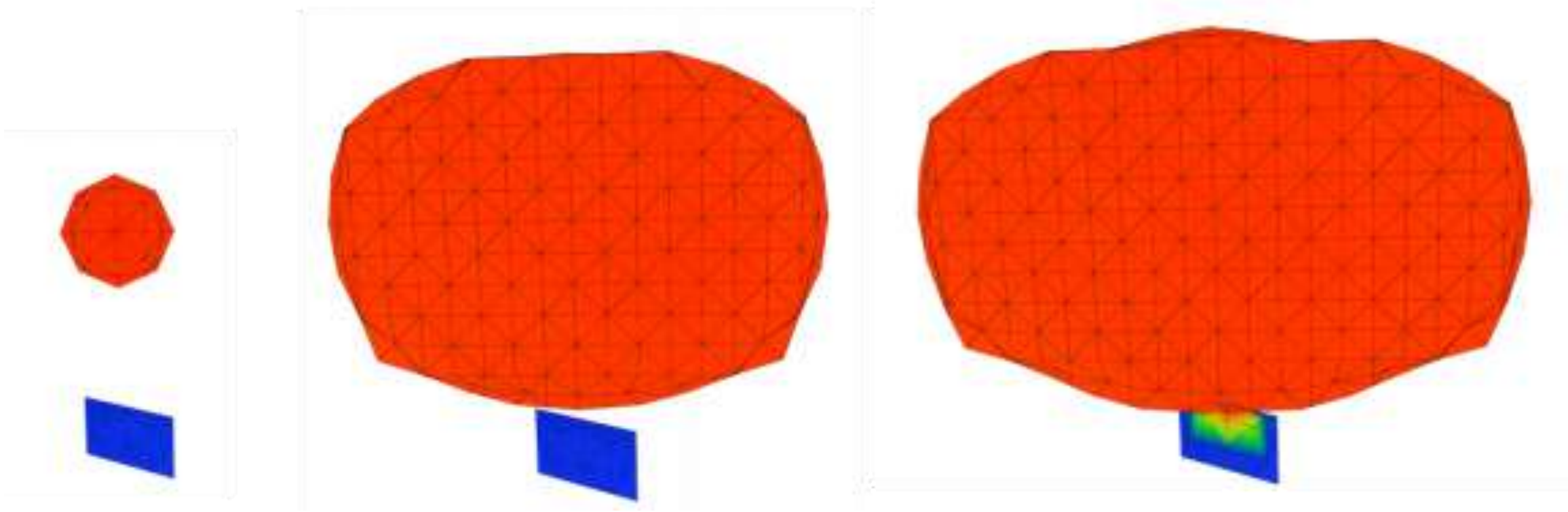

Figure 27: Collision between hydraulic and preexisting fracture from above.

Collision from above is illustrated in the example shown in the figure. Here the hydraulic fracture is above the natural fracture and propagates towards it. As can be seen, there is no pressure communication until the fractures are actually crossed. Note that the hydraulic fracture is allowed to cross the natural fracture in this configuration.

## 10.5 Fracture crossing options

When a hydraulic fracture reaches a preexisting fracture, the crossing criterion is evaluated. The following options are available: 'Always', 'Random', 'Crossing frequency', 'Never', and 'Extended Renshaw-Pollard'.

When the option 'Always' is selected, all hydraulic fractures that reach preexisting fractures are allowed to mechanically cross. If the hydraulic conductivity of the preexisting fracture is sufficiently high, then there can be hydraulic arrest, i.e. all the fluid can flow into the preexisting fracture and the original hydraulic fracture may not noticeably propagate. The only exception is when local termination angle is used. If specified, if the angle between the hydraulic and preexisting fractures is less than the specified value, there is no crossing.

When the option ‘Never’ is selected, all hydraulic fractures that reach preexisting fractures are not allowed to mechanically cross. Note that even when crossing is not allowed, the hydraulic fracture can grow around the preexisting fracture and then close on the other side. This can make the visual appearance of direct crossing.

The option ‘Random’ allows to incorporate random crossing with 50% probability. The option ‘Crossing frequency’ allows to change this probability. Both of these options are still subject to the termination angle logic (if specified), as was described above. In addition, it is possible to specify crossing probability multiplier per preexisting fracture. This multiplier applies for both ‘Random’ and ‘Crossing frequency’ options.

Last but not least option is ‘Extended Renshaw-Pollard’. This is mechanistic option and therefore the logic associated with the termination angle as well as crossing probability multipliers are disabled. The original Renshaw-Pollard criterion (Renshaw and Pollard, 1995) applies when a fracture attempts to cross an orthogonal frictional interface. The criterion can be written as

$$\frac{S_H}{T+S_h} > \frac{0.35+0.35/\mu}{1.06}.$$

Here the fracture is perpendicular to the minimum principal stress $S_h$, the interface is orthogonal to the maximum stress $S_H$, $T$ is tensile strength, while $\mu$ is coefficient of friction. This criterion is derived by calculating the total state of stress comprised of the initial stress as well as the stress field around the approaching fracture tip. If tensile fracture initiation occurs before interface sliding, then the fracture mechanically crosses the interface. However, if sliding (calculated according to Mohr-Coulomb criterion) occurs first, then there is no crossing. The numerical coefficients in the above equation stem from the complex expressions for the near tip stress field and are not empirical constants. Also note that this criterion has been validated using laboratory experiments.

The original Renshaw-Pollard criterion applies when a vertical fracture crosses an orthogonal preexisting fracture. This is rarely the case in practical applications. For this reason, Gu and Weng (2010) extended this criterion for the case of arbitrary oriented vertical preexisting fractures. Since ResFrac can support dipping natural fractures, we made further generalization to arbitrary oriented fractures. Both extensions rely on the same principles as the original work of Renshaw-Pollard (i.e. the competition between interface sliding and tensile fracture initiation). Given lack of simple expressions, please see the corresponding references for more information.

As an illustration, the figure below shows an example simulation made with the ‘Extended Renshaw-Pollard’ option. A single hydraulic fracture is interacting with two dipping faults with slightly different orientations. The parameters are selected such that the fracture crosses one interface and is arrested by another. The general trend is that higher coefficient of friction and cohesion promote crossing, while smaller intersection angles and smaller stress contrasts promote fracture arrest. Also note that effective stresses affect the result, so that the value of Biot coefficient and pore pressure are very important. Finally, Biot coefficient of the preexisting fracture is set to one.

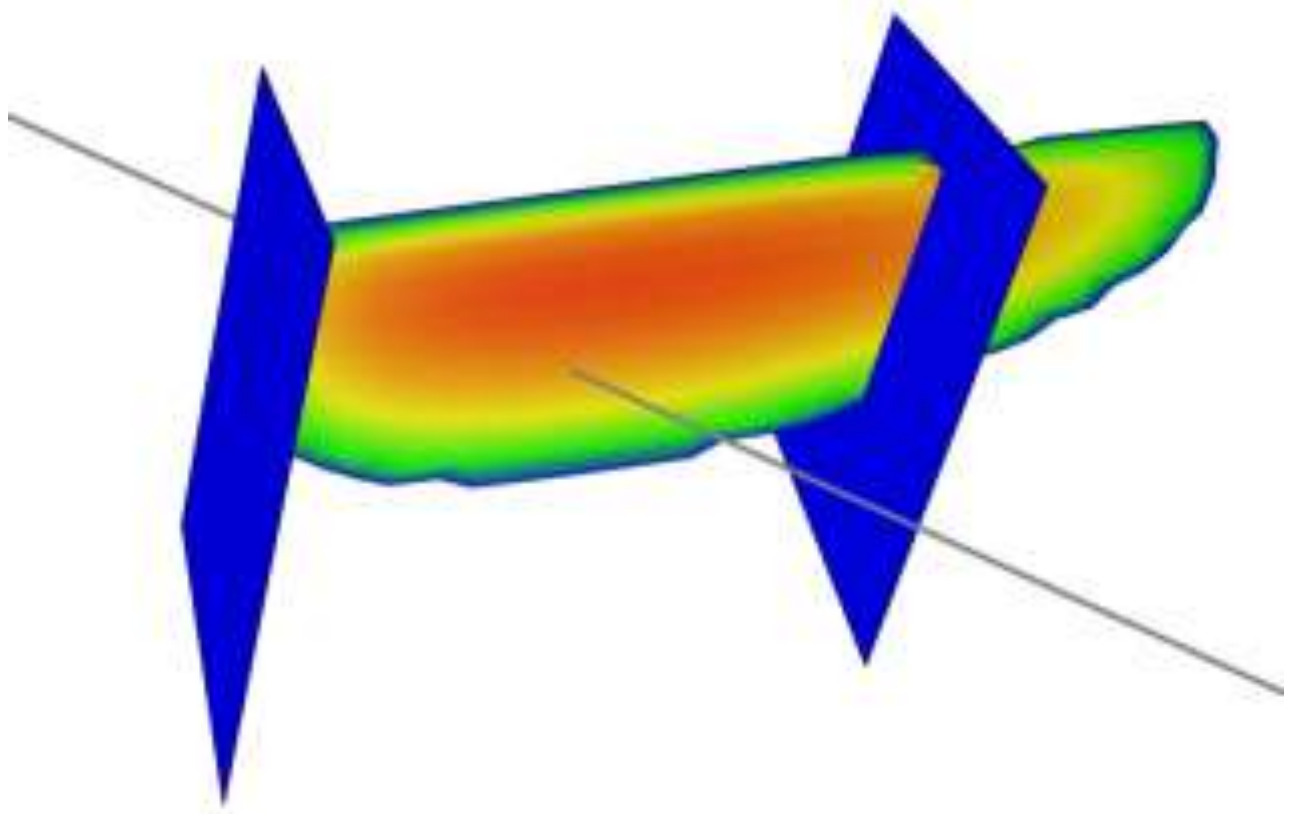

Note that currently the crossing of hydraulic fractures is always allowed and no criteria are checked. This situation can occur when a vertical hydraulic fracture crosses a horizontal fracture or vice versa when a horizontal fracture crosses a vertical hydraulic fracture.

## 10.5 Dipping natural fractures

Some natural fractures and faults are not necessarily vertical. To account for this, the dip angle can be specified for preexisting fractures. If the dip angle is zero degrees, then the preexisting fracture is horizontal, while if the dip angle is 90 degrees, then the preexisting fracture is vertical. The dipping preexisting fractures are rectangular and, if truncated to fit inside the matrix region, remain rectangular (this applies only for continuous front tracking algorithm). The dipping preexisting fractures are not snapped and are meshed using the fracture element length divided by the element aspect ratio along the dip direction and using just the fracture element length along the strike direction. The dipping fracture feature works for both continuous and discrete fracture propagation algorithms. Note that the same fracture crossing logic is used for dipping preexisting fractures as for the vertical fractures. Also, currently, there is no hydraulic fracture initiation from dipping preexisting fractures. Finally, if the dip angle is left blank, then it is assumed that the preexisting fractures are vertical.

Dipping preexisting fractures can open, if stress conditions allow, and carry proppant. Proppant settling velocity is reduced by ‘sin’ of the dip angle. In addition, the total proppant flow rate is affected by the ‘proppant flow reduction factor in horizontal fractures’. Given the dip angle, the proppant flow is reduced by sin(dip)^2 +ProppantFlowReductionFactor*cos(dip)^2. Therefore, if the factor is set to 1 or the fracture is vertical, then there is no effect.

## 10.6 Horizontal hydraulic fractures

Under certain conditions, horizontal hydraulic fractures can form during reservoir stimulation. For instance, Wright et al. (1997) reported observation of horizontal fractures using tiltmeters. At the same time, Alali and Zoback (2018) provided evidence of horizontal fractures in Marcellus shale using microseismic. Horizontal fracture propagation may be suspected when ISIP is greater than the overburden stress.

Horizontal hydraulic fractures are allowed to form in ResFrac if the option “Allow horizontal fractures” is selected. The horizontal fracture initiation condition is checked at perforations, at frictional interfaces, and at user specified vertical interval. The initiation criterion is P > Sv + Tstrv, where P is the fluid pressure, Sv is the total vertical stress, and Tstrv is the effective vertical tensile strength. Note that the initiation criterion for vertical fractures is P > Sh + Tstr, where Sh is the total minimum horizontal stress and Tstr is the effective tensile strength (applies for vertical fractures).

Horizontal fractures propagate in mode I (or opening mode) and their growth is controlled by debonding toughness KIcd, which is expected to be smaller than the vertical and horizontal fracture toughness since the fracture propagation occurs along the bedding.

Proppant transport is expected to be very different in horizontal fractures compared to the vertical ones. First of all, there is no settling. However, particles will accumulate in the lower part of the flow in horizontal fractures and thus will travel slower. To account for this, the parameter 'proppant flow reduction factor in horizontal fractures' is introduced, see also the section on dipping preexisting fractures. As of now, the total proppant flow rate is multiplied by this parameter. Consequently, if it is set to 1, then there is no effect.

Below is an example of horizontal fracture initiation from a frictional interface and from a perforation. The latter is caused by an increased value of Sh due to stress shadow from previous stage. Fracture aperture is shown by the color.

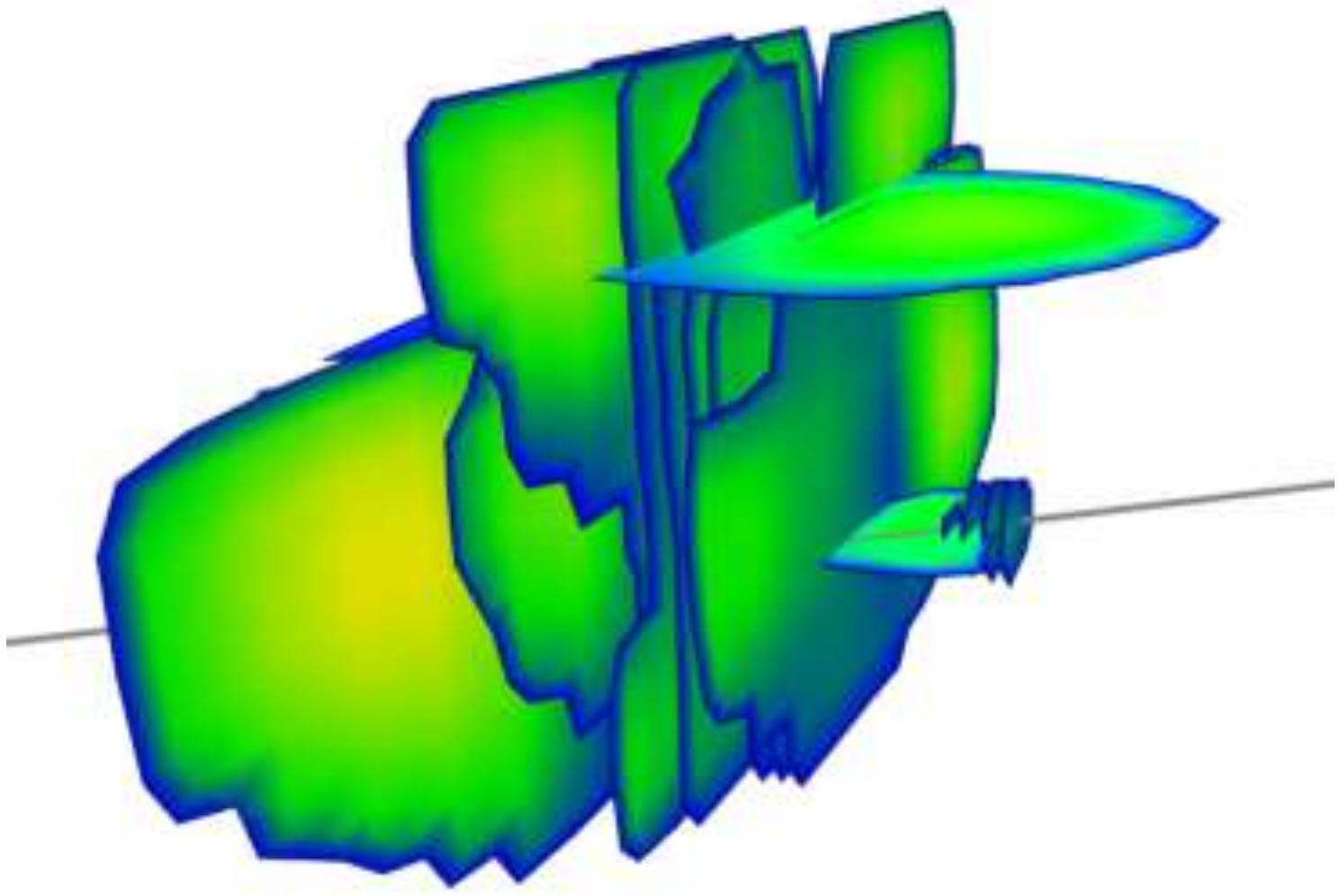

Another example shown below corresponds to the situation in which 'horizontal fracture initiation distance' is specified. In this case, there are many horizontal fractures, some of which grow bigger, while some do not propagate very far. Fluid pressure is shown by the color.

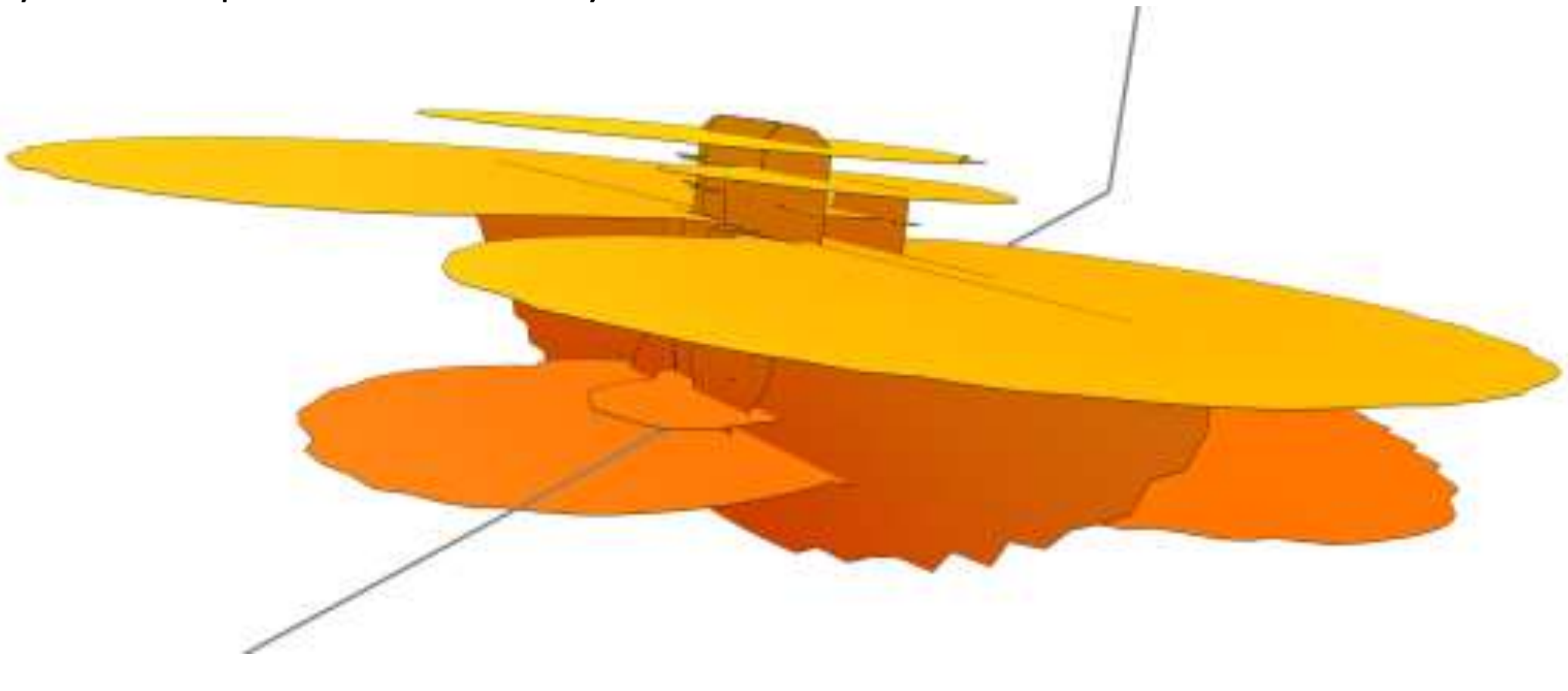

It is also possible to have a well that goes through different facies or formations. In one of the facies, initiation conditions favor horizontal fractures, i.e. Sv+Tstrv < Sh+Tstr. While in another one, the situation is opposite, i.e. Sv+Tstrv > Sh+Tstr, and vertical hydraulic fractures initiate. See such an example below. Fracture aperture is shown by the color.

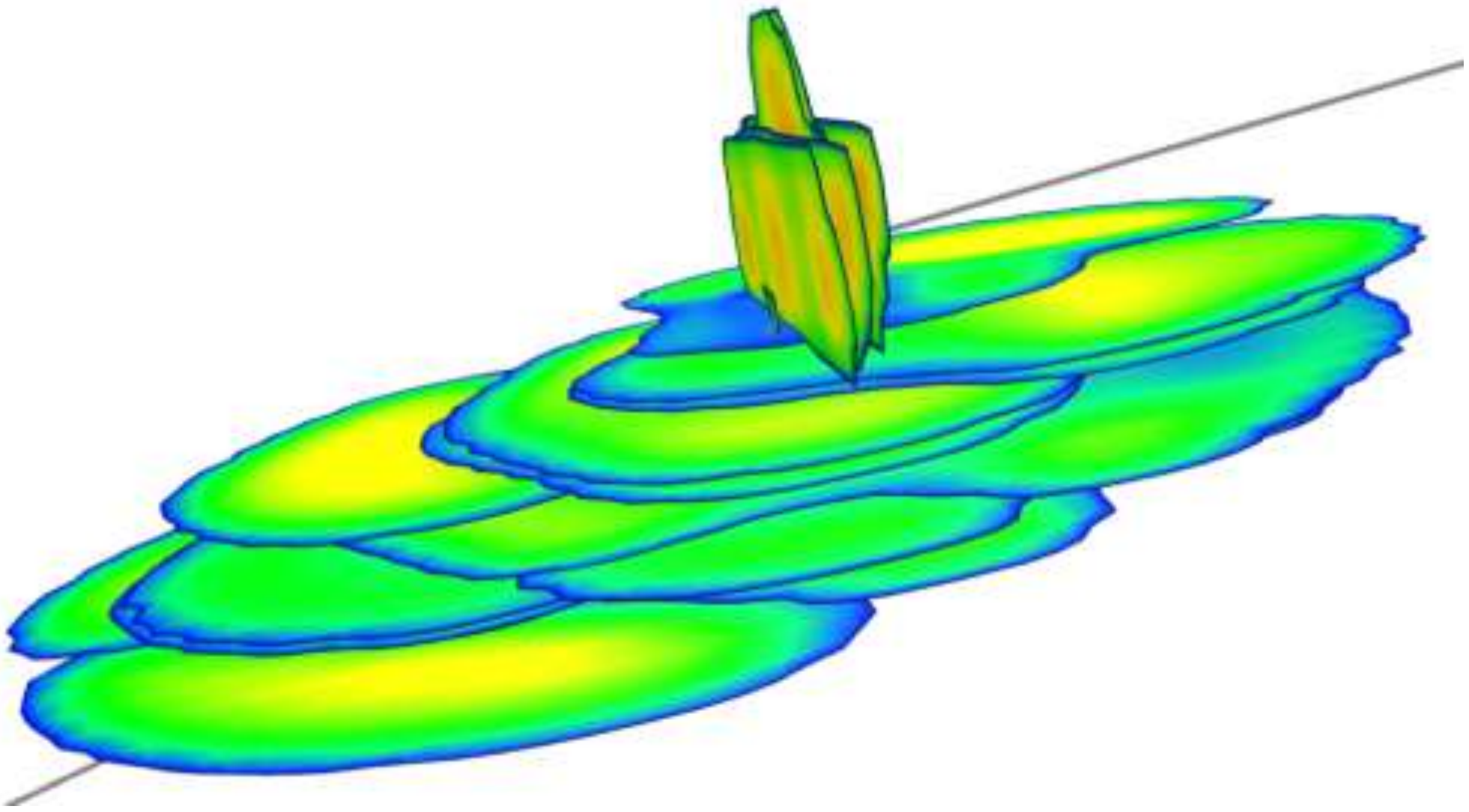

Finally, the example below shows the most complicated situation, in which there are vertical and horizontal hydraulic fractures that interact with two dipping preexisting fractures. Notice how horizontal fractures grow around the faults. Fluid pressure is shown by the color.

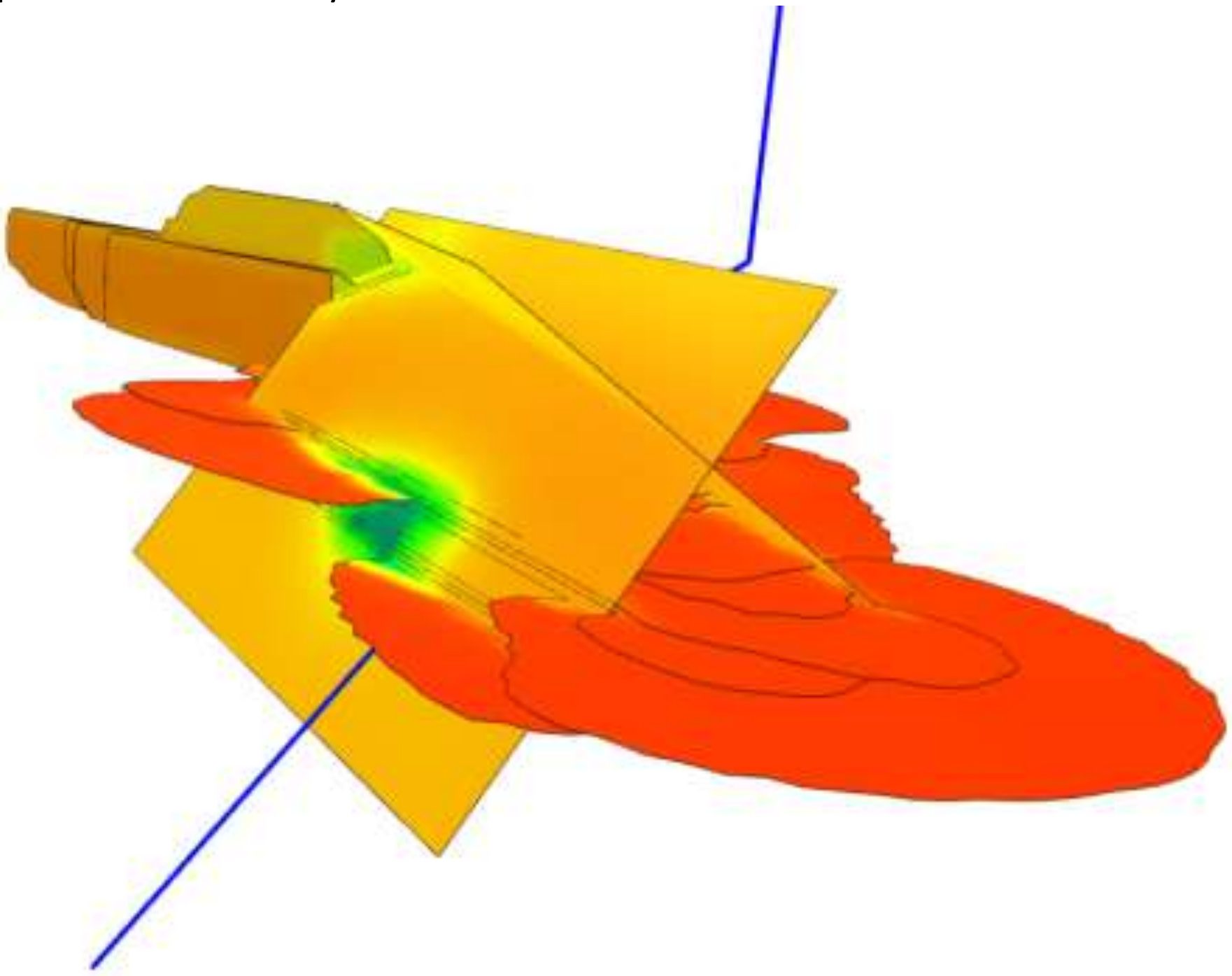

## 10.7 Details of fracture/wellbore flow and flow outside casing

When fractures hit a well, if the well section is openhole, a hydraulic connection always forms between the fracture and the well. If the well is cased, a hydraulic connection forms if the fracture is within a certain distance of a perforation, given by the parameter “Cased well and fracture connection distance.” Also, if you place perforation clusters within this distance, cross-flow will occur between the fractures at each cluster. We most often set this value to around 30 ft; however, in some cases, larger or smaller values may be justified. The idea is that there may be some flow behind casing through an axial, localized hydraulic fracture, or due to an imperfect cement sheath (Ugueto et al., 2019a, 2019b).

For example, let's say that perforation clusters are placed 5 ft apart along a lateral, and you set "Cased well fracture collision distance' to 10 ft. In this case, the hydraulic fracture from one perf cluster will form a hydraulic connection to not only its own perf cluster, but also the two perf clusters on either side (because they are within 10 ft). These connections represent the flow channel outside the casing that would permit fluid crossflow. This matters because of the impact on limited-entry. If the fractures form a flow channel outside the casing that runs along the length of the stage, this effectively defeats limited-entry. Limited-entry forces fluid to distribute across all the perf clusters, which is good for achieving uniform fracturing. However, if fluid can flow along the well in a flow channel between the fractures, this allows fluid to exit the well at one perf cluster, and then flow along the well to another fracture elsewhere. This can allow fluid to localize to a smaller number of fractures, as if limited-entry hadn't been used.

The simulator also permits for flow directly between fracture elements initiating from the same well that are within the "Cased well fracture collision distance". These crossflow terms are restricted by a 'flow barrier' treated like flow in a series (harmonic averaging) and with strength proportional to the distance between the fractures. The strength of the flow barrier is given by the parameter "Connect frac through 'cased well fracture collision distance' transmissibility barrier," which is given in units of md-ft.

There is also an option to specify a "Connect frac through 'cased well fracture collision distance transmissibility multiplier'" to control these crossflow terms. However, this term is no-longer recommended – the default value is 1.0 so that it has no effect. In addition, there is an option to model cross-flow with an additional 'near-wellbore tortuosity' pressure drop term, but this is no longer recommended. The strength of cross-flow between should be tuned using the parameter "Connect frac through 'cased well fracture collision distance' transmissibility barrier."

Because of these crossflow terms, when fluid flows out of a perforation cluster, it can connect into multiple different fracture elements. It is important to have the perforation pressure drop scale appropriately with the *total* flow rate through the perforation cluster (from the well to all the fracture elements), and not to calculate the perforation pressure drop separately for each well-fracture element pair. To do this exactly would be extremely complex because each well-fracture flow term is itself nonlinear, and so it would be necessary to solve a coupled system of nonlinear equations. To simplify, the code makes a reasonable assumption, which is to assume that the flow rate of fracture-well connection scales with the conductivity of the flow term. With this assumption, the flow terms of the fracture-well elements still be calculated independently of each other, albeit with an adjustment to the perforation pressure drop flow term related to the conductivity of *all* the well-fracture connections going through each perforation pressure drop.

A final complication is that a fracture element may connect to the same well element through multiple perforation clusters (because a well element is not restricted to contain only one perforation cluster). To handle this complication, the code independently calculates a well-fracture element flow term for every perforation cluster associated with the connection. In other words, if there are three perforation clusters connecting a well and fracture element, the code calculates three separate flow terms for the same well-fracture element pair, corresponding to each of the perforation clusters.

The figure below shows results from a simulation with 'cased well and fracture collision distance' equal to 50 ft, and cluster spacing equal to 20 ft. The value of the 'transmissibility barrier' is set to 3e6 md-ft. As you can see, there are many clusters that never break down, which is because of the crossflow during stimulation. Also, in this simulation, the third stage (top of the screenshot) is never stimulated. The line plot shows the cumulative production versus time for this unstimulated third stage. This production occurs almost exclusively from cross-flow from the perforations of the third stage to the adjacent stimulated fractures from the second stage.

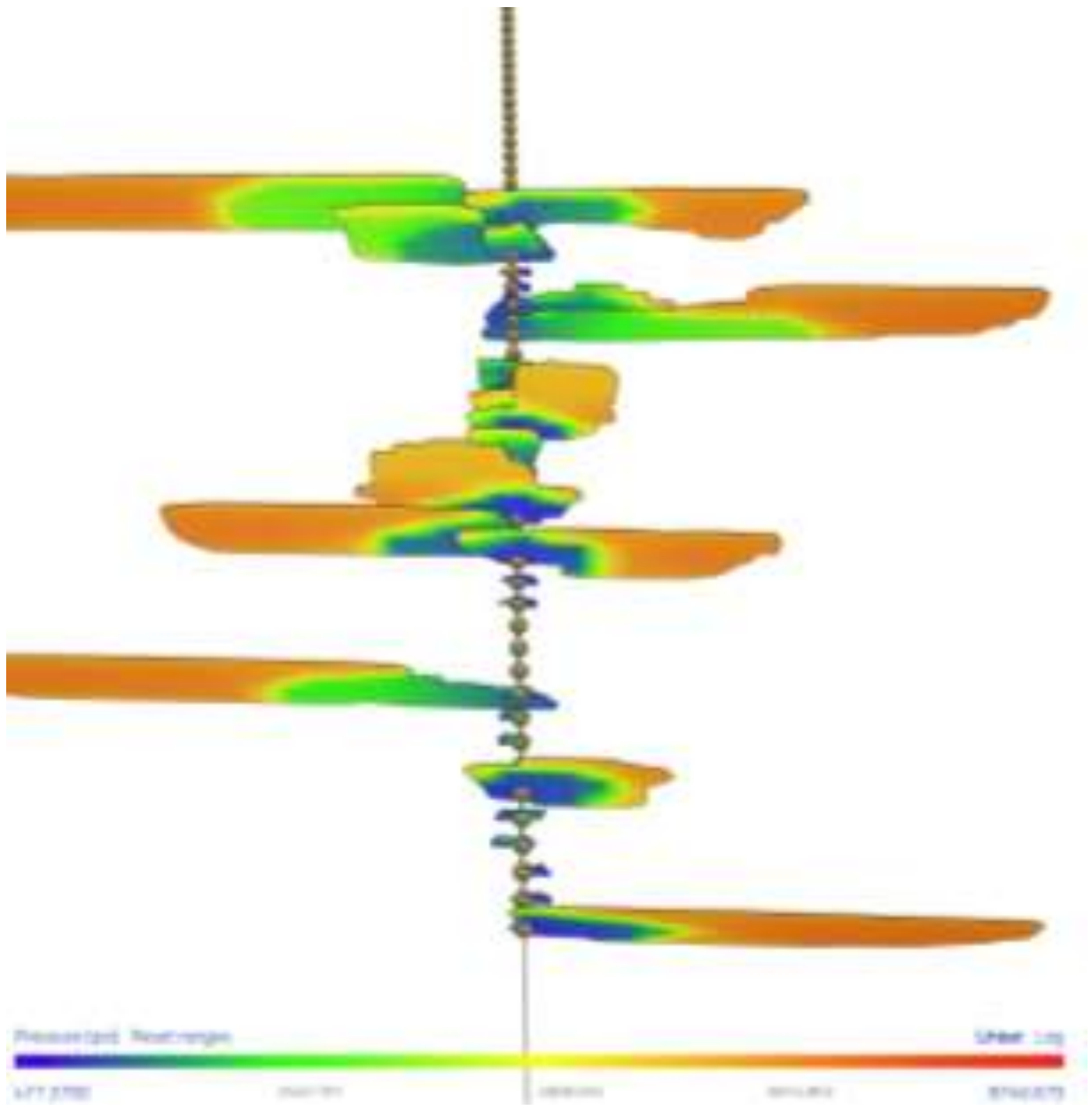

This figure shows the results with the ‘transmissibility barrier’ set to 3e5 md-ft.

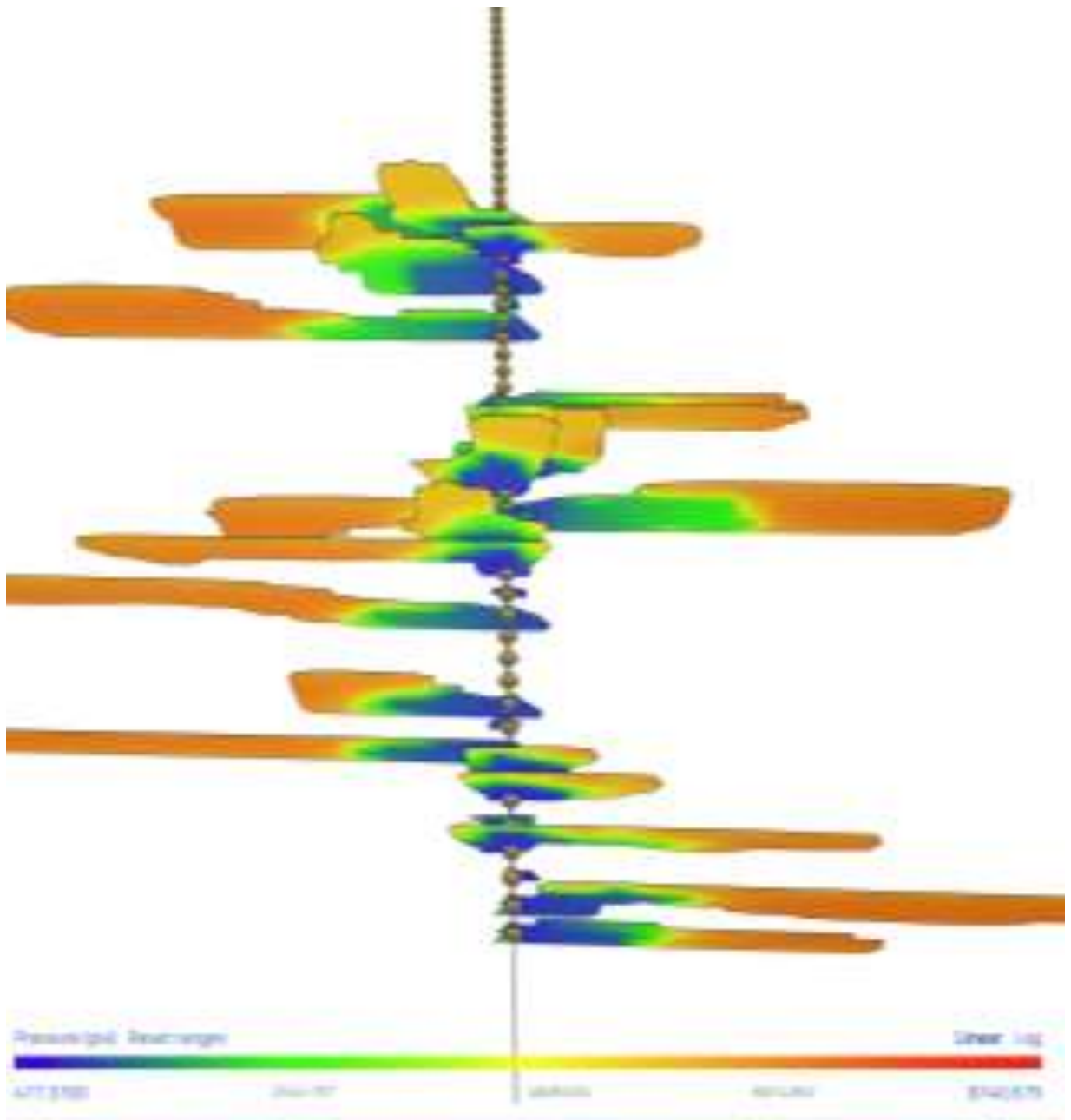

This figure shows the results with the ‘transmissibility barrier’ set to 3e4 md-ft.

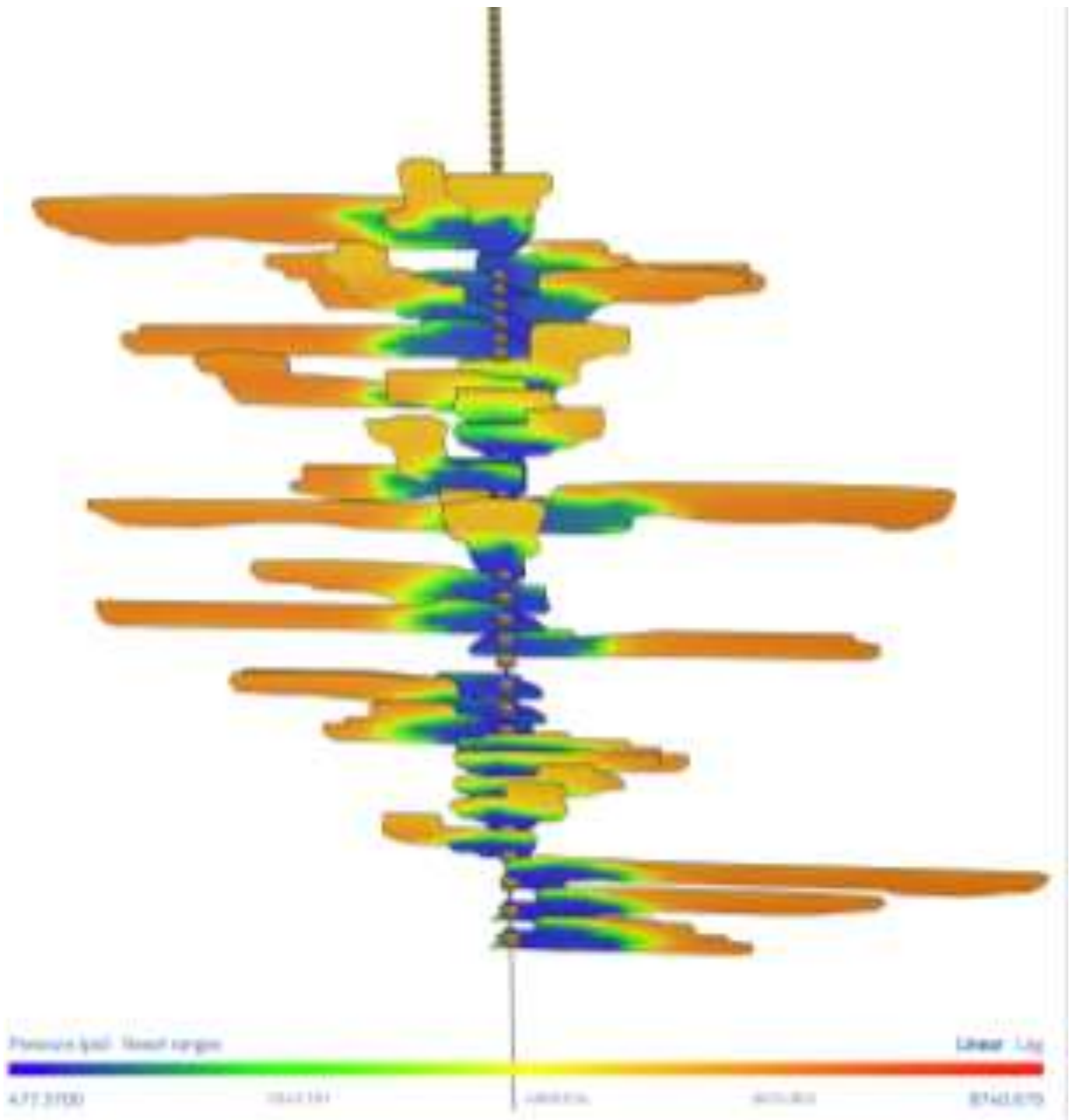

This figure shows the results with the ‘transmissibility barrier’ set to 3000 md-ft.

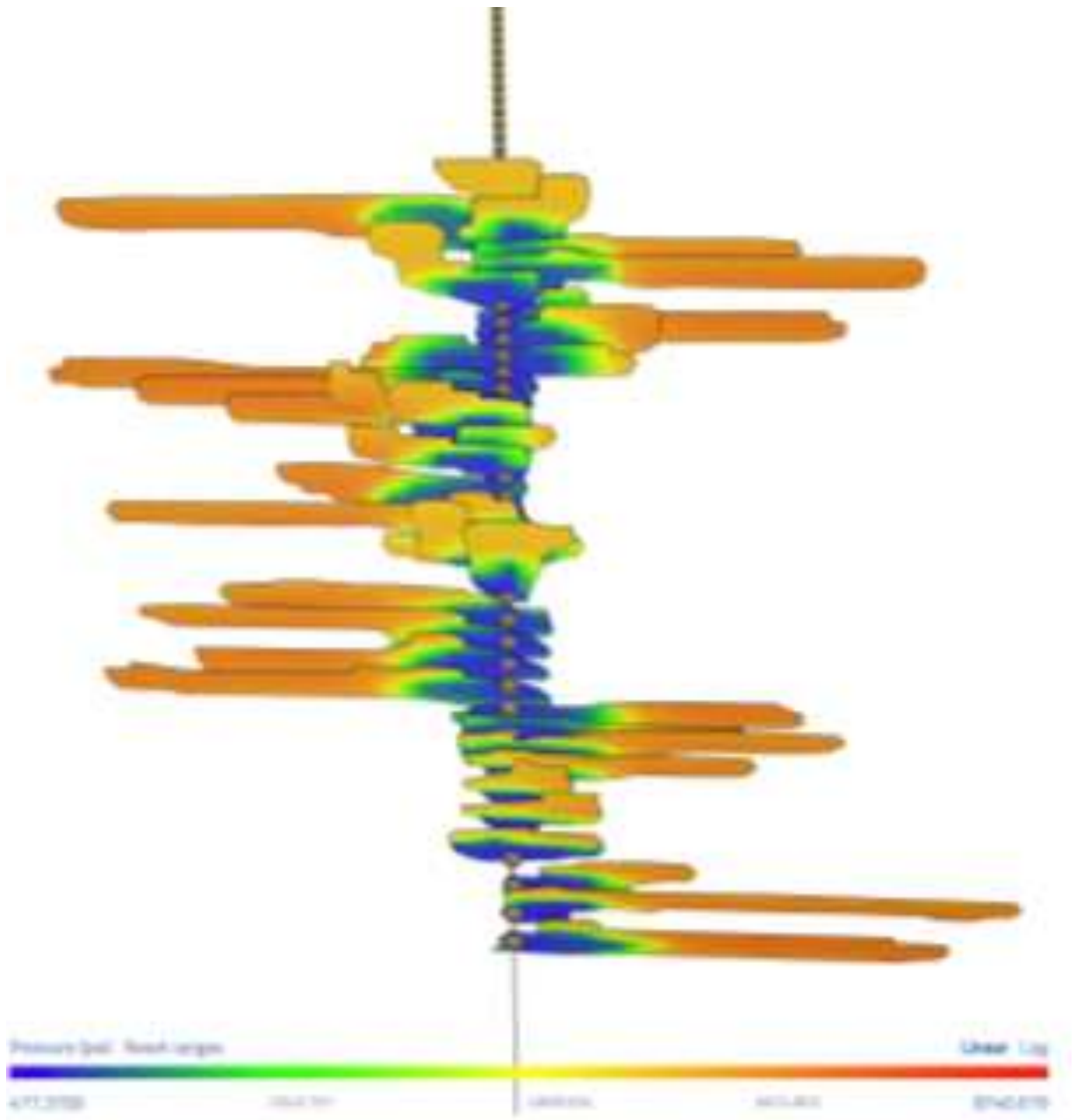

This figure shows the results with the ‘transmissibility barrier’ set to 30 md-ft.

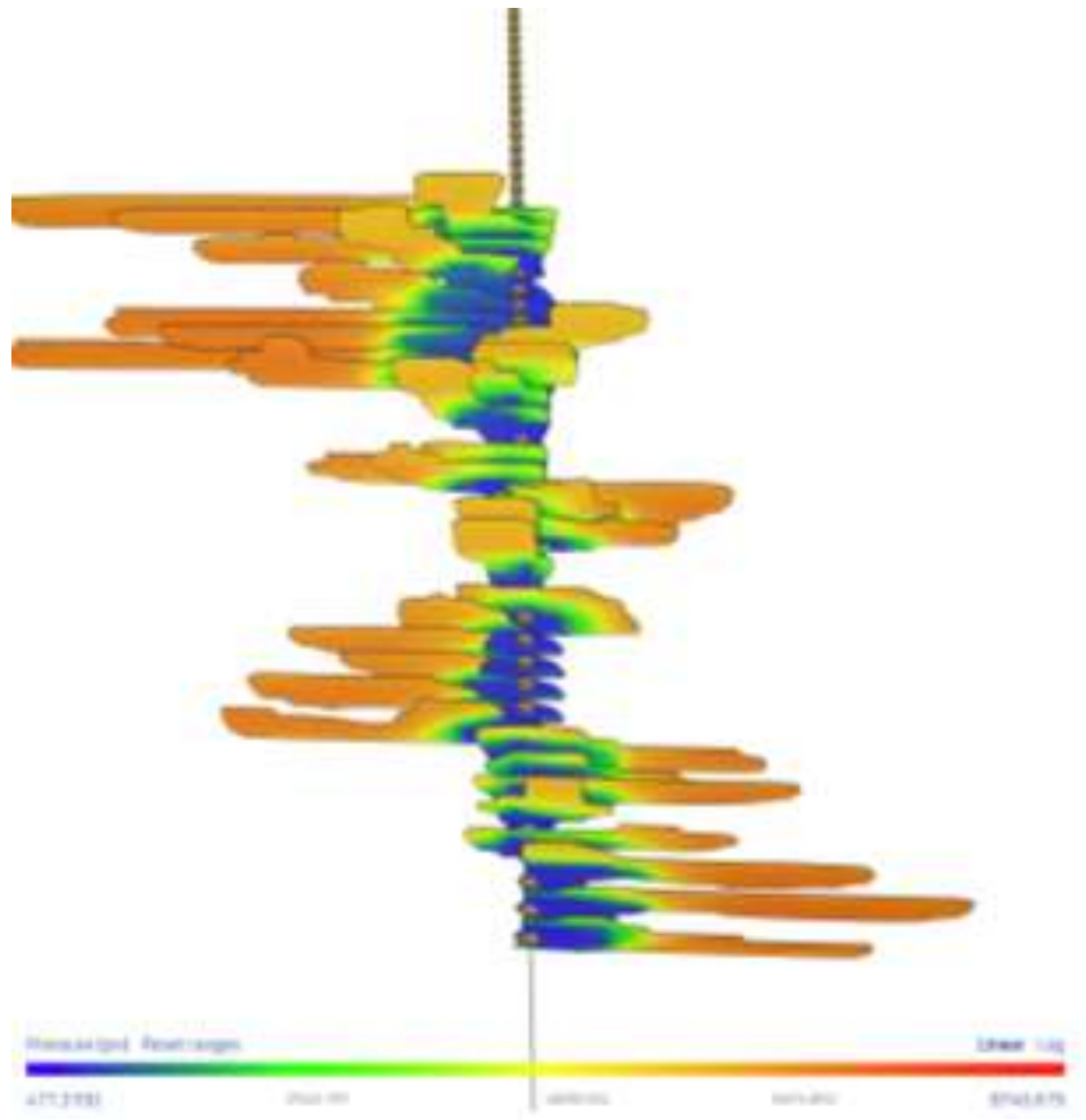

The results indicate that cross-flow during stimulation is significant at 300,000 md-ft, and relatively limited at 30,000 md-ft, and very limited at 3,000 md-ft. During production, flow rates are lower than during stimulation, and so while 3,000 md-ft is enough to prevent cross-flow during stimulation, it does not prevent significant cross-flow during production. On the other hand, going down to 30 md-ft restricts flow sufficiently that cross-flow during production is also significantly reduced.

For a worked example, refer to the ResFrac Blog Post from May 2022, “Automated History Matching and Economic Optimization of an Eagle Ford Refrac.” Note that the settings used for crossflow have changed somewhat over time. Previously, there was an ‘additional near wellbore deltaP’ parameter, but it is now deprecated in favor of the ‘transmissibility barrier’ parameter.

## 10.8 Fracture collisions

Rules-based collision logic to determine when hydraulic fractures collide and form a hydraulic connection. If the fractures had exactly the same orientation, then they wouldn’t be able to collide unless they are perfectly aligned. In reality, they may (or may not) curve inwards towards each other (or outwards away from each other). In ResFrac, there is a parameter called “Fracture collision relative distance (different wells)” that controls this behavior. If you set the parameter to 0.15, and the fracture element length is 50 ft, then fractures will be considered to be ‘collided’ if they propagate within 0.15*50 = 7.5 ft. When fractures ‘collide,’ the simulator forms a hydraulic connection between the elements so that fluid and proppant can flow from one to the other.

After testing, we decided to recommend making the “Fracture collision relative distance (different wells)”a small number, such as 0.01. While the jury is still out (and very likely depends on context), we’ve found so far that data has been best described when we assume that fractures (mostly) do not collide away from the well (ie, a small ‘relative fracture collision distance’).

However, we also recommend setting the ‘cased well and fracture connection distance’ to 10 ft so that collisions do occur between wells and fractures. With this combination of parameters, collisions are not occurring between the wells, but are occurring at the wells.

For one more wrinkle, we recommend making fracture-to-fracture collision more likely when the formation has been depleted. In this case, the stress around the preexisting fracture has been reduced by the depletion, and this could cause the fracture to curve inwards. In testing, we have found that fracture-to-fracture collisions do appear to be more likely to occur in this kind of depleted ‘parent/child’ context. To mimic this in ResFrac, you can define a “Depleted fracture collision distance.” If fractures approach within this distance, and if the rock around one of the fractures has been depleted by an amount equal to the “Fracture collision depletion stress,” then the fractures will ‘collide.’

What about flow ‘outside pipe’ along a well? For this, set “Connect frac through 'cased well fracture collision distance'” to true. With this setting, if two fractures are within ‘cased well and fracture connection distance’ of each other, then they will connect directly to each other, via flow outside the casing. This may be important, for example, because if you place perforation clusters very close-together, then even if limited entry forces fluid to flow out of all of them, fluid may flow outside pipe into a smaller number of dominant fractures. Thus, the flow outside pipe prevents limited entry from creating approximately one propped fracture per perf cluster. The conductivity of these pathways is not well-known, and we do not have a good way of characterizing. Nevertheless, the user does have a way of controlling the conductivity of these ‘outside pipe’ connections, the parameter “Connect frac through 'cased well fracture collision distance' transmissibility multiplier”.

## 10.9 Fracture turning

If you set “straight fractures” to ‘false’, the fractures will be permitted to turn. This capability is only supported with the ‘discrete’ crack propagation algorithm, not the newer ‘continuous’ algorithm.

The propagation direction is calculated using the maximum circumferential stress criterion, as given in Equation 11 from Sheibani and Olson (2013).

The screenshots below show simulation with the new turning option. The first screenshot corresponds to just 100 psi of horizontal stress anisotropy. As expected, there is a large amount of turning and fracture coalescence. The second screenshot shows 1500 psi of stress anisotropy. There is very little fracture turning.

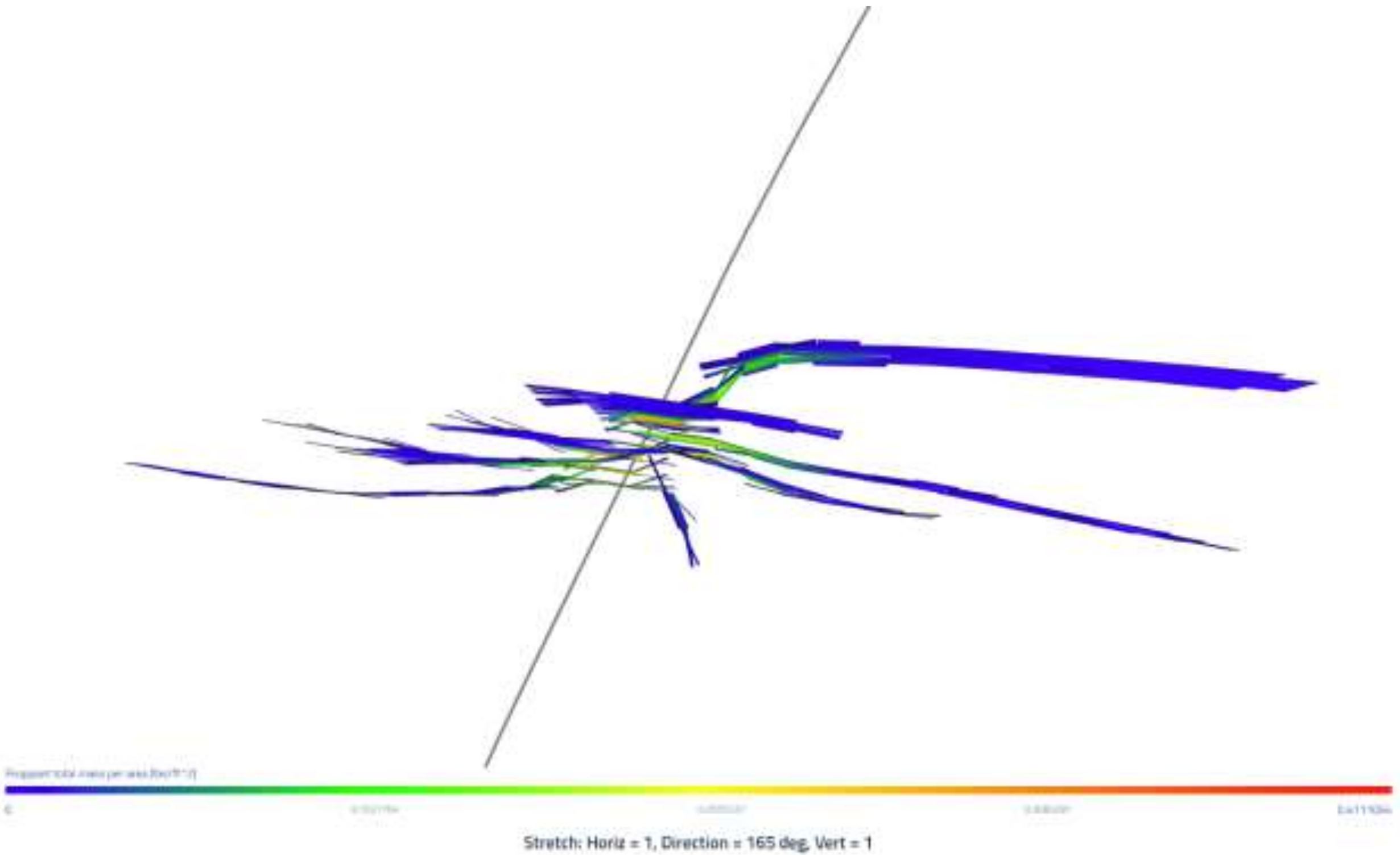

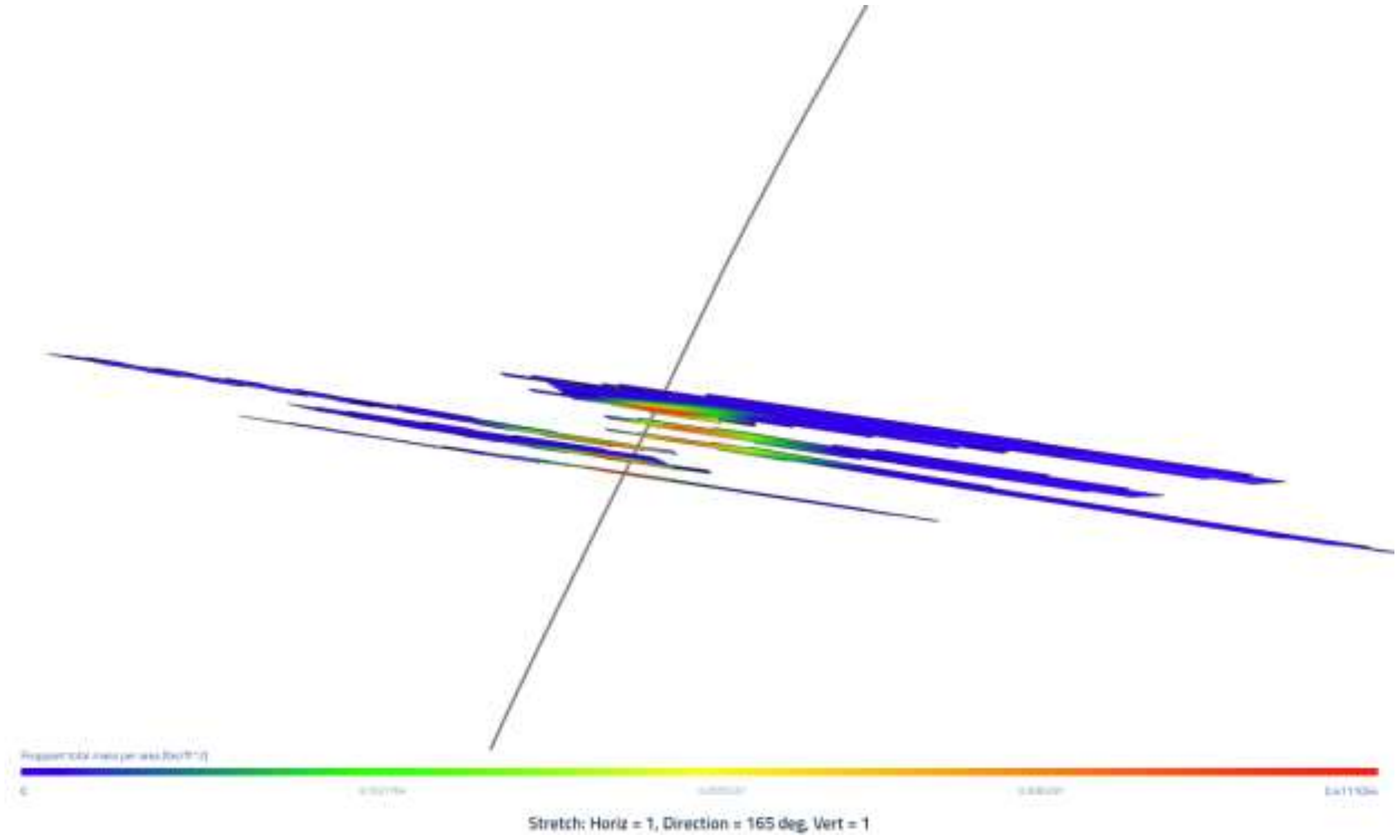

## 10.10 Elastic anisotropy

Transversely anisotropic material is characterized by five independent elastic constants, $E_h$, $E_v$, $\nu_h$, $\nu_v$, and $G_{vh}$. However, the constants $\nu_v$ and $G_{vh}$ are often hard to measure and are typically not available. ResFrac uses $E_h$ (horizontal Young's modulus), $\nu_h$ (Poisson's ratio), and $E_v/E_h$ (vert over horiz modulus) as inputs. This section summarizes the model to calculate $\nu_v$ and $G_{vh}$.

One possibility to generate a three parameter anisotropic material is to upscale a layered material. In particular, if there are two alternating layers with different Young's moduli, the same thickness, and the same Poisson's ratio, then the resultant upscaled anisotropic material is characterized exactly by three elastic constants that can be derived from the properties of individual layers. After some algebraic manipulations, the vertical Poisson's ratio is simply:

$$\nu_v = \nu_h \frac{E_v}{E_h}. \qquad \text{10-4}$$

The expression for the modulus $G_{vh} = C_{44}$ is more complex and is given by:

$$G_{vh} = \frac{1-2\nu_h}{2(1-\nu_h)}\left[\frac{1}{E_v} - \frac{2\nu_h^2}{(1-\nu_h)E_h}\right]^{-1} = \frac{E_h}{2(1+\nu_h)}\left[1 + \frac{(1-\nu_h)}{(1+\nu_h)(1-2\nu_h)}\left(\frac{E_h}{E_v} - 1\right)\right]^{-1}. \qquad \text{10-5}$$

The expression in square brackets is always positive provided that $E_v/E_h \leqslant 1$ and $\nu_h < 0.5$. Note that for an isotropic material (i.e. when $E_v/E_h = 1$) the above expression reduces to $G_{vh} = \frac{1}{2(1+\nu_h)E_h}$. Detailed derivation of the above equations as well as the comparison with other models and laboratory data can be found in arXiv publication Dontsov (2023a).

Once all five elastic constants are known, the next step is to use them to calculate stress interactions between fracture elements. At this point, an approximate solution is used. This approximate handling of anisotropy is used by ResFrac if the 'anisotropic modulus numerics version' is set to 4.

This solution uses stress interaction coefficients calculated using the isotropic solution, but alters the elastic modulus. Most of the fractures are height contained, in which case the relevant effective plane strain modulus is

$$E_v' = 2\left[\frac{C_{33}}{C_{11}C_{33}-C_{13}^2}\left(\frac{1}{C_{44}}+\frac{2}{C_{13}+\sqrt{C_{11}C_{33}}}\right)\right]^{-1/2}, \quad 10\text{-}6$$

where the coefficients $C_{ij}$ can be calculated from $E_h$, $E_v$, $\nu_h$,$\nu_v$, and$G_{vh}$. This modulus applies precisely for a vertically oriented plane strain fracture, see Laubie and Ulm (2014). Thus, this solution works well for the height contained fractures for which the state of stress in the vertical cross-section follows the plane strain condition. As a result, the Young's modulus is set to $E = E_v'(1-\nu_h^2)$ and the Poisson's ratio is $\nu = \nu_h$. In this case, the elastic behavior for height contained fractures is precise, while the result becomes approximate for fractures with substantial height growth.

Al-Shawaf (2023) tested this approach with a PKN-style test problem. He set up a simulation with a single fracture, length much greater than height, zero permeability, and constant pressure injection. The problem setup is shown in the figure below (reproduced from al-Shawaf, 2023). Under these conditions, the crack opens against uniform internal fluid pressure, and can be compared with the plain strain TIV solution from Chertov (2012).

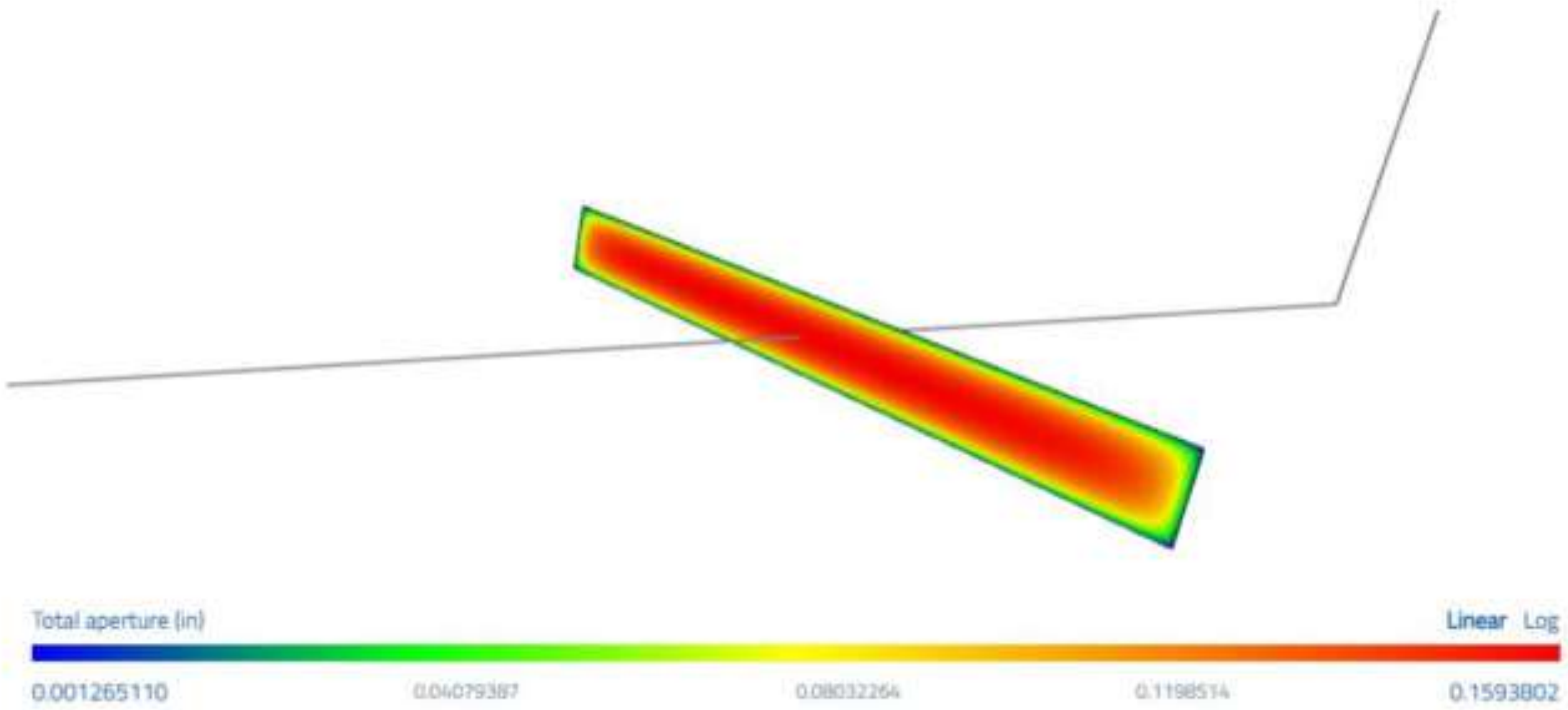

The figure below, reproduced from al-Shawaf, shows the maximum fracture aperture versus the ratio of horizontal and vertical Young's modulus, comparing the analytical solution with the numerical results.

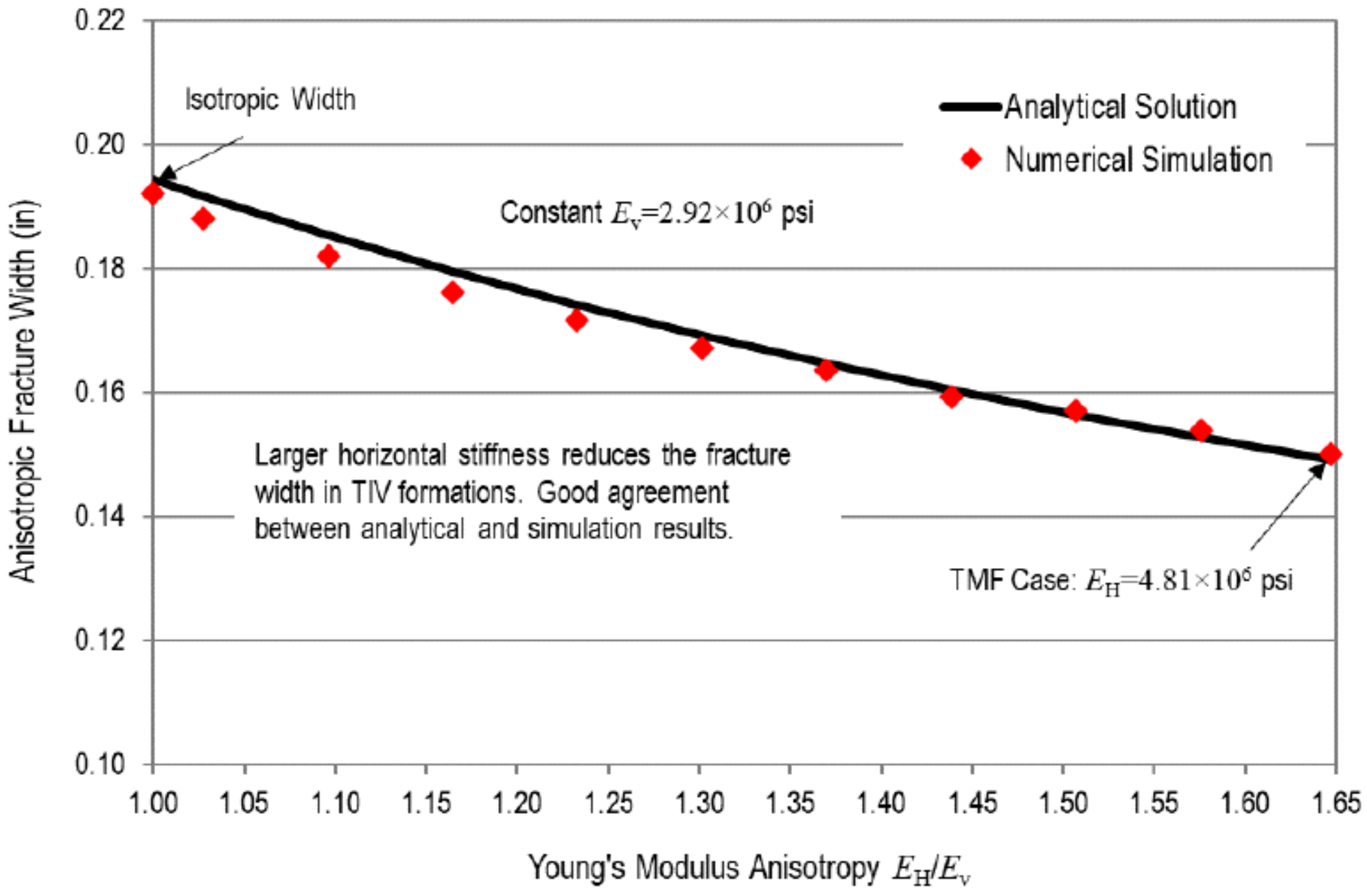

# 11. Poroelastic and thermoelastic stress changes from depletion

## 11.1 General

ResFrac can calculate stress changes in response to pore pressure ($P$) and temperature ($T$) changes in the matrix. These stresses are calculated using the thermoelastic potential function ϕ described in Chapter 2 of Nowacki (1986) or equivalently, the poroelastic potential function $\Phi$ described in Chapter 5 of Wang (2000). The porothermoelastic potential function $\Phi$ is calculated from the equation:

$$\frac{\partial^2 \phi}{\partial x^2} + \frac{\partial^2 \phi}{\partial y^2} + \frac{\partial^2 \phi}{\partial z^2} = \frac{\alpha}{2G}\frac{1-2\nu}{1-\nu}P + \alpha_t \frac{1+\nu}{1-\nu}T, \qquad 11\text{-}1$$

where x, y, and z are the Cartesian coordinates, $\nu$ is the Poisson ratio, $G$ is the shear modulus, $\alpha$ is the Biot coefficient, and $\alpha_t$ is the thermal coefficient of linear expansion. This equation is solved for the thermoelastic potential function ϕ in the matrix elements using the finite difference method, taking into account the distribution of pore pressure ($P$) and temperature ($T$) in every matrix element. The induced stress change in each element is calculated from the second derivatives of the potential function. The resulting changes in stress on fracture elements are calculated with trilinear interpolation onto the fracture using the stresses calculated in the surrounding elements. The porothermoelastic potential method is only exact for homogeneous elastic moduli. ResFrac optionally allows the user to specify spatially variable elastic moduli. The impact of heterogeneity on the porothermoelastic stresses is calculated with an approximate method. The 'source' term in the elastic potential function is calculated using the moduli at the location of pore pressure/temperature change. The changes in stress at any given location are calculated from the elastic potential function using the moduli at that location.

As described in Section 4.10 by Wang (2000), in an infinite homogeneous domain, porothermoelastic stress changes do not affect the total volumetric strain (sum of the diagonal of the strain tensor) outside the region where pore pressure is changed. In other words, if pore pressure is emplaced or injected in one location, there is not a corresponding volumetric strain or poroelastic pressure response at another location. This decouples the flow and deformation problems in the matrix. This is the underlying justification for the typical assumption in conventional reservoir simulators that pore volume in an element can be calculated solely from pore pressure in that element. Note that this is only strictly valid in the special case of a homogeneous, elastic, infinite domain (which is currently the assumption used by ResFrac for these calculations). ResFrac allows the user to specify spatially variable elastic moduli. Nevertheless, in all cases, ResFrac uses the assumption that porothermoelastic stress changes do not affect the nonlocal total volumetric strain.

Even though the volumetric stress/strain (sum of the diagonal of the stress/strain tensor) does not change, the strain and stress in individual directions do change (each individual number in the stress tensor). Therefore, the poroelastic and thermoelastic stresses may strongly affect the normal stress that is resolved on individual fracture elements. Thus, the poroelastic and thermoelastic stress changes are strongly coupled to the fracture flow/deformation, but are decoupled from the pore pressure/pore volume calculation in the other matrix elements.

The boundary conditions for the porothermoelastic stress calculations are ‘no displacement,’ which is equivalent to being a ‘symmetric’ boundary condition. In other words, the boundary condition assumes that the pore pressure/temperature change occurring in the problem domain is also occurring in rock on the other side of the boundary. This boundary condition is very well-suited for sector models, where the edge of the matrix region in the Shmin direction corresponds to the boundary with an adjacent sector (ie, we are modeling a particular stage or group of stages within a lateral, and so it is surrounded by other stages on either side). The stage(s) in the simulation’s problem domain are experiencing similar pressure/temperature changes as the stages on either side, and so the calculation of stresses induced by deformation should account for those changes that are also occurring in the surrounding regions.

For boundaries that do not involve the edges of a ‘sector’ model (ie, the edge of the model in the SHmax direction, or the top/bottom of the matrix region), we recommend that you make the edge of the simulation mesh sufficiently distant from the pore pressure/temperature changes occurring in the simulation that results are unaffected by the boundary.

For the problem formulation to remain valid, at least one boundary (e.g., the top or bottom) should be placed sufficiently far from the changes in pore pressure and temperature occurring in the simulation, so as to effectively represent a boundary located at 'infinity'.

The porothermoelastic potential only allows the computation of irrotational displacement fields. When solving for an injection in an infinite medium, the resulting displacement field is irrotational. Due to the linearity of the porothermoelastic problem, displacement fields from multiple injections can be superimposed. It’s important to note that the superposition of irrotational displacement fields preserves their irrotationality. The presence of planes of symmetry does not affect this property, meaning we can solve the problem by considering only one side of the domain, or one sector in the case of multiple symmetry planes. In the end, while the porothermoelastic potential only enables the computation of irrotational fields, this is precisely what is needed for our solution.

To relate pore pressure and temperature changes to stress changes, users input the Biot coefficient $\alpha$ and thermal coefficient of linear expansion $\alpha_t$.

Poro and thermoelastic stresses are activated using options in the “Startup” panel. The Biot coefficient $\alpha$ and thermal coefficient of linear expansion $\alpha_t$ specified in the “Geologic units (facies list)” table in the “Static Model and Initial Conditions” panel.

While it is common to assume that Biot coefficient $\alpha$ is a constant, in shale, it is strongly sensitive to effective stress (Ma and Zoback, 2017). ResFrac gives users the option to specify a table of Biot coefficient multipliers versus pressure change.

## 11.3 The symmetric boundary condition

The symmetric boundary condition is implemented by imposing zero displacement in direction normal to the boundary, at all the boundary points. In other words, if $\mathrm{u}^T(P) = [u_x \quad u_y \quad u_z]$ is the displacement at a boundary point $P$, and $\mathrm{n}^T(P) = [n_x \quad n_y \quad n_z]$ is a vector normal to the boundary, the boundary condition is

$$\mathrm{u}^T \cdot n = 0, \qquad 11\text{-}2$$

and because the displacement is calculated from the porothermoelastic potential function ϕ by derivation as

$$u_x = \frac{\partial \phi}{\partial x}, \qquad u_y = \frac{\partial \phi}{\partial y}, \qquad u_z = \frac{\partial \phi}{\partial z}, \qquad 11\text{-}3$$

the symmetric boundary condition is simply written as

$$\nabla \phi^T \cdot n = 0. \qquad 11\text{-}4$$

To impose symmetry with a symmetric boundary condition, both the normal displacements and the tractions parallel to the boundary must be set to zero. However, in the current formulation (which utilizes the porothermoelastic potential), setting the normal displacements at the boundary to zero is sufficient to simulate a symmetric boundary condition, as the tractions parallel to the boundary are automatically zero. The proof is provided below.

We now demonstrate that ensuring symmetry in the solution to the porothermoelastic problem with respect to a given boundary --- such as the plane $x_3 = 0$ --- requires only the boundary condition $\partial \phi / \partial x_3 = 0$. In this approach, it is unnecessary to impose that the tractions parallel to the symmetry plane are zero.

### 11.3.1 Implications of assuming a ’smooth’ potential $\phi$

For a function $\phi(\mathrm{x}_1, \mathrm{x}_2, \mathrm{x}_3)$, the mixed partial derivatives, $\frac{\partial^2 \phi}{\partial \mathrm{x_i}\, \partial \mathrm{x_j}}$ and $\frac{\partial^2 \phi}{\partial \mathrm{x_j}\, \partial \mathrm{x_i}}$, will be equal if $\phi$ is continuously differentiable, meaning that $\phi$ belongs to the class of $C^2$ functions (functions with continuous second derivatives).

This result is a consequence of *Schwarz’s theorem* (also known as the *symmetry of second derivatives*), which states that for any smooth (continuously differentiable) function, the mixed partial derivatives are equal if the function is twice differentiable.

On the other hand, as it will be recalled later, the second derivatives are used to compute the stresses from the potential ϕ. Therefore, assuming continuous second derivatives for ϕ implies the continuity of the stress field. Mathematically, for any function ϕ that has continuous second partial derivatives, we have:

$$\frac{\partial^2 \phi}{\partial x_i \partial x_j} = \frac{\partial^2 \phi}{\partial x_j \partial x_i}. \quad 11\text{-}5$$

This condition holds regardless of the specific form or domain of ϕ as long as it satisfies the required smoothness criteria.

In terms of the displacement field $u = \nabla\phi$, these expressions are equivalently written as:

$$\frac{\partial u_i}{\partial x_j} = \frac{\partial u_j}{\partial x_i}. \quad 11\text{-}6$$

## 11.3.2 Strain-Displacement Relation

The symmetry in Eq. 11-6 implies that the strain tensor $\epsilon_{ij}$ is symmetric, as it contains the terms $\partial u_i / \partial x_j$ and $\partial u_j / \partial x_i$. In fact, the strain tensor $\epsilon_{ij}$ is:

$$\varepsilon_{ij} = \frac{1}{2}\left(\frac{\partial u_i}{\partial x_j} + \frac{\partial u_j}{\partial x_i}\right). \quad 11\text{-}7$$

For a displacement field $u = \nabla\phi$ in which the potential ϕ is 'smooth' this becomes

$$\varepsilon_{ij} = \frac{1}{2}\left(\frac{\partial^2 \phi}{\partial x_i \, \partial x_j} + \frac{\partial^2 \phi}{\partial x_j \, \partial x_i}\right) = \frac{\partial^2 \phi}{\partial x_i \, \partial x_j}. \quad 11\text{-}8$$

## 11.3.3 The Stress Tensor Components

The stress tensor $\sigma_{ij}$ is related to the strain tensor by Hooke's law:

$$\sigma_{ij} = 2\mu\epsilon_{ij} + [\lambda\, \epsilon_{kk} - (2\mu + 3\lambda)\alpha_t T - \alpha P]\delta_{ij}, \quad 11\text{-}9$$

where λ and μ are the Lamé constants, $\varepsilon_{kk}$ is the volumetric strain (the trace of the strain tensor), and $\delta_{ij}$ is the Kronecker delta (1 if $i = j$, and 0 otherwise).

Now, substituting the expression for the strain components in terms of the potential ϕ we get, for the diagonal components ($i = j$)

$$\sigma_{ii} = 2\mu \frac{\partial^2 \phi}{\partial x_i \partial x_i} + [\lambda \nabla^2 \phi - (2\mu + 3\lambda)\alpha_t T - \alpha P], \quad 11\text{-}10$$

or, when accounting for Eq. (11-1)

$$\sigma_{ii} = 2\mu\left(\frac{\partial^2 \phi}{\partial x_i \partial x_i} - \nabla^2 \phi\right), \quad 11\text{-}11$$

and for the off-diagonal components ($i \neq j$)

$$\sigma_{ij} = 2\mu \frac{\partial^2 \phi}{\partial x_i \partial x_j} \quad \text{for } i \neq j. \tag{11-12}$$

Thus, the off-diagonal stress components $\sigma_{ij}$ depend directly on the second derivatives of the potential ϕ.

### 11.3. 4 Traction Vector on the Plane $x_3 = 0$ and boundary condition $\partial\phi/\partial x_3 = 0$

The traction vector $t$ on the plane $x_3 = 0$ is given by the Cauchy traction law:

$$t_i = \sigma_{ij} n_j, \tag{11-13}$$

where $\sigma_{ij}$ are the components of the stress tensor, and $n_j$ are the components of the unit normal vector to the plane.

For the plane $x_3 = 0$, the unit normal vector is $n = (0,0,1)$, since the plane is perpendicular to the $x_3$-axis. Substituting $n_1 = 0$, $n_2 = 0$, and $n_3 = 1$ into the traction law, the traction vector becomes:

$$t_i = \sigma_{i3} \quad \text{on the plane} \quad x_3 = 0. \tag{11-14}$$

In Cartesian coordinates $(x_1, x_2, x_3)$, the components of the traction vector are:

- Tangential component in the $x_1$-direction: $t_1 = \sigma_{13}$,
- Tangential component in the $x_2$-direction: $t_2 = \sigma_{23}$,
- Normal component in the $x_3$-direction: $t_3 = \sigma_{33}$.

In terms of the potential function $\phi$, the component of the traction vector on the plane $x_3 = 0$ are

$$t_1 = 2\mu \frac{\partial^2 \phi}{\partial x_3 \, \partial x_1}, \qquad t_2 = 2\mu \frac{\partial^2 \phi}{\partial x_3 \, \partial x_2}, \qquad t_3 = 2\mu \left( \frac{\partial^2 \phi}{\partial x_i \partial x_i} - \nabla^2 \phi \right). \tag{11-15}$$

Because the boundary condition on the plane $x_3 = 0$ is $\partial\phi/\partial x_3 = 0$, differentiating it with respect to $x_1$ and $x_2$

$$\frac{\partial^2 \phi}{\partial x_1 \, \partial x_3} = \frac{\partial^2 \phi}{\partial x_2 \, \partial x_3} = 0. \tag{11-16}$$

allows us to conclude that the traction components $t_1$ and $t_2$ on the plane $x_3 = 0$ are zero:

$$t_1 = 2\mu \frac{\partial^2 \phi}{\partial x_3 \partial x_1} = 0, \qquad t_2 = 2\mu \frac{\partial^2 \phi}{\partial x_3 \partial x_2} = 0 \qquad \text{on the plane } x_3 = 0. \tag{11-17}$$

### 11.3.5 Symmetry of ϕ and boundary condition $\partial\phi/\partial x_3 = 0$

In case the scalar potential ϕ is symmetric with respect to the plane $x_3 = 0$, then:

$$\phi(x_1, x_2, x_3) = \phi(x_1, x_2, -x_3). \qquad 11\text{-}18$$

This equation states that ϕ at a point $x_3$ is equal to its value at $-x_3$, reflecting the symmetry of the system about the plane $x_3 = 0$.

To demonstrate that this is equal to the condition

$$\frac{\partial \phi}{\partial x_3} = 0 \quad \text{on the plane} \quad x_3 = 0, \qquad 11\text{-}19$$

take the partial derivative of both sides of the symmetry equation Eq. (11-18) with respect to $x_3$:

$$\frac{\partial \phi}{\partial x_3} = \frac{\partial \phi}{\partial(-x_3)} \cdot \frac{\partial(-x_3)}{\partial x_3}. \qquad 11\text{-}20$$

On the right-hand side, $\partial(-x_3)/\,\partial x_3 = -1$, so we have:

$$\frac{\partial \phi}{\partial x_3} = -\frac{\partial \phi}{\partial x_3}. \qquad 11\text{-}21$$

The only solution to this equation is:

$$\frac{\partial \phi}{\partial x_3} = 0 \quad \text{on the plane} \quad x_3 = 0. \qquad 11\text{-}22$$

This derivation shows that if ϕ is symmetric about the plane $x_3 = 0$, then its derivative in the $x_3$-direction, $\partial\phi/\,\partial x_3$, must vanish on the plane. The reasoning can be also applied in the opposite direction.

## 11.4 Implementation of the symmetric boundary condition

Two, well-known, mathematical difficulties arise when calculating the porothermoelastic potential function ϕ from Equation (11-1), and the boundary conditions given in Equation (11-4) (Bramble and Hubbard 1965).

- The first is that, given a pore pressure and temperature distributions, a solution that satisfies the imposed zero-displacement boundary condition may not exist. This is, for example, the case in which the pore pressure, the temperature, or both, increase uniformly of a given amount over the whole domain. In this case the body (normally) expands uniformly at each point, including the ones at the boundary. Therefore, in general, the average displacement at the boundary can be any value. This can be immediately verified by applying Green's first identity to Equation (11-1), and by recalling the boundary conditions given in Equation (11-4), thus obtaining:

$$\int_{\Omega}\left(\frac{\alpha}{2G}\frac{1-2\nu}{1-\nu}P+\alpha_t\frac{1+\nu}{1-\nu}T\right)\mathrm{dx\,dy\,dz}=\underbrace{\int_{\partial\Omega}\overbrace{\nabla\Phi^T\cdot n}^{\substack{\text{displacement}\\\text{normal to}\\\text{the boundary}}}\ \mathrm{dS}\ =0}_{\text{From the boundary conditions}}, \qquad \text{11-23}$$

where $\Omega$ indicates the 3D domain, $\partial\Omega$ the domain's boundary, and $\mathrm{d}S$ is an infinitesimal area element of the boundary $\partial\Omega$. The first equality from the left is obtained by applying Green's identity and it is equivalent to require that the average temperature in the domain must be proportional to the average displacement at its boundary. The second equality from the left in (11-23), is simply obtained by substituting the boundary condition given in Equation (11-4). Together, these Equations form a necessary condition for the existence of a solution to Equations (11-1) and (11-4).

- The second difficulty arise because integrating twice Equation (11-1) requires knowing the value assumed by the porothermoelastic potential function ϕ (at least) at one point, and we do not know ϕ at any point. In fact, the zero displacement boundary conditions in Equation (11-4) involve only the first derivatives of the porothermoelastic potential function ϕ, and does not give any information about ϕ. Therefore, the solution is not uniquely determined. However if, on one hand, it is true that in case the solution exists the problem admits infinite many solutions, on the other hand all these differ by a constant value.

These mathematical properties of the system of equations to be solved are also present in the discretized system of equations

$$L\Phi = b, \qquad \text{11-24}$$

where $L$ is the matrix obtained by applying the finite difference method to the left-hand side of Equation (11-1), $\Phi$ is a vector containing the unknown values of porothermoelastic potential center of the matrix elements, and $b$ is a vector containing the source term in Equation (11-1). In fact, from the fundamental theorem of linear algebra (see for example Strang 1986, page 72):

- The linear system (11-24) has a solution only if $b$ is in the range of $L$, or correspondingly $b$ is orthogonal to the nullspace of $L^T$. In other words, in case the nullspace of $L^T$ is not empty (which is the case), the discrete system does not have solution for any source term $b$.
- If the vector $\Phi$ is a solution of the linear system (11-24), then the system has infinite solutions of the form $\Phi + \alpha v$, in which $\alpha$ is a real number and $v$ is the vector in the nullspace of $L$. In fact, by construction, the matrix $L$ is a non-symmetric squared matrix of rank one less than its order and presents a constant nullspace. In other words, one of the eigenvalues is zero and the corresponding eigenvector $v$ is a constant vector such that $v^{\mathrm{T}} = [1 \quad \dots \quad 1]$.

Note that either the two mathematical difficulties presented before, or the corresponding two numerical properties of the discretized system of equations just presented, do not represent an actual obstacle in computing the porothermoelastic stress change.

- An unconstrained uniform expansion, or contraction, of an infinite space does not create a stress change. Because a uniform expansion, (or contraction) is the result of a uniform temperature or pore pressure increment, (or decrement) over the entire space, if the average temperature, or pore pressure, are subtracted from the respective fields, the final stress change will be unaffected, and the necessary condition for the existence of a solution in Equation (11-23) will be satisfied.
- The existence of multiple solutions to the system (11-24) does not affect the calculation of the porothermoelastic stress change because this is obtained from the second derivative of the porothermoelastic potential $\Phi$.

These considerations allow to modify the right-hand side $b$ of the original linear system of equations (11-24), and to solve the system

$$L\Phi = \overline{b}, \tag{11-25}$$

in which $\overline{b}$ is $b$ orthogonalized to the nullspace of $L^T$. Additionally, when solving the linear system (11-25) using a Krylov method, the approximate solution is kept orthogonal to the nullspace $v$ of $L$ ($v^{\mathrm{T}} = [1 \quad \dots \quad 1]$) between different iterations.

# 12. Boundary conditions

## 12.1 General

You can specify a variety of boundary conditions: constant rate production, constant rate injection, constant pressure, or shut-in. With each type of boundary condition, you specify constraints. With constant rate injection, maximum injection pressure is specified. If the maximum pressure is reached, the simulator switches to a constant pressure constraint as long as the rate remains below the specified target injection rate. If constant rate injection is specified but the well produces at the maximum pressure (an unusual but not impossible scenario), then well is shut-in rather than allowing the well to produce fluid. With constant rate production, you specify a minimum production pressure, and it switches to constant pressure production if that is reached. Again, if it (oddly) tries to inject at the minimum production pressure, the well is shut-in. With constant pressure boundary conditions, you specify a maximum injection rate and a maximum production rate.

The well constraints are imposed implicitly so that if a constraint is violated during a timestep, the boundary condition type is changed and the timestep is repeated.

When specifying production rates, you are given the option of specifying water rate, oil rate, gas rate, liquid rate, or total rate. All of these rates are volumetric rates evaluated at surface conditions after the fluid has been put through the surface separation facilities. When you specify 'total rate', this is calculated as STB oil + STB water + Mscf gas. If you are using metric units, it is calculated as m^3 oil + m^3 water + 1000 m^3 gas.

When specifying injection rates, you specify the relative ratios of water, oil, and gas injected (with the black oil model) and water and hydrocarbon (with the compositional model). These are volume ratios, evaluated at standard conditions. Under injection conditions, you specify the composition of injection fluid (which may be pure water or include flash components), concentration of each proppant type, concentration of each type of water solute, and temperature.

Production and injection rates are described in terms of volumes – stock tank barrels (STB) of oil and water and standard cubic feet (scf) of gas. This can be confusing to non-petroleum engineers because 'volume' is not a conserved quantity. There is conservation of mass and moles, but not conservation of volume (because volume is a function of density, which is non-constant). Our petroleum engineering volumes are evaluated at 'standard' conditions. At specified temperature and pressure, density can be considered specified (and constant) and so standard 'volumes' are really proxies for mass or moles (because the volumes can be divided by the reference densities to convert to mass or moles). In reality, density is also a function of composition, and so it is an approximation to assume that densities are defined solely by pressure and temperature.

Section 14.2 describes how reservoir production rates are converted to surface rates (STB and scf) with the black oil model and the compositional model. The calculation accounts for the phase changes occurring in the surface separation units. Section 14.2 also describes how injection volumes (STB and scf) are converted to reservoir conditions.

All well controls are specified in the "Well Controls" panel.

## 12.2 Boundary condition location

The boundary conditions can be specified as 'wellhead,' 'MD', or 'bottomhole' constraints. If specified bottomhole, then the wellbore is removed from the model and the conditions are applied directly on connections between the reservoir and the wellbore. This effectively turns off wellbore storage. The bottomhole constraints can be useful for modeling production. ResFrac does not have a detailed multiphase wellbore flow model, and so during multiphase production, it is most convenient to simply produce at a specified bottomhole pressure.

A drawback to bottomhole constraints is that they can have issues with crossflow within the well. In certain (fairly uncommon) circumstances, fluid may be trying to flow out of the well at some fractures as it tries to flow into the well at others, during overall net production to the surface. During production, ResFrac does not permit that kind of 'outflow' from the bottomhole pressure constraints, which means the crossflow is not being completely realistically handled.

The 'MD' boundary condition type alleviates the crossflow problem, while also avoiding the problem of needing to model artificial lift or multiphase flow in the vertical section of the well. MD constraints can be defined for production sequences, general sequences, or shut-ins.

In an MD constraint, part of the well is included in the model. The user defines the measured depth at which the boundary condition is placed within the well. Downhole from the boundary condition, the wellbore is meshed and fully included in the simulation. This allows for cross-flow, or any other combination of behaviors. Uphole from the boundary condition, the wellbore elements are not included in the model. In practice, you should either define the MD boundary condition at the bottom of the vertical section or within the lateral right next to the matrix region. This way, if you have estimates for bottomhole pressure versus depth, you can still use those to impose on the MD constraint.

For example, in the image below, the grey regions of the well are uphole from the boundary condition. The colored regions downhole from the boundary condition are included in the model.

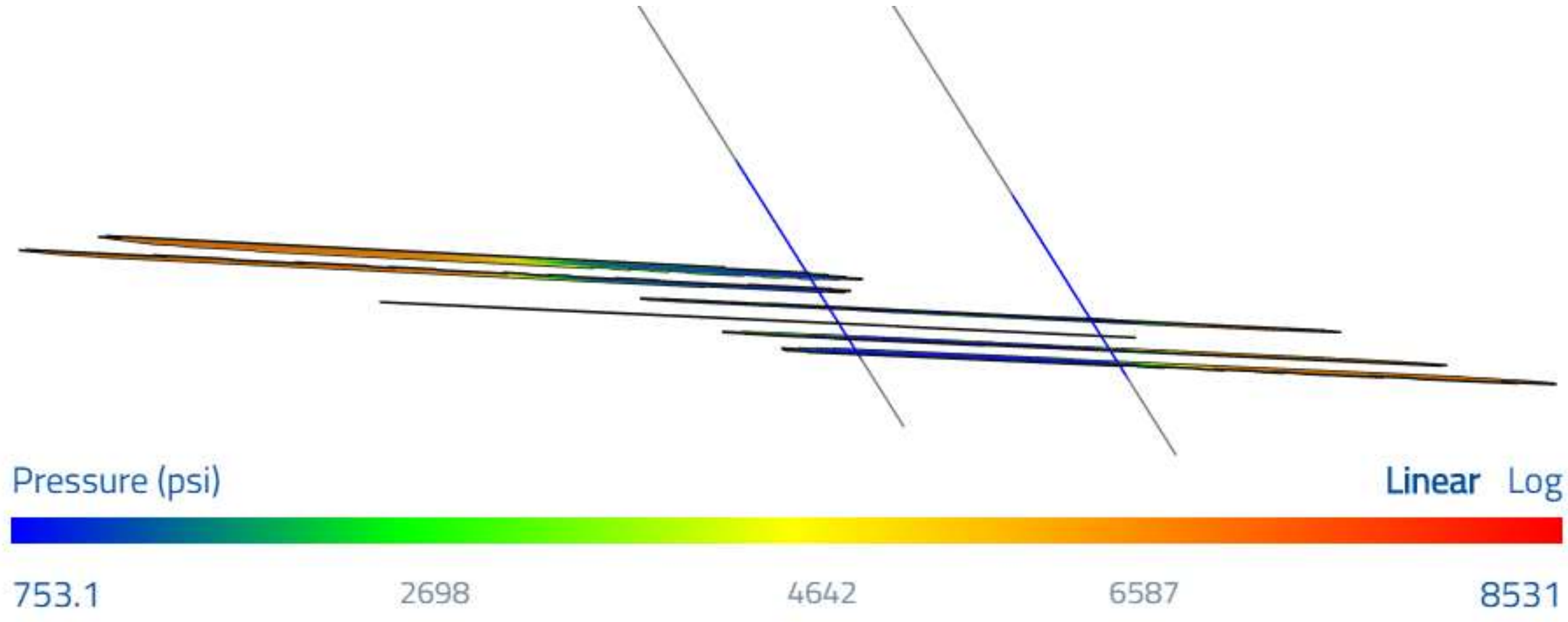

By default, the simulator 'cleans out' the wells at the start of MD production sequences. This removes any proppant from the well, and can cause a partial reset in the pressure distribution. This can be turned off by setting 'clean out well at start of MD production' to false.

## 12.3 Stages

Each wellbore vertex can be assigned to a stage. This effectively divides the wellbore into segments, each assigned to a particular stage number. When you specify boundary conditions, you can specify that all stages are active, some stages are active, or none are active. If all are active, the entire well is open to flow. If some are active, only those stages you specify are open to flow. In the inactive stages, there is no hydraulic connection between the wellbore elements and fracture or matrix elements. In all sections of the wellbore with greater measured depth than the furthest active stage, the wellbore elements are hydraulically isolated from the formation and even from each other. This is intended to capture the isolation created by packers in previously pumped stages.

When the well is shut-in, you can specify a total shut-in by specifying that there are no active stages. If you do this, the simulator acts as if the well does not exist. There is no flow between wellbore elements or between the well and matrix or fracture elements. If you do not specify 'no active stages,' and you do not specify a 'bottomhole' boundary condition constraint, then there can be crossflow through the well. Fluid may enter the well in some places and flow back out of the well in others. If you specify a bottomhole boundary condition constraint and the well is shut-in, then the net fluid flow in/out of the wellbore is imposed to be zero, but crossflow is permitted.

## 12.4 MD shut-ins with storage

A 'storage' shut-in is a special type of MD boundary condition used to model cross-flow between stages within a shut-in well. The idea is that when a frac hit occurs, fluid will flow into the well from the fractures in that stage, flow through the well to the fractures in the other stages, and then flow back out. This impacts the magnitude of the pressure response. If the crossflowing fluid was flowing uphole into a sealed, water saturated wellbore, the storage coefficient of the uphole section would be small, and the pressure response relatively large. However, if the fluid is flowing up into a oil/gas/water filled wellbore section that is connected to a huge number of fractures (which can leak off into the rock), then storage coefficient is far larger, and the pressure response may be far smaller. The MD shut-in with storage mimics this effect.

In this constraint, fluid is permitted to flow in or out of the boundary condition that is placed within the well, kind of like a constant pressure constraint. However, the pressure at the boundary condition changes over time as fluid flows in or out. The pressure changes are calculated assuming that the boundary condition is a constant volume tank. As fluid is added to a tank, conservation of mass requires that the mixture density increase, and so pressure must increase. The converse is true when fluid flows out of a tank. The pressure change for an increment of flow depends on the composition of the mixture in the tank. When the storage shut-in is started, the composition of the 'tank' is initialized as the average composition of the fracture elements connected to the well. The user specifies the volume of the 'tank' uphole from the boundary condition.

Note that fluid that flows in or out of the boundary condition during a 'storage' shut-in is not counted in the 'cumulative' production or injection that is reported by the simulator in the tracking file.

## 12.5 Wellbore cleanouts

Under certain conditions, the simulator performs a 'wellbore cleanout.' In a wellbore cleanout, if there is any proppant in the wellbore, it is removed; the wellbore is filled with water (no dissolved water solutes); and the wellbore pressure is set to hydrostatic. Cleanouts are performed under the following conditions:

a. The user has specified the timing of a cleanout with the “Wellbore cleanouts” setting.
b. A wellbore boundary condition switches between being bottomhole control, wellhead control, or MD control.
c. If “Cleanout well at start of MD production” is set to ‘true’ (the recommended default), then cleanout is performed when you first begin to produce from a BC MD constraint.
d. If “Clean proppant from well at start of production or shut-in” is set to ‘true’ (not the recommended default), then the well is cleaned out any time the well is shut-in or put on production. This is not recommended because the cleanout creates an abrupt pressure change that prevents you from being able to identify ISIP from the shut-in.
e. If you set “Clean out well with isolate wellbore condition” to ‘false’, then when wells are put on ‘isolate wellbore’ conditions, then the well is not cleaned out, regardless of other settings. This setting is global, and also, individual boundary condition control sequences each provide a checkbox to customize this behavior within the control.

Why does the simulator perform cleanouts? Without them, proppant may fill up and clog the wellbore. In reality, if a well became clogged with proppant, a cleanout would be performed. In a ResFrac simulation, the user cannot know in advance whether proppant will clog the well. Thus, we must include logic to clean out the well at reasonable times. As of June 2021, any proppant removed from the well during a cleanout is considered ‘produced proppant’ in the tracking file.

## 12.6 Wellbore skin – formation damage and choked fracture

Using the setting ‘Wellbore properties versus time,’ you can specify two types of skin, for each well, as a function of time.

The formation damage skin is applied for flow between the wellbore elements and matrix elements. It does not affect flow between wellbore elements and fracture elements. It imposes that there is an additional pressure drop that is proportional to flow rate, according to the equation:

$$\Delta P_{skin} = \frac{\mu q}{2\pi k h} s. \quad (12\text{-}1)$$

A skin of 0 (the default) corresponds to no damage. A skin of 10-15 corresponds to significant damage.

The choked fracture skin, following Cinco-Ley and Samaniego (1981), assumes that there is a choke point in the fracture flow pathway between the well and the fracture. The chokepoint has conductivity divided by length equal to:

$$\frac{C_{choke}}{x_{f,choke}} = \frac{k\pi}{s_{choke}}. \quad (12\text{-}2)$$

If the matrix permeability is on the order of nanodarcy, even a small choked fracture skin, such as 1, may be substantial. Consider using values such as 0.01 – 0.1 in low permeability formations.

The choked fracture skin is *only* applied to flow through the proppant bed. It is *not* applied to unpropped conductivity or mechanically ‘open’ fracture conductivity. The intent is that it affects production from propped fractures, but does not affect the process of hydraulic stimulation.

## 13. Initial conditions

Several options are available for specification of initial conditions (at the top of the "Static Model and Initial Conditions" panel). You may specify pressure and saturation in each individual element or by layer, or specify an oil/water contact and allow the simulator to initialize in hydrostatic equilibrium.

Keep in mind, if you use these options, the simulation may not be initialized at hydrostatic equilibrium. In reality, variable pressure/saturation versus depth is kept in equilibrium by capillary forces. However, ResFrac neglects capillarity. Therefore, if you manually set pressure and/or saturation, you might want to consider setting vertical permeability to zero so that fluid flow only occurs horizontally into the fractures/well and not vertically in from above or below.

If you select you choose to initialize the simulation in hydrostatic equilibrium, specify the "Initial water pressure at reference depth" at a "Reference depth" and the "Depth of water-hydrocarbon contact." The code automatically calculates pressure and composition in every element. Because capillary pressure is neglected, the water-hydrocarbon contact is a sharp interface. If it lies within an element, the element fluid saturation is calculated as an average, assuming that the contact is located at the specified depth. Above the contact, the initial water saturation is set to the connate water saturation from the relative permeability function in each facies (so water is immobile).

With the compositional model, you specify the initial composition of the hydrocarbon phases. With the black oil model, the default is to initialize with an oil phase with 'composition' given by the specified initial bubble point. Alternatively, you can tell the model to initialize with 100% gas and no oil. You are not permitted to initialize a model that has two hydrocarbon phases. Natural hydrocarbon deposits are always initially either single phase oil or gas.

The wellbore is always initialized at hydrostatic equilibrium. You are given the option to either initialize wellbore pressure at equilibrium with the reservoir, or to specify a wellhead pressure and initialize in equilibrium with that pressure (the table "Wellbore pressure initiation strategy" in the "Wells and Perforations" panel).

For the thermal initial conditions, you specify the surface pressure and the temperature at a specified "Reference depth." The initial thermal gradient is assumed to be uniform at all depths. The relevant settings are "Surface temperature," "Reference depth," and "Initial temperature at reference depth."

There are several options for specifying the initial stress state (specified at the top of the "Static Model and Initial Conditions" panel). You can specify stress or stress gradient on an element-by-element basis, stress or stress gradient by layer, or specify an overall stress gradient and 'stress deviations' from that background gradient by layer. If you specify "Straight fractures," then fractures do not turn, and the maximum horizontal stress down not impact the simulation results. Otherwise, SHmax affects the tendency of the cracks to turn. To specify the magnitude of the maximum horizontal stress, you specify the parameter "SHmax – Shmin" in the "Static Model and Initial Conditions" panel.

## 14. Fluid properties: the black oil model and the compositional model

Fluid properties can be calculated with either the black oil model or the compositional model. The model type is specified in the "Startup" panel, and then the details of the model are specified in the "Fluid Model Options" panel. The compositional model is more realistic but is more computationally intensive and more complex to set up. With either model, ResFrac assumes that the water and non-water phases are immiscible (ie, no water enters into the oil/gas phases and no hydrocarbons mix into the water phase).

Real hydrocarbons mixtures contain hundreds or thousands of different types of molecules. Hydrocarbons contain alkanes (hydrocarbon chains with only single bonds), alkenes (hydrocarbon chains containing a double bond, also called olefins), aromatics (hydrocarbon chains containing a conjugated ring), and many others. Within these categories, there are all different types of molecules, broadly categorized by how many carbons they contain (propane, butane, pentane, etc.). Finally, each type of molecule has a wide range of isomers. Isomers are molecules with the same chemical composition (the number of each type of atom) but with different arrangement of the atoms within the molecule. For example, n-pentane is an unbranched chain of five carbons where each carbon is bonded to two carbons (except the carbons at the end of the chain, which are bonded to one other carbon). In contrast, isopentane (aka, 2-methylbutane) has the same chemical composition ($C_5H_{12}$), but the second-to-last carbon on the chain is bonded to three carbons. Even though isomers have the same chemical composition, they can have substantially different macroscopic properties (boiling point, etc.).

It is not practical to keep track of thousands of different types of molecules in a reservoir simulator. The black oil model and the compositional model were developed to mimic the phase behavior of real hydrocarbons while keeping track of a much smaller number of components. The black oil model is simpler and runs faster – it keeps track of only three 'components' – oil, water, and gas. The compositional model is more flexible and can more accurately represent the real behavior of mixture. However, compositional simulations run more slowly than black oil simulations.

A rule of thumb is that the conventional black oil model will yield a good description of the reservoir fluid if the initial producing gas-oil ratio is less than around 2000 scf/STB (implying that the fluid is a so-called "black oil" petroleum fluid). Also, the black oil model will be fine if you have only single-phase gas in the reservoir (ie, it is a gas reservoir that does not experience retrograde liquid condensation in the reservoir). A so-called "wet gas" has only single phase gas in the reservoir, but some liquid condensate drops out at the surface. Volatile oils and retrograde condensates have both oil and gas phases in the reservoir, but the initial gas-oil ratio is greater than 2000 scf/STB. Volatile oil and retrograde condensates can be described with a compositional model, or with the 'modified' black oil model. If you are going to run a thermal simulation or if you are going to simulate enhanced oil recovery with gas injection, you should use a compositional model. Strictly speaking, the black oil model should only be used in isothermal calculations. However, ResFrac provides an option to use it in thermal simulations – this is a significant approximation that should only be used the right context.

McCain (1999) and Whitson and Brule (2000) provide a detailed description of the black oil model. Pedersen and Christensen (2007) provide a detailed description of compositional fluid models.

## 14.1 The compositional model

The compositional model keeps track of a set of predefined fluid components (typically no more than fifteen). There are three types of 'components' in a compositional model: defined components, lumped pseudocomponents, and the plus fraction. A defined component is a specific molecule, like methane or CO2. A lumped pseudocomponent is a mixture of components. This could be a mixture like N2-C1 (a mixture of nitrogen and methane), or a mixture of many different types of molecules like C8-C10 (all molecules that have boiling point in the range of n-octane to n-decane). The plus fraction is a pseudocomponent representing all molecules that are larger than a certain amount, like C30+ (all molecules with boiling point greater than an unbranched alkane containing thirty carbons). The word 'pseudocomponent' is used to denote that it is not a real molecule – a pseudocomponent is a hypothetical model that has properties that are averaged between a mixture of real molecules. However, we often use the words 'component' and 'pseudocomponent' interchangeably.

A compositional fluid model consists of: (1) a list of the components in the model, (2) molar mass of components, (3) the pseudocritical temperature and pressure of each component, (4) acentric factor of each component, (5) binary interaction coefficients describing interaction between each component, and (6) other optional parameters, such as a Peneloux volume correction factor. You can also specify parameters for calculating properties such as viscosity.

A different compositional model needs to be built for every real hydrocarbon mixture. You don't need to do it for every well, but if you notice significant differences in GOR between wells in the same formation, you might want to consider having multiple models. Compositional fluid models can be built with a commercial package, such as PVTSim. An excellent practical guide to building compositional models is given by Pedersen and Christensen (2007). You start by taking a fluid sample and measuring the composition. From here, you decide how many pseudocomponents to use. Using more pseudocomponents will yield a model that better matches reality but will result in a simulation that runs more slowly. After you have defined the pseudocomponents, their effective properties are calculated by taking a special type of average of the properties of the underlying components. Finally, regression is used to 'tune' the compositional model the any experimental data you may have available. For example, when you sent your fluid sample to the lab, they may have performed a differential vaporization test, a constant composition expansion, and a separator test.

The saturation, density, and composition of the phases are calculated with a cubic equation of state. ResFrac implements the Peng-Robinson equation of state. There are two slightly different versions of the Peng-Robinson equation of state, and ResFrac implements either (defaulting to use the more recent 1978 version). The algorithms described by Michelsen and Mollerup (2007) are used to calculate phase stability (whether the hydrocarbons form one or two phases), and the saturation of each phase.

ResFrac provides two options for calculating viscosity of the hydrocarbon mixture. The default is to calculate the viscosity of the flash phases from the method of Lohrenz et al. (1964), the "LBC" correlation. This correlation calculates viscosity from the critical molar volume of each component and using several coefficients. ResFrac has default values for these parameters. Alternatively, you can define your own modifications to the LBC correlations and/or define critical molar volumes for each of your components. You may want to specify these modifications if your fluid lab did viscosity measurements and you have used PVTSim (or another comparable code) to calculate the coefficients that best match the experimental data.

Alternatively, ResFrac provides a simplified viscosity model. In this model, you specify a viscosity for each component. The viscosity of a phase is calculated as the mass fraction weighted average of the viscosity of each of the components. This is not as realistic as an LBC model that has been tuned to experimental data, but provides a quick and easy way of specifying viscosity if more detailed characterization information is not available.

When the simulator calculates the production rate, the simulator must convert molar flow rates to surface volumes (STB of oil and water, and Mscf of gas). The conversion is performed as follows. Prior to the simulation, the user specifies the pressure and temperature of the separator and the stock tank. Optionally, the user can specify two separators prior to the stock tank. The simulator takes the produced moles and performs a flash calculation to calculate the phase properties at the separator conditions. The gas phase is removed and sent to 'sales'. The liquid phase is moved to either the second separator or the stock tank. The process is repeated – a flash calculation is performed and gas goes to 'sales'. If there is a second separator, the liquid is again moved to the stock tank. In the stock tank, the liquid phase is also sent to 'sales.' At the end of this calculation, ResFrac has calculated the number of moles of gas sent to sales and also oil sent to sales. The simulator has also calculated the moles of produced water. These three molar quantities are divided by their molar density at standard conditions to calculate STB of oil and water and Mscf of gas.

Injection rates are also specified in terms of standard volumes. These are converted to moles (which is the quantity conserved in the simulator) by calculating the molar density at standard conditions and dividing the specified injection volumes by molar density.

Pure water properties are calculated according to the correlations from the International Association for the Properties of Water and Steam (Cooper, 2007). Viscosity and density are adjusted with correlations to account for the effect of the dissolved solutes (Pedersen and Christensen, 2007). Alternatively, you are permitted to model the water phase as being slightly compressible, in which case you specify the density at a reference pressure and compressibility. You also have the option to specify water viscosity to a constant.

Originally, the correlation for water properties did not permit the formation of steam. More recently, the water properties were extended to allow modeling of steam and supercritical fluid. In addition, while the original version assumed constant heat capacity, you now have the option to have ResFrac calculate fluid enthalpy rigorously as a function of pressure and temperature.

With the new options, the water properties are valid up to 1073.15 K (800 C; 1472 F). To turn on these options, set 'constant water heat capacity' to false, and set 'use steam correlation' to true (both are in the Fluid Properties panel).

There are three correlations used, depending on the pressure/temperature, as shown in the figure below. We did not implement the correlation for region five (which is applicable only above 1073.15 K). The correlations are valid only up to 100 MPa, but we extrapolate up to pressures beyond 100 MPa by assuming constant compressibility from 100 MPa upward. The correlations are close, but not identical, at their boundaries, and so the code must smoothly interpolate between them across the boundary.

The steam/water phase transition is the most challenging and complex part of the implementation. Water has a large heat of vaporization, and so processes at the phase transition line are complex. We use a 'pseudo-two phase' simplification of the phase transition. As the pressure drops to the saturation line, rather than having discontinuous phase change at a single discrete pressure, we interpolate the change in properties across a range of pressures plus/minus 3 MPa above and below the saturation line (with that range narrowing when saturation pressure is less than 3 MPa). The steam/water mixtures within this interpolation range are treated as one 'phase' with a density and enthalpy equal to the average density and enthalpy of the steam/water mixture.

Within the interpolation region around the saturation line, the pressure/temperature/enthalpy relations are inexact and deviate somewhat from the precise values from the correlations. It is important to note – the simulator is rigorously conserving both mass and enthalpy. Effectively within this interpolation region, the simulator can be considered exact on enthalpy and pressure, but inexact in temperature. To avoid confusion, the simulator outputs both the 'simulator internal' temperature (which is approximate within the interpolation region), and the 'exact' temperature that would be back-calculated from a steam table, if specifying the pressure and enthalpy.

The code has been tested up to extreme temperatures and pressures and in complex problems very near the critical point. It has been able to converge efficiently through all of these problems. Steam has an unusually high heat of vaporization. If liquid, high-temperature water enters the bottom of the well, as it moves up the well, pressure drops and it may begin to flash to steam. The heat of vaporization is supplied from thermal energy, and so the temperature drops as enthalpy is conserved. In test problems, ResFrac was able to smoothly simulate through these processes.

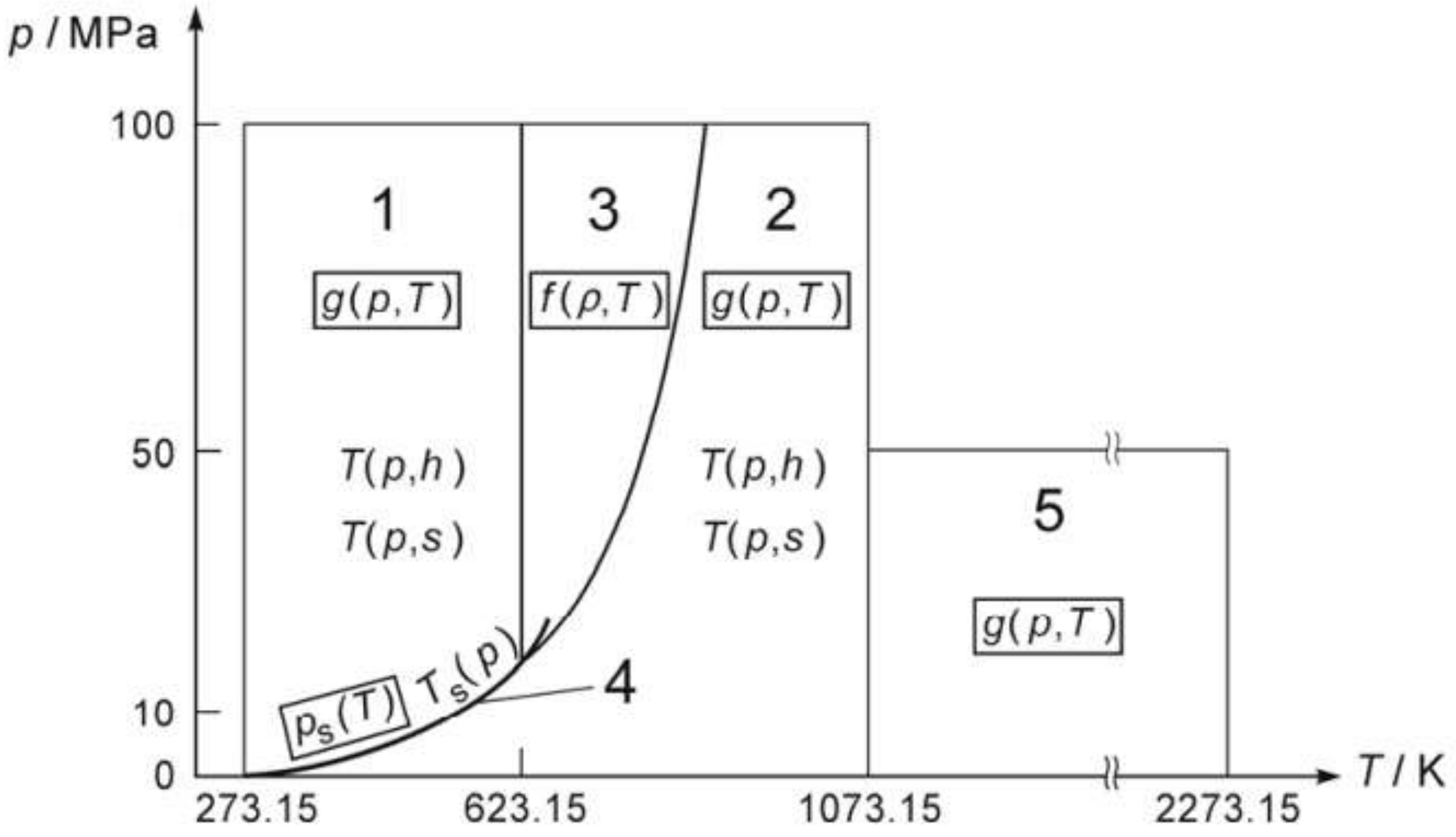

**Fig. 1.** Regions and equations of IAPWS-IF97.

Figure 28: Phase diagram and correlation regions for water/steam properties.

When running simulations with nonconstant water heat capacity, the code outputs the enthalpy and steam volume fraction in the water pseudophase for the 3D visualization, the raw results files, and in the tracking file.

One important thing to note – to regularize numerical challenges, if using this option, the well and fracture elements are assumed to include not only the fluid in the well, but also some of the surrounding rock (which is assumed to be the same temperature as the fluid in the well). This can have minor impact on the results of the temperature transient related to fluid heat exchange with the surrounding rock.

## 14.2 The black oil model

In the black oil model, you specify a table of oil formation volume factor, solution gas-oil ratio, gas formation volume factor, oil viscosity, and gas viscosity as a function of pressure. McCain (1999) and Whitson and Brule (2000) descriptions of the black oil model and how it is built.

Bubble point is not necessarily constant throughout the simulation because it is a function of composition. The water viscosity is assumed constant, and the water density is defined from a reference water formation volume factor at the bubble point pressure, water compressibility, and water specific gravity at standard conditions. Finally, you specify the oil and gas specific gravity at standard conditions.

The black oil model assumes that there are three conserved components – water, oil, and gas. Confusingly, these components are distinct from the phases – water, oil, and gas. In other words, there is an oil 'phase' and an oil 'component' and these are not the same thing. The oil 'component' is defined as 'molecules that will be liquid oil at the surface after going through the surface separation units.' The gas 'component' is defined as 'molecules that will be gas at the surface after going through the surface separation units.' When you bring molecules to the surface, some of the molecules in the liquid (oil) phase vaporize up into the gas phase at standard conditions. Thus, the oil phase in the reservoir contains both oil and gas components. In real hydrocarbon mixtures (specifically, in volatile oils and retrograde condensates), some of the molecules in the gas phase can drop out into the liquid phase at the surface. However, in the standard black oil model, it is assumed that the gas phase in the reservoir contains only gas 'component.'

The key to the black oil table is the table of formation volume factors and solution gas-oil ratio as a function of pressure. These values relate reservoir volumes to surface volumes. Because surface volumes can be converted to moles by dividing by molar density at 'standard conditions,' these expressions of formation volume factor and solution gas-oil ratio are (rather indirect) ways of specifying the composition and density of the phases at reservoir conditions. However, these values are more commonly discussed and understood in terms of the macroscopic behavior that they represent.

The oil, water, and gas formation volume factors are defined as:

$$B_o = \frac{resbbloilproduced}{STBoilproduced}, \quad 14\text{-}1$$

$$B_w = \frac{resbblwaterproduced}{STBwaterproduced}, \quad 14\text{-}2$$

$$B_g = \frac{resbblfreegasproduced}{Mscfgasproduced} freegas. \quad 14\text{-}3$$

A variety of different unit combinations may be used for defining $B_g$.

Some of the liquid oil in the reservoir converts to free gas at the surface. This is expressed with the solution gas-oil ratio:

$$R_s = \frac{scfproducedatsurfacegascomingoutoftheliquidphase}{STBoilproduced}. \quad 14\text{-}4$$

The water produced at the surface comes only from the water phase in the reservoir. The oil produced at the surface comes only from the oil phase in the reservoir (though some of the molecules in the oil phase convert to gas when the liquid comes to the surface). The gas at the surface comes from free gas in the reservoir and gas out of solution from the liquid oil phase in the reservoir. Therefore, the total gas at the surface is:

$$scfgasproducedatsurface = R_s STB + V_g/B_g, \quad 14\text{-}5$$

where STB is the stock tank barrels of oil produced and $V_g$ is the volume of free gas produced (ie, volume evaluated at reservoir conditions).

The key assumption of the black oil model is that the formation volume factors and the solution gas-oil ratio can be written solely as a function of pressure. This is a simplification because in reality these properties are also a function of composition. The values in the black oil table are derived from a complex procedure that involves combining the results from several laboratory experiments designed to mimic different processes occurring in the reservoir and in the surface separation units (McCain, 1999). When the pressure is above the bubble point pressure, the oil phase is assumed to be constant compressibility.

With some manipulations (and by exploiting the assumptions of the black oil model), you can convert a black oil table to a table of molar compositions (defined at the pressure specified in each row). The simulator does not actually enforce these compositions. That would be impossible - the simulator is solving molar balance in each element on the water, oil, and gas components, as well as calculating pressure. Composition is calculated from the molar balance calculation. So how is this handled? The table can be interpreted as showing the composition of the mixture when it has the bubble point equal to the pressure in each row. Thus, fluid properties are calculated as follows:

1. From molar composition, calculate the bubble point of the mixture from the black oil table.
2. If the pressure is above the bubble point:
    a. Evaluate the properties from the black oil table, using the bubble point pressure, not the actual pressure.
    b. Adjust the oil formation volume factor to account for the compression of the liquid as it goes from the bubble point to the actual pressure.
    c. Three options are available (see discussion below). These options are 'Basic,' 'Correlations,' and 'TableOfTables.'
3. If the pressure is below the bubble point:
    a. Read the fluid properties directly from the black oil table at the given pressure.

When pressure is between rows of the table (which is nearly always), a spline is used to accurately interpolate between the values.

ResFrac also provides the option to use the modified black oil model. In the modified black oil model, it is assumed that some stock tank oil is vaporized into the reservoir gas phase. With the modified black oil model, it is possible to simulate retrograde condensate reservoirs, which involve liquid dropping out of the gas phase. When this option is selected, the user is asked to provide $R_v$, the vaporized oil-gas ratio, versus pressure in the black oil table:

$$R_v = \frac{STBoilproduced}{MMscfgasproduced}. \qquad 14\text{-}6$$

The user specifies the dry gas formation volume factor in the table. It is not necessary for the user to also specify the wet gas formation volume factor, which can be derived from the other inputs. The dry gas formation volume factor is:

$$B_{gd} = \frac{resbblfreegasproduced}{Mscfgasproduced}. \qquad 14\text{-}7$$

Three options are available for calculating unsaturated properties (ie, fluid properties when only gas or only oil are present).

1. Basic option
    a. For liquid, oil formation volume factor is calculated at the bubble point pressure. It is adjusted to current pressure assuming a constant (user-specified) oil compressibility. Oil viscosity is assumed to be equal to oil viscosity at the bubble point.
    b. For gas, the formation volume factor and viscosity are read directly from the table of saturated properties.
2. Table of tables
    a. For every row of the 'saturated' black oil table, the user provides a table of properties versus pressure for properties in above/below the dew point. The code interpolates from the tables to use the user-specified values. This can be done with an oil or gas. Note that for retrograde condensates with two dew points, the user specifies two unsaturated tables – for above and below the two dew points. Formatting details are provided in the help content built into the UI.
3. Correlations
    a. For oil: from the composition of the mixture, determine the saturation pressure, Psat. Calculate viscosity and Bo at the saturation point. Adjust oil viscosity to the current pressure using the Standing correlation given by Equation 3.130 from Whitson and Brule (2000). Calculate Bo from Bo at bubble point using the Vazquez and Beggs correlation from Equation 3.108 from Whitson and Brule (2000).

b. For gas: from the composition of the mixture, determine the saturation pressure. If the mixture has two dew points, there may be two saturation pressures corresponding to this composition. Select the upper or lower dew point depending on whether actual pressure is above or below the two dew points.
    i. The Lee and Gonzalez correlation is used to extrapolate gas viscosity as a function of pressure (Equation 3.65 from Whiting and Brule, 2000). First, calculate the constants A2 and A3 (note that the user is always asked to input reservoir temperature, even if an isothermal black oil simulation, so the simulator knows the value of temperature, which is needed in the correlation). To determine the value of A1, calculate A1 such that mug from the correlation is equal to gas viscosity from the table at the saturation pressure. With these constants, then we plug into the correlation to calculate gas viscosity at P.
    ii. Bg is calculated from the definition of gas formation volume factor (Equation 7.12 from Whitson and Brule) and a correlation for the Z-factor – the Hall and Yarborough correlation to the Standing-Katz chart (Equation 3.42 from Whitson and Brule). Critical temperature and pressure are estimated from the Sutton correlation (Equation 3.47 from Whitson and Brule). Molar mass of the stock tank oil (which is needed in the calculation) is estimated from the user-inputted API gravity using the Cragoe (1929) correlation. The correlation is used to calculate Bg at Psat. Then, an adjustment factor is calculated to enforce that the correlation's prediction of Bg at Psat is equal to the value entered in the table. Finally, the correlation, with adjustment factor, is used to calculate Z and then Bg above or below the dew point.

# 15. Data needed to set up a ResFrac simulation

To set up a ResFrac simulation, you need to know (or have reasonable estimates for):

1. Formation properties versus depth: permeability, porosity, initial fluid saturations, initial pressure, and the minimum principal stress. ResFrac uses a 'layer cake' model. It neglects lateral heterogeneity. You can opt to use a model with only a few facies (with uniform properties within each facies) or you can use the "formation properties versus depth" option to put in vertical heterogeneity at finer resolution.
2. Location and geometry of the well. Inner diameter. Location of stages and perforation clusters. Perforation diameter and count per cluster.
3. Relative permeability curves (can be different in each facies, which you define by depth intervals).
4. A fluid model. This could either be a black oil table or a compositional fluid model.
5. Your wellbore boundary conditions. Injection schedule during fracturing (rate, fluid viscosity/type, and proppant type and concentration). During production, an estimate for the producing bottomhole pressure.
6. Orientation of the minimum principal stress.

ResFrac simulations have a lot of secondary parameters that you may not be familiar with or may not be sure what values to use. You can use the default values for these parameters. If you would like more information, refer to the "help" button in the builder next to the parameter. You can press the "suggest" button, and the builder will put in a default value (if feasible).

# 16. Tips for history matching

History matching to data can be time consuming, and we don't recommend that you try get a perfect match. There are a lot of parameters that you could vary, and it can be tedious and not particularly fruitful to be a perfectionist about matching the data. On the other hand, you clearly want to be in the ballpark. This section describes a procedure that is reasonably efficient and effective.

First, try to match the initial shut-in pressure and (if available) fracture length. During injection, there is a complex relationship between the pressure measured in the well (either at the wellhead or even bottomhole) and the pressure in the fracture. Thus, trying to match WHP during pumping will not necessarily going to lead to a much better model. On the other hand, the ISIP is measured after shut-in and is more representative of the true pressure in the fracture during pumping. As long as your minimum stress estimate is reasonable, it should be somewhat below the ISIP. After a full-scale frac job, it might be anywhere from 500-1500 psi lower (with values closer to 500 psi being more likely). The effective fracture toughness controls the net pressure – the difference between the minimum principal stress and the fluid pressure required to propagate the fracture. Fracture toughness is specified in the facies list. We recommend using values between 2500-5000 psi-in$^{1/2}$. But also, you should specify a scale dependent fracture toughness parameter, as described in Section 10.2 (and found in the "Fracture options" tab in the builder). For the scale dependent toughness parameter, we have had success with values between 0.5 and 1.5. Vary this scale dependent fracture toughness parameter until you match ISIP. Find out of if you have any additional information to constrain fracture length. This could be microseismic or reports on the distances where you typically observe frac hits. You'd like your simulated frac length to be in the ballpark of this estimate. If not, you can lengthen the frac by decreasing the scale dependent toughness parameter, or shorten it by doing the opposite. When matching length, make sure you consider how many perf clusters there are in each stage!

Next, you can (optionally) match the wellhead pressure during pumping. This isn't really critical because you already matched the ISIP. The relationship between BHP and WHP is controlled by wellbore friction. You don't really need a perfect wellbore friction calculation to do a good frac simulation. ResFrac has a default correlation for friction, but it certainly isn't always perfect. There are too many different types of fluid out there, and you are unlikely to have perfect information about how they behave when pumped through a well. You can tune the wellbore friction up and down with the "wellbore friction adjustment factor" found in the well controls panel.

The last, and most challenging, part of a history match is the production data. Usually, you want to model long-term production with a constant bottomhole pressure constraint. When you specify a bottomhole constraint, the wellbore is removed from the model and the simulation is performed assuming pressure is uniform in the well. Before you get started, make sure that you have a good estimate for that BHP, which may be changing over time. You probably don't want to vary that to match the data, unless you don't have a good initial estimate.

Alternatively, you may want to specify the injection volumes and try to match BHP. If so, you can still do this as a 'constant BHP pressure' constraint in the builder. Specify 'total rate', which is STB oil + STB water + Mscf gas. The code will calculate the BHP and the relative amount of each phase. You typically will have daily production data. You can specify every single one of those daily rates, but this will slow down the simulation because it will be having to change the boundary condition frequently. Instead, you might want to take 10-day averages and change the BC every 10 days. If you have a large number of these specified rate changes, you might want to consider writing a script to directly modify the simulation's text input file, rather than manually inputting all the controls through the builder.

For history matching, the size of the drainage volume will make a big impact. Look at the proximity of your well to other wells that are nearby. Generally, the drainage volume boundaries should be halfway between the wells. If the neighboring wells are laterals landed at a different depth, you may have to make a judgement call whether or not you think those neighbors are draining the same formation(s) that your well is draining, and whether or not it should be used to specify a drainage volume.

The edges of the matrix region are no-flow boundaries. Therefore, one way to specify the drainage volume is to make the matrix region the same size as the drainage volume. But this is often not the recommended approach.

ResFrac does not allow fractures to grow out of the matrix region. Therefore, you generally want to make the model region large enough that it can contain any fractures that might form. Instead, to modify the drainage volume, use the option “zeropermoutsidecube.” You specify both the size and location of the center of this cube. This cube is your drainage volume – permeability is set to zero outside the cube. You can set the ‘zero permeability’ to turn on at a certain time. For example, you may expect fractures to grow outside the cube, and you’d like fluid to be able to leak off along the entire fracture length. You can set ResFrac so that you don’t ‘turn on’ the ‘zero permeability outside cube’ option until you put the well on production.

When modeling multiple wells, the placement of no-flow boundary conditions can become challenging if the wells have different stage lengths. To handle this, the code also permits the insertion of one or more ‘zero perm inside cubes.’ This gives much greater flexibility to model the desired shape of the matrix domain.

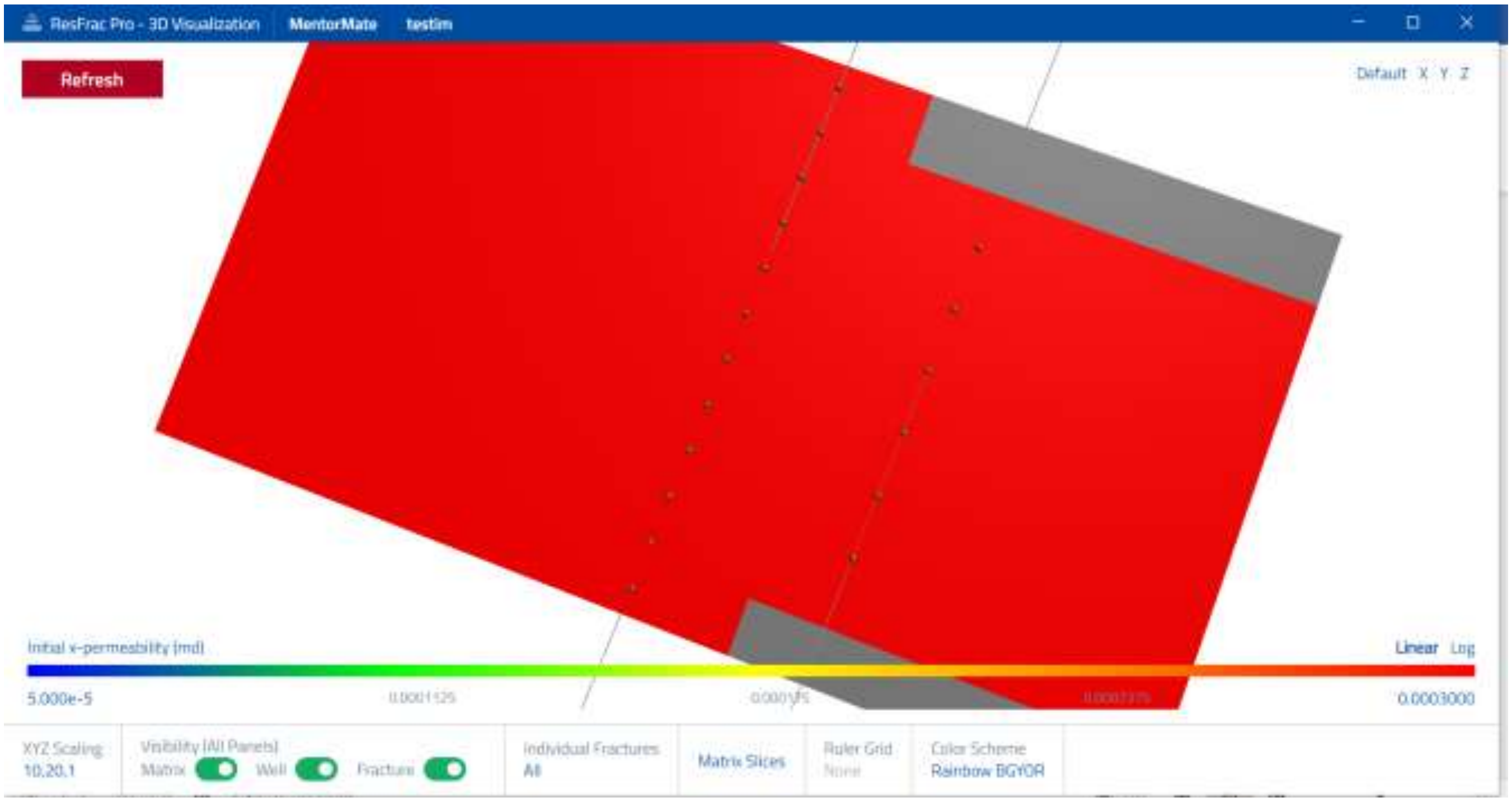

Figure 29: Example of using ‘zero perm inside cube’ constraints to handle mismatched stage lengths.

To vary the relative amounts of the phases, you can vary the relative permeability curves in the facies. To vary the overall pressure drawdown, you could vary either the matrix permeability or the fracture conductivity. To modify the unpropped fracture conductivity, vary “E0max” or “90% closure stress” in the facies list. Higher values result in better unpropped fracture conductivity. E0max controls the overall conductivity, and the 90% closure stress controls the stress sensitivity. I often use values in the vicinity of 0.002 ft for E0max and 500 psi for 90% closure stress. Note that the unpropped fracture conductivity is proportional to the cube of E0max. If you double that value, you will effectively multiply unpropped fracture conductivity by 8x. To modify the propped fracture conductivity, vary k0 and proppant bed compressibility in the list of proppant properties. Propped fracture conductivity is linearly proportional to k0. Proppant bed compressibility affects the sensitivity to effective normal stress.

It is possible to ‘restart’ a ResFrac simulation in the middle, rather than start from scratch. It is even possible to do a restart with different settings than the original simulation. This can be very useful. In a typical simulation, the frac job takes most of the CPU time, and the production period goes quickly at the end. Therefore, if you are matching production data, you can use restarts from the beginning of production, rather than redoing the simulations from scratch.

# 17. Planar fracture modeling and complex fracture network modeling

Planar fracture modeling is the conventional approach to hydraulic fracture modeling, and it is the primary approach used in ResFrac. Hydraulic fractures are assumed to be mostly linear, spatially continuous features. 'Complex fracture network' (CFN) modeling is an alternative approach that has sometimes been used over the past decade (Weng et al., 2011). This approach is started by seeding a network of preexisting natural fractures. Propagating hydraulic fracture are assumed to sometimes terminate against natural fractures, creating branching, zig-zagging flow pathways.

McClure et al. (2020) provides a detailed discussion of this topic. A few key points are repeated in this section.

In-situ observations (from core across studies and offset well fiber) indicate that hydraulic fractures are propagating quite linearly, in a consistent orientation, and in a relatively narrow band (Raterman et al., 2017, 2019; Gale et al., 2018; Ugueto et al., 2019a; 2019b). This is fundamentally at odds with the zig-zagging fracture networks conceptualized by the CFN approach.

In-situ observations show that fractures are complex at small-scale. However, when we zoom out to the reservoir scale, these fractures look like linear, planar features. Core suggests fractures have small-scale bifurcations and jobs, and multiple (subparallel) strands. The effect of these features can be captured using constitutive relations (such as ResFrac's proppant trapping/immobilization model and scale dependent fracture toughness). Complex fracture network models attempt to explicitly represent this small-scale complexity with a DFN. But there is too much complexity to truly reproduce the geometry of the fractures, and so CFN models are also grossly simplifying reality. At the same time, because adding a huge DFN has heavy computational cost, DFN models are forced to sacrifice on physical realism. For example, the model from Weng et al. (2011) is not fully 3D. Not only are CFN models also simplifying reality, they appear to be actually incorrect in most applications. The zig-zagging flow pathways of CFN models are directly contradicted by in-situ observations in major shale plays, which suggest subparallel hydraulic fractures dominate flow. The CFN approach is perhaps most useful in applications like Enhanced Geothermal Systems, or in shallow formations, where there is low stress anisotropy and fractures are less likely to be mineralized shut.

An important result from the core across studies is there are very numerous (subparallel) hydraulic fracture strands. However, Raterman et al. (2019) found that only a small percentage of these fractures contain proppant, and only the propped fractures are associated with pressure depletion a significant distance from the well. Thus, during production, it is reasonable to model production as occurring from a relatively small number of major propped fractures – a planar fracture model. We routinely history match to production in shale with this modeling approach. During fracturing, core indicates that there are many water filled fractures strands. These fracture strands increase the surface area for leakoff, and so cause an accelerated leakoff. In ResFrac, we mimic this increase in leakoff area as an increase in leakoff permeability. This is done with a user-input table of pressure dependent permeability (PDP) multipliers. Thus, during fracturing, leakoff is accelerated by PDP. But production is dominated by the much more sparsely distributed set of propped hydraulic fractures, which is mimicked as the PDP multiplier goes back down to 1.0 as pressure depletes. Occasionally, the history matching process leads us to use a pressure dependent permeability loss to decrease effective permeability as depletion occurs.

# 18. Correlation for multistranded fracture swarms

Inspired by the recent core-through studies, Fu et al. (2020) propose a correlation for modifying constitutive equations to handle the occurrence of multi-stranded fracture swarms. Each individual 'hydraulic fracture' in the model represents a band of multiple hydraulic fractures. This impacts toughness, viscous pressure drop, and leakoff.

This correlation was introduced to ResFrac in May 2020. We had already been modeling elevated toughness and leakoff as being potentially caused by multiple fracture strands. However, we were not modeling a process that could increase viscous pressure drop. Also, the Fu et al. (2020) approach explicitly ties together elevated toughness, viscous pressure drop, through a single parameter, N, which is the number of fracture strands in each swarm. This is specified with the parameter “Fracture strands per swarm” in the “Fracture Options” panel.

The effective toughness should scale with the square root of N, where N is the number of fracture strands. We follow this approach, but also note that our experience finds that it is best to scale toughness with the square root of fracture size. Thus, we also continue to use the ‘scale dependent toughness’ scaling parameter.

Fu et al. (2020) propose to scale viscosity with the square of N. The quadratic viscosity scaling is derived because conductivity scales with the cube of aperture, but also, overall conductivity scales linearly with the number of strands. Their approach is valid if the fluid is Newtonian, but for a non-Newtonian fluid, it is not valid because the viscosity depends on shear rate, which depends nonlinearly on aperture. To adjust, we scale the ‘open fracture’ and ‘closed unpropped’ conductivity inversely with the square of N, and also scale the aperture used to calculate the shear rate and effective viscosity inversely with N. The smaller viscosity of each strand leads to greater shear, and so lower effective viscosity. The conductivity of closed, propped fractures is not scaled with the number of strands. For flow through a propped fracture, the conductivity scales linearly with aperture, rather than scaling with the cube of aperture.

Fu et al. (2020) do not address scaling of leakoff with N, but obviously this would also scale with N, since leakoff is proportional to area times the square root of permeability. We already model this process as ‘pressure dependent permeability,’ and require the user to input a table of pressure dependent permeability multipliers versus pressure change. To incorporate N, we follow the same approach, introducing a PDP multiplier that scales with the square of N from initial pressure to a 1000 psi above initial pressure.

# 19. Miscellaneous features

## 19.1 Accounting for stress shadow from fractures outside the model

Often, when using ResFrac, you would like to run a simulation of just one or a few stages, rather than model every stage in the well. Full well-scale models are useful for certain applications, but many questions can be addressed without needing to simulate 25-50 stages. By running a simulation of fewer stages, you can get faster runtimes and get away with using a finer mesh.

If using ResFrac to model a single stage, you may want to account for the stress shadowing from the previous stages. In other words, the simulation is initialized in a way that accounts for the stress shadow that would be felt by a typical stage along the middle of the well.

This can be accomplished with “External fractures” specified in the Wells and Perforations panel. For each fracture, you specify the location of the center of the fracture (x,y,z), the volume of fluid injected into the fracture, and the net pressure. Based on other input parameters, the code calculates an effective radius of the fracture. For simplicity, these external fractures are assumed circular and leakoff from the fractures is neglected (so that all of the injected fluid remains in the fracture without leaking off). These simplifying assumptions are adequate solely for this purpose of calculate stress shadow from previous stages. The fractures are assumed to be perpendicular to the minimum horizontal stress. Then, the analytical solution from Sneddon (1946) is used to calculate the stresses within the simulation problem domain. This approximate treatment allows the fractures in the simulation to feel the stress shadow from previous stages, without needing to do a detailed simulation of more than one stage.

Stress shadow decreases rapidly with distance, and so the clusters in your model that are closer to the toe may feel substantially more stress shadow than the clusters closer to the heel. How many of these external fractures should you include in the simulation setup? Typically, field experience shows that ISIP trends along wells plateau after 2-3 stages. It is typically recommended that you put in one fracture per well (on the toe side) to capture the aggregated effect of stress shadowing from the fractures from the previous stages.

## 19.2 Diverter

Diverter pills are modeled as a temporary reduction in perforation diameter using the table of "Diverter slugs" in the "Well Controls" panel. The user specifies the timing of pills injected in each well. The transport of the pill down the well is not included – the user specifies the time when the pill reaches the perforations. The amount of diverter injected is specified in a unitless quantity. When the diverter is pill is injected, it is placed at the perforations in an amount proportional to the volumetric flow rate at each perf. The perforation diameter is multiplied by an adjustment factor to account for the blockage of the diverter:

$$D_{pf,adj} = 1 - D_{pf,adj,max} + D_{pf,adj,max}\frac{1}{1+d_m}, \quad 19\text{-}1$$

where $D_{pf,adj}$ is the adjustment factor, $D_{pf,adj,max}$ is the maximum possible adjustment factor (defaults to 0.5), and $d_m$ is the (dimensionless) amount of diverter present at the perf cluster. Thus, if there are four flowing perf clusters with equal flow distribution and one cluster where the fracture has not broken down, and you inject 4 units of diverter, then 1 unit will be placed at each cluster, resulting in $D_{pf,adj}$ equal to 0.75. The perforation pressure drop scales with the fourth power of diameter, so it will increase by 1/0.75^4 = 3.16. This will cause an increase in injection pressure, probably causing the inactive cluster to break down.

The user also specifies a "diverter decay rate," $d_{decay}$, which defaults to 6 hrs^-1. The diverter degrades according to the equation:

$$d\frac{(d|m)}{dt} = -d_m d_{decay}, \quad 19\text{-}2$$

such that:

$$d_m = d_m(t=0)exp\left(-d_{decay}t\right), \quad 19\text{-}3$$

where t is time.

The relevant parameters are "Diverter decay rate," "Minimum perforation diameter with diverter," "Maximum diverter blockage factor," and "Diverter slugs," all in the "Well Controls" panel.

## 19.3 Fracture damage mechanisms

### 19.3.1 Background

Shale wells often experience huge production losses after a frac hit. For example, Figure 22 from King et al. (2017) shows a parent well in the Woodford that experienced a 65% reduction in oil production after it was hit by the fracturing of a neighboring child well. Production damage is not limited to parent wells – child wells tend to significantly underperform parent wells.

A key question: why do parent wells lose so much production? There are reasonably well-understood explanations for why a parent well would lose production, but they appear to be insufficient to explain the severity of production loss at parent wells. The relatively well-understood explanations are: (a) depletion reduces the stress in the formation, and this tends to attract hydraulic fractures (Roussel et al., 2013); (b) fracture asymmetry causes inefficient drainage and loss of depletion efficiency (Cipolla et al., 2018); (c) the wells experience production interference as they compete to produce from the same rock. (Rimedio et al., 2015), and (d) proppant is remobilized and pushed into the wellbore. The severity of the parent/child impact has been found to depend on the age of the parent well (Elliott, 2019), which is consistent with these interpretations. But while these mechanisms are surely having an effect, they do not appear to be sufficient to explain the severity of frac hit production loss in most shales.

After the frac hit, the well may be clogged with proppant and water, and require cleanout with coiled tubing. But even after the cleanout and months of flowback, production may remain heavily depressed.

An important clue: frac hit production loss in parent wells is highly variable by formation. As reviewed by Miller et al. (2016), wells in the Bakken typically do not experience substantial production loss after frac hits. But wells in formations like the Marcellus or Woodford tend to experience significant production loss. Why should this depend so dramatically on formation?

It appears that chemical and/or multiphase flow effects play a major role in causing parent well damage and production loss. These effects depend on the mineralogy of the rock, and the composition of the formation fluid (both hydrocarbon and water phases), which explains why the effect is formation-specific.

Nieto et al. (2018) used swabbing operations to recover black, gunky solid particles from a damaged retrograde gas parent well in the Montney. Analysis indicated that the particles were composed of a mixture of silica (crushed proppant), formation fines, and high molecular weight hydrocarbons. The heavy hydrocarbon components were found to be soluble in an aromatic solvent, but not in a paraffin solvent, suggesting that they were asphaltenes. Importantly, Nieto et al (2018) found that iron oxide formed from the reaction of formation iron with dissolved oxygen in the frac fluid. The iron oxide created nucleation points for asphaltene to agglomerate out of solution, and also cemented together the silica and fines into larger particles. The thick, gooey asphaltene semi-solid helped muck all this together. The parent well experienced more than a 50% reduction in production after the frac hit. Interference tests were performed, and only minor pressure communication was observed between the wells. This all points to a ‘fracture conductivity’ form of frac hit damage.

Rassenfoss (2020) reports on another mechanism that could cause conductivity damage – a so-called ‘gummy bear’ phenomenon in the Woodford. Again, iron from the formation is a culprit. When the cross-linking occurs, the result is a thick gooey mixture of cross-linked gel, ground up proppant, and formation fines. I did a quick informal lit review, and it does appear that the Woodford is unusually high in pyrite, compared to other shales.

The mechanism described by Rassenfoss could occur in a parent well, not just during a frac hit. And in fact, operators in the Woodford do report seeing this type of damage in parent wells. Chemical formulations can be helpful for operators to mitigate these issues.

In contrast to fracture conductivity damage, fracture skin damage is harder to conclusively diagnose. Unlike processes that create physical material that can be pulled out of a well, fracture skin damage involves blockage of flow occurring out in the reservoir. The most obvious potential mechanism for fracture skin damage is ‘water block.’ The idea is that water leaks off into the surrounding rock, and accumulates (rather than flowing back or flowing our further into the formation). The layer of accumulated water could block hydrocarbon flow as it tries to produce into the fracture.

Swanson et al. (2018) report very positive results from pumping chemical remediation treatments in damaged parent wells in the Woodford. They were uncertain about damage mechanism and so threw the kitchen sink – surfactant to reduce interfacial tension and mitigate water block, HCl to dissolve scale, and HF/HCl to dissolve silica/clay mineral scaling/fines, etc. We can't be sure which mechanism was most important (and it might have been the gummy bear mechanism discussed by Rassenfoss). Whatever it was, they got good production recovery from these wells. Surfactant to prevent water block is certainly a reasonable fluid to include in a chemical treatment.

For water block, we can reasonably hypothesize that the problem might be more severe in child fracs than in parent fracs. Elputranto et al. (2018) discuss how capillary end effect can cause water to accumulate along the walls of a hydraulic fracture, blocking flow. In general, capillary end effect occurs when a rock with capillary pressure is opened to a zero-capillary pressure interface (such as the side of core in the lab, or a fracture wall). Capillary end effect is closely related to the process of spontaneous imbibition of a wetting fluid into rock.

Elputranto et al. (2018) did not specifically discuss the topic of frac hits. But we can think through mechanistically why capillary end effect could be more severe after a frac hit than after the original parent frac (kudos to Joe Frantz for pointing this out to me). Elputranto et al. (2018) perform simulations to show that development of capillary end-effect depends on how pressure is available to drive fluid into the fracture. If the pressure gradient is strong enough, it can overcome the capillary end effect, fluid is drawn cleanly into the fracture, and the wetting phase is pulled further out into the formation by capillary pressure, instead of accumulating at the fracture walls. However, if the pressure gradient is weak, then flow still occurs, but capillary end effect is not overcome and water accumulates at the fracture wall.

When a parent frac is performed, the formation fluid pressure is still high, and so capillary end effect is less likely to develop. But, in a frac hit on a parent well, the formation pressure has been drawn down by prior depletion, and so there is less pressure available to overcome the capillary end effect. Thus, we might develop a 'water block' after a frac hit, but not in the original frac.

Preloads are performed by injecting fluid into parent wells prior to an anticipated frac hit. They are, in fact, effective at reducing the amount of frac fluid that flows from the child well to the parent well. But why don't they themselves cause damage – since they also involve injection of water into depleted fractures?

Perhaps because preloads are typically pumped with a different fluid chemistry than frac fluid. They are often pumped with surfactant, iron chelators, or other remediation chemicals that may not be found in a typical frac fluid.

To address these issues, a variety of mechanisms have been implemented to describe frac hit damage mechanisms.

In total, there are four different types of damage that are tracked separately by the code: water block, fracture skin, fracture conductivity damage, and fracture relative permeability damage. You can run a simulation with anywhere from zero, one, two, three, or all four of these different types of damage occurring simultaneously.

### 19.3.2 Fracture damage reactions

Fracture damage reactions can create or remove two different types of fracture damage: (a) fracture conductivity damage, or (b) fracture rel perm damage. Fracture damage reactions are quite flexible – they can be used to describe a variety of different mechanisms.

For each fracture damage reaction, you specify: (a) the name of the water solute that reacts to cause the damage (must be one of the water solutes defined in the water solutes panel), (b) optionally, the name of a second water solute that must be present to react with the first water solute in order to cause damage, (c) a reaction rate constant that controls how rapidly the reaction occurs, (d) a potency constant that determines how much damage is caused (or removed) by the reaction, (e) the type of damage created (or removed) by the reaction, and (f) a Boolean (true or false) that specifies whether the reaction rate is affected by the saturation of the oil phase.

For example, Rassenfoss (2020) described the formation of 'gummy-bear' gunk in fractures because of cross-linking of friction reducer with iron in the formation. This can be described as a fracture damage reaction where you specify only the first water solute, and not the optional second water solute. In reality, this is a reaction between an injection fluid and the formation. This is handled naturally because, by default, fracture damage reactions only occur when fluid is in a fracture, and not when it is in the well.

If the potency constant for a reaction is positive (so that this is a reaction that creates damage), then the reaction progresses according to the equation:

$$\frac{d\overline{m_A}}{dt} = -k_{R,A}\overline{m_A} \qquad 19\text{-}4$$

Where $\overline{m_A}$ is the mass fraction of the water solute in the water phase, and $k_{R,A}$ is the reaction rate constant of the water solute. Damage forms (or is removed) by the reaction according to the equation:

$$\frac{dD_f}{dt} = \frac{d\overline{m_A}}{dt} P_R W S_w \qquad 19\text{-}5$$

Where $D_f$ is the amount of damage (either conductivity or rel perm damage), W is the width of the fracture element, and $P_R$ is the potency constant for the reaction. If the potency constant is positive, then the reaction creates damage, and if the potency constant is negative, then the reaction removes damage. If the user specifies the option that the formation of damage is proportional to the oil saturation, then damage forms according to the equation:

$$\frac{dD_f}{dt} = \frac{d\overline{m_A}}{dt} P_R W S_w S_o \qquad 19\text{-}6$$

For example, to model 'gummy bear' damage, you could specify $k_{R,A}$ equal to 0.001 s^-1, inject friction reducer in the frac fluid at a mass fraction of 0.001 (roughly 8 ppt), set the potency constant for the reaction to 50, specify the damage type to be conductivity damage, and specify that damage is not proportional to the oil saturation. This combination of parameters may lead to formation of 'damage' in the fracture elements in the range of 0.05 inches.

'Damage' is quantified as a variable of volume per area (units of length) in each fracture element – one to keep track of conductivity damage, and another to keep track of rel perm damage. Fracture conductivity damage applies to all types of fracture conductivity: open, closed and unpropped, and closed and propped. It decreases the fracture conductivity according to the relation:

$$C_{damaged} = \frac{C_{undamaged}}{1+100\frac{D_{f,c}}{W}} \qquad 19\text{-}7$$

Relative permeability damage decreases the relative permeability of the oil and gas phases as they flow through the reservoir by increasing the oil and gas Brooks-Corey exponents according to the equations:

$$D_{rp,adj} = \left(1 - exp\left(-10\frac{D_{f,rp}}{W}\right)\right) \quad 19\text{-}8$$

$$n_o = n_{,init} + (n_{,dam} - n_{,init})D_{rp,adj} \quad 19\text{-}9$$

$$n_g = n_{gr,init} + (n_{gr,dam} - n_{gr,init})D_{rp,adj} \quad 19\text{-}10$$

By default, $n_{or,dam}$ and $n_{gr,dam}$ are assigned to a value of 4. This value can be customized with the parameter 'maximumdamagedhydrocarbonbrookscoreyexponent'.

Additionally, relative permeability damage can be made to increase the residual saturation of the water phase through a similar relation:

$$S_{wr} = S_{wr,init} + (S_{wr,dam} - S_{wr,init})D_{rp,adj} \quad 19\text{-}11$$

By default, $S_{wr,dam}$ is zero so this relation has no effect; however, it can be set with the parameter 'maximumdamagedresidualwatersaturation'.

If the potency constant is negative, then this is a reaction that removes damage. In this case, it is a reaction between the water solute and the damage itself. The reaction progresses at a rate that is proportional to the amount of damage per aperture:

$$\frac{d\overline{m_A}}{dt} = -k_{R,A}\overline{m_A}\frac{D_f}{W} \quad 19\text{-}11$$

If you specify a second water solute, then the reaction occurs between two different defined water solutes. The reaction progresses according to the relation:

$$\frac{d\overline{m_A}}{dt} = \frac{M_A}{M_B}\frac{d\overline{m_B}}{dt} = -R_{cond,A,B}\frac{M}{M_B}\overline{m_A m_B} = -R_{cond,A,B}\overline{m_A m_B} \quad 19\text{-}12$$

Where $M_A$ and $M_B$ are the molar masses of A and B, and M is the overall molar mass of the mixture. The molar mass terms are used to maintain stochiometric consistency. The molar mass term $\frac{M}{M_B}$ can be assumed to be rolled into the user-defined constant, $R_{cond,A,B}$. The rate of damage formation can be written in terms of the rate that water solute A is consumed by the reaction, as shown above.

Fracture damage reactions are specified in the table of “Fracture damage reactions” in the “Water Solutes” panel.

### 19.3.3 Fracture skin and water block

Fracture skin and water block are damage types that block flow into (and out of) fractures from the matrix. Fracture skin blocks flow of water, oil, and gas. Water block only blocks the flow of oil and gas. As discussed in the blog post, these damage types are intended to mimic the effect of capillary end effect in blocking production after water leakoff. They occur when water leaks off into the formation, but only if the formation fluid pressure has been depleted below the initial formation fluid pressure. Fracture skin and water block form as water leaks off, but they do not dissipate during production. They only decrease if you define a ‘Water block reduction reactions’ or ‘Fracture skin reduction reactions’, as discussed below, to mimic the effect of chemical treatment.

The amount of water block and fracture skin are quantified with a ‘water block thickness’ and a ‘fracture skin thickness.’ These values are defined between each individual fracture-matrix element connection. In other words, if a fracture element is connected to four different matrix elements, the thickness of the damage zones can be different for the fracture element’s connections to each of the four matrix elements. In the visualization tool, the simulator outputs the 'average' damage zone thickness for each fracture element, averaged over all of its connections.

To activate the formation of fracture skin or water block, set 'Fracture skin pressure reduction threshold' or 'Water block pressure reduction threshold' to a number between 0 and 1. If they are not set (which is the default), these damage mechanisms do not occur.

For example, let's say that the initial formation fluid pressure is 8000 psi. If you set 'Water block pressure reduction threshold' to 0.3, then water block will only form if fluid leaks off into the formation after the formation has been depleted to a pressure lower than 8000*(1-0.3) = 5600 psi. Then, the thickness of the water block layer grows according to the equation:

$$\Delta d_{wb} = 0.5V_L\left(1 - \frac{P_m}{P_{init}(1-P_{wb,rel,thresh})}\right) \quad 19\text{-}13$$

Where $\Delta d_{wb}$ is the increase in the thickness of the water block layer (on each side of the fracture), $V_L$ is the volume of fluid leaked off per area, $P_m$ is the pressure in the matrix, $P_{init}$ is the initial pressure of the matrix element, and $P_{wb,rel,thresh}$ is the 'Water block pressure reduction threshold'. The implication is that the lower the matrix pressure has become, the more rapidly leakoff forms a damage layer. To avoid mesh effects, $P_m$ is evaluated from the 1D submesh method as being the matrix pressure at a distance of 1 m from the fracture.

The skin layer grows according to a similar equation:

$$\Delta d_{skin} = 0.5V_L\left(1 - \frac{P_m}{P_{init}(1-P_{skin,rel,thresh})}\right) \quad 19\text{-}14$$

Optionally, you can define a maximum allowed thickness for these layers: 'Fracture skin maximum zone thickness' and 'Water block maximum zone thickness.'

The layers act like a filtercake – a thin, very low permeability layer blocking flow in or out of the fracture from the matrix. The skin layer affects all phases, and the water block layer affects only flow of oil and gas. The layer permeabilities are set by the parameters 'Fracture skin permeability reduction' and 'Water block permeability reduction'. These parameters are set relative to the in-situ permeability. For example, if the matrix permeability is 100 nd, and the 'water block permeability reduction' parameter is set to 0.001, then the permeability of the water block layer is 0.1 nd. In the simulator, flow is calculated using a thickness-weighted harmonic permeability average, which is the appropriate type of permeability average for flow in series.

To mimic the effect of surfactant and other chemical treatment, you can define one or more 'Water block reduction reactions' or 'Fracture skin reduction reactions'. For each you define: (a) the name of one of the water solutes defined in the 'water solutes' list, and (b) a reference composition. If water containing that water solute leaks off, then it removes the skin or water block damage.

Specifically:

$$\Delta d_{wb} = -0.5V_L\frac{\overline{m_{ws}}}{\overline{m_{ws,wb,ref}}} \quad 19\text{-}15$$

Where $\Delta d_{wb}$is the change in the thickness of the water block layer, $V_L$ is the volume of water leaked of per area, $\overline{m_{ws}}$ is the mass fraction of the water solute in the water phase as it leaks off, and $\overline{m_{ws,wb,ref}}$ is the reference mass fraction. A lower reference composition causes the damage layer to dissipate more rapidly. A similar equation is used for removing skin.

You can use damage reduction reactions either to remove skin that has already formed, or to prevent it from forming. As an example of the latter, during the frac job, if you include a water solute in the frac fluid that is defined as reducing damage, then as water leaks off, potentially causing damage, the water solute is simultaneously preventing it from forming.

The relevant parameters are “Water block pressure reduction threshold,” “Water block permeability reduction,” “Water block zone thickness,” “Water block reduction reactions,” “Fracture skin pressure reduction threshold,” “Fracture skin permeability reduction,” “Fracture skin maximum zone thickness,” and “Fracture skin reduction reactions.”

## 19.4 Water banking

With ResFrac, we usually use a fairly coarse matrix mesh. To avoid discretization error, we use the ‘1D submesh’ method, which automatically submeshes matrix elements containing fractures and calculates leakoff/production on a much finer reduced-order grid (discussed in Section 2). The mesh refinement of the 1D submesh method is necessary to capture nonlinearity created by sharp pressure gradients near the fracture. The 1D submesh does not solve a full multiphase flow problem, and so several approximate methods are used to capture multiphase effects. The parameter ‘adjust submesh for 1D flow’ handles gas/oil effects on GOR (discussed in Section 2). This section discusses the ‘water bank’ method used to handle water flowback.

As water leaks off from fracture elements, the code tracks how much water has leaked off and assigns it to a ‘water bank,’ which is assumed to be a region of high water saturation around a fracture. As water flows back, the water bank is depleted. Typically, the matrix element thickness is much greater than the water bank thickness – ~10 ft versus a fraction of an inch. Because the total matrix element is relatively large, the leakoff does not cause a significant change in saturation or rel perm in the matrix element itself. Instead, to capture elevated water cut during flowback, the ‘water bank’ treatment increases the water rel perm (and optionally decreases the hydrocarbon rel perm) during flowback in order to account for the effect of the water bank.

The setting ‘Water bank option’ allows you to select an option: ‘Original’ (the old way), ‘None’ (no special treatment for the water bank), or to use the newest (and recommended) method: ‘Dec2020Update’ or ‘Jan2021Update’. The latter two work similarly, with small differences discussed below.

For water flowback into the fracture, the water rel perm is calculated as:

$$k_{rw} = \gamma_{wb,w} + (1 - \gamma_{wb,w})k_{rw,orig} \qquad 19\text{-}16$$

$$\gamma_{wb,w} = min\left(1.0, \frac{d_{wb}}{d_{wb,ref,w}}\right) \qquad 19\text{-}17$$

Where $d_{wb}$ is the thickness of the water bank, $d_{wb,ref,w}$ is the ‘Water bank ‘rel perm increase’ scaling thickness’, and $k_{rw,orig}$ is the water rel perm in the matrix element overall (not considering the water bank). As water leaks off and the bank thickness grows, the interpolation factor $\gamma_{wb}$ approaches and eventually reaches 1.0. When $\gamma_{wb}$ is equal to 1.0, then the rel perm for water flowback is 1.0. When $\gamma_{wb}$ is zero, the water bank adjustment has no effect. You can make the water bank adjustment stronger by specifying a lower value for $d_{wb,ref}$.

Note that the adjustment above has increased water relative permeability but has not decreased relative permeability of the hydrocarbon phases. Based on the behavior we see in field data, we think that this is a reasonable assumption, at least as a starting point. This counterintuitive observation may be due to mixed-wettability of the formation.

Regardless, you may optionally choose to also reduce the relative permeability of the hydrocarbon phases. Specify '$d_{wb,ref,h}$', the 'Water bank 'rel perm decrease' scaling thickness'. Then:

$$k_{rh} = (1 - \gamma_{wb,h})k_{rh,orig} \quad \text{19-18}$$

$$\gamma_{wb,h} = min\left(1.0, \frac{d_{wb}}{d_{wb,ref,h}}\right) \quad \text{19-19}$$

Where $k_{rh}$ is the relative permeability of the hydrocarbon phase(s). The parameter $d_{wb,ref,h}$ is optional. If you do not specify it, then $\gamma_{wb,h}$ is always set equal to zero.

The two options 'Dec2020Update' and 'Jan2021Update' work similarly, with minor differences. The Dec2020Update option calculates the leakoff rate separately for each pair of matrix and fracture elements. The Jan2021Update version calculates leakoff rate using an average leakoff rate of the entire fracture element and its neighbors, and then assigns leakoff proportionally to each fracture-matrix pair based on their relative contribution to leakoff. Either is fine, but the latter may be slightly more robust because it avoids localized mesh artifacts.

Refer to the 'water bank' option in the advanced section of the 'Fracture Options' panel, as well as the parameters 'Water bank 'rel perm increase' scaling thickness' and 'Water bank 'rel perm decrease' scaling thickness'.

## 19.5 Pressure-dependent near-wellbore tortuosity

In the full calculation of flow rate between fracture and well elements, the simulator solves the equation:

$$\Delta P = AQ + BQ^2 + \Delta P_{nw} \quad \text{19-20}$$

Where A is a term related to Darcy pressure gradient, and B is a term that lumps together Forchheimer pressure gradient and perforation pressure drop.

We have now implemented an alternative way of modeling the near-wellbore pressure drop, based on a pressure-dependent near-wellbore pressure drop (or tortuosity) transmissibility.

The new relation is:

$$\Delta P_{nw} = \frac{\mu Q}{T_{nw}} \quad \text{19-21}$$

Where $\mu$ is the viscosity of the fluid, $T_{nw}$ is a 'transmissibility'. The transmissibility is calculated from the equation:

$$T_{nw} = T_{nw,min} + T_{nw,ref}10^{\frac{0.5(P_f+P_w)-P_{nw,ref}}{P_{nw,scale}}} \quad \text{19-22}$$

Where $T_{nw,min}$, $T_{nw,ref}$, $P_{nw,scale}$, and $P_{nw,ref}$ are user-specified constants. For example, let's say that Shmin = 8000 psi. You could set $T_{nw,min}$ equal to 0.01 md-ft, $T_{nw,ref}$ equal to 1000, $P_{nw,ref}$ equal to 9200 psi, and $P_{nw,scale}$ equal to 250 psi. If the pressure in the fracture is 8200 psi, and the wellbore pressure is 10,200 psi, then $\frac{0.5(P_f+P_w)-P_{nw,ref}}{P_{nw,scale}}$ equals zero, and $T_{nw}$ will be equal to 1000.01 md-ft. If pressure in the well drops to 8200 psi and the pressure in the fracture remains the same, then $\frac{0.5(P_f+P_w)-P_{nw,ref}}{P_{nw,scale}}$ will equal -4, and $T_{nw}$ will equal 0.11 md-ft. If the pressure in the well and fracture both equal 7200 psi, then $T_{nw}$ will equal 0.01001 md-ft.

Instead of using $0.5(P_f + P_w)$, there is an option to alternatively use the minimum of $P_f$ and $P_w$, which can lead to better convergence behavior.

Relative to the original near-wellbore pressure drop treatment, the pressure-dependent near-wellbore pressure drop treatment causes the 'near-wellbore' period of the transient to stretch out for longer. This can be useful for matching field-data.

If you specify any of the four constants in the pressure-dependent near-wellbore model, you must specify all of them. If these constants are specified, the parameters used for the conventional near-wellbore pressure drop treatment are disabled, and not used. In other words, the near-wellbore pressure drop is calculated using one method or another, and not both at the same time.

Here is a DFIT simulation with $T_{nw,min} = 0.01\ md - ft$, $T_{nw,ref} = 100\ md - ft$, $P_{nw,scale} = 1000\ psi$, and $P_{nw,ref} = 9300\ psi$. Below, are simulations with lower values of $P_{nw,scale}$. The progression shows how the PD NW pressure drop causes an extended duration of the near-wellbore tortuosity pressure drop, and weakens the indication of fracture closure in the pressure transient.

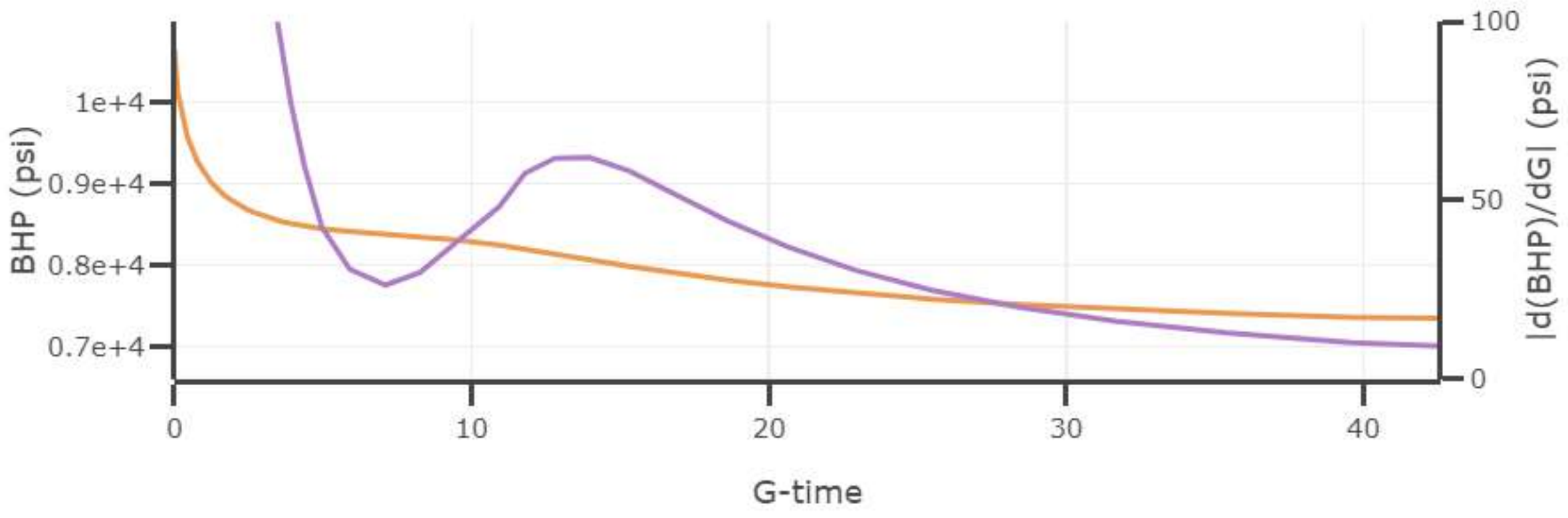

This simulation has $P_{nw,scale} = 800\ psi$.

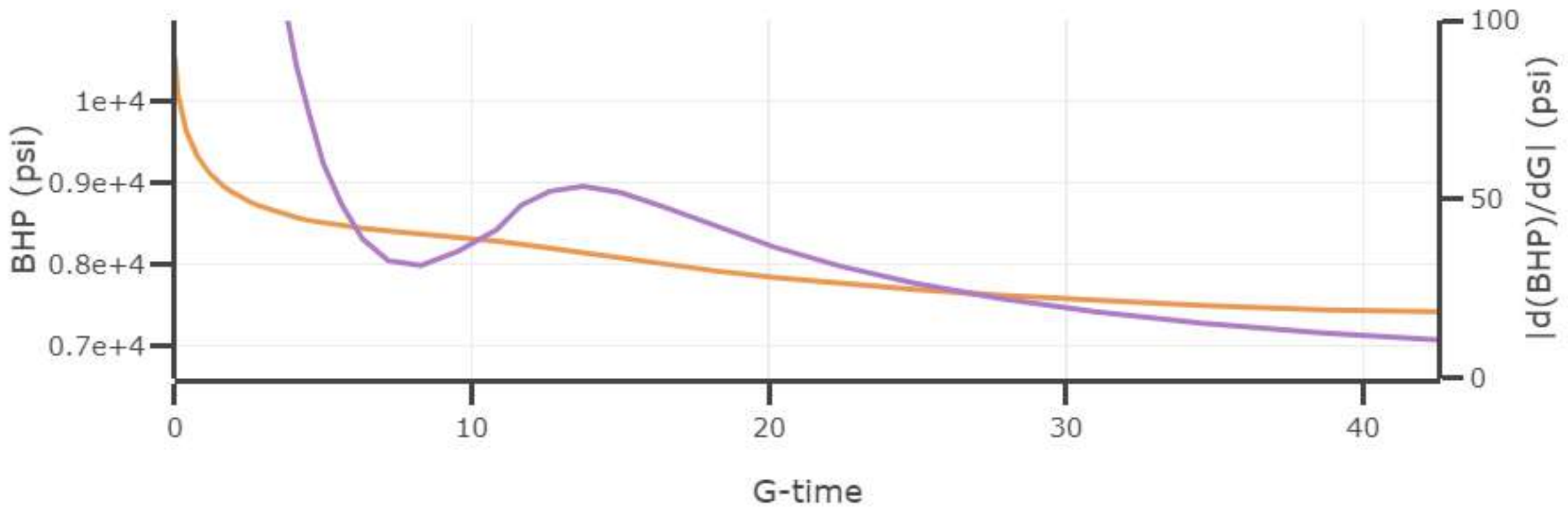

This simulation has $P_{nw,scale} = 650\ psi$.

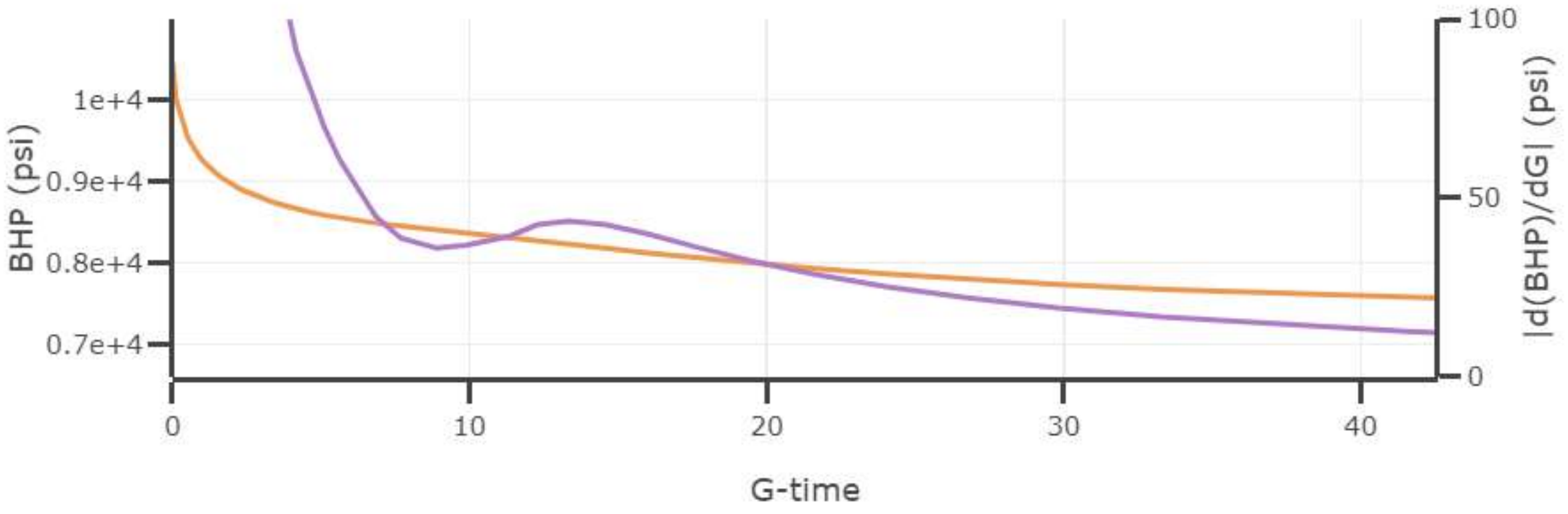

This simulation has $P_{nw,scale} = 500\ psi$.

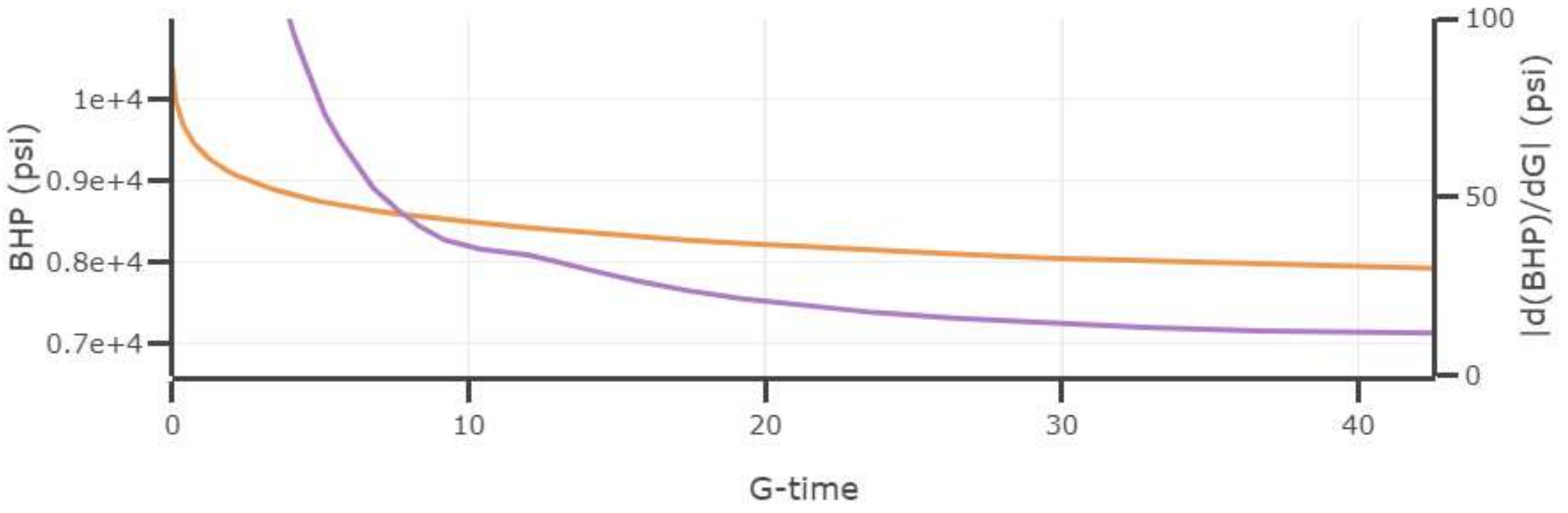

This simulation has $P_{nw,scale} = 250\ psi$.

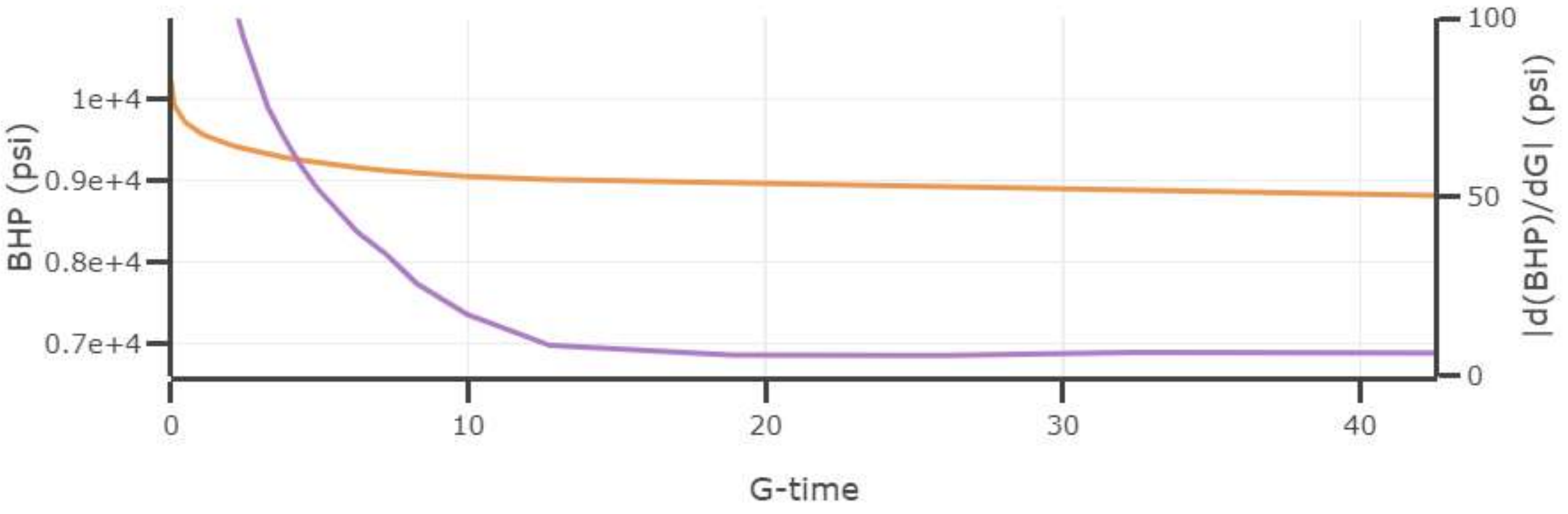

Refer to the four 'PD NW' parameters in the advanced section of the 'Wells and Perforations' panel.

## 19.6 Proppant embedment

Proppant embedment prevents some of the proppant in the fracture from creating a propping effect. Figure 4 from Alrahami and Sundberg (2012) shows that embedment is typically anywhere from 0 to 0.9 mm. The magnitude of embedment increases with the effective normal stress and decreases with Young's modulus (and clay content). For a rock with typical Young's modulus of 4e6 psi, then their Figure 5 shows that embedment at 5000 psi will be around 0.2 mm. If the proppant grain specific gravity is 2.5, then 0.2 mm of embedment corresponds to 0.5 kg/m^2, or 0.1 lbs/ft^2.

In-situ, the relationship between embedment and conductivity is not well-known. Core-through studies suggest that proppant deposition is highly nonuniform. Thus, even if the average proppant deposited was equal to 0.1 lbs/ft^2, and this was the same as the embedment, this would not necessarily mean that all of the proppant will embed. If the proppant is concentrated into patches covering 25% of the total area, then the equivalent thickness of these clumps will be 4x higher. Thus, we should be cautious about applying these embedment values too literally. While we can include embedment in the model, it should not be considered well-constrained.

In ResFrac, you can specify the amount of embedment on a layer-by-layer basis. The default behavior is for embedment to be constant and not a function of stress. However, you have an option to make embedment a function of the square root of effective normal stress (roughly matching the shape of Figure 4 from Alrahami and Sundberg, 2012)). This option is activated by specifying the “Proppant embedment reference stress.” The amount of embedment is then modeled as being equal to:

$$Embedment = (Specifiedembedment)\sqrt{\sigma_n'/\sigma_{n'_{embedment,ref}}}. \quad \text{19-23}$$

Embedment is modeled as causing a reduction in the amount of proppant available to increase conductivity. The conductivity is calculated as a function of proppant mass per area. The amount of embedment is subtracted from the ‘proppant mass per area’ before the conductivity calculation is performed.

## 19.7 Time-dependent proppant pack conductivity

By default, proppant pack conductivity is function of: effective normal stress (with an option to make changes irreversible), proppant concentration, formation properties (such as embedment), and ‘conductivity damage’ related to frac hits and chemical damage mechanisms.

There is also a capability to degrade proppant pack conductivity over time. This is motivated by: (1) observations from laboratory experiments suggesting time-dependent conductivity (Duenckel et al., 2017; Pearson et al., 2020), and (2) experience with history matching suggesting that time-dependent conductivity loss may help with matching late-time EUR observations and RTA curvature.

The treatment of time-dependent conductivity must be able to handle a general set of situations. What if different well are fractured at different times, and proppant of different ages mix together? What if a well is refractured? Because of these nuances, we cannot merely define ‘proppant conductivity as a function of time.’ Instead, we must define a ‘time-dependent conductivity damage’ variable, define an equation to relate that damage variable to conductivity, and define a relation to describe how this variable evolves over time.

ResFrac offers two options:

*Option 1:*

$$D_{td,fac,j} = 1 - \frac{D_{td,j}}{m_j} \quad \text{19-24}$$

*Option 2:*

$$D_{td,fac,j} = \left(1 - \frac{D_{td,j}}{m_j}\right)^3 \quad \text{19-25}$$

Where $m_j$ is the mass per surface area of proppant *j* in the fracture.

In both cases, an overall factor $D_{td,fac}$ is calculated as the volume fraction weighted average of $D_{td,fac,j}$ for each type of proppant:

$$D_{td,fac} = \sum D_{td,fac,j} \frac{C_{pr,j}}{C_{pr}} \qquad \text{19-26}$$

The proppant pack conductivity is multiplied by the overall $D_{td,fac}$. In other words, if $D_{td,fac}$ is equal to 0.01, then proppant pack conductivity has reduced by 99%.

For each type of proppant, the user specifies a rate constant that expresses how quickly conductivity loss evolves over time, $k_{D,td,j}$. For the two different options, the expressions are:

*Option 1:*

$$\frac{d\left(D_{td,j}/m_j\right)}{dt} = \frac{k_{D,td,j}}{ln(10)} 10^{\frac{-D_{td,j}}{m_j}\frac{1}{k_{D,td,j}}} \qquad \text{19-27}$$

*Option 2:*

$$\frac{d\left(D_{td,j}/m_j\right)}{dt} = k_{D,td,j}\left(1 - 36\left(\frac{\left(\frac{D_{td,j}}{m_j}\right)^2}{2} - \frac{\left(\frac{D_{td,j}}{m_j}\right)^3}{3}\right)\right)^2 \qquad \text{19-28}$$

For the particular case that $m_j$ is constant, Option 1 implies that for t > 0:

$$\frac{D_{td,j}}{m_j} = k_{D,td,j} * log_{10}(t) \qquad \text{19-29}$$

Thus, Option 1 scales conductivity loss with the logarithm of time. This relation is consistent with prior literature, such as from Duenckel et al. (2017) and Pearson et al. (2020). However, experience with history matching suggests that it can lead to unusual results. The relation does not cause conductivity to asymptotically approach zero. Instead, it can reach zero rather abruptly. This causes unrealistic-looking kinks on RTA plots derived from simulation using this relation.

Option 2 was designed to address these problems. With Option 2, the conductivity loss over time occurs smoothly, and the conductivity asymptotically approaches zero over time, resulting in smoothly curving RTA plots, consistent with real observations.

The figure below shows the evolution of the conductivity multiplier as a function of time, using Option 2, on a log scale.

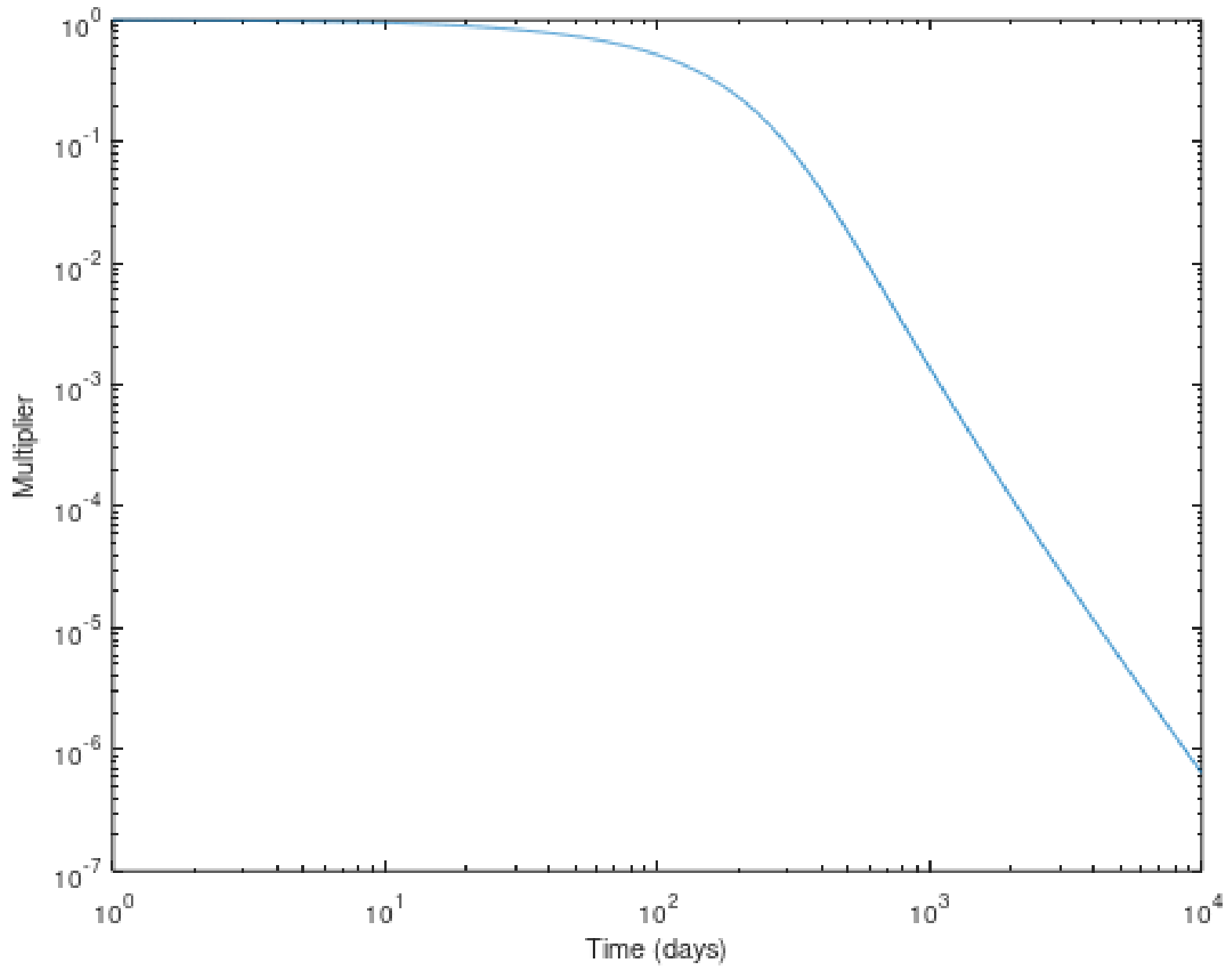

Figure 30: Conductivity multiplier versus time using time-dependent conductivity loss with Option 2.

The figure below shows an example of an RTA plot from a simulation using time-dependent conductivity loss with Option 2. The RTA curvature is primarily caused by the time-dependent conductivity loss.

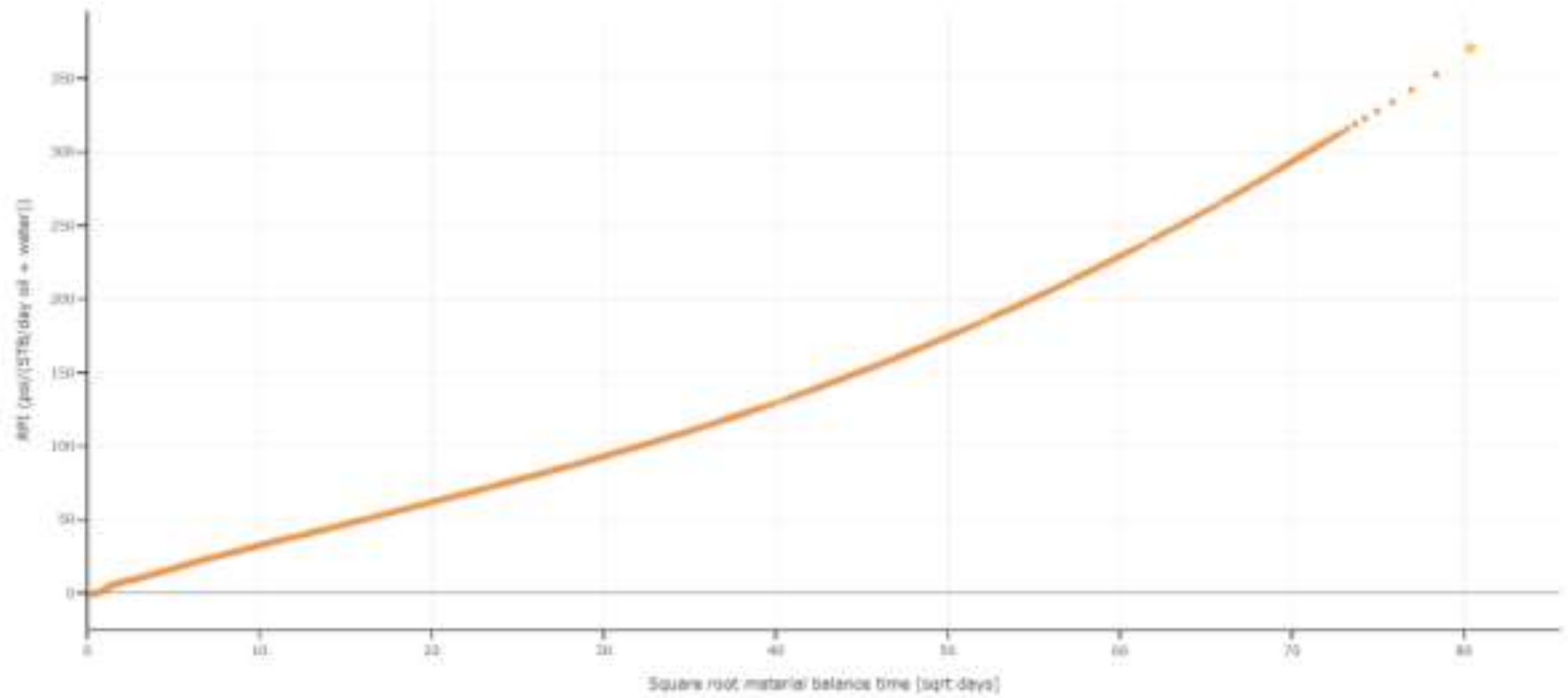

Figure 31: Rate-transient analysis plot from a simulation using time-dependent conductivity loss Option 2. The upward curvature is caused primarily by the conductivity loss.

There is an option to make the rate parameter ($k_{D,td,j}$) increase with effective normal stress. Specifically, you specify a reference stress $\sigma'_{n,tdref,j}$. The rate parameter is multiplied by the parameter $\left(1 + \frac{\sigma'_n}{\sigma'_{n,tdref,j}}\right)^2$, where effective normal stress is always greater than or equal to zero.

## 19.8 Economics module

The *Economic Models* option in ResFrac allows users to perform economic analysis of simulation models results and compare between realizations with different economic assumptions. The economic models can be defined in the Output Options tab of the UI.

In the economic model, the user can select to use either the ‘Oil and gas’ application (the default), or the ‘Geothermal’ application, or both. By selecting both ‘Oil and gas’ and ‘Geothermal’ applications in the same economic model, the user can calculate economic outputs for an oil and gas site, converted to a geothermal energy plant.

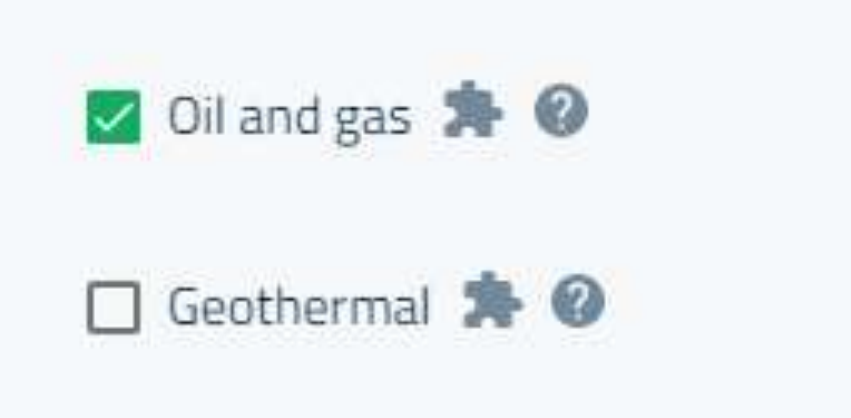

The user can define more than one economic model and each model generates a separate csv file containing the following outputs:

1. Time series for time discounted cost, revenue, cashflow, CAPEX and OPEX, and production.
2. Breakdown of the time discounted costs into time discounted CAPEX and OPEX at the end of the simulation.
3. Net present value (NPV), ratio of NPV to the total cost, Discounted Return on Investment/Invest Efficiency (DROI) and Internal Rate of Return (IRR) at the end of the simulation.
4. Total water, oil and gas production (for the ‘Oil and gas’ application) at the end of the simulation, overall for all wells and for each specified group of wells (well groupings, optionally specified in the Output Options panel).

The economic model calculates the time discounted incremental cost and revenue at the end of each timestep t in the simulation using the following expressions:

$$DiscountedIncrementalCost = \frac{IncrementalCost}{(1+r)^y} \qquad \text{19-30}$$

$$DiscountedIncrementalRevenue = \frac{IncrementalRevenue}{(1+r)^y} \qquad \text{19-31}$$

where $y$ is the time in years since the start and $r$ is the user input annual $discount\ rate$.

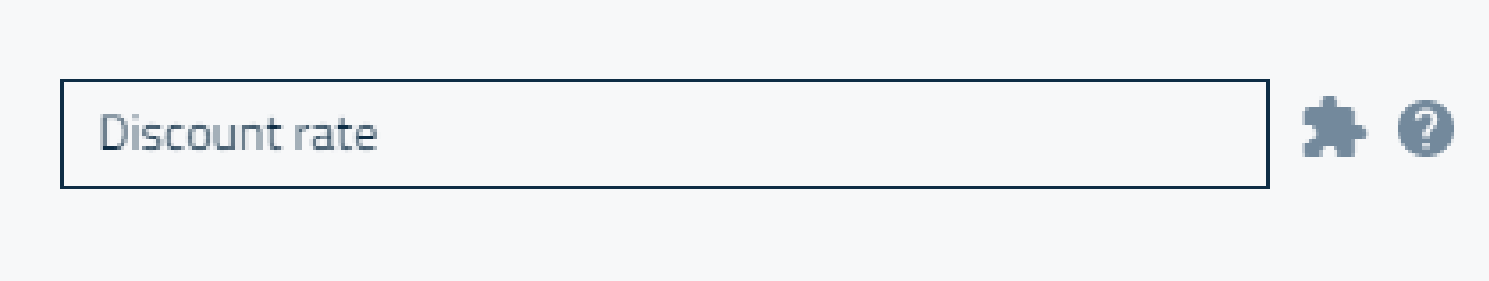

The cumulative cost and revenue at the end of each time step are computed as:

$$CumulativeDiscountedCost = \sum_{j=0}^{t} DiscountedIncrementalCost_j \qquad \text{19-32}$$

$$CumulativeDiscountedRevenue = \sum_{j=0}^{t} DiscountedIncrementalRevenue_j \qquad \text{19-33}$$

where t is the current time step. The cumulative discounted cost is a sum of the $CumulativeDiscountedCapex$ and the $CumulativeDiscountedOPEX$. Utilizing the $CumulativeDiscountedCost$ and $CumulativeNetDiscountedRevenue$, the NPV, DROI and the IRR are calculated with the following expressions:

$$NPV = CumulativeDiscountedRevenue - CumulativeDiscountedCost \quad 19\text{-}34$$

$$DROI = \frac{DiscountedCumulativeRevenue - DiscountedCumulativeOPEX}{DiscountedCumulativeCAPEX} \quad 19\text{-}35$$

$$IRR = DiscountrateatwhichNPVis0. \quad 19\text{-}36$$

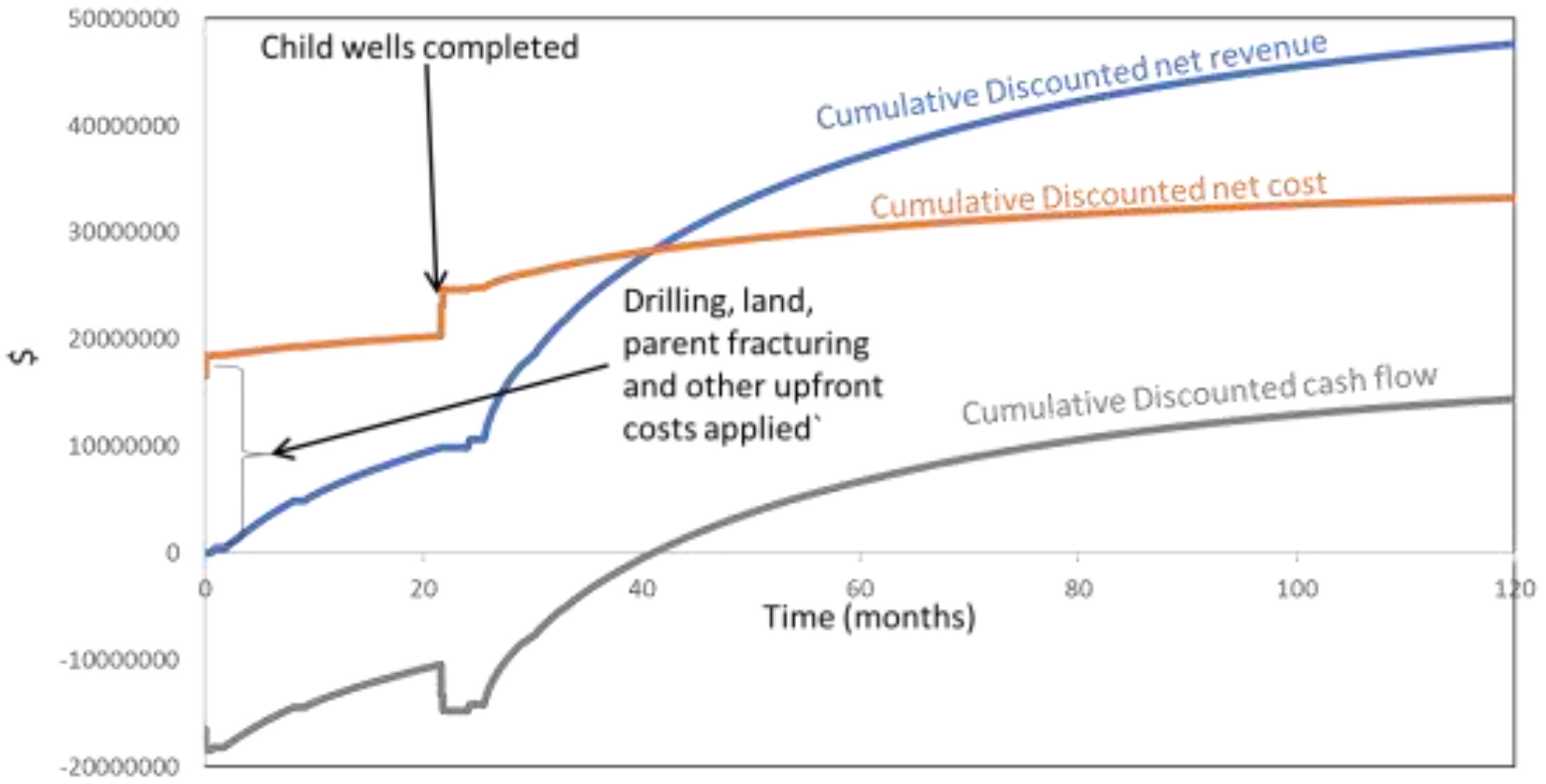

*Figure 32: Results from an example economic model with the 'Oil and gas' application. The time-series for the cumulative discounted net revenue, cumulative discounted net cost and cumulative discounted cashflow are exported in the economics summary csv file along with the NPV, IRR and DROI at the end of the simulation. The example shows the application of the parent and child well completion costs at different stages in the simulation.*

In the following, the description is given for the cost and revenue inputs, common to the 'Oil and gas' and 'Geothermal' applications, and those specific to 'Oil and gas'. Note that the parameters, related to oil, gas, and NGL, are accessible only when 'Oil and gas' application is selected. For the list of the inputs, specific to the 'Geothermal' application, refer to the section "19.8.2 Geothermal economic model (new)" in the end of this section.

The user can click the Add button to create a new economic model.

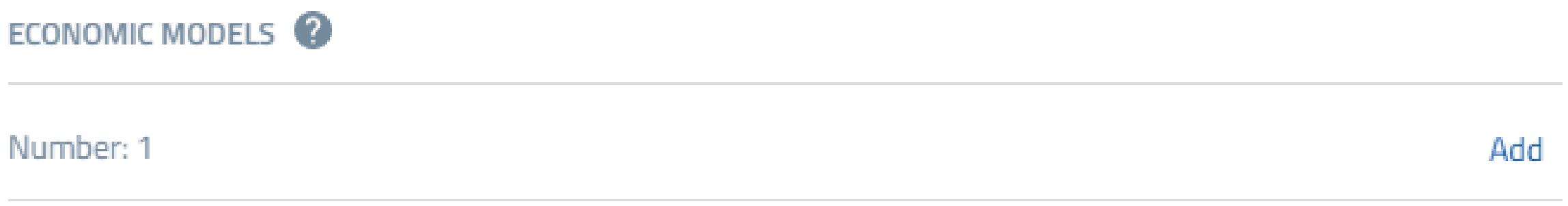

The users can specify which wells they want to include in the model calculations.

INCLUDE IN MODEL

| | Well name | Include in model |
|---|---|---|
| 1 | Well_Child1 | ☑ |
| 2 | Well_Parent | ☑ |
| 3 | Well_Child2 | ☑ |
| 4 | Well_Child3 | ☑ |

The cost parameters include costs that are always applied at the beginning such as the land cost, costs that are applied during specific events such as drilling costs, fracturing costs and cost that depend on the simulation time series results such as produced water costs. The users can also input their working interest in the CAPEX or OPEX as a fraction. These are the fraction of the operating or capital expenses paid by your company. All the cost inputs in the different categories are multiplied by the respective working interest fractions. The land cost, miscellaneous up-front cost, drilling cost per well, drilling cost per measured depth, fixed fracturing cost per well, completion cost per stage, injection water and proppant cost per well and facilities cost per well are considered as CAPEX. The produced water cost, produced oil cost and produced gas cost are considered as OPEX.

All the costs are multiplied by working interest in CAPEX/OPEX and one plus the overhead rate. At the start of the simulation, the land cost and the miscellaneous upfront cost is applied. The total land cost is calculated by multiplying the input cost/acre with the surface area (user-specified well spacing times lateral length). When a well is first drilled, a drilling cost is applied. Note that the wells can be drilled during a simulation by defining the drilling times option. When you first inject into a well, an upfront fracturing cost and the completion cost per stage are applied for that well for all the stages (by default). The user has an option to apply the upfront fracturing cost and the completion cost per stage to selected wells at specified fracturing and refrac times, by unchecking the box 'Apply fracturing cost at the start of injection' and filling-in the table 'Fracturing start times by well'.

☐ Apply fracturing cost at the start of injection

FRACTURING START TIMES BY WELL

| | Well name | Fracturing start time by well period one [days] | Fracturing start time by well period two [days] | Fracturing start time by well period three [days] |
|---|---|---|---|---|
| 1 | Well_Demo | | | |

As the injection is performed, cost is applied for each bbl of water and lb of proppant injected, as well as for each bbl of oil and Mscf of gas injected.

There is an option for applying the cost of injected and produced fluids to the net injected volume or the net produced volume at a moment of time. It can be used in simulations that involve reinjection of produced fluids, such as for an EGS system with long-term circulation, waterflood, or enhanced oil recovery with gas reinjection. This option is activated by using the table 'Net-out instantaneous injection and production start and end times by well'. For each well, the user can specify up to five 'instantaneous net-out' periods, during which the well will be included in the cost for net instantaneous injection and production. In any given timestep, the overall injection and production volumes will be netted-out among all the wells, that are active in an 'instantaneous net-out' period. This applies to each fluid separately (water, oil, and gas). If more fluid is produced than injected, the produced fluid cost ('produced water cost', 'produced gross oil cost', or 'produced gross gas cost') is applied to the net produced volume. If more fluid is injected than produced, the injection fluid cost ('injection water cost', 'injected oil cost', or 'injected gas cost) is applied to the net injected volume. If you include the 'Geothermal' application in the economic model, then the 'replacement water cost' will be used instead of the 'injection water cost' in some situations; refer to the description of the geothermal economic model in the section "19.8.2 Geothermal economic model (new)".

NET-OUT INSTANTANEOUS INJECTION AND PRODUCTION START AND END TIMES BY WELL

| | Well name | (Optional) Start time of period one to apply cost for net instant... | (Optional) End time of period one to apply cost for net instantan... | (Optional) Start time of period two to apply cost for net instant... | (Optional) End time of period two to apply cost for net instantan... |
|---|---|---|---|---|---|
| 1 | Well_1 | | | | |
| 2 | Well_2 | | | | |
| 3 | Well_3 | | | | |

For the EOR simulations with gas reinjection, users can opt to use the dedicated EOR economics module, instead of filling-in the table 'Net-out instantaneous injection and production start and end times by well'. The EOR economics module has more parameters, specific to gas reinjection; see section "19.8.1 EOR economics module".

When a well first goes on production it accumulates operations costs, defined as an annual cost per well. Since the operational costs and production costs may change over time, the user has the option to specify up to three different phases of the life of the well with different annual operations costs and production costs.

While the taxes are calculated from the revenue generated, they are added to the cost, specifically the OPEX.

For an 'Oil and gas' economic model (the default), the severance and ad-valorem tax rates are input by the user. These rates are multiplied by the net revenue to get the respective taxes in $ amounts. The users can also specify a miscellaneous tax rate in the parameter 'Miscellaneous taxes' as a fraction. This is also calculated from the revenue and added to the OPEX. Optionally, the user can include taxes and royalties in this parameter instead of the more detailed tax and net revenue interest parameters.

For an economic model with the 'Geothermal' application, a 'geothermal tax and royalty rate' is specified in the table 'Geothermal economic model inputs' and is multiplied by the geothermal revenue; see the detailed description of the geothermal economic model in the section "19.8.2 Geothermal economic model (new)".

The cost inputs required in the model are (in the same order as the UI):

1. *Drilling cost per well* ($): This is the cost of drilling each well. This is applied at the time of drilling of each well defined by the 'Wellbore drilling times' input. This input is required if you do not specify 'Drilling cost per measured depth'.

2. *Drilling cost per measured depth* ($/ft): This is the cost of drilling per length unit of the measured well depth. This is applied at the time of drilling of each well defined by the 'Wellbore drilling times' input, in addition to 'Drilling cost per well'. This input is required if you do not specify 'Drilling cost per well'.
3. *Land cost* ($/hectare): This is the upfront land cost from a leasing perspective and is applied at the beginning of the simulation.
4. *Fixed fracturing cost per well* ($): This is the fixed cost of fracturing each well and is applied at the time of first injection into the well. If the setting 'Apply fracturing cost at the start of injection' is set to false, then this cost is applied at the times, specified in the table 'Fracturing start times by well'.
5. *Fracturing proppant cost* ($/lb): This is the cost of each pound injected and is applied at each timestep with proppant injection.
6. *Injection water cost* ($/bbl): This is the cost of each barrel injected and is applied at each timestep with injection. Note that, if you use the table 'Net-out instantaneous injection and production start and end times by well' to specify 'instantaneous net-out' periods, the 'injection water cost' will not be applied to each barrel of water injected into wells, that are active in 'instantaneous net-out' periods. In addition, if you include the 'Geothermal' application in the economic model, a related parameter 'replacement water cost', specified in the table 'Geothermal economic model inputs', will be used for the cost of injected water in some situations. Refer to the description of the table 'Net-out instantaneous injection and production start and end times by well' for more details.
7. *Overhead rate*: Fraction of the spending that goes to overhead. Overhead rate times the cost is added to the cost at each timestep.
8. *Produced water cost* ($/bbl): This is the cost of treatment of produced water and is applied along with the water production volumes. Note that, if you use the table 'Net-out instantaneous injection and production start and end times by well' to specify 'instantaneous net-out' periods, the 'produced water cost' will not be applied to each barrel of water produced from wells, that are active in 'instantaneous net-out' periods. Refer to the description of the table 'Net-out instantaneous injection and production start and end times by well' for more details.
9. *Produced gross oil cost* ($/bbl): This is the cost of surface facilities and transport for the produced oil. Note that, if you use the table 'Net-out instantaneous injection and production start and end times by well' to specify 'instantaneous net-out' periods, the 'produced gross oil cost' will not be applied to the total oil produced from the wells, that are active in 'instantaneous net-out' periods. Refer to the description of the table 'Net-out instantaneous injection and production start and end times by well' for more details.
10. *Produced gross gas cost* ($/Mscf): This is the cost of surface facilities and transport for the produced gas. Note that, if you use the table 'Net-out instantaneous injection and production start and end times by well' to specify 'instantaneous net-out' periods, the 'produced gross gas cost' will not be applied to the total gas produced from the wells, that are active in 'instantaneous net-out' periods. Refer to the description of the table 'Net-out instantaneous injection and production start and end times by well' for more details.
11. *Miscellaneous upfront costs* ($): Upfront costs applied the beginning of the simulation.
12. *Facilities cost per well* ($/well): Fixed one-time capital expense imposed at the beginning for each well. This accounts for the cost of surface facilities such as separators.
13. *Operations annual cost per well*: This is applied from the beginning of production of each well. Since operational costs can change with time, this parameter can be defined separately for up to three phases with the phase one and phase two durations as inputs.
14. *Oil severance tax*: Tax rate paid on produced oil.
15. *NGL severance tax*: Tax rate paid on produced NGL.
16. *Gas severance tax*: Tax rate paid on produced gas.

17. *Ad valorem tax*: The ad valorem tax is calculated on revenue after the payment of severance taxes. For example, ad valorem tax paid = ((Gross oil sale revenue)*(1 - Oil severance tax) + (Gross NGL sale revenue)*(1 - NGL severance tax) + (Gross oil sale revenue)*(1 - Gas severance tax))*(Ad valorem tax).
18. *Miscellaneous taxes*: The fraction of revenue that goes to miscellaneous payments. These are added to the OPEX. This is applied as a fraction of the gross revenue. Optionally, this can be used to lump in taxes and royalties. Alternatively, you can use the more detailed parameters such as 'net working interest', gas and oil severance and ad valorem tax, etc.

Additional optional cost inputs include the following parameters:

19. *Completion cost per stage* ($): This is the fixed completion cost per stage, applied to the total number of stages in each well. The number of stages in each well is calculated from the user inputs and corresponds to the scaling, in economic calculations, of the injection and production volumes up to the 'full well'. This cost is applied at the time of first injection into the well. If the setting 'Apply fracturing cost at the start of injection' is set to false, then this cost is applied at the times, specified in the table 'Fracturing start times by well'.
20. *Injection oil cost* ($/bbl): This is the cost of each barrel of oil injected and is applied at each timestep with injection. Note that this parameter cannot be specified simultaneously with specifying the values for '*Cost of purchased oil'* in the EOR economics module, activated by checking the box 'Include EOR inputs'. Refer to the description of the EOR economics module.
21. *Injection gas cost* ($/Mscf): This is the cost of each Mscf of gas injected and is applied at each timestep with injection. Note that this parameter cannot be specified simultaneously with specifying the values for '*Cost of purchased gas'* in the EOR economics module, activated by checking the box 'Include EOR inputs'. Refer to the description of the EOR economics module.
22. *Produced water cost phase two* ($/bbl): This is the treatment cost of produced water, for phase two of the life of a well.
23. *Produced water cost phase three* ($/bbl): This is the treatment cost of produced water, for phase three of the life of a well.
24. *Produced gross oil cost phase two* ($/bbl): This is the cost of the surface facilities and transport for produced oil, for phase two of the life of a well.
25. *Produced gross oil cost phase three* ($/bbl): This is the cost of the surface facilities and transport for produced oil, for phase three of the life of a well.
26. *Produced gross gas cost phase two* ($/Mscf): This is the cost of the surface facilities and transport for produced gas, for phase two of the life of a well.
27. *Produced gross gas cost phase three* ($/Mscf): This is the cost of the surface facilities and transport for produced gas, for phase three of the life of a well.
28. *Fixed fracturing cost per well per maximum injection rate* ($/bpm): This optional parameter allows you to specify marginal cost associated with the maximum injection rate. This cost is applied at the time of first injection into the well. If the setting 'Apply fracturing cost at the start of injection' is set to false, then this cost is applied at the times, specified in the table 'Fracturing start times by well'.

Note that, if you use the table 'Net-out instantaneous injection and production start and end times by well', the 'injection oil cost' and the 'injection gas cost' will be applied to the netted-out injection from the wells, that are active in an 'instantaneous net-out' period. Additionally, for each netted-out well in phase two, the 'produced water cost phase two' will be applied to its contribution to the netted-out water production, calculated proportionally to its contribution to the overall (not netted-out) production from all the netted-out wells. Similar calculations apply to the use of the above parameters for the produced water, oil, and gas, for phases two and three, when the table 'Net-out instantaneous injection and production start and end times by well' is filled-in.

The revenue from production of hydrocarbons is calculated by specifying a price for oil, gas and natural gas liquids (NGL). The oil and gas volumes are multiplied by the Gross oil and gas shrinkage fractions if specified. The amount of natural gas liquids yielded per standard volume of natural gas Gross NGL yield (STB/Mscf) can be specified optionally by the users. Note that this is already calculated by the simulator if using the modified black oil or compositional models, in these cases this parameter can be used to assume that there is some amount of NGL produced from the standard volume of gas additional to what is reported from the simulator. All the revenue is computed pre-taxes. The net revenue is computed multiplying with the net revenue interest (accounting for the royalty and sharing of revenue between different companies with joint interest in the wells).

The revenue inputs in the model are:

- *Price of oil* ($/bbl): Gross sale price of oil
- *Gross oil shrinkage*: Fraction of oil lost to shrinkage. For example, if you specify 0.01, then 1% of the produced oil is lost and not counted towards gross production.
- *Price of gas*($/Mscf): Gross sale price of gas.
- *Gross gas shrinkage*: Fraction of oil lost to shrinkage. For example, if you specify 0.01, then 1% of the produced gas is lost and not counted towards gross production.
- *Price of NGL* ($/bbl): Gross sale price of NGL.
- *Gross NGL yield*: The amount of natural gas liquids yielded per standard volume of produced gas. Please note that if you use the 'modified black oil model' or the 'compositional model', then the simulator should already be calculating the amount of 'liquids' from the gas produced from the reservoir. Thus, using this parameter is options (you may opt to make it 0). Regardless of the type of fluid model, this parameter is used to assume that there is some ADDITIONAL amount of NGL produced from the 'standard volume' of gas that is reported from the simulator.
- *Net revenue interest in oil*: Fraction of the oil revenue that is received. This accounts for royalty payments and the sharing of revenue between different companies with joint interest. This is also used for the NGL revenue.
- *Net revenue interest in gas*: Fraction of the gas revenue that is received. This accounts for royalty payments and the sharing of revenue between different companies with joint interest.

Optionally, the user can specify time-dependent prices of oil, gas, and NGL by filling-in the table of oil, gas, and NGL prices versus time. The values, specified in this table, will override the values, specified for the parameters 'Price of oil', 'Price of gas', and 'Price of NGL'.

OIL, GAS, AND NGL PRICES

| | (Optional) Prices cumulative time [days] | (Optional) The sale price of oil [dollars/bbl] | (Optional) The sale price of gas [dollars/Mscf] | (Optional) The sale price of NGL [dollars/bbl] |
|---|---|---|---|---|
| 1 | | | | |

As simulations usually include a few stages, the users can specify the ‘Well scaling factor’ for each well, so that the production and injection volumes and costs are scaled up to a full well. For example, if the simulation has two stages, and the well has 50 total, then the production and injection costs are scaled 25x.

WELL SCALING FACTOR

| | Well name | Well scaling factor |
|---|---|---|
| 1 | Well_1 | 25 |
| 2 | Well_2 | 25 |
| 3 | Well_3 | 25 |

Alternatively, the users can specify the number of stages in the wells, to scale the production and injection volumes and costs up to a full well. However, this option is no longer recommended.

STAGES PER WELL (THIS SETTING IS NO LONGER RECOMMENDED)

| | Well name | Stages per well (this setting is no longer recommended) |
|---|---|---|
| 1 | Well_1 | 50 |
| 2 | Well_2 | 50 |
| 3 | Well_3 | 50 |

If the user has specified injection and production volume multipliers by using the table 'Well scaling factor', or the table 'Stages per well', and, additionally, has specified 'Well flow scalers' in the 'Well controls' panel, it is recommended to turn on the option 'Do not apply volume multipliers during 'well flow scalers' in the economic models'. This option is located under the 'Advanced options' of the 'Output Options' panel. When this option is set to 'true' (the default), the volume multipliers, calculated in the economic models, are not applied during a period when the well has an active 'well flow scaler'.

☑ Do not apply volume multipliers during 'well flow scalers' in the economic models

The users also need to specify average well spacing and lateral length. These are used to calculate the surface footprint of each well in order to calculate the land cost.

LATERAL LENGTH PER WELL [FT]

| | Well name | Lateral length per well [ft] |
|---|---|---|
| 1 | Well_Child1 | 3280.83989501312 |
| 2 | Well_Parent | 3280.83989501312 |
| 3 | Well_Child2 | 3280.83989501312 |
| 4 | Well_Child3 | 3280.83989501312 |

AVERAGE WELL SPACING PER WELL [FT]

| | Well name | Average well spacing per well [ft] |
|---|---|---|
| 1 | Well_Child1 | 3280.83989501312 |
| 2 | Well_Parent | 3280.83989501312 |
| 3 | Well_Child2 | 3280.83989501312 |
| 4 | Well_Child3 | 3280.83989501312 |

There is an option to provide baseline values for the economics analysis. This optional input allows the economics values to be calculated as the difference between the results of the current simulation and a set of baseline results. For example, this could be useful when simulating a refrac. First run a simulation without the refrac. Then, run a simulation with the refrac and input the production from the initial no-refrac simulation as the 'baseline' for the economics calculation. Outputs such as 'net present value' will be based on the incremental production and incremental cost relative to the initial simulation - providing a true 'NPV' for the refrac.

BASELINE VALUES

| | (Optional) Baseline cumulative time [days] | (Optional) Baseline cumulative cost [$] | (Optional) Baseline cumulative water production [STB] | (Optional) Baseline cumulative oil production [STB] | (Optional) Baseline cumulative gas production [Mscf] | |
|---|---|---|---|---|---|---|
| 1 | 0 | 4000000 | 0 | 0 | 0 | |
| 2 | 0.0000014389534 | 4000000 | 0 | 0 | 0 | |
| 3 | 0.000002895192 | 4000000 | 0 | 0 | 0 | |
| 4 | 0.000005756849 | 4000000.1 | 0 | 0 | 0 | |
| 5 | 0.000011287357 | 4000000.3 | 0 | 0 | 0 | |

Namely, the new $CumulativeCost_j$ at timestep $j$ is computed as:

$$CumulativeCost_j = CumulativeCost_j(Design2) - CumulativeCost_j(Design1) \quad \text{19-37}$$

Similarly, direct subtraction of the baseline results is performed for the incremental undiscounted cost and for the cumulative and incremental production values for water, oil, and gas. The baseline undiscounted revenue and baseline taxes are calculated from the baseline production values and are subsequently subtracted from the new undiscounted revenue and baseline taxes, such as:

$$IncrementalRevenue_j = IncrementalRevenue_j(Design2) - IncrementalRevenue_j(Design1) \quad \text{19-38}$$

Using these new values for the undiscounted cumulative and incremental cost and revenue, from which the baseline values were subtracted, the calculations of the other economics outputs (the discounted cost, the discounted revenue, the NPV, DROI and the IRR) are performed using the same equations, as those used for the case without baseline values, i.e. using the equations (19-30) – (19-36) above.

To fill-in the table of baseline values, the user needs to input the baseline values for the cumulative cost and the cumulative production of water, oil and gas, versus cumulative time. Time values must start from 0 (zero) and grow monotonically from one row to the next one. Cost and production values must be non-negative and must not decrease from one row to the next one. Rows must have all boxes filled-in.

To generate the baseline values from a previous simulation ("Design 1"), the utilized units (field or metric) must be the same as in the new simulation ("Design 2"). Two cases can be considered:

- The simulation "Design 1" was run without filling-in the table of baseline values.
  In this case, after the computations have finished for the simulation "Design 1", the user can open the Economics Summary Output file, navigate to the time series, and copy the columns for cumulative time, cumulative cost, cumulative water production, cumulative oil production, and cumulative gas production. These columns can be pasted directly into the table of baseline values to be used in the new simulation "Design 2".

- The simulation "Design 1" was run with the filled-in table of baseline values.
  In this case, after the computations have finished for the simulation "Design 1", the user can open the Economics Summary Output file, navigate to the time series, and copy the columns for cumulative time, "cumulative cost without subtracting baseline", "cumulative water production without subtracting baseline", "cumulative oil production without subtracting baseline", and "cumulative gas production without subtracting baseline". These columns can be pasted directly into the table of baseline values to be used in the new simulation "Design 2".

If the copied baseline values do not start from time zero or from a sufficiently small time, the user can add a new row to the top of the baseline table and fill it with zeros.

If the baseline values were provided, some outputs will be printed as nan (not a number). This is done for quantities, for which the baseline results cannot not be computed from the baseline cost and production.

### 19.8.1 EOR economics module

There is an option to specify additional input parameters for enhanced oil recovery. This option is activated by checking the box ‘Include EOR inputs’.

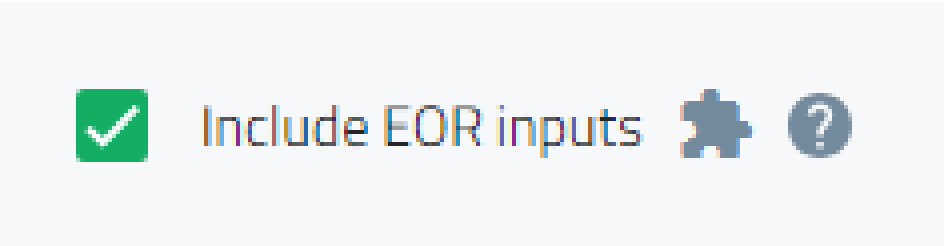

When this setting is set to true, assumptions, specific to the EOR economic analysis, are applied. For a simulation with gas injection, produced gas is split into gas, used for reinjection ('recycled gas'), and gas that goes to the sales ('sales gas'). The 'sales gas' is further split into the 'reservoir gas', considered as native gas, and the 'EOR sales gas', considered as gas that has been previously injected. The taxes and the royalties do not apply to the revenue from the 'EOR sales gas'. Similar logic is applied when there is oil injection, the difference being whether it's liquid (oil) or vapor (gas) at standard conditions. The input parameters are:

- (Optional) *OPEX cost per day per well applied during gas injection (liquid or vapor)* ($/well/day). This is an optional fixed daily OPEX cost per well, applied during gas or oil injection.
- (Optional) *OPEX cost per unit of oil injected* ($/bbl), and (Optional) *OPEX cost per unit of gas injected* ($/Mscf). These are additional optional OPEX amounts applied for each barrel of oil or Mscf of gas in-jected.
- *Compressor fuel fraction of throughput volume*. The compressor fuel gas volume is calculated by multi-plying the injected gas (vapor) volume by this value. The cost of the compressor fuel is added to the cost in addition to the cost of the injected gas. The cost per Mscf of the compressor fuel gas is the same as the cost per Mscf of injected gas and is specified in the column 'Cost of purchased gas' in the table 'EOR time-dependent inputs'.

- (Optional) *Reservoir GOR*. When specified, the simulator will use the reservoir GOR to calculate the produced reservoir gas (vapor) volume from the produced reservoir oil volume. The rest of the sales gas is considered as the EOR sales gas, and taxes and royalties do not apply to it. If the 'reservoir GOR' is not specified, then, after the start of gas injection, all sales gas is considered as the EOR sales gas, until all of the injected gas volume has been produced. See the equations below.
- (Optional) *Oil differentials* ($/bbl). If specified, this value is multiplied by the produced oil volume and is deducted from the oil revenue.
- *Royalties*, and *Working interest in revenue*. These parameters are specified as separate parameters in the EOR inputs section. After the payment of taxes and royalties, working interest in revenue is the fraction of revenue that is received. Working interest in revenue accounts for the sharing of revenue between different companies with joint interest in the wells. Note that the resulting net revenue interest will override the values of 'Net revenue interest in oil' and 'Net revenue interest in gas', specified elsewhere in the economic model.

Users can additionally specify the values of several inputs that can vary with time, in the table 'EOR time-dependent inputs'. They are:

- *Cost of purchased oil* ($/bbl). This is the cost of a unit of the purchased injection oil. Note that, when you fill-in the table 'EOR time-dependent inputs', you cannot simultaneously specify the parameter 'injection oil cost' elsewhere in the economic model.
- *Cost of purchased gas* ($/Mscf). This is the cost of a unit of the purchased injection gas. Note that, when you fill-in the table 'EOR time-dependent inputs', you cannot simultaneously specify the parameter 'injection gas cost' elsewhere in the economic model.
- *Fraction of field gas in the total gas injected*. This fraction of the injected gas volume is considered as field gas, and the 'cost of purchased gas' does not apply to it. In addition, if there is gas production and gas injection at the same timestep, then a part of the produced gas volume is considered as the reinjected field gas (recycled gas) and is not used in the revenue calculations. The same fraction is applied to the injected oil.
- *Gas differentials* ($/Mscf). This value is multiplied by the sales gas volume (excluding the recycled gas) and is deducted from the gas revenue.
- *Compressor CAPEX* ($), and *Compressor OPEX* ($). These are dollar amounts, applied at specified times, that can account for the purchase of or renting a compressor. Users can enter negative values for 'Compressor CAPEX', to account for a residual value of the compressor.
- *Additional EOR CAPEX* ($), and *Additional EOR OPEX* ($). These are dollar amounts, applied at specified times, that can account for other facilities costs.

(OPTIONAL) EOR TIME-DEPENDENT INPUTS

| | Time [days] | Cost of purchased oil [dollars/bbl] | Cost of purchased gas [dollars/Mscf] | Fraction of field gas in total gas injected |
|---|---|---|---|---|
| 1 | | | | |

New Row Resize Table

If the table ‘EOR time-dependent inputs’ is not filled-in, the following default values will be used in the calculations with the EOR economics module:

- *‘Cost of purchased oil’* is defaulted to the value of ‘injection oil cost’, specified elsewhere in the economic model.
- *‘Cost of purchased gas’* is defaulted to the value of ‘injection gas cost’, specified elsewhere in the economic model.
- *‘Fraction of field gas in the total gas injected’* is defaulted to zero. In this case, all the injected gas and all the injected oil are considered as purchased, and there is no recycled gas or recycled oil.
- *‘Gas differentials’* is defaulted to zero.

*Revenue calculation for the case with gas injection (no oil injection)*

In any given timestep, produced gas is split into gas, used for re-injection ('recycled gas'), and gas that goes to the sales ('sales gas'):

$$sales\ gas = produced\ gas - recycled\ gas$$

$$recycled\ gas = \min(injected\ gas \cdot f, produced\ gas)$$

where $f$ is the value of the 'Fraction of field gas in the total gas injected', specified in the table 'EOR time-dependent inputs', and the following assumptions apply:

- If there is no gas injection in this timestep, all produced gas is considered as the 'sales gas'. The 'recycled gas' volume is zero.
- If there is gas injection and gas production, the volume of the 'recycled gas' is calculated from the overall injected gas volume as shown in the equation above. The 'recycled gas' volume is capped by the overall produced gas volume in the current timestep.
- The parameters 'produced gross gas cost', 'produced gross gas cost phase two', and 'produced gross gas cost phase three', specified elsewhere in the economic model, are not applied to the 'recycled gas'.

The 'sales gas' is further split into the 'reservoir gas', considered as native gas, and the 'EOR sales gas', considered as gas that has been previously injected. The taxes and the royalties do not apply to the revenue from the 'EOR sales gas'. In particular:

- Prior to gas injection, all produced 'sales gas' is considered as 'reservoir gas'.
- After EOR gas injection has started, if the user has specified the optional parameter 'Reservoir GOR', the simulator calculates the 'reservoir gas' produced volume by multiplying the produced oil volume by the 'Reservoir GOR'. The rest of the 'sales gas' is considered as 'EOR sales gas'. If you leave the parameter 'Reservoir GOR' blank (nan), all of the ‘sales gas’ is considered as 'EOR sales gas'.

$$reservoir\ gas = produced\ oil \cdot reservoir\ GOR$$

$$EOR\ sales\ gas = sales\ gas - reservoir\ gas$$

- After all of the injected gas volume has been produced, which includes 'recycled gas' and 'EOR sales gas', all of the ‘sales gas’ is considered as 'reservoir gas' for economic calculations.
- The revenue is generated from the 'reservoir gas', the 'EOR sales gas', and from oil production.
- If the user has specified a non-zero value for the parameter ‘Gross NGL yield’ elsewhere in the economic model, it will be used to calculate the NGL volumes, produced from the ‘reservoir gas’ and from the ‘EOR sales gas’. The revenue will be generated from this NGL production as well.
- 'Oil differentials' value, if specified, is multiplied by the produced oil volume, and the result is deducted from the oil revenue. The current value of 'Gas differentials' is multiplied by the 'sales gas' volume, and the result is deducted from the 'reservoir gas' revenue.
- The taxes and the royalties do not apply to the revenue from the 'EOR sales gas'. If the user has specified a non-zero value for ‘Gross NGL yield’, the taxes and the royalties are also not applied to the revenue from the NGL produced from the 'EOR sales gas'. The taxes and the royalties are applied to all other revenue after the oil and gas deducts have been subtracted from it.
- The net revenue is calculated by applying the 'working interest in revenue' to all the revenue generated, after the application of taxes and royalties.
- Taxes are added to the cost. Taxes parameters are specified elsewhere in the economic model, including ‘Oil severance tax’, ‘NGL severance tax’, ‘Gas severance tax’, ‘Ad valorem tax’, and ‘Miscellaneous taxes’.

*Revenue calculation for the case with oil and gas injection*

The calculations are similar to those in the case above. Produced gas is split into 'recycled gas' and 'sales gas' in the same way. Similarly, produced oil is split into oil, used for re-injection ('recycled oil'), and oil that goes to the sales ('sales oil'):

$$sales\ oil = produced\ oil - recycled\ oil$$

$$recycled\ oil = \min(injected\ oil \cdot f, produced\ oil)$$

If there is no oil injection, all produced oil is considered as 'sales oil'. The parameters 'produced gross oil cost', 'produced gross oil cost phase two', and 'produced gross oil cost phase three', specified elsewhere in the economic model, are not applied to the 'recycled oil'.

The 'sales oil', and the total produced oil in the current timestep, are split into the reservoir volumes (coming from the reservoir) and the EOR volumes (coming from the previously injected volume), as follows:

- Prior to oil injection, all produced oil is considered as 'reservoir oil' (equal to the 'reservoir sales oil').
- After oil injection has started, all of the 'sales oil' is considered as 'EOR sales oil', and all of the produced oil in the current timestep is considered as 'EOR oil', until all of the injected oil volume has been produced. After that, all the produced oil in the current timestep is considered as 'reservoir oil' for economic calculations (and equal to the 'reservoir sales oil').

The 'sales gas' is split into the 'reservoir gas' and the 'EOR sales gas':

- Prior to gas injection, all produced 'sales gas' is considered as 'reservoir gas'.

- After gas injection has started, if you specify the optional parameter 'Reservoir GOR', the simulator calculates the 'reservoir gas' volume by multiplying the 'reservoir oil' volume by the 'Reservoir GOR'. The rest of the 'sales gas' is considered as 'EOR sales gas'. If you leave the parameter 'Reservoir GOR' blank (nan), all of the sales gas is considered as 'EOR sales gas'.

$$reservoir\ gas = reservoir\ oil \cdot reservoir\ GOR$$

$$EOR\ sales\ gas = sales\ gas - reservoir\ gas$$

- After all of the injected gas volume has been produced, which includes 'recycled gas' and 'EOR sales gas', all of the sales gas is considered as 'reservoir gas' for economic calculations.

The rest of the revenue structure is the same as in the case above, except that:

- The revenue is generated from 'reservoir gas', 'EOR sales gas', 'reservoir sales oil', 'EOR sales oil', and from NGL production.
- If you specify 'Oil differentials', its value is multiplied by the 'sales oil' volume, and the result is deducted from the 'reservoir sales oil' revenue.
- The taxes and the royalties are not applied to: the revenue from the 'EOR sales gas', the revenue from the 'EOR sales oil', and the revenue from the NGL produced from the 'EOR sales gas'.

*Cost*

Users specify the injection gas cost as time-varying data in the column 'Cost of purchased gas' in the table 'EOR time-dependent inputs'. The cost of the purchased injection gas is then calculated at any given timestep as:

$$cost = injected\ gas\ volume \cdot (1 - f) \cdot Cost\ of\ purchased\ gas$$

where $f$ is the current value of the 'Fraction of field gas in the total gas injected', and the injected gas volume is computed by ResFrac.

Similar calculation is applied for the cost of the purchased injection oil, where the parameter 'Fraction of the field gas in the total gas injected' also applies to oil injection, and the parameter 'Cost of purchased oil' is specified in the table 'EOR time-dependent inputs'.

There are several parameters related to the cost of the compressor usage:

- In the table 'EOR time-dependent inputs', you can specify dollar amounts for 'Compressor CAPEX' and 'Compressor OPEX', applied at defined points in time. These amounts can account for the compressor purchase or for renting a compressor. In the column for 'Compressor CAPEX', you can enter negative values, to account for a residual value of the compressor.
- Purchased compressor fuel gas is calculated as a fraction of the overall injected gas volume, defined by the parameter 'Compressor fuel fraction of throughput volume'. At any given timestep, the compressor fuel gas volume is calculated by multiplying the overall injected gas volume by the parameter 'Compressor fuel fraction of throughput volume'. The compressor fuel cost, applied to the compressor fuel gas volume, is the same as the cost of the purchased injection gas and is specified in the column 'Cost of purchased gas' in the table 'EOR time-dependent inputs'.

In the table 'EOR time-dependent inputs', you can also specify dollar amounts for 'Additional EOR CAPEX' and 'Additional EOR OPEX', applied at defined points in time. You can additionally specify fixed OPEX values, applied during gas injection (liquid or vapor), per well or per unit of oil or gas injected.

### 19.8.2 Geothermal economic model (new)

The economics model has been extended to include an option for geothermal. Previously, there was a separate 'geothermal' economics module, specified in the Output Options panel, under the Advanced options, and described in Section 19.9 Geothermal economic module. This option is still available, but in addition, the primary economic model option has been extended to include geothermal. This allows you to run geothermal economics calculations with a broader set of features, and even to run analyses on simulations that include both oil and gas and geothermal at the same time.

For an EGS simulation, select the 'Geothermal' application in the economic model, and uncheck the 'Oil and gas' application. When both 'Oil and gas' and 'Geothermal' applications are selected, the economic analysis will take into account both oil and gas, and geothermal, economic input parameters.

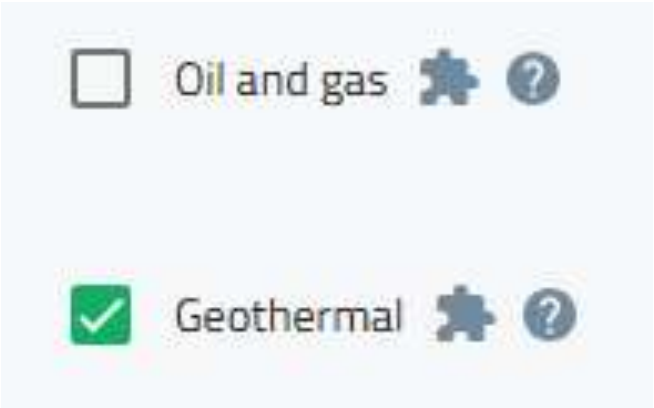

You can choose the geothermal end-use to be either for electricity or for direct heat usage.

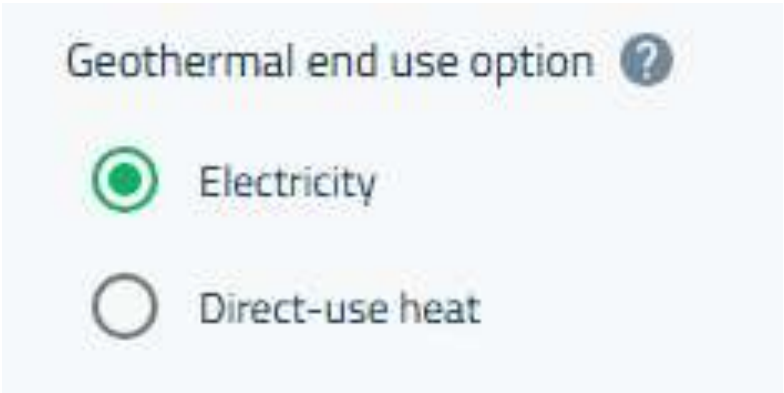

The model calculates heat production, and gross and net electricity production for the electricity end-use, from the enthalpy and flow rate of water at the surface. The thermal production rate, and the gross and net electricity production rates are calculated in the same way as in the existing Geothermal economic module, see the equations in Section 19.9. For the electricity end-use, the revenue is calculated from the net electricity production, and the levelized cost of electricity (LCOE) is calculated. For the direct-heat end-use, the revenue is calculated directly from the heat generation, and the levelized cost of heat (LCOH) is calculated. The heat generation, the gross and net electricity generation for the electricity end-use, as well as the LCOE or LCOH, are printed to the tracking file, in addition to other outputs from the economic model.

Namely, the LCOE and the LCOH are calculated as follows:

$$LCOE = \frac{DiscountedCumulativeCost}{DiscountedCumulativeNetElectricityGenerated}$$

$$LCOH = \frac{DiscountedCumulativeCost}{DiscountedCumulativeHeatGenerated}$$

Note that, in the economic model, the taxes on the geothermal revenue are added to the cost and will affect the values of the LCOE and the LCOH. This is contrary to the calculation of the LCOE in the previously existing Geothermal economics module, where the taxes are subtracted from the revenue. Therefore, the value of the LCOE, calculated in the economic model with the 'Geothermal' application, will be different from that calculated in the previously existing Geothermal economics module.

The table 'Geothermal economic model inputs' is accessible when the 'Geothermal' application is selected, and includes the following parameters:

GEOTHERMAL ECONOMIC MODEL INPUTS

| | Price of electricity [$/kW-hr] | (Optional) Cost of electricity for pumping power [$/kW-hr] | Time when surface plant cost is applied [days] | Surface plant cost [$] |
|---|---|---|---|---|
| 1 | | | | |

- *Price of electricity* ($/kW-hr). This price is used to calculate the revenue from the net electricity sold. It will also be used to calculate the cost of electricity, purchased for the pumping power, if you do not specify the parameter 'Cost of electricity for pumping power'.
- *Cost of electricity for pumping power* ($/kW-hr). For the electricity end-use, this price will be applied to calculate the cost of electricity purchased for the pumping power, when the gross electricity output is not yet large enough to cover the pumps power consumption. For the direct-heat end-use, this price will be applied to calculate the cost of the pumping power at all times during the simulation.
- *Price of heat* ($/MW-hr) (accessible if the end-use is direct use heat). This is the heat price for direct-use heat sales.
- *Time to start counting geothermal cost* (days). This input is accessible only if both 'Oil and gas' and 'Geothermal' applications are selected, and defines the time of application of several geothermal economic parameters. This can be useful when the calculations are performed for an oil and gas site, converted to a geothermal energy surface plant. It allows the user to exclude from the calculation of the geothermal-related cost and CAPEX any cost spent during the oil-and-gas phase. The 'Fixed geothermal operational cost per year' is applied on a per year basis, starting from this time, and the 'Investment tax credit (ITC)' is applied to the CAPEX, calculated starting from this time. The 'Geothermal facilities cost per well' is applied at this time, or at the time of drilling of each well, whichever is later. In addition, the calculation of the LCOE or the LCOH does not take into account any spending prior to this time.
- *Time when surface plant cost is applied* (days). This is the time when the surface plant cost is applied. Prior to this time, there will be no generation of power or heat. If this input is not specified, it defaults to time zero, in case if the 'Oil and gas' application is not selected, and to 'Time to start counting geothermal cost', if the 'Oil and gas' application is selected.
- *Surface plant cost* ($). This is an upfront, lumped-together capital cost for surface plant equipment.
- *Surface plant cost per gross capacity (electric)* ($/kWe), and *Installed gross plant capacity (electric)* (MWe). These inputs are accessible if the end-use is electricity. They allow you to account for the additional surface plant cost, based either on the peak gross electricity output, or on the specified 'Installed gross plant capacity'. If you do not specify the 'Installed gross plant capacity', the simulator will multiply the 'Surface plant cost per gross capacity (electric)' by the calculated peak gross electricity generation rate, achieved over the lifetime of the power plant. If you specify the 'Installed gross plant capacity', it will cap the gross electricity generation rate, and the simulator will multiply it by the 'Surface plant cost per gross capacity (electric)', instead of using the peak gross electricity generation rate.
- *Additional surface plant cost per net capacity (electric)* ($/kWe). This input is accessible if the end-use is

electricity. It allows you to account for the additional surface plant cost, based on the peak net electricity output, such as, for example, the interconnection cost. The simulator will multiply the 'Surface plant cost per net capacity (electric)' by the calculated peak net electricity generation rate, achieved over the lifetime of the power plant.

- *Surface plant cost per capacity (thermal)* ($/kWth), and *Installed plant capacity (thermal)* (MWth). These inputs are accessible if the end-use is direct use heat. They are used similarly to the parameters ‘Surface plant cost per gross capacity (electric)’ and ‘Installed gross plant capacity (electric)’: if you do not specify the 'Installed plant capacity (thermal)', the simulator will multiply the 'Surface plant cost per capacity (thermal)' by the calculated peak thermal generation rate, achieved over the lifetime of the plant, etc.
- *Time to start calculating peak output for surface plant cost* (days). Starting from this time, the simulator will calculate the peak generation rates, in order to apply the surface plant cost per capacity (gross and net electric, or thermal).
- *Geothermal facilities cost per well* (dollars/well). This is a fixed one-time capital expense that accounts for costs at a geothermal site that scale with the number of wells, such as pipeline or pumps. It is imposed at the time of drilling of each well. If both the ‘Oil and gas’ and the ‘Geothermal’ applications are selected, this expense is imposed at the time of drilling of each well or at the time specified in 'Time to start counting geothermal cost', whichever is later.
- *Fixed geothermal operational cost per year* ($/year). This is the operational cost applied continuously throughout the simulation, on a per year basis.
- *Surface plant OPEX per year as a fraction of up-front cost*. This is the fraction of the surface plant CAPEX, which is added to the operational cost, on a per year basis. It allows the user to specify an operational cost that scales with the size (CAPEX) of the surface plant. For example, if set to 0.05, it adds 5% of the surface plant CAPEX to the annual OPEX cost. The surface plant CAPEX combines the 'Surface plant cost', the 'Surface plant cost per gross capacity (electric)', and the 'Additional surface plant cost per net capacity (electric)', when the end-use is 'Electricity'. When the end-use is 'Direct-use heat', the surface plant CAPEX combines the 'Surface plant cost', and the 'Surface plant cost per capacity (thermal)'.
- *Investment tax credit (ITC)*. The ITC is incorporated as a one-time revenue, equal to the specified fraction of the total CAPEX. The investment tax credit is applied one year from ‘Time when surface plant cost is applied’.
- *Production tax credit (PTC)*, and *Number of years to apply PTC*. These inputs are accessible if the end-use is electricity, and cannot be specified simultaneously with an investment tax credit (ITC). The PTC is incorporated as an annual revenue, calculated as the PTC value multiplied by the annual net electricity generated. The PTC is applied annually, starting one year after ‘Time when surface plant cost is applied’, and for a number of years, specified by ‘Number of years to apply PTC’.
- *Tax and royalty rate on geothermal revenue*. This is the fraction of revenue from electricity or heat sales that is lost to taxes and royalties.
- *Replacement water cost* ($/bbl). Water is produced, and then reinjected. However, some water is lost long-term to the formation. This is the cost of the additional water that must be injected to make up for water lost to leakoff. In addition, if the selected end-use is 'Electricity', and you specify a nonzero value for 'Fraction of produced water lost in power plant', this cost will be applied to the additional water injected to make up for the water consumed in the power plant. Note that, the 'replacement water cost' will be applied only during the 'instantaneous net-out' time periods, specified in the table 'Net-out instantaneous injection and production start and end times by well'. In any given timestep, injection and production from all the wells, that are active in an 'instantaneous net-out' period, is netted-out. Either the 'injection water cost', or the 'replacement water cost' is applied to the net injected water, as follows:
  - If you include the 'Geothermal' application and do not include the 'Oil and gas' application in the economic model, the 'replacement water cost' is applied to the net injected volume.
  - If you include both the 'Oil and gas' and the 'Geothermal' applications, the 'injection water cost' is applied to the net injected volume prior to 'time when the surface plant cost is applied', and the 'replacement water cost' is applied after 'time when the surface plant cost is applied'.
  - For any well, that is not netted-out, the 'injection water cost' is applied.

NET-OUT INSTANTANEOUS INJECTION AND PRODUCTION START AND END TIMES BY WELL

| | Well name | (Optional) Start time of period one to apply cost for net instant... | (Optional) End time of period one to apply cost for net instantan... | (Optional) Start time of period two to apply cost for net instant... | (Optional) End time of period two to apply cost for net instantan... |
|---|---|---|---|---|---|
| 1 | Well_1 | | | | |
| 2 | Well_2 | | | | |
| 3 | Well_3 | | | | |

- *Fraction of produced water lost in power plant (for the electricity end-use)*. This is a fraction of produced water that is lost in the power plant due to evaporation. When you specify a nonzero value for it, the 'replacement water cost' is applied to the additional water, injected to make up for the water lost in the power plant. In addition, the 'produced water cost' is not applied to the evaporated fraction of produced water.
- *Lower cutoff temperature* (F). Once the produced fluid goes below this temperature, it is no longer used to generate electricity, is not counted in the 'thermal output', and is not included in the revenue calculation.
- There are several other parameters, which are used in the same way, as in the existing Geothermal economic module, specified in the Output Options panel, under the Advanced options. They are: *Optional net power plant efficiency* (for the electricity end-use), *Injection pump inlet pressure*, *Injection pump efficiency*, *Production pump outlet pressure*, and *Production pump efficiency*. Refer to Section 19.9 Geothermal economic module for the description of their use.

## 19.9 Geothermal economic module (old)

The *Geothermal Economic Modules* option in ResFrac allows users to perform economic analysis of geothermal simulation models results and compare between realizations with different economic assumptions. This section describes the original Geothermal Economic Module. A new geothermal option has since been added to the primary economic model section of ResFrac (details on which can be found in Section 19.8 Economics module). This newer version allows you to run geothermal economics calculations with a broader set of features, and even run analyses on simulations that include both oil and gas and geothermal at the same time. Refer to Section 19.8.2 Geothermal economic model (new) for the description of this alternative geothermal economics model.

The original geothermal economic models can be defined in the Advanced Options section of the Output Options tab of the UI. The user can define more than one geothermal economic model and each model prints the following parameters in the tracking file:

1. Time series for cumulative revenue, cumulative time discounted revenue, cumulative cost, cumulative discounted cost.
2. Net present value (NPV), Discounted Return on Investment (DROI).
3. Thermal production rate.
4. Gross electricity generation rate.
5. Net electricity generation rate.

The Thermal production rate is calculated as

$$Thermalproductionrate = massflowrate * fluidenthalpy \quad (19\text{-}39)$$

The Gross electricity generation rate is computed as

$$Grosselectricitygenerationrate = \eta * fluidenthalpy \quad (19\text{-}40)$$

where the efficiency $\eta$ can be defined in different ways.

1. It can be set to a user specified constant in the geothermal economic module

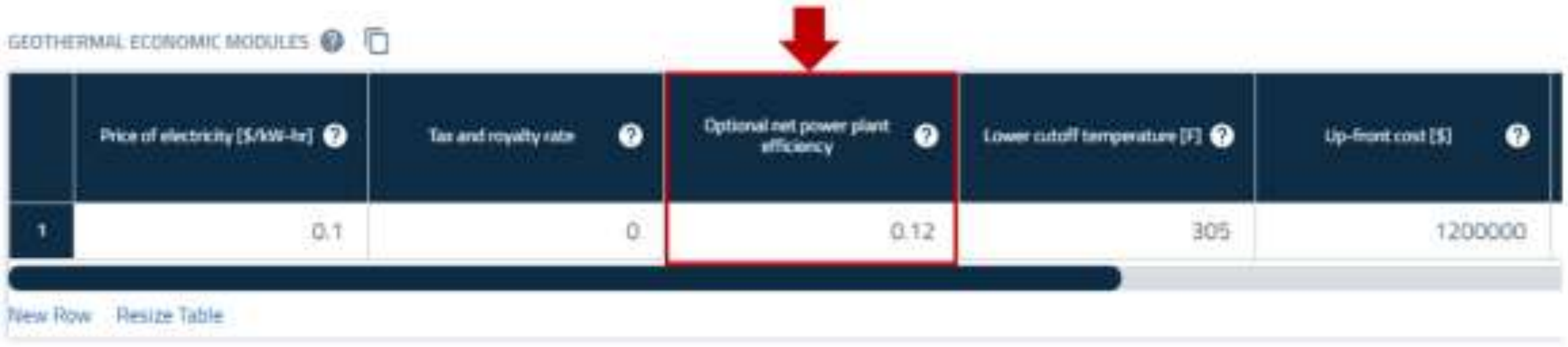

GEOTHERMAL ECONOMIC MODULES

| | Price of electricity [$/kW-hr] | Tax and royalty rate | Optional net power plant efficiency | Lower cutoff temperature [F] | Up-front cost [$] |
|---|---|---|---|---|---|
| 1 | 0.1 | 0 | 0.12 | 305 | 1200000 |

New Row Resize Table

In this case, this is a net efficiency $\eta$ considering all parasitic losses. For example, water at 150 C (423 K) and 2 MPa (300 Psia) has enthalpy of 633.2 kJ/kg (see Table 3). So, 1 kg/s of water at those conditions yields 633.2 kW thermal, and at 0.1 net efficiency, would give Gross electricity generation rate = 63.2 kW.

2. In case the power plant efficiency is left blank in the geothermal economic module

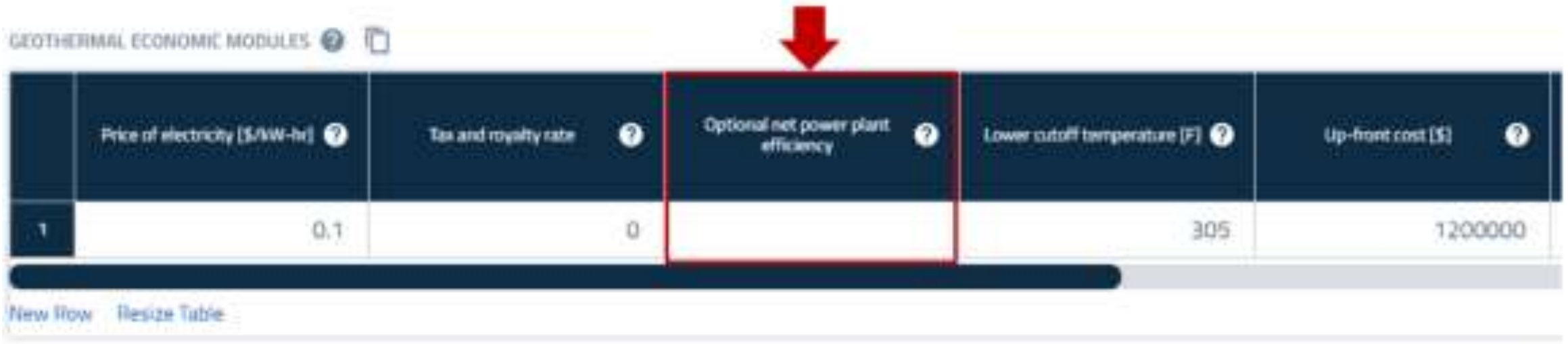

GEOTHERMAL ECONOMIC MODULES

| | Price of electricity [$/kW-hr] | Tax and royalty rate | Optional net power plant efficiency | Lower cutoff temperature [F] | Up-front cost [$] |
|---|---|---|---|---|---|
| 1 | 0.1 | 0 | | 305 | 1200000 |

New Row Resize Table

then ResFrac will use the correlation from Figure 13 from Zarrouk and Moon (2014), 'Efficiency of geothermal power plants: A worldwide review'. This correlation is based on actual results from geothermal power plants, and calculates efficiency as a function of the produced fluid enthalpy.

3. The user can define the efficiency $\eta$ as function of temperature by inputting the corresponding tabulated data in the Brine efficiency versus temperature table. The table is in the Advanced Options section of the Output Options tab of the UI.

BRINE EFFICIENCY VERSUS TEMPERATURE

| | Temperature [F] | Brine efficiency | |
|---|---|---|---|
| 1 | 100 | 0 | |
| 2 | 200 | 0.01 | |
| 3 | 340 | 0.05 | |
| 4 | 400 | 0.12 | |

New Row Resize Table

Note that, through this table, the user can alternatively specify its own correlation between efficiency $\eta$ and enthalpy. This is the case in which the enthalpy can be assumed to be a function of the sole temperature, and not of temperature and pressure. For example, Table 3 provides a correlation between water temperature and enthalpy at pressure of 2.07 Mpa (300 Psia).

Table 3: Water enthalpy as a function of temperature at the pressure of 300 Psia (2.06843 MPa) from Wagner and Pruß, (2002).

| Temperature (K) | Density (kg/m3) | Enthalpy (kJ/kg) | Temperature (K) | Density (kg/m3) | Enthalpy (kJ/kg) | Temperature (K) | Density (kg/m3) | Enthalpy (kJ/kg) |
|---|---|---|---|---|---|---|---|---|
| 283.16 | 1000.6 | 44.077 | 330.66 | 985.32 | 242.49 | 378.16 | 955.62 | 441.76 |
| 285.66 | 1000.4 | 54.542 | 333.16 | 984.05 | 252.94 | 380.66 | 953.76 | 452.3 |
| 288.16 | 1000 | 65 | 335.66 | 982.75 | 263.4 | 383.16 | 951.87 | 462.86 |
| 290.66 | 999.6 | 75.452 | 338.16 | 981.41 | 273.85 | 385.66 | 949.95 | 473.42 |
| 293.16 | 999.1 | 85.899 | 340.66 | 980.03 | 284.31 | 388.16 | 948.01 | 484 |
| 295.66 | 998.55 | 96.342 | 343.16 | 978.63 | 294.77 | 390.66 | 946.03 | 494.58 |
| 298.16 | 997.93 | 106.78 | 345.66 | 977.18 | 305.24 | 393.16 | 944.03 | 505.17 |
| 300.66 | 997.25 | 117.22 | 348.16 | 975.71 | 315.71 | 395.66 | 942 | 515.77 |
| 303.16 | 996.52 | 127.66 | 350.66 | 974.2 | 326.18 | 398.16 | 939.95 | 526.38 |
| 305.66 | 995.73 | 138.09 | 353.16 | 972.66 | 336.66 | 400.66 | 937.87 | 537.01 |
| 308.16 | 994.9 | 148.53 | 355.66 | 971.09 | 347.15 | 403.16 | 935.76 | 547.64 |
| 310.66 | 994.01 | 158.96 | 358.16 | 969.49 | 357.64 | 405.66 | 933.62 | 558.29 |
| 313.16 | 993.07 | 169.4 | 360.66 | 967.86 | 368.13 | 408.16 | 931.46 | 568.95 |
| 315.66 | 992.09 | 179.84 | 363.16 | 966.2 | 378.63 | 410.66 | 929.27 | 579.62 |
| 318.16 | 991.07 | 190.28 | 365.66 | 964.51 | 389.13 | 413.16 | 927.05 | 590.31 |
| 320.66 | 990 | 200.71 | 368.16 | 962.79 | 399.64 | 415.66 | 924.8 | 601.01 |
| 323.16 | 988.89 | 211.16 | 370.66 | 961.04 | 410.16 | 418.16 | 922.53 | 611.73 |
| 325.66 | 987.74 | 221.6 | 373.16 | 959.26 | 420.69 | 420.66 | 920.23 | 622.46 |
| 328.16 | 986.55 | 232.04 | 375.66 | 957.46 | 431.22 | 423.16 | 917.9 | 633.21 |

The Net electricity generation rate is computed as

$$Netelectricitygenerationrate = Grosselectricitygenerationrate - pumpspowerconsumption \quad (19\text{-}41)$$

in which the pumps power consumption is

$$pumpspowerconsumption = productionpumppowerconsumption + injectionpumppowerconsumption \quad (19\text{-}42)$$

The production pump power consumption is computed as

$$productionpumppowerconsumption = productionpumpeffciciency * (productionpumpoutletpressure - productionpumpinletpressure) \quad (19\text{-}43)$$

and similarly, the injection pump power consumption is computed as

$$injectionpumppowerconsumption = injectionpumpeffciciency * (injectionpumpoutletpressure - injectionpumpinletpressure) \quad (19\text{-}44)$$

The production/injection pump outlet pressure is the fluid pressure immediately downstream of the production/injection pump (in the flow direction). Similarly, the production/injection pump inlet pressure is the fluid pressure immediately upstream of the production/injection pump. The production/injection pump efficiency is also called 'wire-to-water efficiency' as it calculates the power input to the motor in relation to the power output from the production/injection pump. While ResFrac calculates the production pump inlet pressure and the injection pump outlet pressure, the remaining pressure values, as well as the pump efficiencies, can be optionally input by the user in the geothermal economic module.

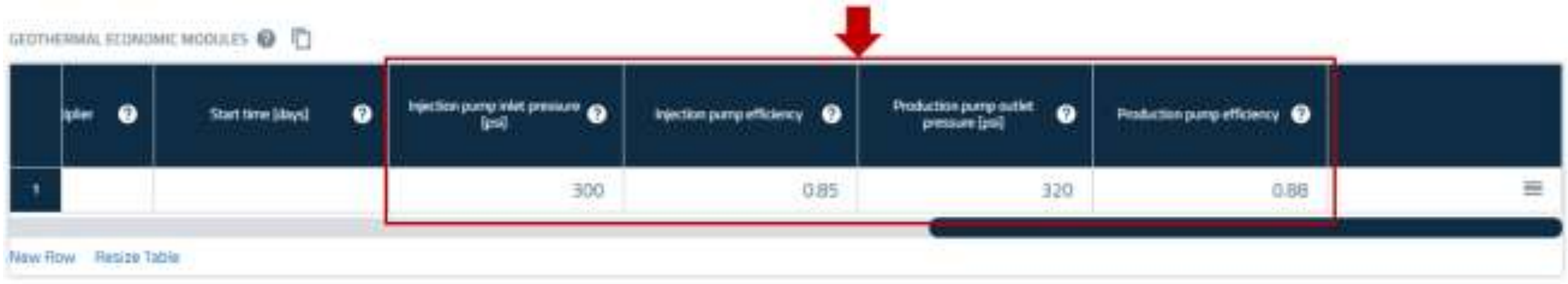

In case these values are left blank, the Net electricity generation rate coincides with the Gross electricity generation rate. The production/injection pump power consumption can be accounted for independently from the injection/production pump power consumption by specifying the inlet/outlet pressure and pump efficiency while leaving blank the others.

## 19.10 Black oil tracers and black oil fluid regions

### 19.10.1 Black oil tracers

Black oil tracers can be used to get an exact allocation of production by layer. Typically the black oil model has only three fluid ‘components’ – water, oil, and gas. The oil and gas ‘components’ may both be soluble in the oil and gas phases. With the ‘black oil tracers’ capability, you can define additional components in the black oil model. The simulator fully tracks mass balance for these components, just like any other component. For purposes of calculating fluid properties, the black oil tracer components are treated as either identical to the ‘oil’ or ‘gas’ components.

This capability gives you the ability to make a fully exact allocation of production by layer into each individual well. First, you define the names of the ‘black oil tracer components’. These components can be considered identical to either the ‘gas’ or ‘oil’ component. You can initialize the concentration of the black oil tracers in each layer of the model. When this simulation is run, we can plot the fraction of each tracer in the produced fluid of each well, giving us a flow allocation by layer.

For example, in the simulation below, the ‘formation2’ tracer is in one layer (shown in lower right), and the ‘formation3’ tracer is in another layer (shown in the lower left). The plot in the upper left shows that during production about 71% of the production is coming from the layer with the ‘formation3’ tracer, and the other 29% is coming from the formation with the ‘formation2’ tracer.

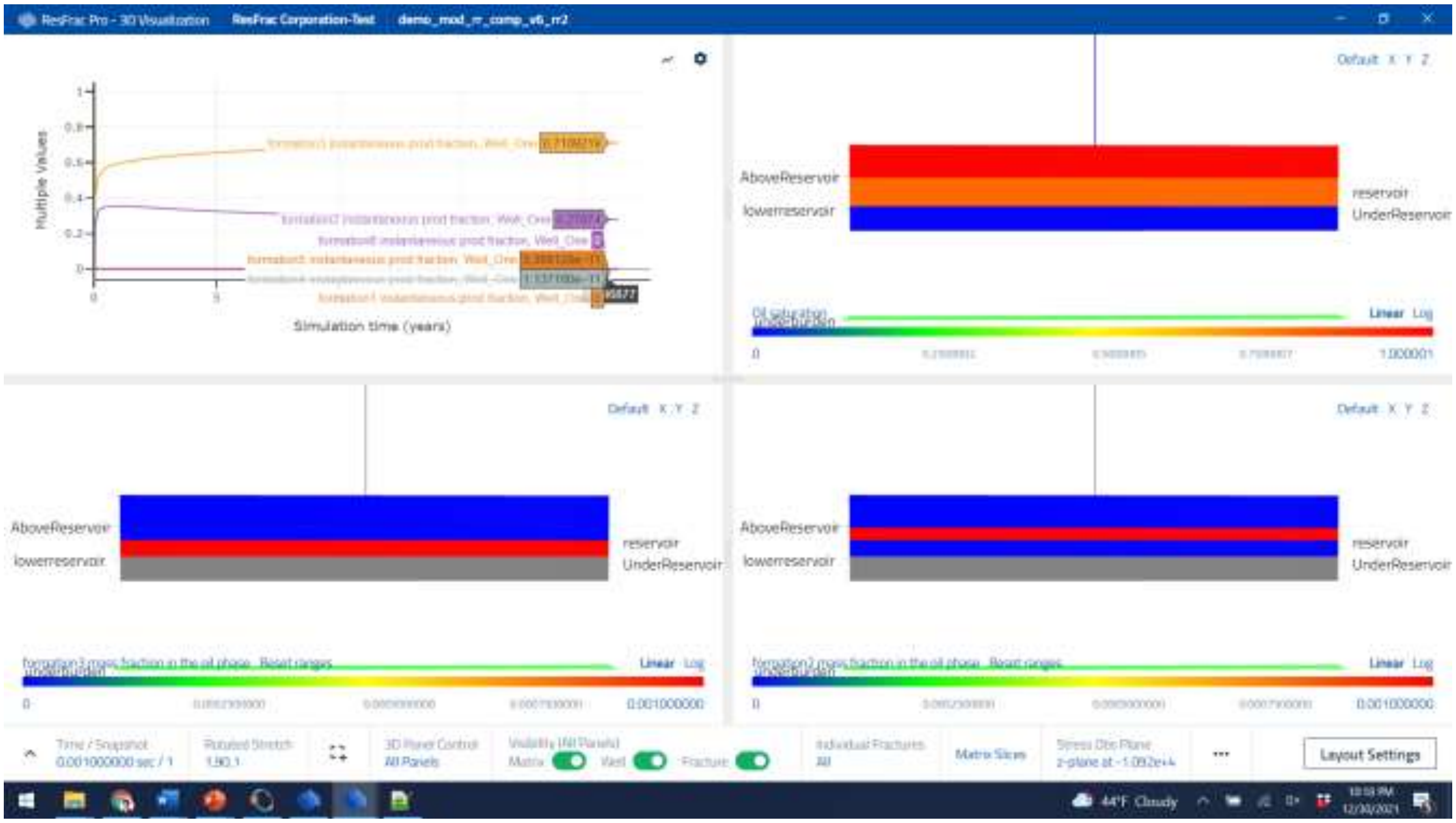

Figure 33: Example of flow allocations using the black oil tracer capability.

### 19.10.2 Black oil fluid regions

The code supports black oil 'fluid regions' in which different black oil tables are used in different sections of the reservoir. There are two options – 'SpatialOnly' and 'SeparateComponents.'

In the 'SpatialOnly' option, the global system of equations continues to have three components – oil, water, and gas – but a different black oil table can be used to calculate fluid properties from those component, depending on location in the reservoir. This option is the simplest, but is physically problematic, and is only recommended for situations where the separate layers will have little or no mixing. Otherwise, a unit of oil and gas can flow from one layer to another and abruptly experience a change in properties or phase behavior.

In the 'SeparateComponents' option, a separate oil and gas component are defined for each fluid region. These components are defined in the global system of equations and mass balance is tracked for each in *all* elements, regardless of which 'fluid region' they belong to. In elements that have mixtures of oil and gas components from different fluid regions, the properties are calculated as the mole fraction weighted average.

For example, let's say we have three fluid regions. An element has: $z_w = 0$, $z_{o1} = 0.1$, $z_{o2} = 0.1$, $z_{o3} = 0.3$, $z_{g1} = 0.05$, $z_{g2} = 0.15$, and $z_{g3} = 0.3$, where z is mole fraction of each component.

Then for each pair of components: ($z_{o1}$, $z_{g1}$), ($z_{o2}$, $z_{g2}$), and ($z_{o3}$, $z_{g3}$), a separate black oil calculation is performed. For each pair of components ($z_{o1}$, $z_{g1}$), the separate black oil models are used to calculate the oil and gas phase mole fractions of the oil and gas phases corresponding to those components: $L_1$, $G_1$, $L_2$, $G_2$, $L_3$, and $G_3$.

The density, molar density, composition, and density of these pseudophases is calculated using the black oil model corresponding to each of the fluid regions.

The oil and gas phases are considered fully miscible so that the mole fraction of the overall liquid and gas phase mole fractions are: $L = L_1 + L_2 + L_3$, etc.

The fluid properties of the mixture phases are calculated with a mole fraction weighted average. Finally, the mixture phase saturations are mole fraction divided by molar density.

## 19.11 Fracture swarm fractal scaling

The code has a capability to represent flow effects from 'fracture swarm fractal networks.' This phenomenon is also sometimes called 'anomalous diffusion,' though we do not favor this term, because it is not an accurate description of what is actually happening in the reservoir.

As outlined in "Rate Transient Analysis of Fracture Swarm Fractal Networks" by Jorge Acuna (2020), fractal fracture spacing is a plausible explanation for (part of) why RTA plots bend upwards in shale formations. Core through studies show erratically spaced productive hydraulic fractures (Raterman et al., 2019). Evidently, fluid flows outside casing from beyond the perf clusters, and hydraulic fractures kick off from the well at erratic locations. Thus, even there are perf clusters spaced uniformly, the actual spacing between producing fractures is highly variable. Maybe sometimes, they are only 5 ft apart, and other times, perhaps they are 50 ft apart.

If the producing fractures have such a distribution of spacing, then we would expect that they experience production interference from their neighbors at all different times. This should create a gradual upward bend in the RTA plot. In fact, we do usually see such a bend in the RTA plots from shale wells.

Also, Raterman et al. (2019) proposed that early time production may be increased by unpropped hydraulic fractures near the well, which quickly close off. This could also contribute to an upward bend in the RTA plot.

To mimic these effects, ResFrac permits the user to specify a ‘fractal dimension’ (referred to as *D*; following Acuna 2020). If you set to 0, it has no effect. If you set to 0.3, there is a mild effect. If set to 0.5-0.7, the effect is fairly strong, and if set to 1.0 or greater, the effect is very strong.

The behavior is implemented by modifying the values in the 1D submesh. The size of the 1D submesh connecting each fracture and matrix element is limited by the size of the matrix element. Thus, if you use the ‘fractal D’ setting, you need to make sure that the simulation mesh is relatively coarse in the direction perpendicular to the fractures, so that the submesh is sufficiently large. It is recommended that the elements are at least 20 ft wide.

The setting is implemented by increasing the flow ‘volume’ and ‘area’ in the vicinity of the fracture, with the effect diminishing with distance, until 20 ft from the fracture face, at which point there is no longer any adjustment.

Larger values of D increase the early-time production rate more, and also increase the early-time leakoff rate. So, if using the ‘fractal D’ setting, you probably do not need to use ‘reversible pressure dependent permeability’ to increase leakoff during injection.

Let’s say, for example, that the baseline 1D submesh is defined by a function of distance from the fracture, A(x), that defines the cross-sectional area for flow. If you specify ‘submesh fractal D,’ then the function is modified to a new function, $Ⱥ(x)$, according to the equations:

$$Ⱥ(x) = A(x) \text{ if x > 20 ft}$$
$$Ⱥ(x) = A(x)\left(\frac{x}{20}\right)^{-D} \text{ if x > 0.1 ft and x < 20 ft} \qquad \text{19-45}$$
$$Ⱥ(x) = A(x)\left(\frac{0.1}{20}\right)^{-D} \text{ if x < 0.1 ft}$$

The length of the submesh may be truncated to ensure that it does not ‘overcount’ the total volume of the element. For example, let’s say that the total volume of the original, unmodified submesh is:

$$V = \int_0^L A(x)dx, \qquad \text{19-46}$$

where *L* is the length of the submesh from the fracture face. The modified submesh must have the same volume, and so the submesh is truncated at a distance $Ƚ$, such that:

$$\int_0^L A(x)dx = \int_0^{Ƚ} Ⱥ(x)dx. \qquad \text{19-47}$$

The size of the unmodified submesh is limited by the matrix element width perpendicular to the fracture faces. This is why it is important to use a reasonably coarse mesh, with 20 ft as the recommended minimum.

The figures below show the effect of using ‘submesh fractal D,’ using values of 0, 0.3, and 0.6. The ‘submesh fractal D’ setting increases early-time production. It is equivalent to assuming that each individual ‘crack’ represented in the simulator is actually a swarm of fractures. With higher ‘fractal D,’ each swarm has a larger number of fractures. As a result, simulations with higher value have more production. However, with more fractures, interference between fractures occurs more rapidly, which is why there is more curvature in the RTA curves.

Higher values of ‘fractal D’ increase fluid leakoff, and so simulations with very high ‘fractal D' value may have noticeably smaller fractures, because of the increased leakoff. Typically, if you use the ‘submesh D’ parameter, there is no need to also use ‘reversible pressure dependent permeability’ increase to accelerate leakoff after fracturing, since this is already implicitly included in the ‘fractal D’ calculation. GOR generally increases with higher values of ‘fractal D,’ because the pressure in the formation is being drawn down more aggressively by the larger number of cracks throughout the rock mass.

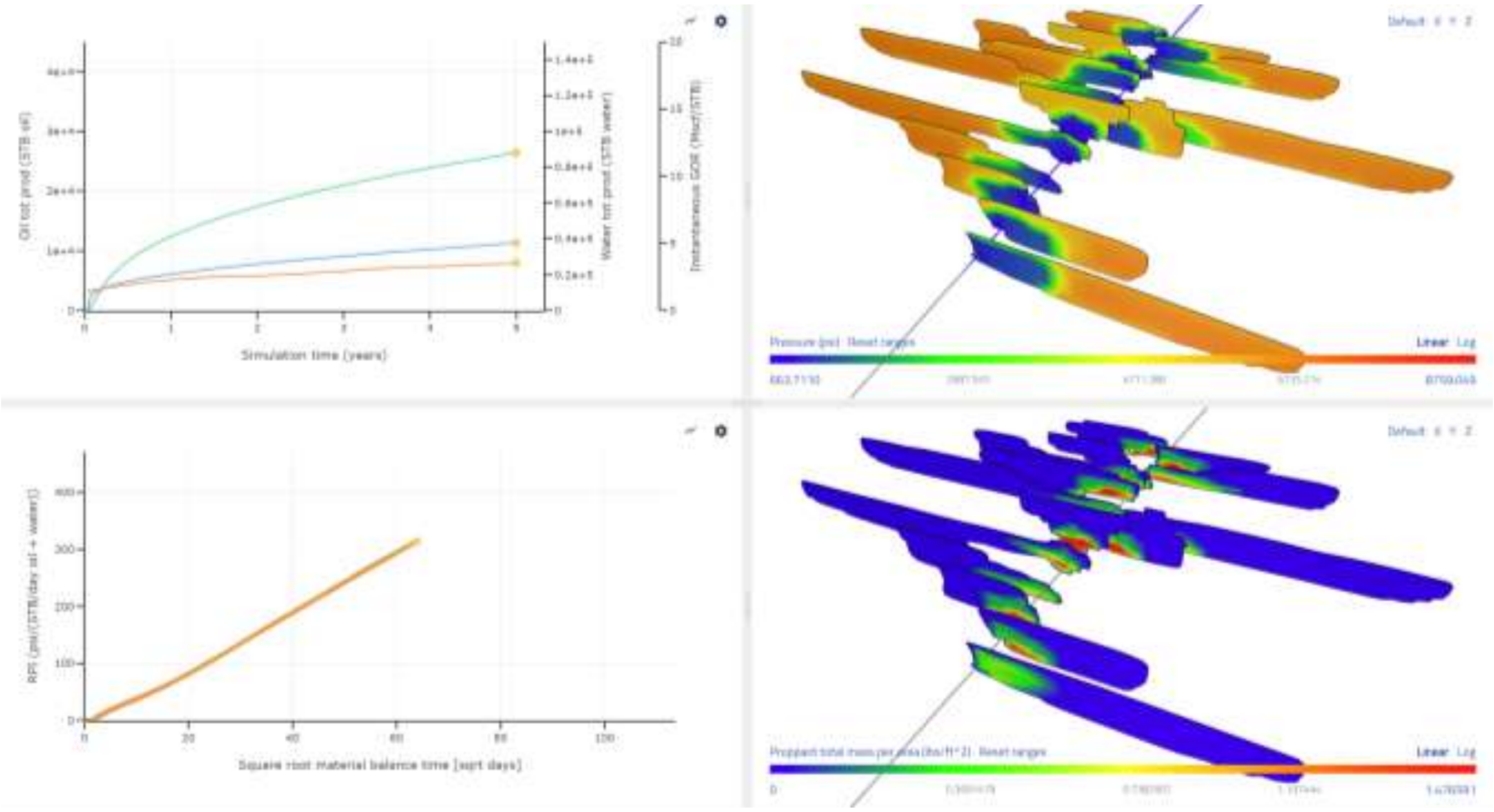

*Figure 34: Example simulation that does not use the ‘submesh fractal D’ setting.*

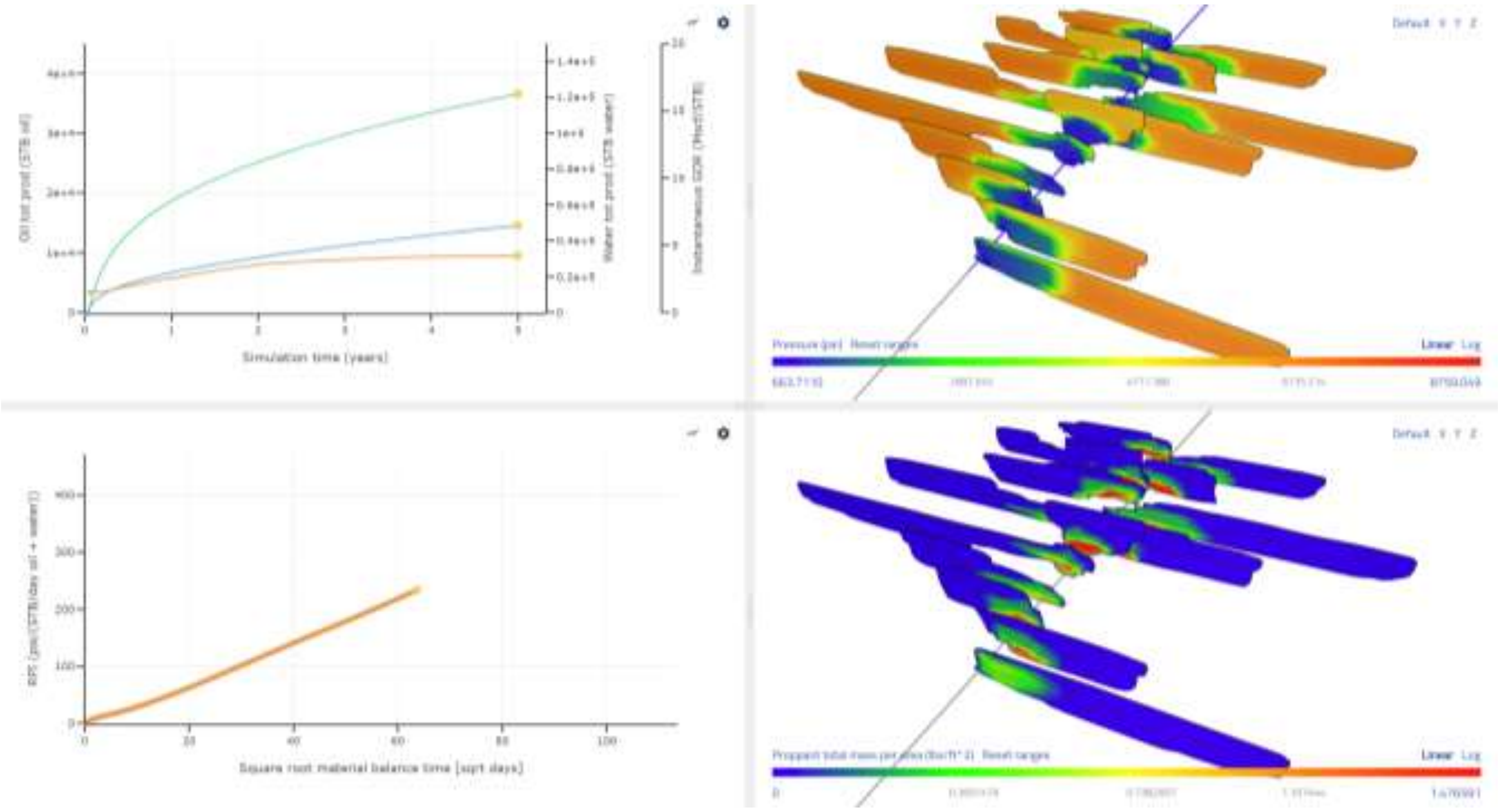

*Figure 35: Example simulation with ‘submesh fractal D’ set to 0.3.*

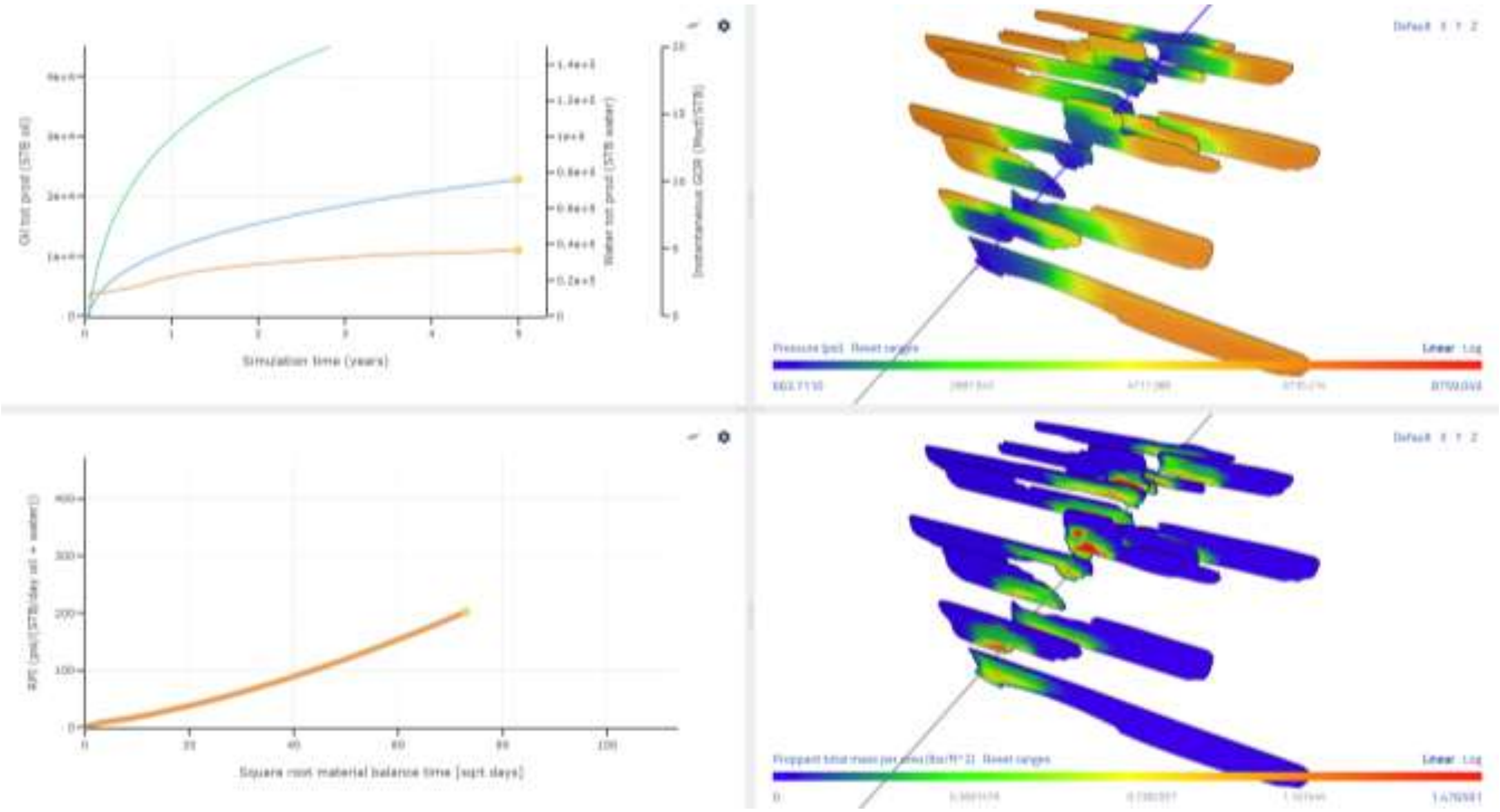

*Figure 36: Example simulation with ‘submesh fractal D’ set to 0.6.*

When matching the RTA curvature in real data, you have a variety of tools – relative permeability loss with drawdown, pressure dependent permeability loss, pressure/stress dependent conductivity loss, and time-dependent conductivity loss. The ‘fractal D’ setting gives you one more tool that can be used to match RTA curvature. It may often be useful (and realistic) to include multiple simultaneous simulations. For example, the figure below shows the same simulation as shown above with ‘fractal D’ set to 0.6, but also with some time-dependent conductivity loss (Section ‘19.7 Time-dependent proppant pack conductivity’). The simulation exhibits even stronger RTA curvature, and also experiences more muted increase in GOR.

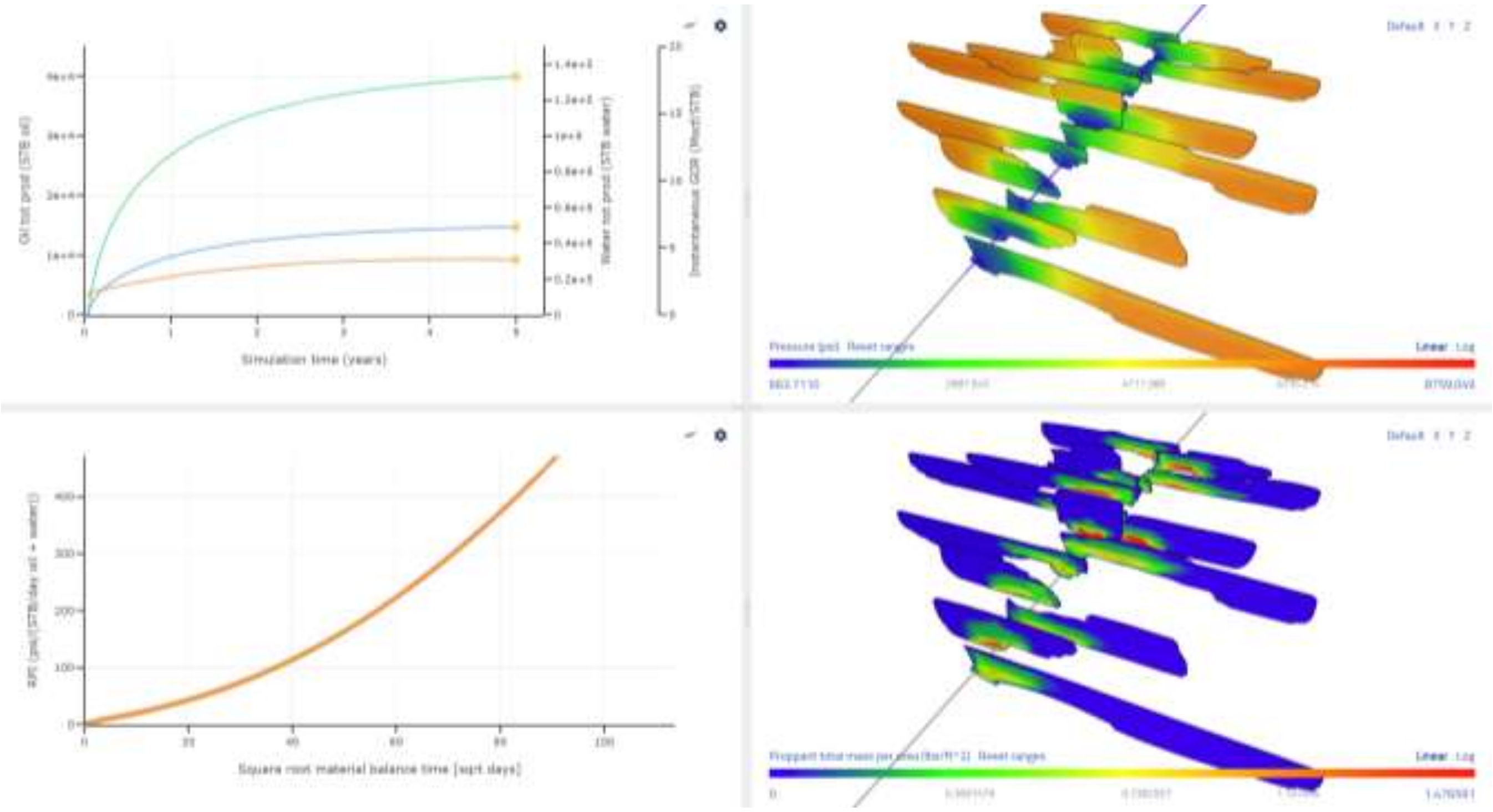

*Figure 37: Example simulation with ‘submesh fractal D’ set to 0.6, and also, time-dependent conductivity loss (type 2) with a time constant of .0004 days^-1.*

# 20. Sensitivity analysis

ResFrac’s sensitivity analysis tools enable the user to create and run batches of simulations that vary systematically, to help understand the effects of changing simulation inputs. To use the sensitivity analysis tools, the user starts with a ‘base simulation’ around which to run sensitivity analysis. The base simulation inputs specify all the values needed to run a ResFrac simulation, and the user selects a subset of input parameters of the base simulation to vary in the sensitivity analysis.

Representing a single modeling idea, such as changing the cluster spacing, often entails creating simulations that differ from each other for several simulation input parameters. In other words, to express one idea, the simulations often need to be changed in more than one place. Additionally, it is natural to consider a few different modeling ideas at once, each of which is expressed by varying several parameters, and so the user might need to vary dozens or more parameters at a time.

In mathematical terms, the user wants to vary inputs in a low dimensional input space (a few modeling ideas) and have that variation be reflected in the high dimensional simulation parameter space (systematically change many parameters embodied in the simulation input and settings files). To enable this, ResFrac’s sensitivity analysis tools are built around dimension reduction of the input space via the concept of ‘parameter groups.’ One parameter group, corresponding to a single modeling idea, represents a collection of simulation input parameters that vary together.

After a base simulation is chosen and a suitable dimension reduction is defined, the user then specifies a sampling scheme to generate simulations to run.

## 20.1 Dimension reduction of input parameters using parameter groups

ResFrac simulations live in a high dimensional (thousands of dimensions) space represented as $X$, where $X$ is the set of all possible simulations. A single simulation $x \in X$ is a point in that space. The sensitivity analysis dimension reduction procedure consists of a mapping $T: Z \rightarrow X$, where $Z$ is a low dimension (typically $dim(Z)$1to 10) space. The map $T$ is defined such that for every $z \in Z$ there is a corresponding unique point $x = T(z)$. In other words, for each point in the low dimension parameter space $Z$ there is a corresponding unique simulation input and settings file. Note that the transformation function $T$ is in general not invertible.

The user specifies $T$ and $Z$ by populating the parameter groups and parameters tables in the ResFracPro user interface. The number of parameter groups defines the dimension of $Z$, with each row of $z$ representing the position of the point along the dimension of the corresponding one parameter group. Each row of $z$ takes on a scalar value between $-1$ and $+1$, with $-1$ indicating the 'left' end along that dimension, $+1$ indicating the 'right' end, and 0 indicating the 'center.' Collectively, the values in the parameter groups and parameters tables specify the individual transformation functions that make up $T$.

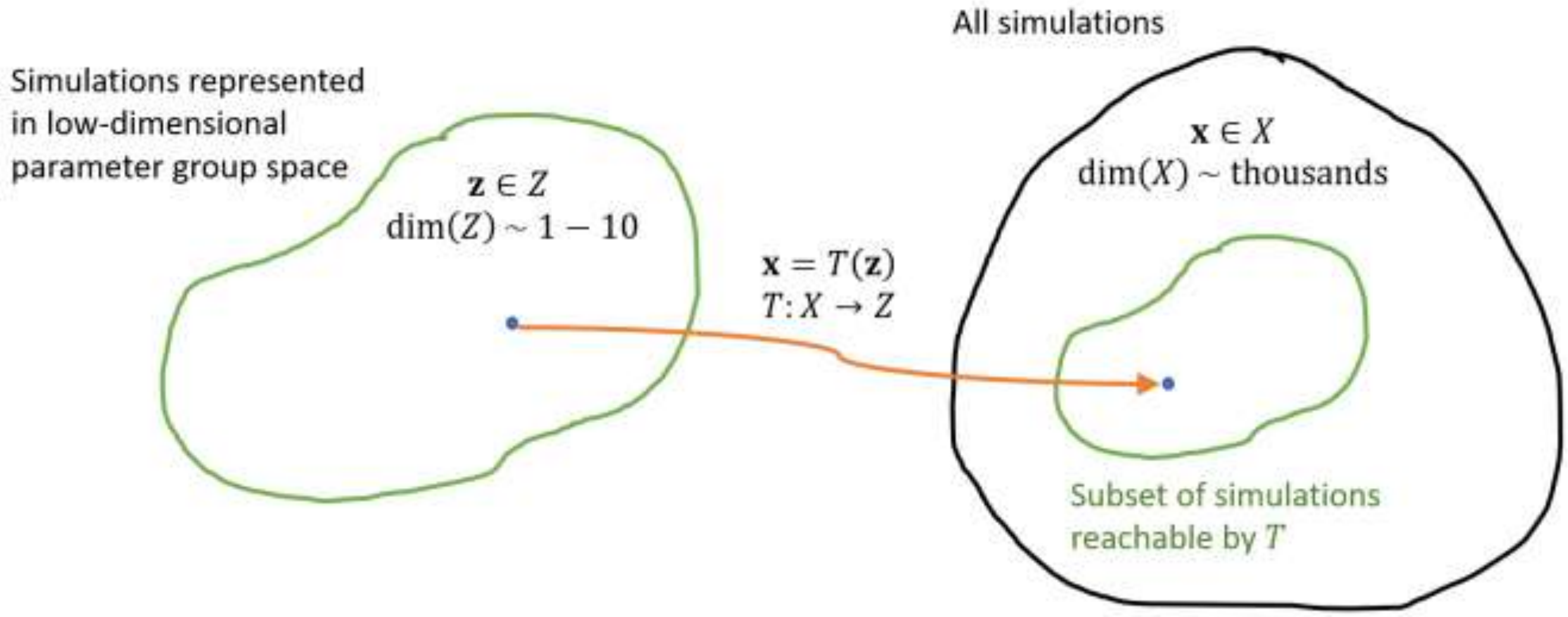

Three different transformation functions from parameter group space to parameter space are supported: linear adder, linear multiplier, and logarithmic multiplier. It is possible to use different types of transformations for different parameters, so the full map $T$ can incorporate a combination of the three types of transformations.

The sensitivity analysis tools allow the user to select any physically meaningful numerical field for use in a sensitivity study. Non-numerical fields such as binary and categorical values are not allowed to be selected, nor are fields with no physical meaning such as maximum simulation wall clock time. Integer numerical fields, such as number of shots in a perforation cluster, are handled by rounding: The transformation calculation treats the integer numerical field as a floating-point number, and then the post-transformation floating-point values are rounded to the closest integer.

The three different types of transformation functions are as follows:

Linear adder transformation

$$x_j = T(z_i) = x_{j,0} + \begin{cases} T_{ij,c} - (T_{ij,l} - T_{ij,c})z_i, \wedge z_i < 0 \\ T_{ij,c}, \wedge z_i = 0 \\ T_{ij,c} + (T_{ij,r} - T_{ij,c})z_i, \wedge z_i > 0 \end{cases}$$

20-1

Linear multiplier transformation

$$x_j = T(z_i) = x_{j,0} \begin{cases} T_{ij,c} - (T_{ij,l} - T_{ij,c})z_i, \wedge z_i < 0 \\ T_{ij,c}, \wedge z_i = 0 \\ T_{ij,c} + (T_{ij,r} - T_{ij,c})z_i, \wedge z_i > 0 \end{cases}$$

20-2

Logarithmic multiplier transformation

$$x_j = T(z_i) = x_{j,0} \begin{cases} T_{ij,c}^{1+z_i} T_{ij,l}^{-z_i}, \wedge z_i < 0 \\ T_{ij,c}, \wedge z_i = 0 \\ T_{ij,c}^{1-z_i} T_{ij,r}^{z_i}, \wedge z_i > 0 \end{cases}$$

20-3

In equations 20-1 – 20-3, $z_i$ is the value of the ith parameter group, $x_j$ is the value of the jth parameter (and this parameter is a member of the ith parameter group), $x_{j,0}$ is the center value of the jth parameter, $T_{ij,l}$ is the left transformation adder or multiplier coefficient for the ith parameter group and jth parameter, $T_{ij,r}$ is the right transformation adder or multiplier coefficient, and $T_{ij,c}$ is the transformation adder or multiplier center value.

The transformation adder or multiplier center value, $T_{ij,c}$, is usually equal to zero for linear adder transformations and equal to one for linear or logarithmic multiplier transformations. In the special cases of linear adder transformation with left and right adders both less than zero, or both greater than zero, applying an adder center value of zero is nonsensical, so in these cases $T_{ij,c}$ takes on nonzero value. Similarly, for linear or logarithmic multipliers, $T_{ij,c}$ takes on nonunity value if the left and right multipliers are both greater than one, or both less than one. In these special cases, $T_{ij,c}$ is specified as the center value of the adder or multiplier interval, so that the parameters are symmetric across the left and right bounds, according to the following relations:

Linear adder transformation

$$T_{ij,c} = \begin{cases} \frac{T_{ij,l} + T_{ij,r}}{2}, \wedge (T_{ij,l} > 0 \text{ and } T_{ij,r} > 0) \text{ or } (T_{ij,l} < 0 \text{ and } T_{ij,r} < 0) \\ 0, \wedge \text{ otherwise} \end{cases}$$

20-4

Linear multiplier transformation

$$T_{ij,c} = \begin{cases} \frac{T_{ij,l} + T_{ij,r}}{2}, \wedge (T_{ij,l} > 1 \text{ and } T_{ij,r} > 1) \text{ or } (T_{ij,l} < 1 \text{ and } T_{ij,r} < 1) \\ 1, \wedge \text{ otherwise} \end{cases}$$

20-5

Logarithmic multiplier transformation

$$T_{ij,c} = \begin{cases} \exp\left(\frac{logT_{ij,l} + logT_{ij,r}}{2}\right), \wedge (T_{ij,l} > 1 \text{ and } T_{ij,r} > 1) \text{ or } (T_{ij,l} < 1 \text{ and } T_{ij,r} < 1) \\ 1, \wedge \text{ otherwise} \end{cases}$$

20-6

The parameter center value $x_{j,0}$ defaults to the value in the base simulation. The user can override the base simulation value by specifying a different value in the parameters table. If the parameter center value is left blank (i.e., set to NaN) in the parameters table, the value in the base simulation is applied. For parameters that represent more than one row in the simulation, the base simulation value is used for $x_{j,0}$ wherever the value in the base simulation for that value is not blank (not NaN). If the base simulation value is blank (NaN), and the user specifies a non-NaN parameter center value in the parameters table, then the user specified non-blank (non-NaN) value is used for $x_{j,0}$.

## 20.2 Sampling schemes

Sampling schemes are used to generate sets of simulations to run in the sensitivity analysis. Essentially a sampling scheme is a method of defining a set of points $\{z_1, z_2, z_3, \dots, z_n\}$ that will be used to generate simulations. The types of sampling schemes available are full factorial, Sobol sequence, Latin hypercube, centered composite, Box Behnken, Plackett Burman, fractional factorial, uniform random, one-at-a-time, and user-defined.

Given the array of options, what do we recommend as the 'best' to use? Sobol sampling has been shown to be one of the most reliable and effective sampling schemes, and it conveniently can be used with any number of simulations – so it easily scales up or down depending on the amount of detail that you want. Sobol sequences provide a 'quasi-random' design, which places points uniformly throughout the sample space, maximizing the average 'distance' between each point. However, a limitation is that Sobol sequences do not sample up-against the edges of the problem domain, and in some applications, that can be an issue. So, to get points right up against the edges, you might supplement by using Sobol in addition to Box Behnken or centered composite.

When a sensitivity analysis is run, the center point (z = 0) is always included automatically. If more than one sampling scheme is specified for a sensitivity analysis, then the set of points used is the superset of all points specified by any of the sampling schemes, with duplicates removed.

## 20.3 Monte Carlo uncertainty quantification

Sensitivity analysis workflows can be used to perform Monte Carlo uncertainty quantification. In the 'Decision support' panel of a sensitivity analysis setup, the user can select either 'Sensitivity analysis' or 'Uncertainty quantification.'

The concept is to specify a statistical distribution on model inputs, representing uncertainty. For example, you may place a normal distribution on porosity and saturation, and a log-normal distribution on permeability. Then, you randomly draw from these distributions and generate a group of simulations. You run them, and then create a histograms of the results, such as oil production. This allows you to quantify the effect of these uncertainties on quantities of interest.

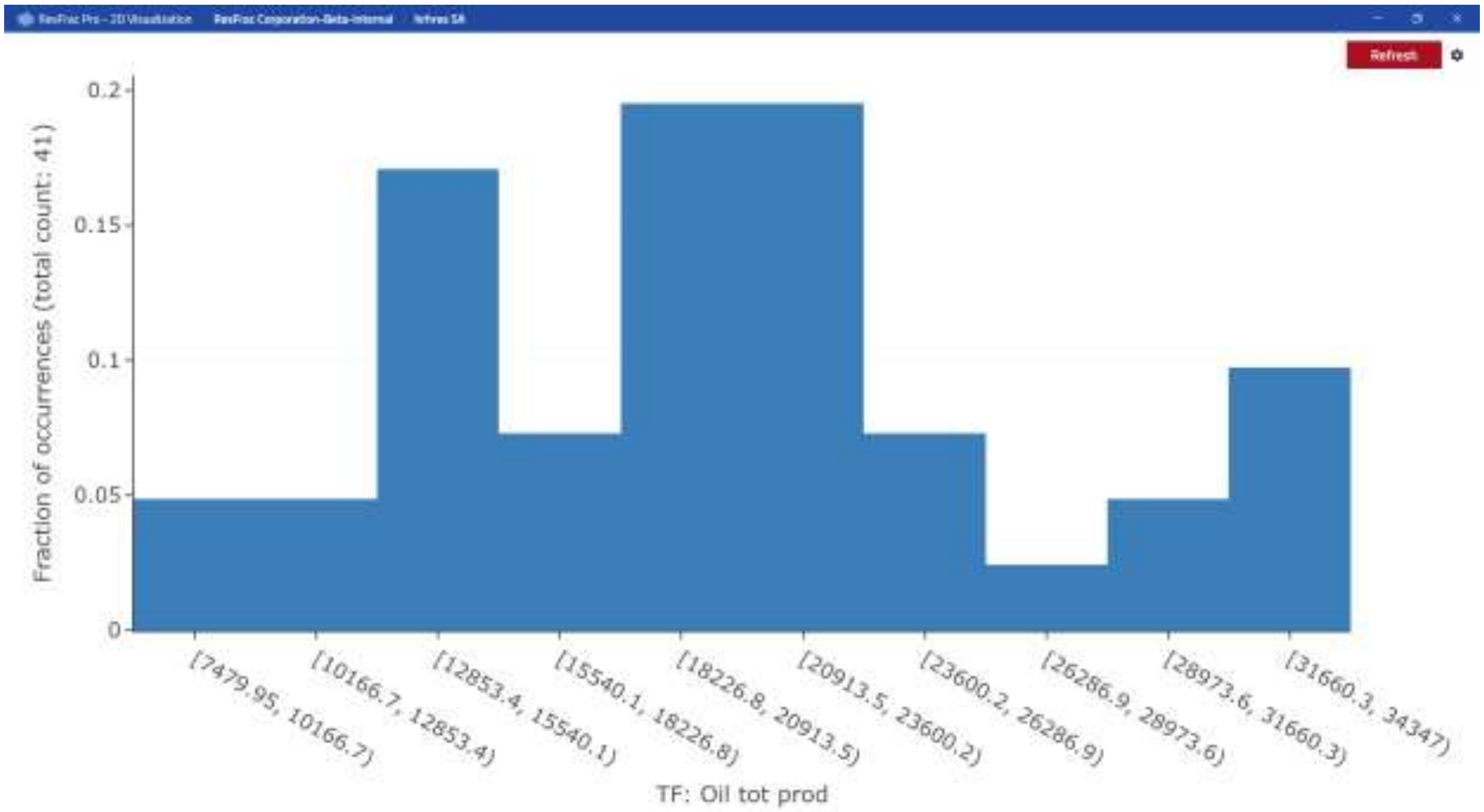

*Figure 38: Example of a histogram of results from an uncertainty quantification workflow.*

For each parameter group, the user selects whether the parameter will be uniformly distributed, univariate Gaussian, or multivariate Gaussian.

If uniform, then a random 'parameter group interpolator' will be selected for the parameter group, randomly located between -1 and 1. For each parameter in the parameter group, the user must specify a 'minimum' and 'maximum' value. If the PG interpolator is equal to -1, for example, then each parameter in the group will take on its minimum value. If it is 0, then each parameter will take on the midpoint between the minimum and maximum value.

If univariate Gaussian, then the PG interpolator will be selected with a random draw from a Gaussian distribution with mean of zero, and standard deviation of 0.2. The standard deviation is set to 0.2 so that virtually all the parameter group values will lie in the range -1 to 1, the same as the range used for the uniform distribution. Next, for each individual parameter in the parameter group, the PG interpolator will be scaled to pick a particular value. For example, let's say that the PG interpolator for a particular Monte Carlo draw is 0.1 (ie, to the right of the mean by 0.5 times the standard deviation). For each parameter in the group, the user must specify an 'average' multiplier, a standard deviation, and min/max values. For example, let's say that permeability is 2 md, and the user has selected that the 'average' multiplier is 1.0, and the standard deviation is 0.15. In that case, for a parameter group interpolator of 0.1, the multiplier will be calculated as: 1.0 + 0.15*(0.1/0.2) = 1.075, and the permeability for that draw will be equal to 2.15 md. The factor 0.15*0.1/0.2 is used to scale from the standard deviation used in the parameter group draw (which is always 0.2) to the standard deviation for this particular parameter (which was set to 0.15). What if the multiplier exceeded the min/max values specified by the user? In that case, the value would be decreased/increased to be equal to that min/max value that was exceeded. So as a hypothetical example, if the 'maximum' multiplier was set to 1.05, then if the output of the scaled normal distribution was 1.075, the value would be set back to 1.05.

The user does not have to specify only 'multipliers.' They can specify 'adders' or 'logarithmic' multipliers. The logarithmic multipliers can be useful if values are varying over orders of magnitude. For example, if you would like to vary permeability multipliers from 0.01 to 100.0. You could specify a logarithmic multiplier, and instead of making the multiplier uniform (or normally distributed) between those bounds, the *logarithm* of the multiplier would be uniform or normally distributed.

We specify 'multipliers' and 'adders' instead of the values themselves because each row in the 'parameters' table can actually encompass a group of values. For example, let's say that the list of properties versus depth defines 100 layers. You may select the permeability of 90 of these layers and add them to the list of parameters in the table. It would be cumbersome to list 90 parameters separately in the table, and so instead, they are all combined into a single row of the 'parameters' table. Because they can each have different permeability values, we must specify adders and multipliers.

If the user specifies multivariate Gaussian, then they must specify a matrix of correlation coefficients that relate the parameter group interpolators for any other parameter groups specified to be multivariate Gaussian. Correlation near 1.0 suggests that the interpolators are nearly identical. Correlation near -1 suggests that they have nearly perfect anticorrelation. Correlation of zero means that they have no relationship to each other. For each draw, the parameter group interpolators are selected with a draw from the multi-Gaussian distribution, using the user-specified correlation coefficients.

## 20.4 Postprocessing

The ResFracPro user interface includes a postprocessing tool to help interpret the results of a sensitivity analysis. The postprocessing tool consists of two main sets of features: target functions and plotting tools. The recommended workflow is to first define target functions, and then use these target functions in the plotting tools.

A target function is a quantity of interest by which to assess the simulation. Any column in the sim_track_xxx.csv results file in a ResFrac simulation can be used to define a target function. A target function can be for the value in the column at a point in time, or the maximum, minimum, or average value over a range of time. For example, someone might define one target function as the maximum water cut for a specific well over the production period of the simulation, a second target function as the cumulative overall oil production at the end of the simulation, and a third target function as the average rate of water production from another well starting from one month after the well has started production.

Presently, the user interface provides two sensitivity analysis-specific plot types: spider plot and scatterplot matrix. A spider plot, intended to be used along with one-at-a-time sampling, plots target function values on the vertical axis and parameter group values on the horizontal axis, for points that have at most one non-zero value in $z$. The scatterplot matrix creates a grid of scatterplots with target functions on the vertical axis and parameter groups on the horizontal axis, and plots all of the points with results.

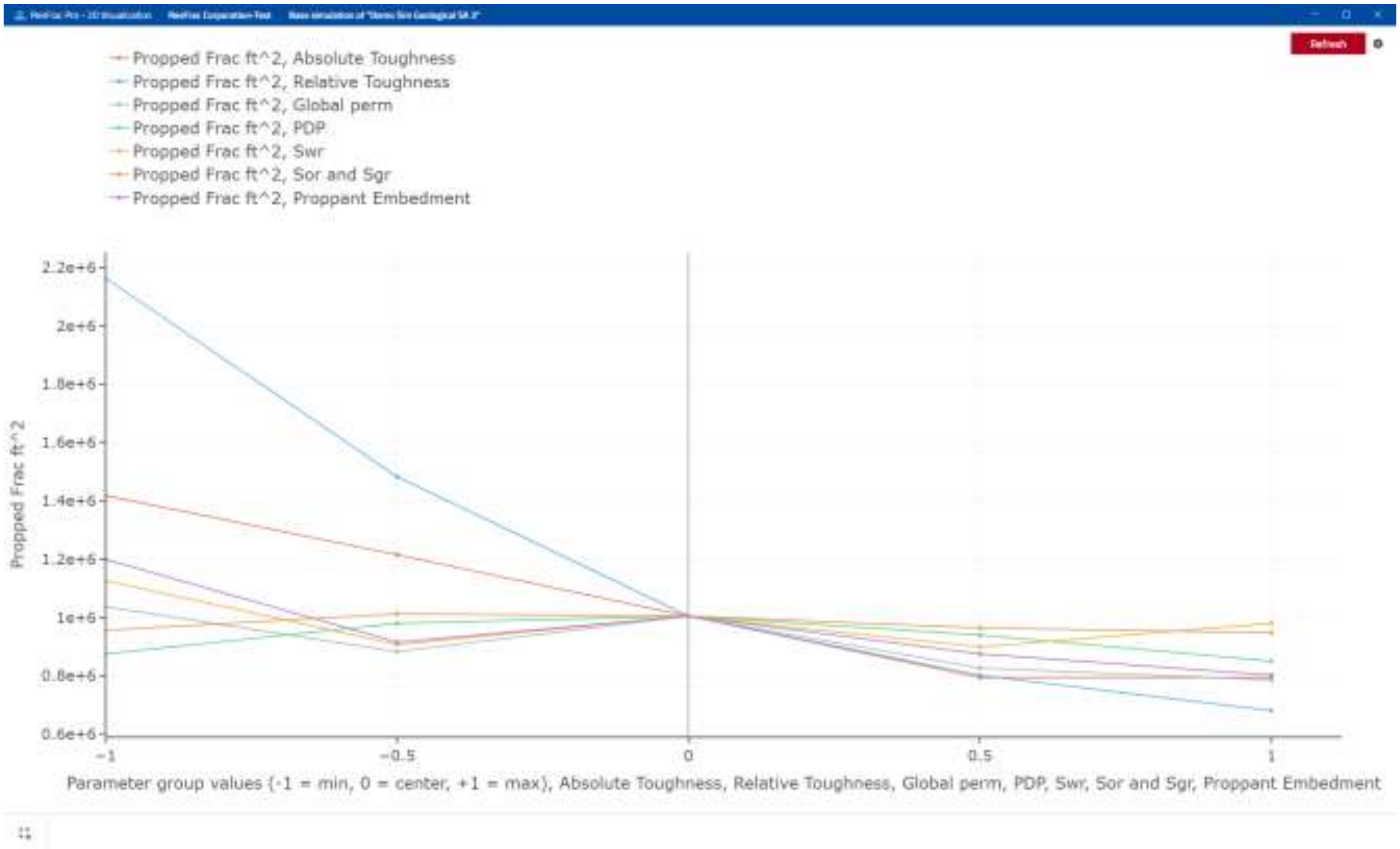

Figure 39: Example of a spider plot from the postprocessing tool.

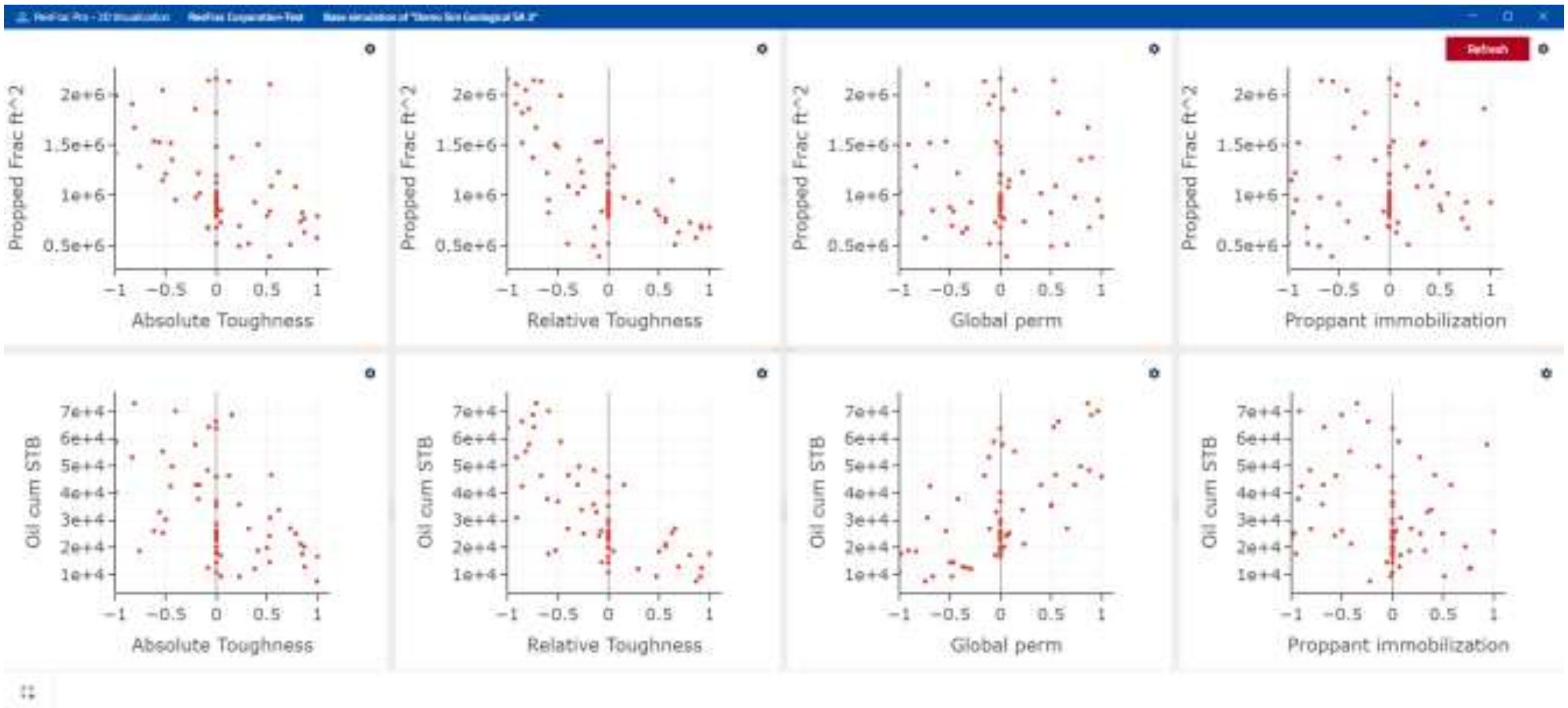

Figure 40: Example of scatterplots from the postprocessing tool.

# 21. External libraries

ResFrac uses several publicly available libraries to perform calculations. Petsc is used as an iterative linear solver (Balay et al., 1997, 2016). Hypre is used to construct and apply the preconditioners used with the iterative linear solver (Falgout et al., 2006). MUMPS (Amestoy et al., 2001, 2019) and Eigen SparseLU (Guennebaud and Jacob, 2010) are used as direct linear solvers. The Template Numerical Toolkit (TNT) and the JAMA/C++ Linear Algebra Package are used for a handful of specialized numerical operations (Pozo, 1997). The Poros distribution of Metis is used as a mesh partitioner (Karypis and Kumar, 1999). The lower-level numerical algebra packages GotoBLAS (Blackford et al., 2002; Xianyi et al., 2012) and LAPACK (Anderson et al., 1999) are used. Finally, we make use of MPICH. We use an implementation of the modified Cholesky decomposition from Fang and O'Leary (2008).

# 22. Validation simulations

The numerical accuracy of the simulator is checked with a test suite of problems that have known solution. The test suite also includes a variety of simulations designed to test the ability of the simulator to converge through difficult problems. To ensure that ongoing development does not inadvertently introduce errors, a script is used to automatically run the full test suite and compare the results against known solutions. This appendix briefly reviews some of the suite problems that are designed to test accuracy. All results shown below have close match with the known benchmark solution.

1. Sneddon (1946) Solution for Stress Around an Open Crack

Injection is performed at constant pressure (greater than the minimum principal stress) into a circular preexisting fracture in an impermeable formation. The injection rate goes to zero as the pressure in the fracture becomes uniform. A grid of stress observation points is defined, and the calculated stresses are compared against the Sneddon (1946) solution.

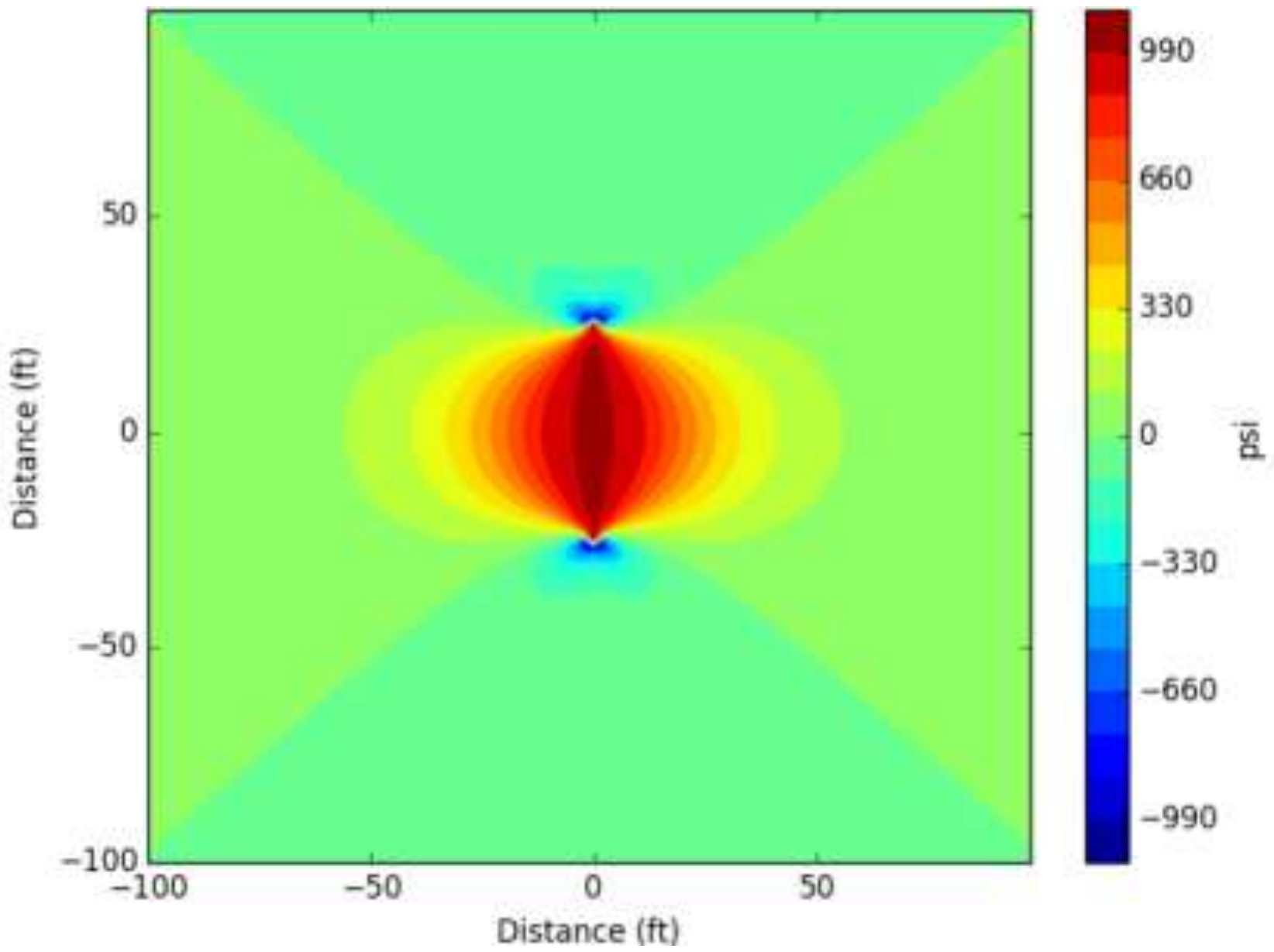

Figure 41: Numerically calculated change in the $\sigma_{xx}$ stress around a circular crack opening with uniform internal pressure.

2. SPE1 Comparison Problem

The SPE1 comparison problem is a black oil simulation with gas injection. The simulated gas and oil production rates are compared with the solution provided by (Odeh, 1981).

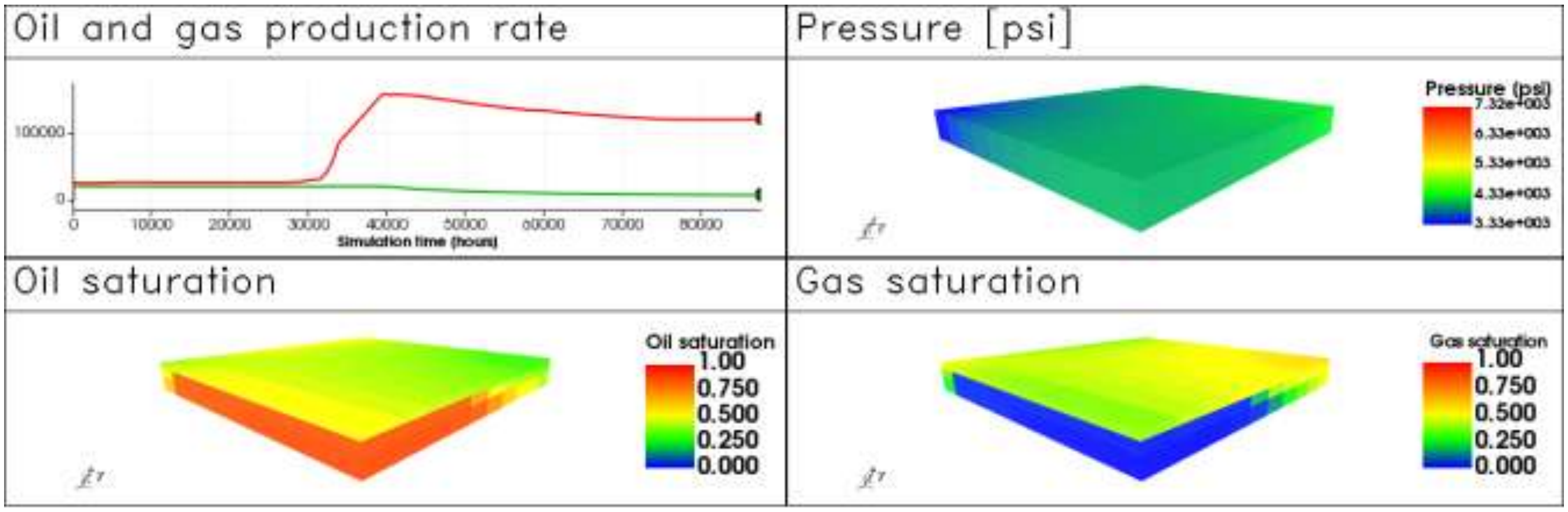

Figure 42: Calculated oil and gas production rate, and the final distribution of pressure, oil saturation, and gas saturation from the solution to the SPE1 comparison problem.

3. SPE3 Comparison Problem

The SPE3 comparison problem is a compositional simulation with gas injection. The reservoir fluid is a retrograde condensate. Liquid produced at the surface is sent to sales. The produced gas is reinjected to volatilize and sweep out the components dropped out as liquid in the reservoir. The simulated oil production rates are compared with the solution provided by (Kenyon and Behie, 1987). The problem is solved with the ARCO fluid model.

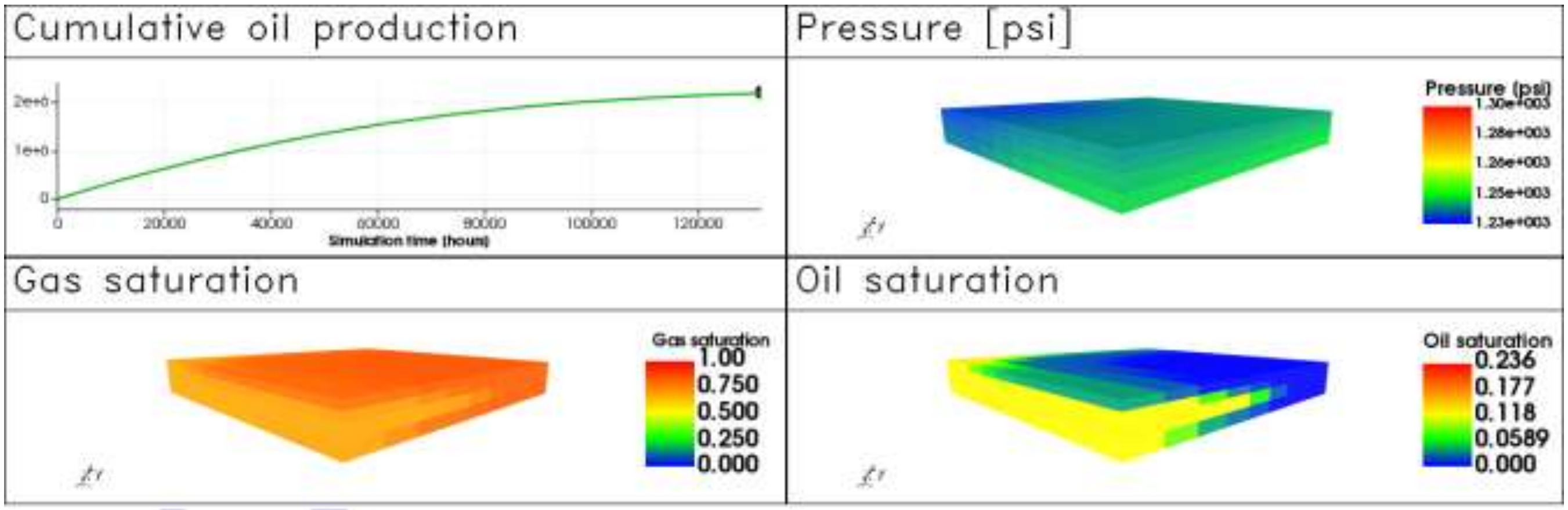

Figure 43: Calculated oil production rate and the final distribution of pressure, gas saturation, and oil saturation from the solution to the SPE3 comparison problem.

In addition, the SPE3 problem has been used to validate our implementation of the modified black oil model. The SPE3 simulation was run with a modified black oil model table constructed to mimic the original compositional fluid model. The results are close, though not identical (which is to be expected, because the MBO is a significant simplification of a compositional model).

4. Radial Crack Propagation

Radial crack propagation problems are solved for the limiting cases of low toughness and high toughness with no leakoff. To match the analytical solutions, the stress gradient is assumed to be zero, wellbore storage is neglected, non-Darcy pressure drop is neglected, and gravity is neglected. With low toughness, the solution is (Equation 6C-5 from Economides and Nolte, 2001):

$$R = 0.52\sqrt[9]{\frac{YQ^3}{(1-\nu^2)\mu}}t^{\frac{4}{9}}. \qquad \text{(A1)}$$

where R is radius, Y is Young's modulus, Q is volumetric flow rate, $\nu$ is Poisson's ratio, $\mu$ is viscosity, and t is time.

With high toughness, the solution can be derived by assuming uniform pressure in the crack and combining the solutions for volume and stress intensity factor of a circular crack:

$$V = Qt = \frac{16R^3(1-\nu^2)}{3Y}\Delta P, \qquad \text{(A2)}$$

$$K_{Ic} = 2\Delta P\sqrt{\frac{R}{\pi}}, \qquad \text{(A3)}$$

$$R = \sqrt[2.5]{\frac{3QtY}{8(1-\nu^2)\sqrt{\pi}}}. \qquad \text{(A4)}$$

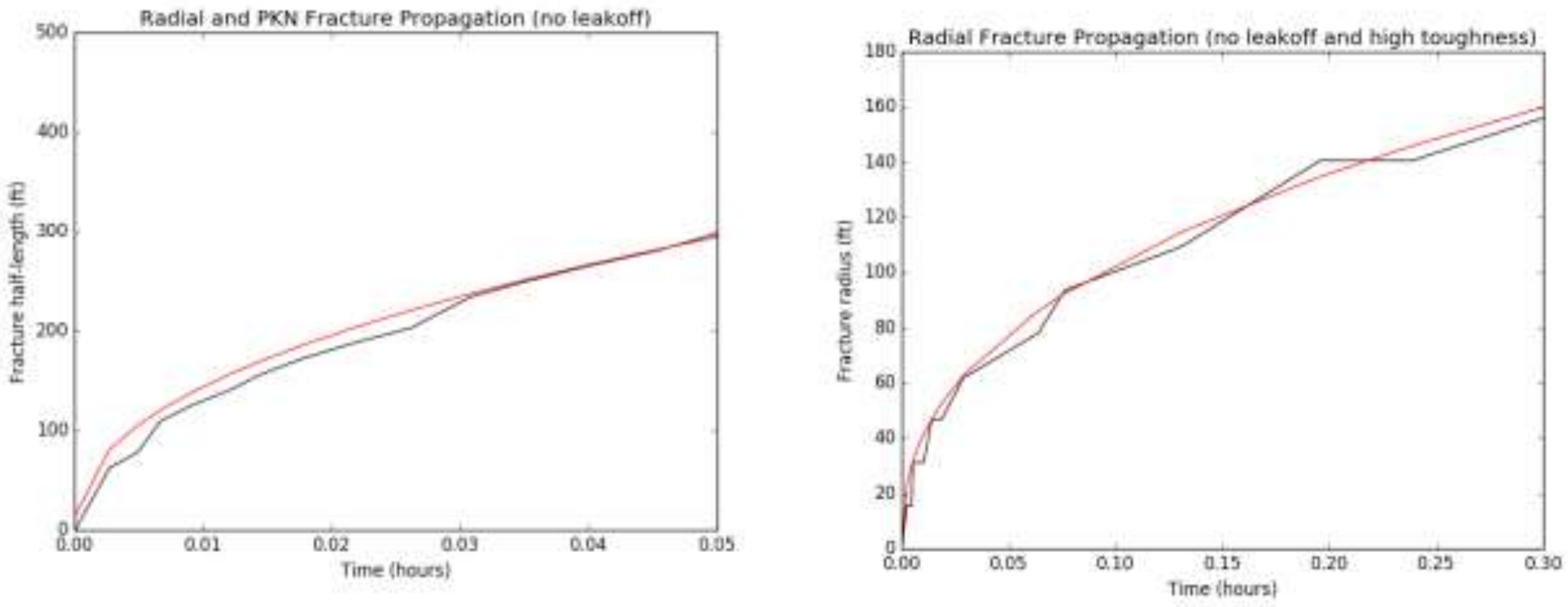

Figure 44: Radius versus time for radial crack propagation with no leakoff and low and high toughness. Black lines show numerical result and red lines show the analytical solutions.

5. PKN Crack Propagation with and without Leakoff

The assumptions of the PKN model are not exactly reproduced by a 3D simulator. However, in the limiting case of perfect height confinement and low toughness, a 3D simulator should be expected to approximately match the PKN solution. To match the analytical solution, the stress gradient is assumed to be zero, wellbore storage is neglected, non-Darcy pressure drop is neglected, and gravity is neglected. The analytical solution with leakoff is given by Equations 9.41 and 9.42 from Valko and Economides (1995). The analytical solution without leakoff is given by Equation 9.13 from Valko and Economides (1995). Two simulations of the PKN with leakoff problem are performed. One uses a highly refined mesh toward the fracture. The second uses a coarse mesh and the 1D subgrid method for calculating leakoff (McClure, 2017). In these simulations, the crack initially propagates radially before reaching a maximum height of 328 ft. Once the length is substantially greater than 328 ft, the propagation should be approximately PKN.

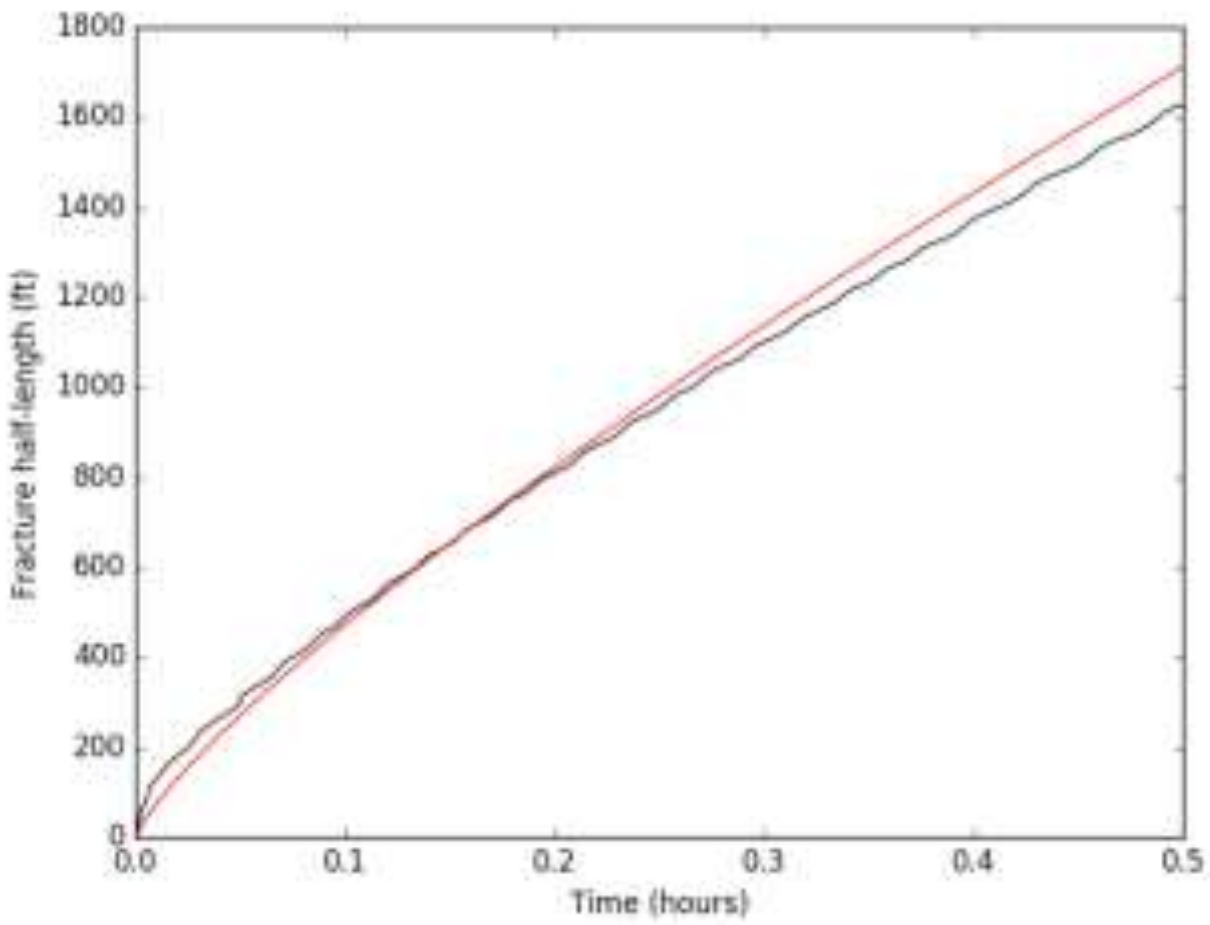

Figure 45: Approximate PKN propagation (at later time) with no leakoff. The black line shows the numerical result and the red line shows the analytical result.

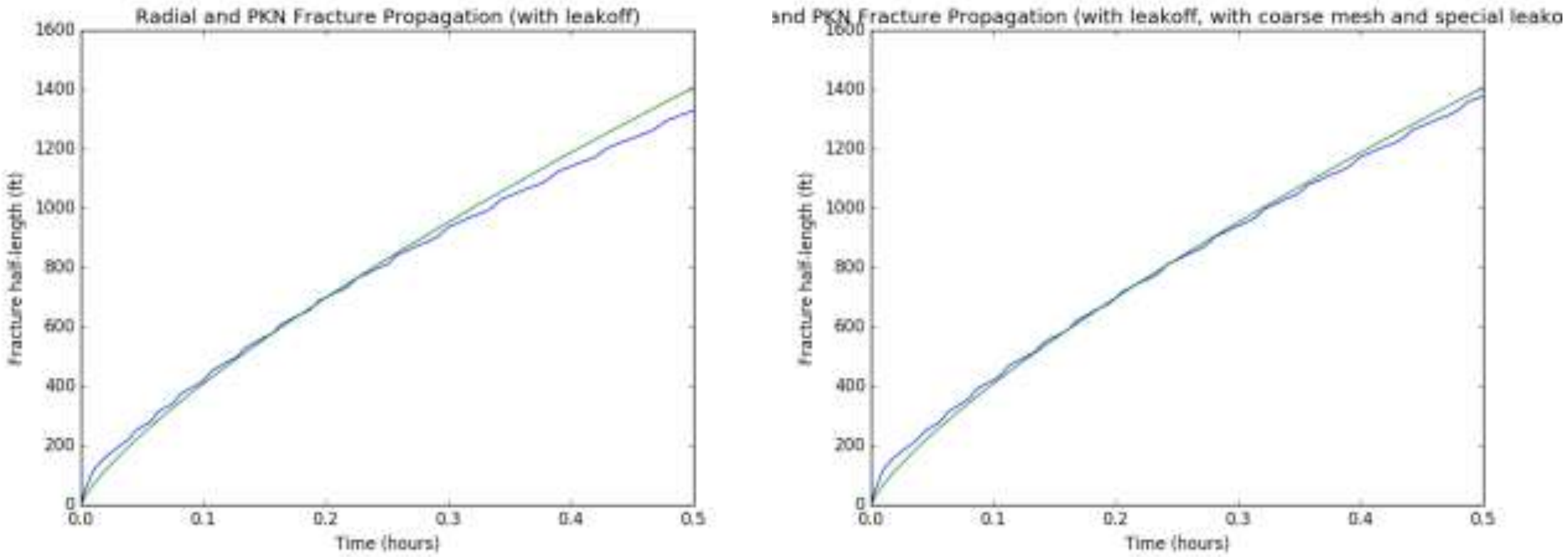

Figure 46: Approximate PKN propagation (at later time) with leakoff. The blue lines show the numerical result and the green lines shows the analytical result. The solution on the left uses a refined mesh towards the matrix. The solution on the right uses the 1D subgrid method and a coarse mesh.

6. Thermal conduction into a Crack

Gringarten et al. (1975) provide an analytical solution for produced temperature in a scenario with water flow between two wells through a crack embedded in a zero permeability medium. Injection is performed at constant temperature and rate. The water is heated by conduction as it flows through the crack between the well. The problem domain is sufficiently long perpendicular to the fracture that the problem domain is effectively infinite in the direction perpendicular to the fracture. The Gringarten et al. (1975) solution assumes 1D heat conduction, and so the boundaries of the matrix mesh do not extend beyond the wells.

Figure *47* shows a 3D visualization, with a cross-section cut through the matrix. Figure *48* shows the simulated and analytical solutions for two cases: high refined mesh towards the fracture and a coarse mesh using the 1D subgrid method (McClure, 2017) for heat conduction.

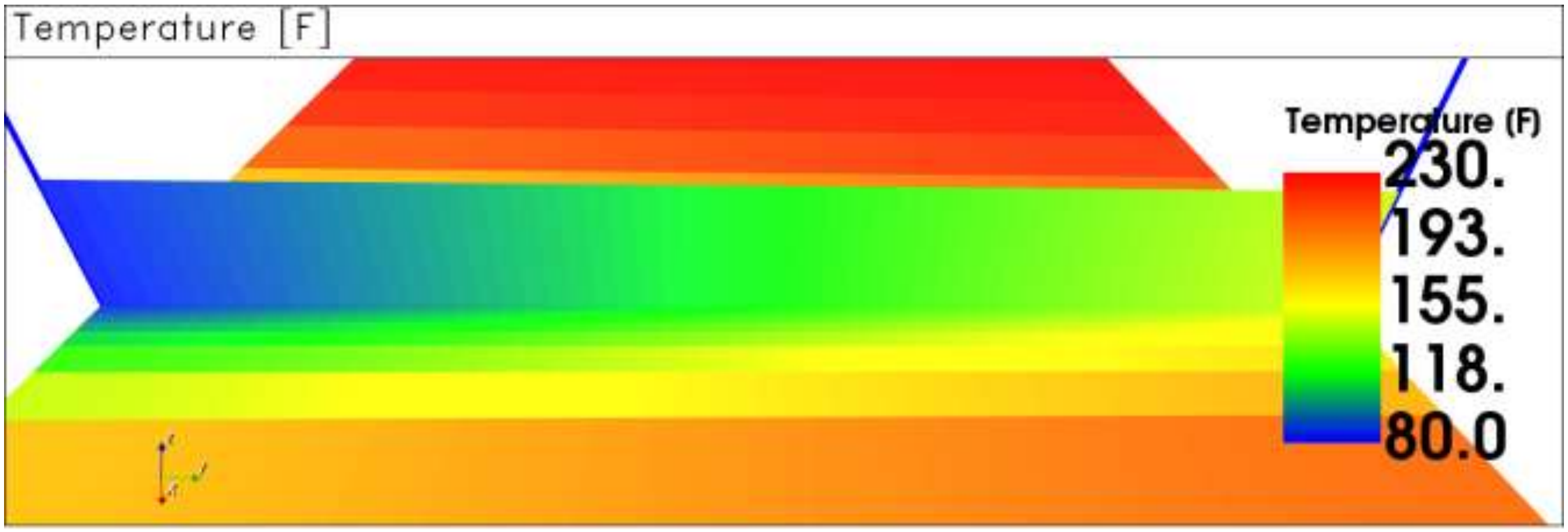

Figure 47: Spatial distribution of temperature at the end of the Gringarten simulation. A horizontal cross-section is cut through the matrix.

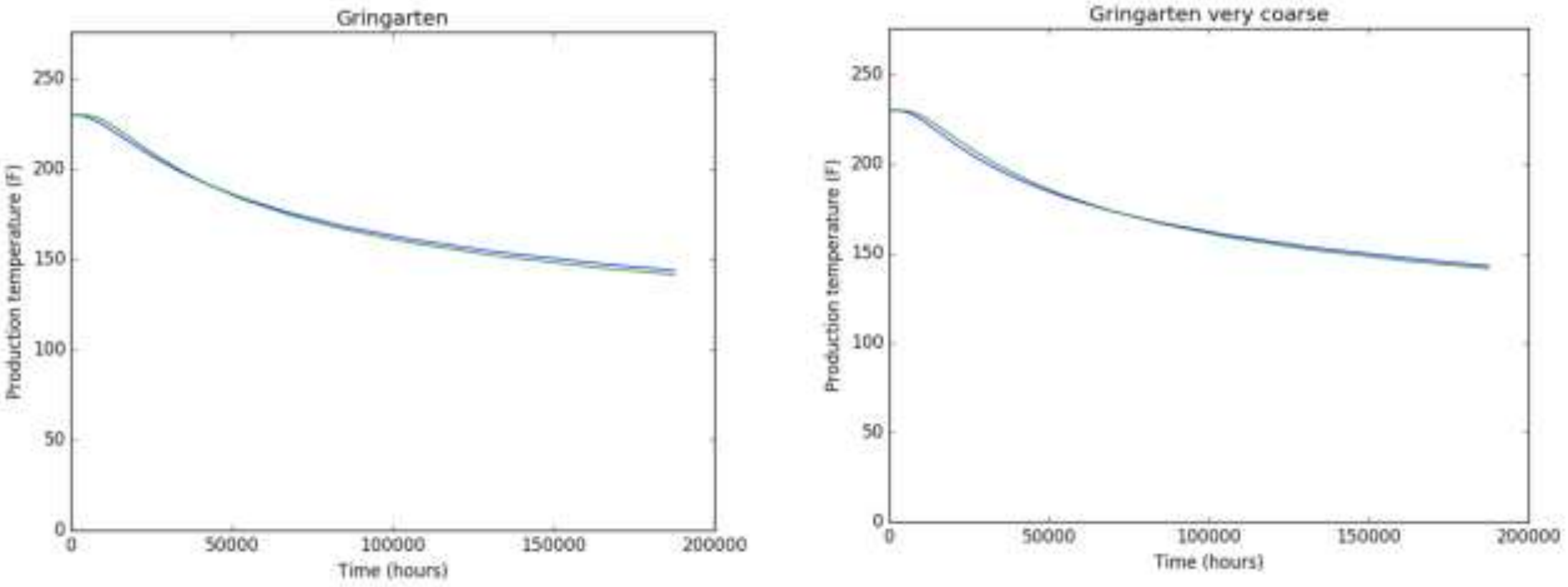

Figure 48: Production temperature versus time in the solution to the Gringarten problem. The blue lines show the simulated results, and the green lines show the analytical solution.

7. Estimating Leakoff Coefficient from a Diagnostic Fracture Injection Test

Under ideal conditions, the leakoff coefficient can be estimating from the preclosure transient after shut-in from a diagnostic fracture injection test (Nolte, 1979). In the test, fluid is injected, creating a hydraulic fracture, and then pressure is monitored after shut-in.

To test, we set up an idealized simulation of a diagnostic fracture injection test (single phase and single component, very small residual aperture after closure, no non-Darcy pressure drop, no gravity, perfect height confinement). Figure *49* shows the simulated shut-in transient. Using the equations for calculating leakoff coefficient from the derivative of pressure with respect to G-time (summarized by Marongiu-Porcu et al., 2014), the leakoff coefficient can be estimated from the simulation as 9.8e-6 $m/s^{1/2}$. The exact value is 1e-5 $m/s^{1/2}$.

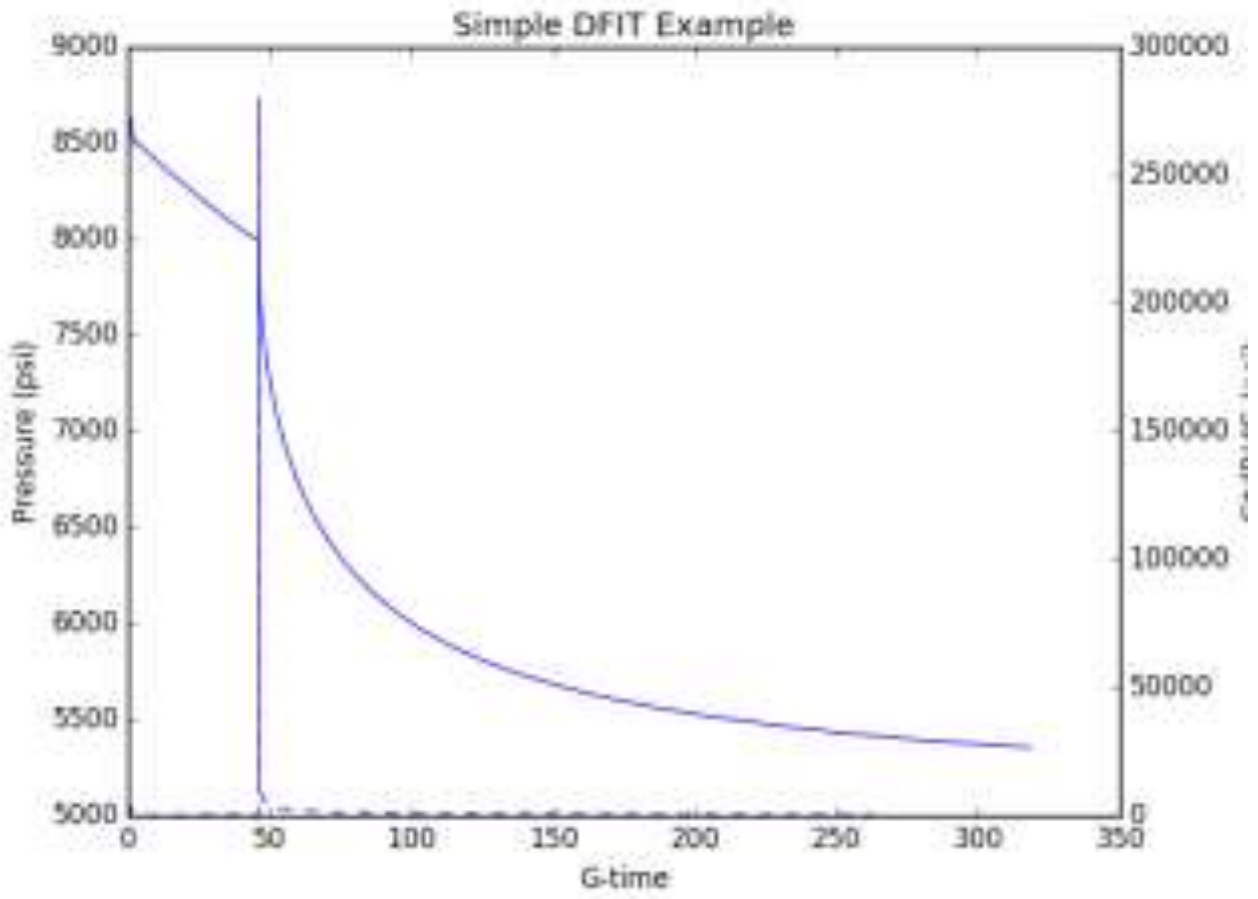

Figure 49: Shut-in transient from an idealized DFIT simulation.

8. Poroelastic Stresses Around a Cuboid of Constant Pressure Change

Page 73 from Nowacki (1986) provides an analytical solution for the stresses induced outside a rectangular cuboid of uniform change in pressure (or temperature). To match this solution, a simulation is set up with the special condition that permeability is zero outside a cube in the center of the problem domain. Then, injection is performed at constant pressure until the pressure is uniform within the cuboid. The stresses induced by the deformation are calculated numerically (shown in Figure *50*) and compared with the analytical solution.

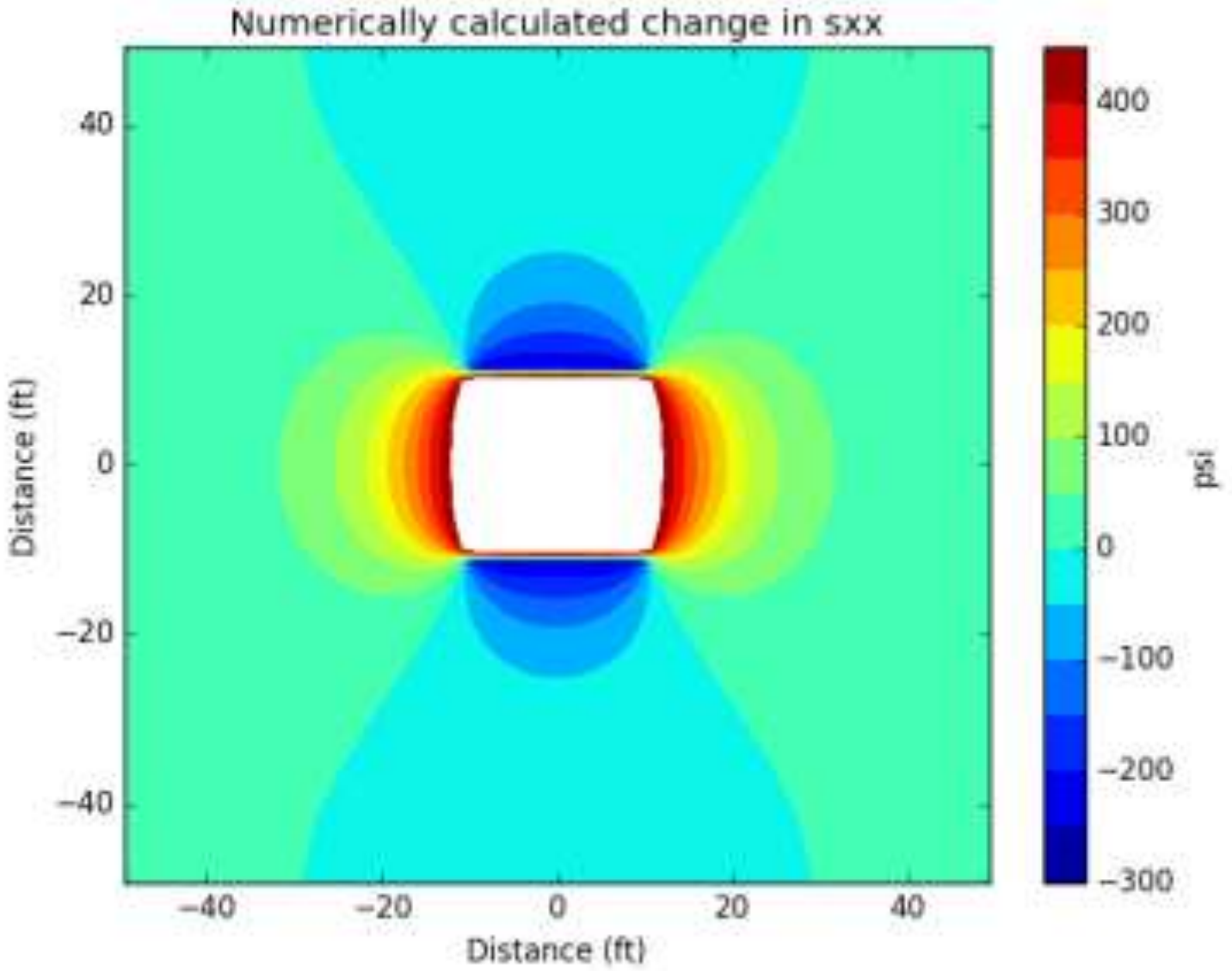

Figure 50: Numerically calculated change in $\sigma_{xx}$ due to a uniform change in pressure within a cube of rock embedded in an infinite domain.

9. Norne Comparison Problem

The Norne comparison problem is a corner point simulation with water injection. The Norne benchmark case is a black oil model for an oil field in the Norwegian Sea. The grid is a faulted corner point grid, with heterogenous and anisotropic permeability. Features used include dissolved gas, vaporized oil, transmissibility multipliers, and pressure-dependent porosity, the capillary pressure in the original model was removed because capillary pressure is not currently supported in ResFrac. The model is made openly available by The Open Porous Media Initiative (2021).

The Norne comparison problem has been used to validate the ResFrac implementation of corner point gridding. The Norne simulation was modified in order to remove features that are not yet supported by ResFrac, such as non-physical well paths. The results from ResFrac were compared with the results from the opensource reservoir simulator, Flow (The Open Porous Media Initiative, 2021). The results are compared in Figure *52* and show a very good match between the two different simulators.

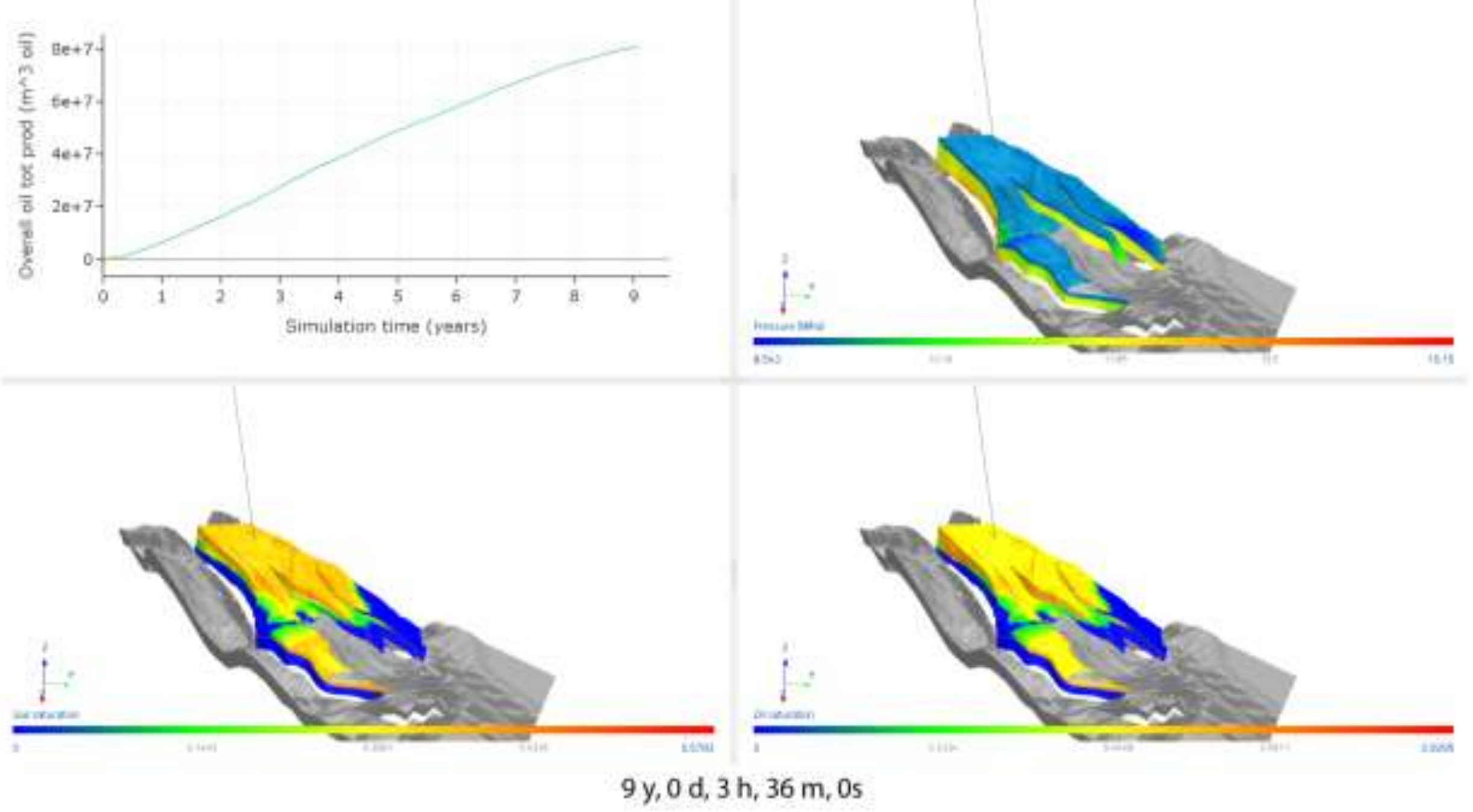

Figure 51: Calculated oil production rate and the final distribution of pressure, gas saturation, and oil saturation from the Norne comparison problem.

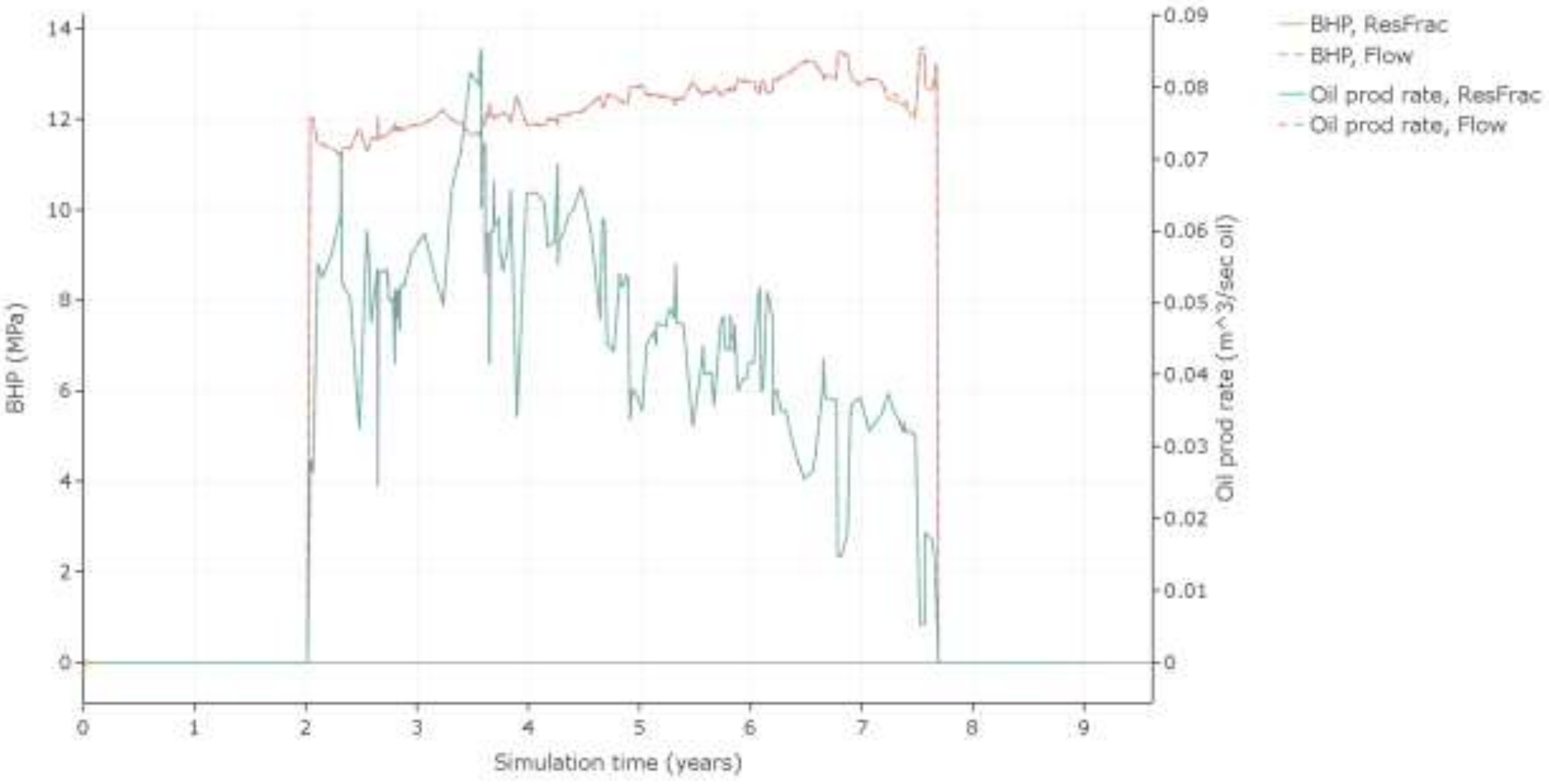

Figure 52: Comparison between flow and the ResFrac simulation of the Norne benchmark case displaying the oil production rate and the bottom hole pressure from Well_E2H. The purple dashed line coincides with the flow simulated data, the orange solid line refers to the well head pressure of the ResFrac simulation, and the green line the ResFrac oil production rate.

## 10. Corner point to Rectilinear Comparison Problem

As a simple validation for the corner point grid capability, we set up a corner point grid simulation to have an identical mesh as an existing simulation built with the rectilinear capability. While the simulations are identical, the grids are specified differently and handled differently internally to the simulator. The simulations were compared to confirm consistency.

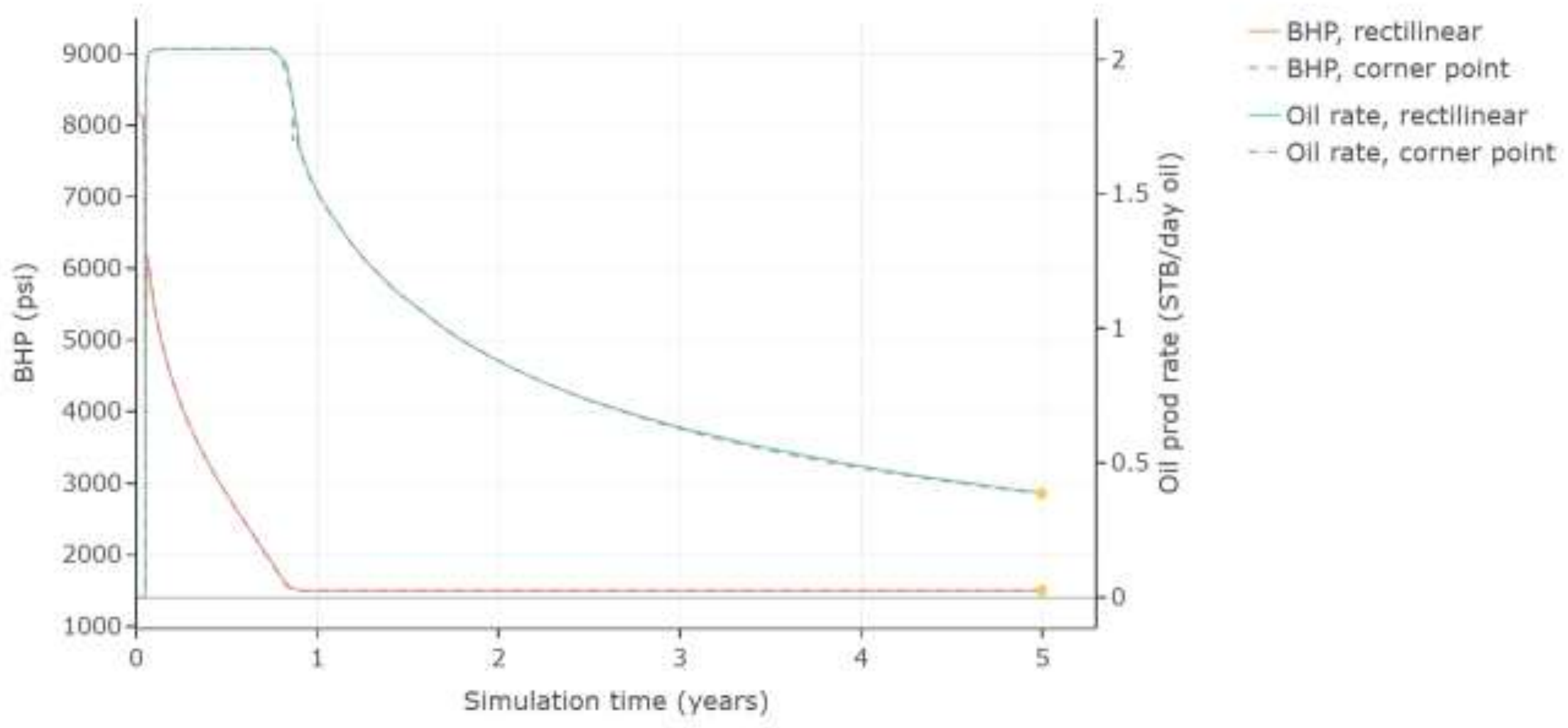

Figure 53: Comparison of rectilinear simulation using a corner point mesh implementation of rectilinear and a true rectilinear.

11. Leakoff from a single hydraulic fracture

Carter's leakoff law analytically models the volume loss of fracturing fluids into a porous formation, where fluid velocity is inversely proportional to the square root of time. It utilizes a leakoff coefficient to measure the rate of fluid loss, commonly used to calculate fluid consumption, filter cake buildup, and fracture volume efficiency. It is used to validate the fluid flow from fracture to matrix element in ResFrac with a series of simulations featuring a single, confining homogeneous layer containing a single hydraulic fracture perforated at the center. Validation simulations include a pure water injection case without any filtercake formation, a “pure” filtercake case where filtercake forms as soon as leakoff starts and reservoir has an effectively infinite permeability, and a slickwater injection case where leakoff is under combined influence of filtercake and reservoir. Simulation results show that the numerically calculated flow from fracture to matrix element aligns closely with the leakoff calculated by Carter’s leakoff law. In the subsequent figures, numerical cumulative leakoff volume per area from ResFrac is plotted in red, and that from Carter’s leakoff law is plotted in blue. The y-axis is cumulative leakoff volume per area, and x-axis is timestep number.

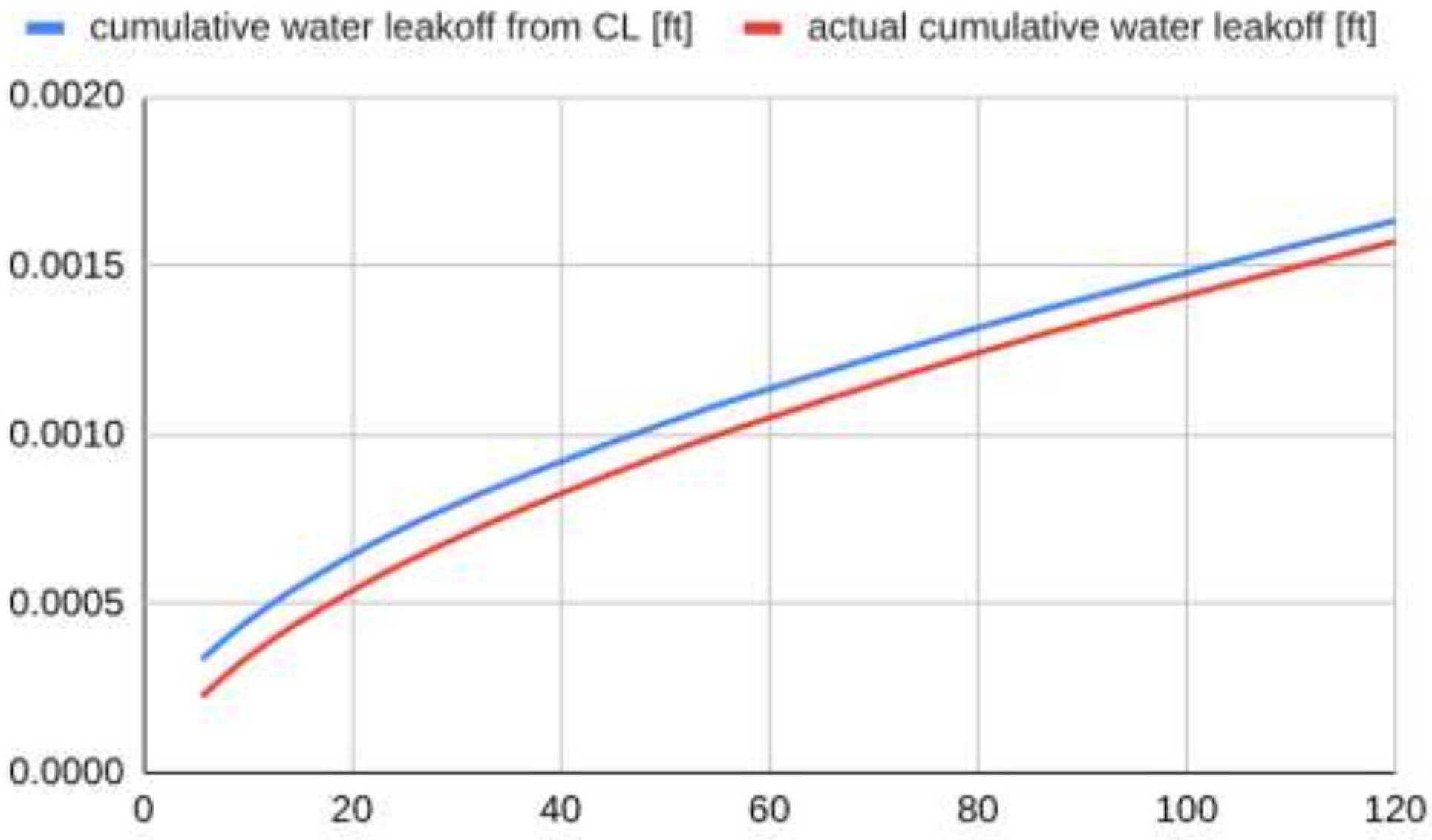

Figure 54: Cumulative leakoff volume per area versus timestep in the pure water case.

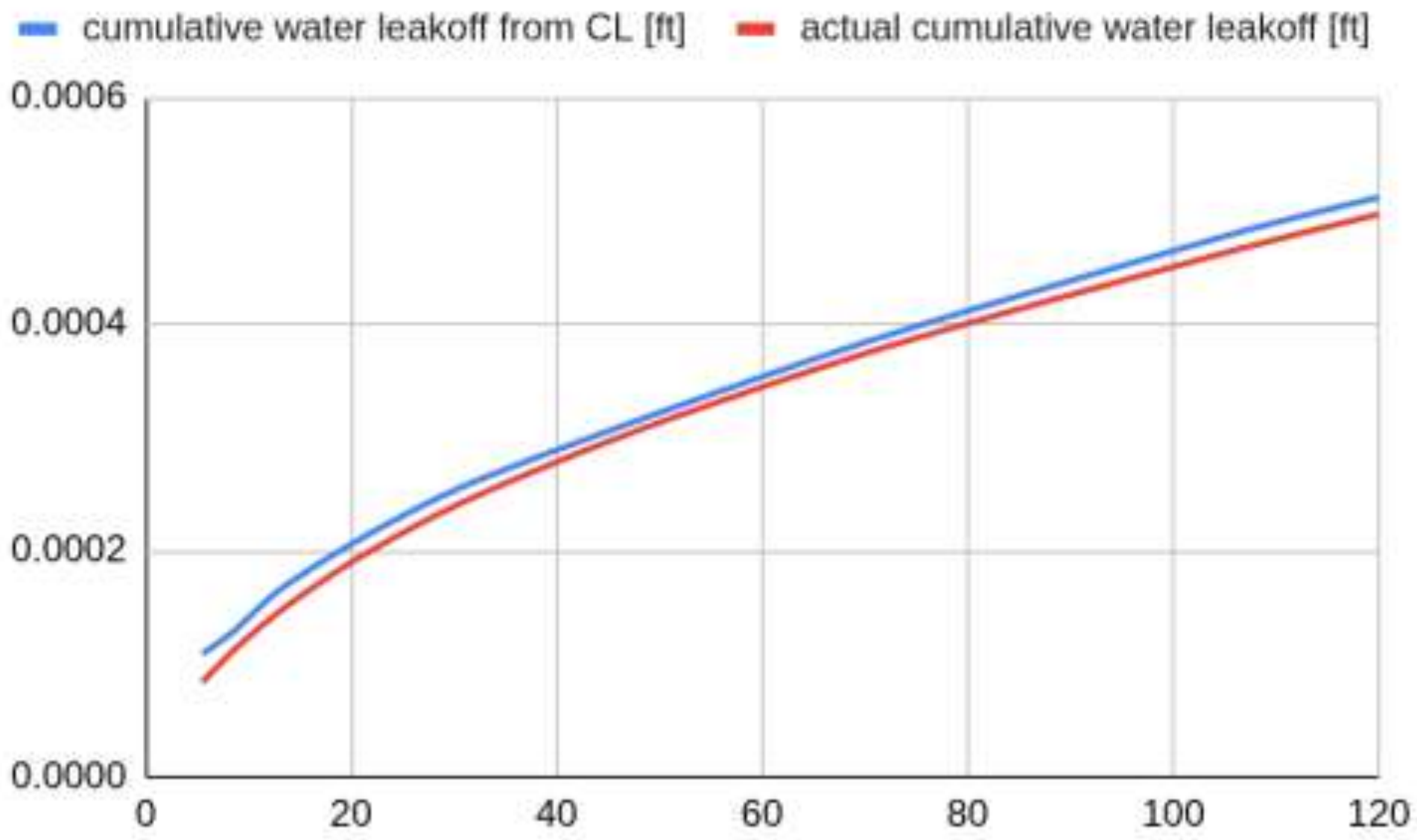

Figure 55: Cumulative leakoff volume per area versus timestep in the “pure filtercake leakoff” case.

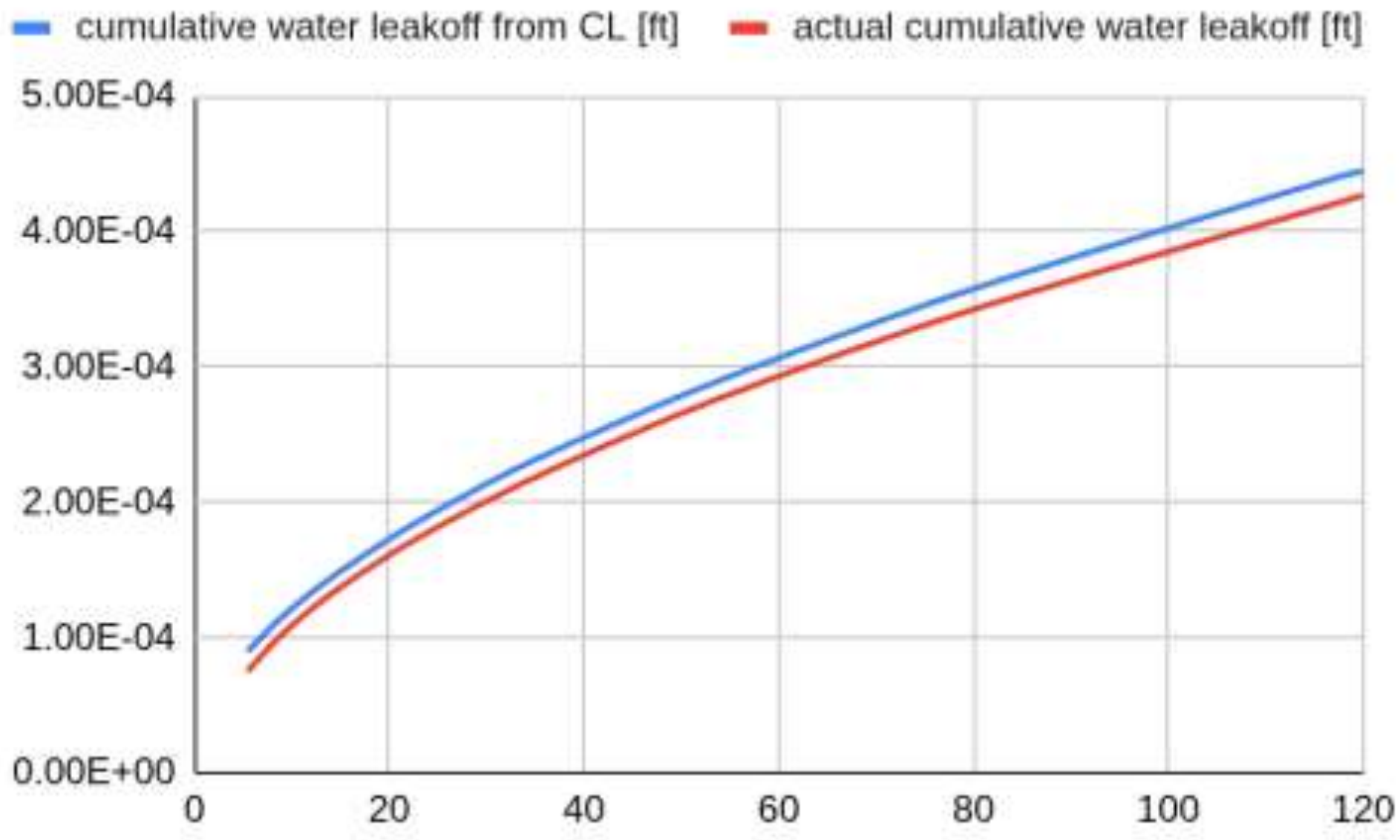

Figure 56: Cumulative leakoff volume per area versus timestep in the slickwater injection case.

# 23. Log interpretation equations in the UI –Properties versus depth Wizard

## 23.1 Rock solid matrix density

The rock-solid matrix density $\rho_{grain}$, or grain density, is the density of the solid mineral matrix in the rock, excluding any pores. This can be computed from the dry rock bulk density $\rho_{dry\ bulk}$ (i.e. the mass of a dry porous rock sample divided by its total volume, including the pore space) as

$$\rho_{grain} = \frac{\rho_{dry\ bulk}}{1-n}, \qquad \text{23-1}$$

where $n$ is the porosity. Alternatively, it can be estimated from the rock bulk density $\rho_{bulk}$ (i.e. the mass of a porous rock sample divided by its total volume, including the pore space and the mass of any fluids filling the pores) as

$$\rho_{grain} = \frac{\rho_{bulk} - n\,\rho_f}{1-n}, \qquad \text{23-2}$$

where $\rho_f$ is the density of the fluid that fills the pores, and $n$ is the rock porosity.

Parameters needed:
- rock porosity [unitless],
- dry rock bulk density [kg/m^3]
OR
- rock bulk and fluid densities [kg/m^3]

## 23.2 Vertical stress

This method is not exposed to the user. It is used only internally by many other methods.
The vertical total stress $\sigma_v$, at depth $D$, can be computed from log data as

$$\sigma_v(D) = \sigma_{v0} + \int_{D_0}^{D} \rho_{bulk}(z) g \, dz, \qquad \text{23-3}$$

where $\sigma_{v0}$ is the vertical total stress at depth $D_0$, $\rho_{bulk}$ is the rock bulk density as a function of depth, $z$, and $g = 9.80665\ m/s^2$ is the standard acceleration of gravity.

Parameters needed:
- rock bulk density as a function of depth [m],
- vertical total stress at the shallowest depth where formation bulk density is provided [MPa].

## 23.3 Isotropic Poisson's ratio

From the equations of elastodynamic, the dynamic Poisson's ratio $\nu_{dyn}$ (Fjær, Holt, Horsrud, & Raaen, 2008),

$$\nu_{dyn} = \frac{0.5\left(V_p/V_s\right)^2 - 1}{\left(V_p/V_s\right)^2 - 1}, \qquad \text{23-4}$$

with:

$$V_p = 1/\Delta t_c, \qquad \text{23-5}$$
$$V_s = 1/\Delta t_s, \qquad \text{23-6}$$

$\Delta t_c$, and $\Delta t_s$ are respectively the 'compressional slowness', and the 'shear slowness' (Clark & Weber, 1967). Dynamic and static values of Poisson's ratio are assumed to coincide.

Parameters needed:
- p-waves velocity [m/s] as a function of depth [m],
- s-waves velocity [m/s] as a function of depth [m],

OR
- compressional slowness [s] as a function of depth [m],
- shear slowness [s] as a function of depth [m].

## 23.4 Isotropic Young's modulus

From the elastodynamic equations, the Shear Modulus $G_{dyn}$ can be computed as (Fjær, Holt, Horsrud, & Raaen, 2008)

$$G_{dyn} = \rho_{dry\ bulk} V_s^2, \quad \text{23-7}$$

the dynamic Youngs's modulus $E_{dyn}$ is then computed as

$$E_{dyn} = 2G_{dyn}(1 + \nu_{dyn}), \quad \text{23-8}$$

and $\rho_{dry\ bulk}$ is the dry rock bulk density. The static value of the Young's modulus $E_{st}$ can be obtained from empirical correlations such as (Brotons, et al., 2016)

$$E_{static} = 11.531\ \rho_{dry\ bulk}^{-0.457} E_{dyn}^{1.251}, \quad \text{23-9}$$

or alternatively

$$E_{static} = 3.97\ 10^6\ \rho_{dry\ bulk}^{-2.090} E_{dyn}^{1.287} n^{-0.116}, \quad \text{23-10}$$

where $E_{static}$ is the static Young's modulus in [GPa], $E_{dyn}$ is the dynamic modulus expressed in [GPa], $\rho_{dry\ bulk}$ is the dry rock bulk density [kg/m^3], $n$ is the total porosity [%]. Both correlations (23-9 – *one parameter relation*) and (23-10 – *two parameter relations*) are available in ResFrac.

Parameters needed:
- dry rock bulk density [kg/m^3] as a function of depth [m],
- p-waves velocity [m/s] as a function of depth [m],
- s-waves velocity [m/s] as a function of depth [m],
-Optional, total porosity [%] as a function of depth [m],
OR
- Instead of p-waves velocity, it could use the compressional slowness [s].
- Instead of s-waves velocity, it could use the shear slowness [s].
- Instead of p-waves and s-waves velocities, it could use the dynamic shear modulus [GPa], and the Poisson's ratio [unitless].
- Instead of p-waves and s-waves velocities, it could use the dynamic Young's modulus [GPa].

## 23.5 Minimum horizontal stress / frac gradient

### 23.5.1 Matthews and Kelly's formula, or "Effective stress and K0 coefficient"

At a given depth $z$, the minimum horizontal stress $\sigma_{h\ min}$ can be estimated by summing the pore pressure $p_p$ with the minimum horizontal effective stress $K_0(\sigma_v - p_p)$

$$\sigma_{h\ min} = K_0(z) \times (\sigma_v - p_p) + p_p. \quad \text{23-11}$$

The minimum horizontal stress is obtained from the overburden stress $\sigma_v$, and the pore pressure, by multiplying the vertical effective stress $(\sigma_v - p_p)$ by a coefficient $K_0(z)$ varying with depth $z$ (Matthews & Kelly, 1967).

Parameters needed:
- vertical total stress [MPa] as a function of depth [m],
- horizontal over vertical stress coefficient [unitless].
- pore pressure [MPa].
OR
- Instead of pore pressure, see parameters needed for calculating it.
- Instead of vertical total stress, see parameters needed for calculating it.

### 23.5.2 Eaton's formula, or "Effective stress and Poisson's ratio"

For a linear, elastic, and homogeneous material vertically loaded with stress $(\sigma_v - p_p)$, and that cannot expand laterally, the value $K_0 = \nu/(1-\nu)$ (Eaton, 1969).

$$\sigma_{h\,min} = \frac{\nu}{1-\nu}(\sigma_v - p_p) + p_p. \quad \text{23-12}$$

Parameters needed:
- vertical total stress [MPa].
- Poisson's ratio [unitless].
- pore pressure [MPa].
OR
- Instead of pore pressure, see parameters needed for calculating it.
- Instead of Poisson's ratio, see parameters needed for calculating it.
- Instead of vertical total stress, see parameters needed for calculating it.

### 23.5.3 Hard rock formula

An estimate for the minimum horizontal stress $\sigma_{h\,min}$ is derived by solving the linear poroelasticity equations for a homogeneous isotropic material, with vertical stress set equal to the overburden, and horizontal strains set to zero. In this configuration, the only deformation occurs along the vertical direction, and the minimum horizontal stress $\sigma_{h\,min}$ is (Andreson, Ingram, & Zanier, 1973)

$$\sigma_{h\,min} = \frac{\nu}{1-\nu}(\sigma_v - \alpha\, p_p) + \alpha\, p_p, \qquad (not\ used) \quad \text{23-13}$$

where $\alpha$ is the Biot coefficient, $\sigma_v$ is the overburden stress, $p_p$ is the pore pressure, $\nu$ is Poisson's ratio. Because the linear isotropic poroelastic model did not seem to adequately describe existing stress conditions in hard-rock areas, a different model (Newberry, Nelson, & Ahmed, 1985)

$$\sigma_{h\,min} = \frac{\nu}{1-\nu}\left(\sigma_v - \alpha\, p_p\right) + \ p_p, \qquad (the\ actual\ 'Hard\ Rock\ formula') \qquad 23\text{-}14$$

was derived by setting the Biot's coefficient $\alpha = 1$ in the direction of the least (horizontal) principal stress. This model accounts for the presence of microcracks that are supposed to be mostly aligned to the least principal stress.

Parameters needed:
- Poisson's ratio [unitless].
- vertical total stress [MPa].
- Biot's coefficient [unitless].
- pore pressure [MPa].
OR
- Instead of Poisson's ratio, see parameters needed for calculating it.
- Instead of vertical total stress, see parameters needed for calculating it.
- Instead of pore pressure, see parameters needed for calculating it.

### 23.5.4 Extended Eaton Model from Blanton & Olson 1999

The "Blanton & Holson 1999" method accounts for the tectonic stress and temperature gradient (Blanton & Olson, 1999). The equations for the horizontal stress in $x$ and $y$ directions

$$\sigma_{h\,x} = \frac{\upsilon}{1-\upsilon}\left(\sigma_V - \alpha\, p_p\right) + \alpha\, p_p + \frac{E}{1-\upsilon^2}\left(\varepsilon_x + \nu\,\varepsilon_y\right) + \frac{E\,\alpha_T}{1-\upsilon}\Delta T, \qquad 23\text{-}15a$$

$$\sigma_{h\,y} = \frac{\upsilon}{1-\upsilon}\left(\sigma_V - \alpha\, p_p\right) + \alpha\, p_p + \frac{E}{1-\upsilon^2}\left(\varepsilon_y + \nu\varepsilon_x\right) + \frac{E\,\alpha_T}{1-\upsilon}\Delta T, \qquad 23\text{-}15b$$

are simplified by assuming that one of the two tectonic strains $\varepsilon_x$, $\varepsilon_y$ is zero while the other is set equal to $\epsilon_{tect}$ (plane strain assumption). This implies that there are two alternative expressions for $\sigma_{h\,min}$ depending on the actual sign of $\epsilon_{tect}$:

$$\begin{aligned} \sigma_{h\,min} &= C_1\epsilon_{tect} + C_2 \qquad if\ \ \epsilon_{tect} < 0 \quad extensional\ tectonic\ strain, \\ \sigma_{h\,min} &= \nu C_1\epsilon_{tect} + C_2 \quad\ if\ \ \epsilon_{tect} > 0 \quad compressional\ tectonic\ strain. \end{aligned} \qquad 23\text{-}16$$

where the parameters $C_1$, $C_2$ are defined as

$$C_1 = \frac{E}{1-\nu^2}, \qquad 23\text{-}17$$

$$C_2 = \frac{\nu\sigma_v + (1-2\nu)\alpha\, p_p + E\alpha_T\Delta T}{1-\nu}, \qquad 23\text{-}18$$

where $\alpha_T$ is the thermal coefficient of expansion, $\Delta T = T - T_{surface}$ is the difference between the temperature at a given depth $T$, and the surface temperature $T_{surface}$. The value of the tectonic strain $\epsilon_{tect}$ can be obtained from direct measurements of $\sigma_{h\,min}$ as

$$\epsilon_{tect} = \frac{\sigma^*_{h\,min} - C^*_2}{C^*_1} \qquad if\ \ \epsilon_{tect} < 0\ \ \ extensional\ tectonic\ strain, \qquad 23\text{-}19a$$

$$\epsilon_{tect} = \frac{\sigma^*_{h\,min} - C^*_2}{\nu C^*_1} \qquad if\ \ \epsilon_{tect} > 0\ \ \ compressional\ tectonic\ strain, \qquad 23\text{-}19b$$

where asterisks indicate that these terms are associated with the depth at which the minimum horizontal stress has been determined. $\sigma^*_{h\,min}$ is the measured (or estimated) value of the minimum horizontal stress. Note that, in case the Poisson's ratio $\nu < 0$, it can be unclear which formula (Eq. 23-19a or Eq. 23-19b) is to be considered. This is unlikely to happen as rocks do not usually exhibit negative values for Poisson's ratio. However, if this is the case, ResFrac sets $\epsilon_{tect} = 0$, and returns a warning to the user.
In case the user specifies N>1 values for $\sigma^*_{h\,min}$, the value of $\epsilon_{tect}$ is obtained by minimizing the squared error $\mathrm{E}(\epsilon_{tect})$

$$\mathrm{E}(\epsilon_{tect}) = \sum_{i=1}^{=N} (C_{1i}\epsilon_{tect} + C_{2i} - \sigma^*_{h\,\min i})^2 \qquad if\ \ \epsilon_{tect} < 0\ \ \ extensional\ tectonic\ strain,$$

or,

$$\mathrm{E}(\epsilon_{tect}) = \sum_{i=1}^{=N} (\nu C_{1i}\epsilon_{tect} + C_{2i} - \sigma^*_{h\,\min i})^2 \qquad if\ \ \epsilon_{tect} > 0\ \ compressional\ tectonic\ strain,$$

which can also be expressed as

$$\mathrm{E}(\epsilon_{tect}) = A\epsilon^2_{tect} + B\epsilon_{tect} + C \qquad 23\text{-}20$$

in which

$$\mathrm{A}(\epsilon_{tect}) = \begin{cases} \mathrm{A}^- = \sum_{i=1}^{=N} C^2_{1i} & if\ \epsilon_{tect} < 0 \\ \mathrm{A}^+ = \sum_{i=1}^{=N} \nu^2_i C^2_{1i} & if\ \epsilon_{tect} > 0 \end{cases}$$

$$\mathrm{B}(\epsilon_{tect}) = \begin{cases} \mathrm{B}^- = 2\sum_{i=1}^{=N} [C_{1i}(C_{2i} - \sigma^*_{h\,\min\ i})] & if\ \epsilon_{tect} < 0 \\ \mathrm{B}^+ = 2\sum_{i=1}^{=N} [\nu_i C_{1i}(C_{2i} - \sigma^*_{h\,\min\ i})] & if\ \epsilon_{tect} > 0 \end{cases}$$

$$C = \sum_{i=1}^{=N} (C_{2i} - \sigma^{*}_{h\,\min\,i})^2.$$

Because the A coefficient in Eq. 23-20 is always positive, the equation admits always a value $\epsilon_{tect}$

$$\epsilon_{tect} = -\frac{\mathrm{B}}{2\mathrm{A}} \qquad \text{23-21}$$

for which the squared error $\mathrm{E}(\epsilon_{tect})$ is minimum. However, because the A and B values depend on the actual value of $\epsilon_{tect}$, it is possible to estimate two values

$$\epsilon^{+}_{tect} = -\frac{\mathrm{B}^{+}}{2\mathrm{A}^{+}}, \qquad and \qquad \epsilon^{-}_{tect} = -\frac{\mathrm{B}^{-}}{2\mathrm{A}^{-}}$$

The final value for $\epsilon_{tect}$ is to be decided on the sign of the estimates for $\epsilon^{-}_{tect}$ and $\epsilon^{+}_{tect}$, as well as the values $\mathrm{E}(\epsilon^{-}_{tect})$ and $\mathrm{E}(\epsilon^{+}_{tect})$ of the squared error. The results are summarized in the following table 23.1:

| | $\epsilon^{-}_{tect} > 0$ | $\epsilon^{-}_{tect} < 0$ | $\epsilon^{-}_{tect} = 0$ |
|---|---|---|---|
| $\epsilon^{+}_{tect} > 0$ | $\epsilon_{tect} = \epsilon^{+}_{tect}$ | $\epsilon_{tect} = \min[\mathrm{E}(\epsilon^{+}_{tect}), \mathrm{E}(\epsilon^{-}_{tect})]$ | $\epsilon_{tect} = \min[\mathrm{E}(\epsilon^{+}_{tect}), \mathrm{E}(\epsilon^{-}_{tect})]$ |
| $\epsilon^{+}_{tect} < 0$ | $\epsilon_{tect} = 0$ $= \min[\mathrm{E}(\epsilon^{+}_{tect}), \mathrm{E}(\epsilon^{-}_{tect})]$ | $\epsilon_{tect} = \epsilon^{-}_{tect}$ | $\epsilon_{tect} = \min[\mathrm{E}(\epsilon^{+}_{tect}), \mathrm{E}(\epsilon^{-}_{tect})]$ |
| $\epsilon^{+}_{tect} = 0$ | $\epsilon_{tect} = \min[\mathrm{E}(\epsilon^{+}_{tect}), \mathrm{E}(\epsilon^{-}_{tect})]$ | $\epsilon_{tect} = \min[\mathrm{E}(\epsilon^{+}_{tect}), \mathrm{E}(\epsilon^{-}_{tect})]$ | $\epsilon_{tect} = 0$ |

Table 23.1. Optimal value for the tectonic strain $\epsilon_{tect}$.

Parameters needed:
- Poisson's ratio [unitless]
- vertical total stress [MPa]
- Biot's coefficient [unitless]
- pore pressure [MPa]
- Young's modulus [MPa]
- temperature at a given depth minus surface temperature [℃]
- thermal expansion coefficient [$℃^{-1}$]
- known values of minimum horizontal stress [MPa] at specific depths [m].
OR
- Instead of Poisson's ratio, see parameters needed for calculating it.
- Instead of vertical total stress, see parameters needed for calculating it.
- Instead of Young's modulus, see parameters needed for calculating it.
- Instead of pore pressure, see parameters needed for calculating it.

### 23.5.5 Non-isotropic extended Eaton's method

The equations for the horizontal stress in $x$ and $y$ directions considered in the "Blanton & Holson 1999" method

$$\sigma_{h\,x} = \frac{\upsilon}{1-\upsilon}\left(\sigma_V - \alpha\, p_p\right) + \alpha\, p_p + \frac{E}{1-\upsilon^2}\left(\varepsilon_x + \nu\,\varepsilon_y\right) + \frac{E\,\alpha_T}{1-\upsilon}\Delta T \quad \text{23-15a}$$

$$\sigma_{h\,y} = \frac{\upsilon}{1-\upsilon}\left(\sigma_V - \alpha\, p_p\right) + \alpha\, p_p + \frac{E}{1-\upsilon^2}\left(\varepsilon_y + \nu\varepsilon_x\right) + \frac{E\,\alpha_T}{1-\upsilon}\Delta T \quad \text{23-15b}$$

can be extended to the case of a Vertical Transverse Isotropic (VTI) rock (e.g. Zhang et al 2023)

$$\sigma_{x\,VTI} = \frac{E_h \upsilon_V}{E_V(1-\upsilon_h)}\left(\sigma_V - \alpha_V\, p_p\right) + \alpha_h\, p_p + \frac{E_h}{1-\upsilon_h^2}\left(\varepsilon_x + \nu_h \varepsilon_y\right) + \frac{E_h \alpha_{Th}}{1-\upsilon_h}\Delta T \quad \text{23-22a}$$

$$\sigma_{y\,VTI} = \frac{E_h \upsilon_V}{E_V(1-\upsilon_h)}\left(\sigma_V - \alpha_V\, p_p\right) + \alpha_h\, p_p + \frac{E_h}{1-\upsilon_h^2}\left(\varepsilon_y + \nu_h \varepsilon_x\right) + \frac{E_h \alpha_{Th}}{1-\upsilon_h}\Delta T \quad \text{23-22b}$$

where $E_V$ and $\upsilon_V$ are Young's modulus and Poisson's ratio in the vertical directions, respectively; $E_h$ and $\upsilon_h$ are Young's modulus and Poisson's ratio in the horizontal directions, respectively; $\alpha_h$ and $\alpha_V$ are Biot's coefficients in the horizontal and vertical directions, respectively; $\alpha_{Th}$ is the thermal expansion coefficient in the horizontal direction. The 4 elastic constants ($E_V$, $E_h$, $\upsilon_V$, $\upsilon_h$) required for the method can be expressed using the elements of the anisotropic stiffness tensor (in Voigt notation)

$$\begin{pmatrix} \sigma_1 \\ \sigma_2 \\ \sigma_3 \\ \sigma_4 \\ \sigma_5 \\ \sigma_6 \end{pmatrix} = \begin{pmatrix} C_{11} & C_{12} & C_{13} & 0 & 0 & 0 \\ C_{12} & C_{11} & C_{13} & 0 & 0 & 0 \\ C_{13} & C_{13} & C_{33} & 0 & 0 & 0 \\ 0 & 0 & 0 & C_{44} & 0 & 0 \\ 0 & 0 & 0 & 0 & C_{44} & 0 \\ 0 & 0 & 0 & 0 & 0 & C_{66} \end{pmatrix} \begin{pmatrix} \varepsilon_1 \\ \varepsilon_2 \\ \varepsilon_3 \\ \varepsilon_4 \\ \varepsilon_5 \\ \varepsilon_6 \end{pmatrix} \quad \text{23-23}$$

in which the axis of symmetry coincides with $x_3$. The Young's moduli and Poisson's ratios are

$$E_v = C_{33} - \frac{C_{13}^2}{C_{11} - C_{66}}, \qquad v_v = \frac{C_{13}}{2(C_{11} - C_{66})}$$

$$E_h = C_{11} + \frac{C_{13}^2(C_{11} - 4C_{66}) - C_{33}(C_{11} - 2C_{66})^2}{C_{33}C_{11} - C_{13}^2}, \qquad v_h = \frac{C_{33}(C_{11} - 2C_{66}) - C_{13}^2}{C_{33}C_{11} - C_{13}^2}$$

Equations 23-22a, 23-22b are then simplified by assuming that one of the two tectonic strains $\varepsilon_x$, $\varepsilon_y$ is zero while the other is set equal to $\epsilon_{tect}$ (plane strain assumption). The other steps mirror what's done for the Blanton & Olson method.

Parameters needed:
- 2 Poisson's ratios [unitless]
- vertical total stress [MPa]
- 2 Biot's coefficients [unitless]
- pore pressure [MPa]
- 2 Young's moduli [MPa]
- temperature at a given depth minus surface temperature [℃]
- thermal expansion coefficient [$℃^{-1}$]
- known values of minimum horizontal stress [MPa] at specific depths [m].

OR
- Instead of 2 Poisson's ratios and 2 Young's moduli, could use C11, C13, C33, C66.
- Instead of vertical total stress, see parameters needed for calculating it.
- Instead of pore pressure, see parameters needed for calculating it.

### 23.5.6 Viscoplastic Stress Relaxation

The "Hard-Rock formula", and the "Constant-strain" formulas are variations of the well-known Eaton equation (Eaton, 1969). The main limitation for these equations is that they are derived assuming linear elastic, and time independent rock behavior. A time dependent viscoplastic relation can be used to estimate the minimum horizontal stress (Sone & Zoback, 2014)

$$\sigma_{h\,min,\ VSR} = \sigma_v - E_{horiz}\frac{t^{-n}}{(1-n)}\frac{\varepsilon_0}{k}, \tag{23-24}$$

and it proved to be accurate in normal or normal/strike-slip faulting environments. In this relation:
- $E_{horiz}$ is the horizontal Young's modulus.
- $n$ is a coefficient calculated as a function of $E_{horiz}$ using the linear equation (Ma & Zoback, 2020),

$$n = \frac{70 - E_{horiz}}{700}, \qquad with\ E_{horiz}\ [GPa], \qquad 23\text{-}25$$

or, alternatively, using the power law relation (Mandal, Saraut, Rezaee, & Finkbeiner, 2022), (Pratim, Sarout, & Rezaee, 2023):

$$n = 0.0934 \frac{1}{E_{horiz}^{0.334}}, \qquad with\ E_{horiz}\ [GPa]. \qquad 23\text{-}26$$

These relations (equations 23-25, and 23-26) are both empirical. They are computed using a database of $E_{horiz}$ and $n$ values computed in laboratory creep experiments using a varied lithology across US (linear relation - Eq. 23-25), and Australian (power law - Eq. 23-26) shale reservoirs.

- $t$ is the time scale at which stress relaxation has occurred, in other words it is the age of the formation. This can be estimated as the time since the last major tectonic uplift occurred. Practically, due to the power law characteristics the sensitivity of the result to time is insignificant beyond a few million years, i.e., the estimates assuming 10 million years or 20 million years will be nearly the same, (see Sone & Zoback (2014) for a discussion).

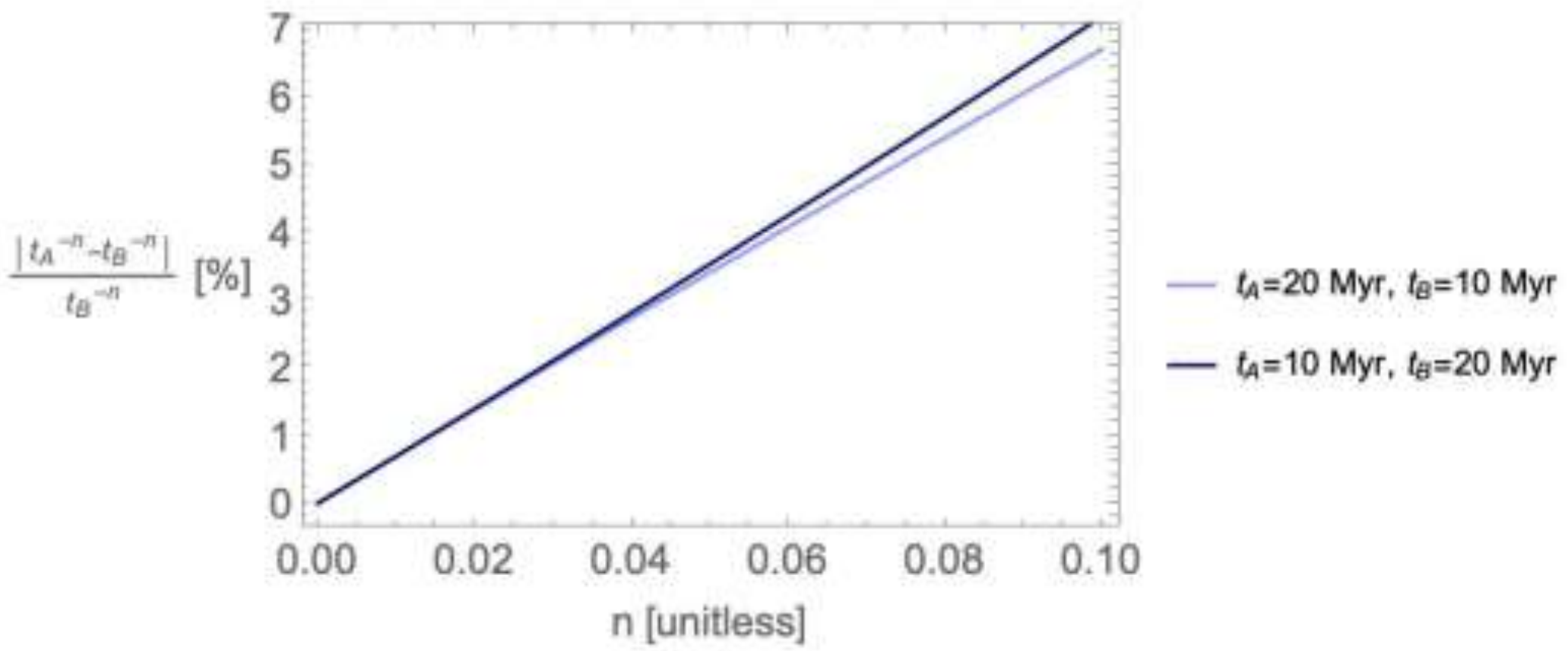

Figure 1. *Relative difference of the factor* $t^{-n}$ *considering* $t_A = 20 \cdot 10^6$ *years,* $t_B = 10 \cdot 10^6$ *years, and vice versa.*

- $\varepsilon_0/k$ is a parameter to be tuned by fitting Eq. 23-24 at specific depths for which $\sigma_{h\,min}$ is known from e.g., DFIT. The calibrated values of $\varepsilon_0/k$ are assumed to be the same throughout the depths of interest. In case the user specifies N>1 values of $\sigma_{h\,min}$, the value of $\varepsilon_0/k$ is obtained by minimizing the squared error $\mathrm{E}(\varepsilon_0/k)$

$$\mathrm{E}(\varepsilon_0/k) = \sum_{i=1}^{=N} \left( C_{1i} \frac{\varepsilon_0}{k} + C_{2i} - \sigma_{h\,\min i}^{*} \right)^2, \qquad 23\text{-}27$$

where the coefficients $C_{1i}$, and $C_{2i}$are

$$C_{1i} = -E_{horiz\,i} \frac{t^{-n_i}}{(1 - n_i)}, \qquad C_{2i} = \sigma_{v\,i}$$

and the minimization is performed by following the same procedure as used for Eq. 23-20.

In the presence of a normal faulting environment, the minimum horizontal stress computed by the viscoplastic stress relaxation method $\sigma_{h\,min,\ VSR}$ (Eq. 23-24) cannot be less than the one predicted by frictional equilibrium $\sigma_{h\,min,\ FE}$

$$\sigma_{h\,min,\ FE} = \frac{\sigma_v - p_p}{\left[\sqrt{\mu^2+1}+\mu\right]^2} + p_p \qquad \leq \sigma_{h\,min,\ VSR}\,, \tag{23-28}$$

where $\mu$ is the frictional coefficient. This coefficient varies between $0.4 - 0.8$ for all major unconventional reservoir formations, and 0.6 is usually a good value to assume (Zoback & Kohli, 2019). For less clay rich rocks, a range from $0.6 - 1.0$ may be more applicable.

Parameters needed:
- vertical total stress [MPa]
- horizontal Young's modulus [MPa]
- age of the formation [years]
- known values of minimum horizontal stress [MPa] at specific depths [m]
- frictional coefficient [unitless]
- optional, pore pressure [MPa].
OR
- Instead of vertical total stress, see parameters needed for calculating it.

## 23.6 Pore pressure / pore pressure gradient

### 23.6.1 Pore pressure gradient from resistivity measurements

The pore pressure gradient $p_p\,/z$, at depth $z$ can be estimated as (Eaton (1972), Eaton (1975))

$$\frac{p_p}{z} = \frac{\sigma_v}{z} - \left(\frac{\sigma_v}{z} - p_{ng}\right)\left(\frac{R}{R_n\,e^{bz}}\right)^n, \tag{23-29}$$

where $p_{ng}$ is the hydrostatic pore pressure gradient (normally 1.03 MPa/km, dependent on water salinity); $R$ is the resistivity obtained from well logging; $R_n$ is the resistivity at hydrostatic pressure; $n$ is the exponent varied from 0.6 to 1.5 (Zhang, 2011); $z$ is the depth; $b$ is the slope of logarithmic resistivity normal compaction trendline.

Parameters needed:
- vertical total stress gradient [MPa/km]
- hydrostatic pore pressure gradient [MPa/km]
- resistivity obtained from well logging
- resistivity at hydrostatic pressure
- constant $n$
- constant $b$ [1/m]
- depth [m]

### 23.6.2 Pore pressure gradient from transit time

The pore pressure gradient $p_p\,/z$ can be also estimated using the empirical formula (Eaton, 1975)

$$\frac{p_p}{z} = \frac{\sigma_v}{z} - \left(\frac{\sigma_v}{z} - p_{ng}\right)\left(\frac{\Delta t_n}{\Delta t_c}\right)^3, \qquad \text{23-30}$$

where the compressional slowness at hydrostatic pressure, $\Delta t_n$ , is a function of depth $z$, and it can be expressed using the general formula (Zhang, 2011)

$$\Delta t_n = \Delta t_m + (\Delta t_{ml} - \Delta t_m)e^{-c\,z}, \qquad \text{23-31}$$

in which c, $\Delta t_m$ , and $\Delta t_{ml}$ are constant. In particular, $\Delta t_m$ is the compressional slowness in the matrix (with zero porosity), and $\Delta t_{ml}$ is the mudline compressional slowness. See Zhang, (2011) for more details on how to determine these constants.

Parameters needed:
- vertical total stress gradient [MPa/km]
- hydrostatic pore pressure gradient [MPa/km]
- constants $\Delta t_m$ [s/m], $\Delta t_{ml}$ [s/m], $c$ [1/m]
- p-waves slowness [s/m]

## 23.7 Vertical permeability

Vertical permeability $k_v$ is calculated from horizontal permeability $k_h$ via the anisotropy factor ω as

Parameters needed:
- horizontal permeability [m^2]
- anisotropy permeability factor [unitless].

$$k_v = \omega\, k_h\,. \tag{23-32}$$

## List of variables

$A_{nw}$: coefficient for the near wellbore complexity pressure drop correlation (Pa/(m^3/s)^$\alpha_{nw}$; psi/bpm^$\alpha_{nw}$)
$A_{leakoff}$: leakoff area ($m^2$; $ft^2$)
$b$: PDL leakoff coefficient exponent ($MPa^{-1}$; $psi^{-1}$)
$c_{acc}$: actual total mass fraction of water solutes in the same pipe friction table (dimensionless)
$c_{min}$: minimum mass fraction recorded in a pipe friction table (dimensionless)
$C_P$: PDL leakoff coefficient multiplier (dimensionless)
$C_{P,max}$: PDL maximum leakoff coefficient multiplier (dimensionless)
$C_{pf}$: perforation discharge coefficient
$C_{pr}$: volume fraction of proppant
$C_{pr,c}$: effective volume fraction of proppant at closure
$C_{pr,max}$: maximum possible volume fraction of proppant in a packed bed
Cw: leakoff coefficient ($ft/min^{1/2}$)
$c_{b,\phi}$: proppant bed porosity compressibility ($Pa^{-1}$; $psi^{-1}$)
$c_{gel}$: mass concentration of gel solute ($g/m^3$; ppg)
D: wellbore diameter
D: fractal dimension in the fracture swarm correlation
$D_{pf}$: perforation diameter (m; ft)
$D_{pf,adj}$: perforation diameter diverter adjustment factor (dimensionless)
$D_{pf,adj,max}$: maximum possible perforation diameter diverter adjustment factor (dimensionless)
$D_{spurt}$: dimensionless spurt loss metric (dimensionless)
$D_t$: tube diameter (m; ft)
d: proppant grain diameter
$d_m$: dimensionless amount of diverter
$d_{decay}$: diverter decay rate
E: fracture aperture (m; ft)
$E_0$: portion of proppant-free fracture aperture at closure that is dependent on aperture (m; ft)
$E_{0,max}$: maximum value of $E_0$ (m; ft)
$E_b$: “proppant bed” portion of the fracture aperture (m; ft)
$E_{cr}$: “crack” portion of the fracture aperture (m; ft)
$E_{open}$: portion of the fracture aperture in excess of the aperture at mechanical closure (m; ft)
$E_{pr}$: portion of the mechanically closed fracture aperture in excess of $E_0$ and $E_{res}$ (m; ft)
$E_{res}$: residual fracture aperture at infinite normal stress (m; ft)
$E_{res,max}$: maximum residual fracture aperture at infinite normal stress (m; ft)
f: Fanning friction factor
$f_b$: factor used in calculating $k_b$
$f_{blend}$: overall blended friction factor of fluid in well
$f_i$: mass fraction of water solute $i$
$f_N$: Fanning friction factor assuming Newtonian fluid
$f_{PL}$: Fanning friction factor assuming power law fluid
$f_{table}$: friction factor calculated from pipe friction table
$F_{brk}$: breaker aggressiveness factor of a ‘Modified Power Law vT’ water solute (dimensionless)
$f_{fis}$: PDL fissure reopening pressure fraction (dimensionless)
$F_{PDL}$: permeability multiplier from pressure-dependent leakoff (dimensionless)

$F_{PDL,con}$: permeability multiplier from pressure-dependent leakoff for a fracture–matrix connection (dimensionless)
g: gravitational constant (m/s$^2$; ft/s)
h: sigmoidal averaging function
$h_f$: fracture height (m; ft)
$ID_{table}$: pipe inner diameter in pipe friction table records (m, ft)
$ID_{actual}$: actual pipe inner diameter (m, ft)
$K_{Ic}$: Fracture toughness (MPa-m$^{1/2}$; psi-in$^{1/2}$)
$K_{Ic,init}$: Initial (scale independent) fracture toughness (MPa-m$^{1/2}$; psi-in$^{1/2}$)
$K_{Ic,fac}$: Fracture toughness scaling factor (m$^{-1/2}$; ft$^{-1/2}$)
K: viscosity constant from the power law or modified power law (Pa-s$^n$; cp/s$^{n-1}$)
$K'$: power law consistency index from laboratory measurements (Pa-s$^{n'}$; lbf-s$^{n'}$/ft$^2$)
$K_{imm}$: rate constant for proppant immobilization (min^-1)
$K_{washout}$: rate constant for proppant washout
k: permeability (m$^2$; md)
$k_{0,b}$: factor used for calculating $k_b$
$k_b$: proppant bed permeability (m$^2$; md)
$(k_{eff})_{ij}$: effective permeability for flow between element i and element j (m$^2$; md)
$k_{fc}$: filtercake permeability (m$^2$; md)
$k_{fc,i}$: filtercake permeability of water solute $i$ (m$^2$, md)
$k_{fc,con}$:effective filtercake permeability at a connection (m$^2$, md)
$k_{fc,con,eff}$: filtercake permeability at a connection after the pressure-dependent permeability factor is applied (m$^2$; md)
$k_{fc,mix}$: filtercake permeability of the fluid mixture(m$^2$, md)
$k_m$: matrix permeability in a dual porosity model (m$^2$; md)
$k_{rp,b}$: relative permeability of phase p in flow through a closed, proppant filled fracture
$k_{rp,cr}$: relative permeability of phase p in flow through a proppant-free crack
$k_{rp,cr,max}$: maximum relative permeability of phase p in flow through a proppant-free crack at the maximum residual saturations of the other phases
$k_{rp,cro}$: relative permeability of phase p in flow through an open fracture
$k_{rp,ij}$: relative permeability of phase p for flow from element i to element j
$k_s$: scaling constant used in the "Filtercake permeability wizard" (m$^2$, md)
$L_{eff}$: fracture length scale, either length or height, whichever is smaller (m; ft)
$l_i$: distance from center of element i to the interface (m; ft)
$M$: molar mass (g/mol; lbs/lbmol
$M_{p,b}$: absolute mobility factor for phase p for flow through a closed, proppant-filled crack (m$^3$/(Pa-s); md-ft/cp)
$M_{p,cr}$: absolute mobility factor for phase p for flow through a proppant-free crack (m$^3$/(Pa-s); md-ft/cp)
$\overline{m}_{gel}$: gel mass fraction
$m_{leakoff}$: leakoff mass of the fluid mixture (kg; lbs)
$m_{solute,i}$: mass fraction of water solute $i$ (dimensionless)
$m_i$: mass of proppant per area (kg/m$^2$; lbs/ft$^2$)
$m_{i,imm}$: mass of immobile proppant per area (kg/m$^2$; lbs/ft$^2$)
$m_{tot,imm,max}$: maximum allowed mass of immobile proppant per area (kg/m$^2$; lbs/ft$^2$)
$m_{pr,a}$: mass of proppant per fracture area (kg/m$^2$; lbs/ft$^2$)
$m_{spurt,i}$: spurt loss mass per unit area of water solute $i$ (kg/m$^2$; lbs/ft$^2$)
N: number of fracture strands in a swarm
$N_c$: number of components in the simulation
$N_{gc}$: dimensionless number expressing the tendency for gravitational bulk slurry convection
$N_{pf}$: number of perforations in a cluster
$N_{pr}$: number of proppant types in the simulation

$N_s$: number of water solute components in the simulation
$N_{sh,crit}$: critical Shields number
n: exponent from the power law or modified power law model
$n'$: power law flow behavior index from laboratory measurements (dimensionless)
$n_{p,cr}$: Brooks-Corey power law relative permeability exponent for flow through a crack
P: pressure (MPa or Pa; psia)
$P_f$: fluid pressure in a fracture element (MPa or Pa; psia)
$P_{fis}$: fissure reopening pressure (MPa or Pa; psia)
$P_{init}$: initial pore pressure of a matrix element (MPa or Pa; psia)
$P_m$: fluid pressure in a matrix element (MPa or Pa; psia)
$p_i$: is the spurt loss potency of water solute $i$ (dimensionless)
$P_L$: Langmuir pressure (MPa or Pa; psia)
$P_p$: fluid pressure of phase p (MPa or Pa; psia)
Q: total volumetric flow rate ($m^3$/s; bpm)
$Q_d$: volumetric flow per element volume in the dual porosity model (1/s; 1/s)
$Q_{pr}$: total volumetric flow rate of proppant ($m^3$/s; bpm)
$q_{c,ij}$: molar flow rate of component c from element i to element j (moles/s; lbmoles/s)
$q_p$: volumetric flow rate of phase p ($m^3$/s; bbl/day)
$q_{p,c}$: volumetric flow rate of phase p in a mechanically closed fracture ($m^3$/s; bbl/day)
$q_{p,o}$: volumetric flow rate of phase p in a mechanically open fracture ($m^3$/s; bbl/day)
R: relative buoyancy factor used in the Ferguson and Church (2006) correlation
Re: Reynolds number
$Re_t$: particle Reynolds number in a Newtonian fluid
$Re_t$': particle Reynolds number in a power law fluid
$Re_{MPL}$': particle Reynolds number in a modified power law fluid
$S_p$: saturation of phase p
$S_{pr,cr}$: residual saturation of phase p in a proppant-free fracture
$S_{pr,cr}$: residual saturation of phases other than p in a proppant-free fracture
s: exponent used for proppant jamming adjustment
$T_{ij}$: transmissibility factor for flow from element i to element j ($m^3$; md-ft)
t: time (s; hrs)
$T$: temperature (°C; °F)
$t_a$: age of a 'Modified Power Law vT' water solute, the time since the fluid was mixed (s; min)
$t_D$: dimensionless age of a 'Modified Power Law vT' water solute (dimensionless)
$t_{kick}$: breaker kick-in time of a 'Modified Power Law vT' water solute (s; min)
$t_{max}$: maximum aging time in the parameter versus time tables at a given temperature (s; min)
$u$: overall mixture Darcy velocity (volumetric flux) (m/s; ft/s)
$v_{true}$: actual fluid velocity (m/s; ft/s)
$V_{leakoff}$: spurt loss volume per unit area of a fluid mixture (m, ft)
$u_p$: Darcy velocity (volumetric flux) of phase p (m/s; ft/s)
$V_{t,\infty}$: terminal settling velocity of an isolated particle (m/s; ft/s)
$V_{t,adj,clust}$: settling velocity adjustment for clustered settling
$V_{t,hind}$: hindered particle settling velocity (m/s; ft/s)
v: superficial flow velocity (volumetric flux) of the mixture in the wellbore (m/s; ft/s)
$v_{c,homogeneous}$: velocity at which a flowing proppant slurry in a pipeline becomes homogeneous (m/s; ft/s)
$v_D$: critical deposition velocity for proppant settling in the wellbore (m/s; ft/s)
$v_{D,0}$: critical deposition velocity for proppant settling in the wellbore, with dilute proppant concentration (m/s; ft/s)
$v_a$: gas volume of adsorption (m^3; ft^3)
$v_a$: Langmuir volume (m^3; ft^3)

$v_{pt}$: reference value of wellbore superficial velocity used for calculating proppant holdup as due to inertia as it flow out of the well (m/s; ft/s)
W: width of the flowing region within a fracture (height, etc.) so that cross-sectional area for flow is equal to this width times the aperture (m; ft)
$w_{fc}$: thickness of filtercake layer on one side of the fracture wall (m; ft)
$w$: local mass fraction of a water solute (dimensionless)
$w_{min}$: lowest measured concentration, as mass fraction, in the parameter versus time tables at the local temperature (dimensionless)
$w_p$: mass fraction of a 'Modified Power Law vT' water solute (dimensionless)
$w_{tr}$: mass fraction of the age tracer of a 'Modified Power Law vT' water solute (dimensionless)
$x_{base}$: mass fraction share of all water solutes without a designated friction table (dimensionless)
$x_k$: mass fraction share of water solutes utilizing friction table $k$ (dimensionless)
$X_{ws1,ws2}$: first order reaction rate constant between water solute 1 and 2 (s; hrs)Z: exponent used in the Garside and Al-Dibouni (1977) hindered settling correlation
z: depth (m; ft)
$z_{c,p,ij}$: molar fraction of component c in phase p for flow from element i to element j
$\alpha$: parameter in the Ellis fluid model
$\alpha_d$: shape factor in the dual porosity model (m$^{-2}$; ft$^{-2}$)
$\alpha_c$: parameter in the Cannella equation for shear rate during flow through porous media
$\alpha_k$ is the low-concentration transition multiplier for pipe friction table $k$ (dimensionless)
$\alpha_{nw}$: exponent for near wellbore complexity
$\beta$: Forchheimer coefficient (1/m; 1/ft)
$\beta_b$: single phase Forchheimer coefficient for flow through a closed, proppant-filled crack (1/m; 1/ft)
$\beta_{cr}$: single phase Forchheimer coefficient for flow through a proppant-free crack (1/m; 1/ft)
$\beta_{cro}$: single phase Forchheimer coefficient for flow through an open crack (1/m; 1/ft)
$\beta_{rp,cr}$: relative Forchheimer coefficient for flow of phase p through a proppant-free crack
$\beta_{rp,cro}$: relative Forchheimer coefficient for flow of phase p through a closed, proppant-filled crack
$\dot{\gamma}$: shear rate (s$^{-1}$; s$^{-1}$)
$\dot{\gamma}_{1/2}$: transition shear rate from Newtonian and power law behavior in the modified power law (s$^{-1}$; s$^{-1}$)
$\gamma_b$: weighting factor indicating the fraction of the roughness dominated part of the fracture that is filled with proppant at closure
$\gamma_f$: weighting factor indicating whether the fracture flow is mechanically open or closed
$\dot{\gamma}_s$: shear rate of a settling particle (s$^{-1}$; s$^{-1}$)
$\Delta L$: fluid travel length in pipe (m, ft)
$\Delta P_{fric}$: pressure drop due to pipe friction (Pa; psia)
$\Delta P_{pf}$: pressure drop across perforations (Pa; psia)
$\Delta P_{nw}$: pressure drop between the well and a fracture due to near wellbore complexity (Pa; psia)
$\Delta t$: timestep duration (s; hrs)
$\Delta t_{age}$: aging increment of a 'Modified Power Law vT' water solute in one timestep (s; hrs)
$\Delta w_{tr}$: change in the age tracer mass fraction in one timestep (dimensionless)
$\Delta x$: flow distance (m; ft)
$\Delta \Phi_p$: hydraulic potential difference driving flow of phase p (Pa; psi)
$\eta_i$: relative filtercake permeability multiplier for water solute $i$
$\theta_r$: proppant angle of repose
$\theta_w$: wellbore angle from vertical
$\lambda$: layer specific spurt loss multiplier (dimensionless)
$\mu$: viscosity (Pa-s; cp)
$\mu_0$: viscosity parameter from the modified power law (Pa-s; cp)
$\mu_{0,adj}$: adjustment factor used in calculating $\mu_0$ at non-standard conditions
$\mu_{0,target}$: zero-shear viscosity multiplier that reproduces the measured $K'$ of a 'Modified Power Law vT' water

solute (dimensionless)
$\mu_a$: apparent viscosity of a flowing non-Newtonian fluid (Pa-s; cp)
$\mu_{p,ij}$: viscosity of phase p for flow from element i to element j (Pa-s; cp)
$\mu_{p,b}$: viscosity of phase p in flow through a proppant bed (Pa-s; cp)
$\mu_{p,cr}$: viscosity of phase p flowing through a proppant-free fracture (Pa-s; cp)
$\mu_{p,s}$: viscosity of phase p in slurry with entrained proppant (Pa-s; cp)
$\mu_{r,N}$: viscosity adjustment factor for the effect of proppant concentration in a Newtonian fluid
$\mu_{r,PL}$: viscosity adjustment factor for the effect of proppant concentration in a power law fluid
$\mu_w$: viscosity of pure water (Pa-s; cp)
$\xi$: maximum asperity height parameter from the Chen et al. (2015) correlation (m; ft)
$\xi_L$: layer-specific filtercake permeability multiplier
$\overline{\rho}$: average slurry density including all phases and proppant types (kg/m$^3$; lbs/ft$^3$)
$\rho_f$: fluid density (kg/m$^3$; lbs/ft$^3$)
$\rho_{fc}$: filtercake density (kg/m$^3$; lbs/ft$^3$)
$\sigma_{hmin}$: minimum principal stress (Pa; psia)
$\sigma_{hmin,init}$: initial minimum horizontal stress of a matrix element (Pa; psia)
$\rho_{M,p,ij}$: molar density of phase p for flow from element i to element j (moles/m$^3$; lbmoles/ft$^3$)
$\rho_{\text{mixture}}$ is the fluid mixture density (kg/m$^3$; lbs/ft$^3$)
$\rho_p$: density of phase p (kg/m$^3$; lbs/ft$^3$)
$\rho_{pr}$: proppant density (kg/m$^3$; lbs/ft$^3$)
$\overline{\rho}_{pr}$: average proppant density (kg/m$^3$; lbs/ft$^3$)
$\rho_{p,s}$: density of phase p in slurry with proppant (kg/m$^3$; lbs/ft$^3$)
$\sigma_n$: fracture normal stress (Pa; psia)
$\sigma_n'$: fracture effective normal stress (Pa; psia)
$\sigma_{n,ref}$: fracture 90% closure stress (Pa; psia)
$\tau_{1/2}$: shear stress at the transition from Newtonian to power law behavior (Pa; psia)
$\tau_{tr}$: internal timescale of the age tracer of a 'Modified Power Law vT' water solute (s; hrs)
$\Phi$: phase hydraulic potential (Pa; psia)
$\Phi_{p,i}$: hydraulic potential of phase p in element i (Pa; psia)
$\phi$: porosity
$\Phi$: the porothermoelastic potential function
$\chi$: blocking function used for modeling proppant bridging in the fracture
$\omega$: moles adsorbed per unit mass of rock